\documentclass[a4paper,11pt]{article}
\usepackage{graphicx}
\usepackage{bm}
\usepackage{bbm}
\usepackage{here}
\usepackage{ascmac}
\usepackage{amsmath,amssymb,amsfonts}
\usepackage{adjustbox}
\usepackage{mathtools}
\usepackage{ulem}
\usepackage{mathrsfs}
\usepackage{appendix}
\usepackage{longtable}
\usepackage{CJKutf8}
\usepackage{jheppub}                
\title{
A Lattice U(1) Chern--Simons Theory \\via Lattice Deligne--Beilinson Cohomology
}
\author[]{Yo Ikeda}
\affiliation[]{Institute of Physics, University of Tokyo,\\Komaba, Meguro-ku, Tokyo 153-8902 Japan}
\emailAdd{ikeda-yo721@g.ecc.u-tokyo.ac.jp}

\preprint{UT-Komaba/26-2}

\abstract{We define Deligne--Beilinson (DB) cohomology on a cubic lattice and use it to formulate and analyze lattice $U(1)$ Chern--Simons theory at even levels. The continuum DB cohomology provides a refined mathematical framework for continuum $U(1)$ connections constructed in a patchwise manner. The lattice DB cohomology we construct retains many essential properties of the continuum DB cohomology and naturally incorporates a notion of self-linking number.
The lattice $U(1)$ Chern--Simons action formulated using the lattice DB cohomology is expressed as a simple quadratic form via the star product, which naturally exhibits level quantization. Framed Wilson lines respecting staggered symmetry are defined in a gauge-invariant manner, and their expectation values are shown to be given by the self-linking number, as follows from completing the square.
Using the lattice Hodge decomposition, we explicitly characterize the DB cohomology on a three-dimensional cubic toroidal lattice and present a gauge-fixed, rigorous path integral for the lattice Chern--Simons theory. To regulate divergences in the lattice Chern--Simons path integral arising from staggered symmetry, we introduce a small Maxwell term. The resulting error is controlled by the linear order in the small Maxwell coupling.}
\arxivnumber{}

\begin{document}
\maketitle
\newpage
\section{Introduction}\label{sec100}
\subsection*{Chern--Simons theory}
In 1974, Shiing-Shen Chern and James Simons discovered Chern--Simons differential form~\cite{Chern:1974ft}. It was the beginning of the long history of Chern--Simons theory. In 1978, A.S. Schwarz first formulated Chern--Simons theory and pointed out the relationship between Chern--Simons theory and a topological invariant, which is called Ray-Singer torsion~\cite{Schwarz:1978cn}. This is now called the first example of TQFT.

Chern--Simons theory was first formulated as a Lagrangian QFT around 1980~\cite{Schonfeld:1980kb,Deser:1981wh}. In 1988, a connection between $U(1)$ Chern--Simons theory and the linking number of knots was pointed out~\cite{Polyakov:1988md,Frohlich:1988}. It is so famous that Witten discovered the relationship between the $SU(2)$ Chern--Simons theory and the Jones polynomial in 1989~\cite{Witten:1988hf}.

The Chern--Simons theory still plays a very important role at the cutting edge of physics today, but while the Chern--Simons theory can be formulated as an axiomatic TQFT with mathematical rigour, the continuum Chern--Simons theory as a quantum field theory of path integral still is not mathematically rigorous. From this perspective, establishing lattice Chern--Simons theory is an extremely important task.
\subsection*{History of lattice formulations of Chern--Simons theory}
The definition of Chern--Simons theory on a three-dimensional space-time lattice has been under discussion for about 30 years, but a satisfactory formulation has only been achieved recently. In this section, we provide an overview of the process leading to the establishment of lattice Chern--Simons theory.

Chern--Simons theory on lattices began to be formulated in the early 1990s. The model known today as Dijkgraaf-Witten theory was introduced as something close to a lattice Chern-Simons theory~\cite{Dijkgraaf:1989pz}. In the same year, a na\"{i}vely discretized Chern--Simons action on a triangular lattice using the cup product of a simplicial complex was proposed in~\cite{Kavalov:1989kg}. A similar approach has been taken to formulate lattice $U(1)$ Chern--Simons theory using the $U(1)$ cochain formalism, namely a method that directly uses the $U(1)$ link variables on the triangulation~\cite{DeMarco:2019pqv}. However, the action defined in this work has a rounding operation to the nearest integer, which makes the action a discontinuous function.

In 1992, a Chern--Simons theory on a discrete 2-dimensional + continuum 1-dimensional lattice was introduced~\cite{Eliezer:1992sq}, and the discussion of Wilson lines with nontrivial homology was advanced, but in~\cite{DeMarco:2019pqv}, it is pointed out that the theory becomes nonlocal.

The zero modes of lattice Chern--Simons theory induce difficulties for the path integral formulation. These zero modes were pointed out in~\cite{Berruto:2000dp}, and it was also pointed out that these zero modes can be eliminated by the Maxwell term. Lattice Maxwell Chern--Simons theory was also a model that had already been discussed at that time. In 1993,~\cite{Diamantini:1993iu} pointed out topological excitations as strings connecting monopoles and antimonopoles in lattice Maxwell Chern--Simons theory. For this reason, the formulation of lattice Chern--Simons theory without zero modes has also been studied. For example, a lattice Chern--Simons theory without zero modes as a determinant of the lattice Ginsparg-Wilson fermion~\cite{So:1984nf,Coste:1989wf,Bietenholz:2002mt} should be mentioned.

In 2023, a research offering a new perspective on the zero-mode problem in lattice Chern--Simons theory appeared in~\cite{Jacobson:2023cmr}. Jacobson and Sulejmanpasic constructed a lattice $U(1)$ Chern--Simons theory using a method called the modified Villain formalism. While the zero modes associated with the staggered symmetries had previously been regarded as something to be removed, Jacobson and Sulejmanpasic reinterpreted them as a mechanism that imposes a framing on Wilson lines. The paper also discusses the linking number and the 't Hooft anomaly. However, the convergence of the path integral remains unclear, and their discussion is restricted to even Chern--Simons levels.

Although the work of Jacobson and Sulejmanpasic seemed at first to provide a beautiful solution to the zero-mode problem, it is pointed out that a Maxwell term is still necessary for the convergence of the path integral~\cite{Xu:2024hyo}. Xu and Chen studied a model where a Maxwell term is added to the modified Villain $U(1)$ Chern--Simons theory, showed that the partition function is finite, and gave its expression. They also analyzed several detailed properties of the theory, and discussed the case of odd Chern--Simons level $k$. They also consider Wilson lines with nontrivial homology. However, their formulation of Wilson lines differs from~\cite{Jacobson:2023cmr}, and their discussion is in infinite volume and focuses on classical solutions, so it is not fully general.
\subsection*{Deligne--Beilinson cohomology}
Deligne--Beilinson cohomology, which at the time was called ``multiplicative de Rham complex'' was originally introduced by Pierre Deligne to incorporate a concept called the intermediate Jacobian into cohomology theory~\cite{Deligne:1971}. The term ``multiplicative de Rham complex'' is now called Deligne--Beilinson cohomology, named after the paper published by Alexander Beilinson in 1985~\cite{Beilinson:1985}.

On the other hand, the definition of Deligne--Beilinson cohomology adopted in many physics papers, including this one, is different but equivalent to Deligne's. This definition is related to the patchwise definition of gauge fields in classical field theory. For example, consider a $U(1)$ connection on a manifold $M$. A $U(1)$ connection with a non-trivial Chern class can always be described by a $1$-form which satisfies the gluing condition on the good cover of $M$. The space obtained by enumerating $U(1)$ gauge configurations that satisfy the gluing condition and dividing them by gauge equivalence is the Deligne--Beilinson $2$-cohomology. If we enumerate $k$-forms instead of $1$-forms and divide them by gauge equivalence, we obtain the Deligne--Beilinson $(k+1)$-cohomology. This construction is similar to a method called \v{C}ech--de Rham complex, and appears to differ from the method first proposed by Deligne. For this reason, this construction is sometimes called smooth Deligne cohomology to distinguish it from the method by Deligne. Today, it is more commonly referred to as Deligne--Beilinson cohomology without making much of a distinction.

Thus, considering a Deligne--Beilinson $2$-cohomology on a manifold is equivalent to enumerating all possible $U(1)$ connections on $M$, which is the gauge-fixed path integral domain in $U(1)$ gauge theory. In this sense, it is quite natural to consider a path integral on a Deligne--Beilinson 2-cohomology in $U(1)$ Chern--Simons theory as well. In recent years, the formulation and properties of continuum $U(1)$ Chern--Simons theory using the Deligne--Beilinson cohomology have discussed in~\cite{Guadagnini:2008bh,Guadagnini:2014mja,Thuillier:2015vma}, and the relationship between Deligne--Beilinson cohomology and a doubled lattice $U(1)$ Chern--Simons theory on the lattice is pointed out in~\cite{Chen:2019mjw}.

Deligne--Beilinson $(k+1)$-cohomology is also known in physics. The construction of B-fields in string theory is well known and has also been discussed in the context of generalized symmetry~\cite{Kapustin:2014gua,Kapustin:2014zva}. Deligne--Beilinson cohomology of rank three or higher is often referred to as a gerbe with connection in some contexts.

From these perspectives, it seems useful to consider Deligne--Beilinson cohomology even in the formulation of lattice $U(1)$ gauge theory. In 2021, Norton defined Deligne--Beilinson cohomology on simplicial complexes~\cite{Norton:2021psm}, showing that this is well-defined as a Deligne--Beilinson cohomology in discretized spacetime and that it can be discussed in the same way as continuous Deligne--Beilinson cohomology. The publication of this paper has opened the door to the formulation of lattice models using Deligne--Beilinson cohomology.
\subsection*{Outline of this paper}
The main objectives of this paper are as follows:
\begin{itemize}
    \item $\,$To define the lattice Deligne--Beilinson cohomology for cubic lattices, following the (discrete) Deligne--Beilinson cohomology for triangular lattices defined in~\cite{Norton:2021psm}.
    \item $\,$To define and analyze the lattice $U(1)$ Chern--Simons theory on the $3$d cubic toroidal lattice, referring to the method of the continuum $U(1)$ Chern--Simons theory pointed out in~\cite{Guadagnini:2008bh,Thuillier:2015vma}.
\end{itemize}

This paper is divided into three parts.
Part~\ref{part1} is a review part. In Section~\ref{sec200}, we briefly review the Deligne--Beilinson formulation of continuum $U(1)$ Chern--Simons theory discussed in~\cite{Guadagnini:2008bh,Thuillier:2015vma}, and in Section~\ref{sec300}, we briefly review the modified Villain formulation of lattice $U(1)$ Chern--Simons theory discussed in~\cite{Jacobson:2023cmr,Xu:2024hyo}.

In Part~\ref{part2}, we give a definition of the lattice DB cohomology. In Section~\ref{sec400} and \ref{sec500}, we define the Deligne--Beilinson cohomology for the $d$-dimensional toroidal lattice. This discussion shows that the methodology used in~\cite{Norton:2021psm} to discuss the Deligne--Beilinson cohomology for the triangulated lattice can be similarly applied to the $d$-dimensional cubic toroidal lattice. In Section~\ref{sec600}, we define the star product, which is an important operation in Deligne--Beilinson cohomology theory. In Section~\ref{sec700}, we define gauge-invariant integrals in Deligne--Beilinson cohomology. The gauge-invariant half-integer integral introduced in Section~\ref{sec700} shows that the definition of Wilson line in~\cite{Jacobson:2023cmr} can also be defined within lattice Deligne--Beilinson theory. In Section~\ref{sec800}, we analyze aspects of the Deligne--Beilinson cohomology as a differential cohomology, which can be used to define gauge-fixed path integrals. In Section~\ref{sec900}, we discuss the relationship between star products and knot theory in the three-dimensional cubic toroidal lattice. The mod $2k$ linking number is a particularly interesting topic, because it is where the geometric ideas of knot theory and Deligne--Beilinson cohomology come together.

Part~\ref{part3} is one of the main topics of this paper. In Section~\ref{sec1000}, we define $U(1)$ Chern--Simons theory on a three-dimensional toroidal lattice using the lattice Deligne--Beilinson cohomology and discuss its properties. This discussion shows that the method of continuum Chern--Simons theory~\cite{Guadagnini:2008bh,Thuillier:2015vma}, formulated in terms of continuous Deligne--Beilinson cohomology, can also be applied to Chern--Simons theory on a lattice. In Section~\ref{sec1100}, we demonstrate that lattice Maxwell--Chern--Simons (MCS) theory has a finite partition function, and the expectation values of Wilson lines appear as linking numbers and self-linking numbers, and have the properties expected from continuum Chern--Simons theory. While the expectation values calculated in lattice Maxwell--Chern--Simons theory contain some error, we show that the error can be controlled in the limit $\epsilon\to 0$, where $\epsilon$ denotes the coefficient of the Maxwell term.

Finally, we discuss some applications. In Section~\ref{sec1200}, we briefly review the non-invertible chiral transformation defect in continuum massless QED, and then discuss how the same defect can be constructed using lattice Maxwell--Chern--Simons theory. In Section~\ref{sec1300}, we discuss other applications, and we summarize our results and give the outlook in Section~\ref{sec1400}.
\newpage
\part{Review}\label{part1}
\section{Continuous Chern--Simons theory formulated using DB cohomology}\label{sec200}
In this section, we define continuous Deligne--Beilinson (DB) cohomology and provide a very brief review of the properties of $U(1)$ Chern--Simons theory defined from it. The content of this section is based on~\cite{Guadagnini:2008bh,Thuillier:2015vma}.
\subsection{DB $2$-cohomology on $3$-manifolds}\label{sec210}
For simplicity, we will only discuss DB 2-cohomology in this review. For a more general definition, see~\cite{Guadagnini:2008bh,Thuillier:2015vma}.

Let $M$ be a compact three-dimensional manifold. We fix a finite good open cover of $M$ $\{U_\alpha\}_{\alpha\in I}$. We introduce an abbreviation
\begin{align}\label{equ2010}
    &U_{\alpha\beta}=\left\{\begin{array}{cl}U_\alpha\cap U_\beta&(\alpha\neq\beta)\\\emptyset&(\mathrm{otherwise})\end{array}\right.&
    &U_{\alpha\beta\gamma}=\left\{\begin{array}{cl}U_\alpha\cap U_\beta\cap U_\gamma&(\alpha\neq\beta\neq\gamma\neq\alpha)\\\emptyset&(\mathrm{otherwise})\end{array}\right..&
\end{align}
The Deligne--Beilinson 2-cocycles on $M$, denoted $Z^2_{\mathrm{DB}}(M)$,\footnote{Since the top form is the $1$-form, it is sometimes written by $Z^1_{\mathrm{DB}}(M)$, but this is inconsistent with the star product described below. In this paper, we respect consistency with the product and write it as $Z^2_{\mathrm{DB}}(M)$.} are given by
\begin{equation}\label{equ2020}
    Z^2_\mathrm{DB}(M)=\Bigl\{\bigl(\{A_{\alpha}\}_{\alpha\in I}\;,\;\{\Lambda_{\alpha\beta}\}_{\alpha,\beta\in I}\;,\;\{n_{\alpha\beta\gamma}\}_{\alpha,\beta,\gamma\in I}\bigr)\Bigr\},
\end{equation}
where $A,\Lambda,n$ are
\begin{align}\label{equ2030}
    &A_\alpha\in\Omega^1(U_\alpha),&&\Lambda_{\alpha\beta}\in\Omega^0(U_{\alpha\beta}),&&n_{\alpha\beta\gamma}\in\Omega^{-1}(U_{\alpha\beta\gamma}),
\end{align}
that satisfy
\begin{align}\label{equ2040}
    &(\delta A)_{\alpha\beta}=A_\beta-A_\alpha=d\Lambda_{\alpha\beta},&&(\delta \Lambda)_{\alpha\beta\gamma}=\Lambda_{\beta\gamma}+\Lambda_{\gamma\alpha}+\Lambda_{\alpha\beta}=-d_{-1}n_{\alpha\beta\gamma}.
\end{align}
$\Omega^n(U)$ is the set of differential $n$-forms on $U$, and we interpret a $(-1)$-form that belongs to $\Omega^{-1}(U_{\alpha\beta\gamma})$ as an assignment of an integer
to $U_{\alpha\beta\gamma}$. The operator $d_{-1}$ is defined to regard such a $(-1)$-form as a $\mathbb{Z}$-valued function on $U_{\alpha\beta\gamma}$. We define the addition operation on $Z^2_\mathrm{DB}(M)$ as the sum of each component. For simplicity, we omit the subscripts and denote
\begin{equation*}
    \bigl(\{A_{\alpha}\}_{\alpha\in I}\;,\;\{\Lambda_{\alpha\beta}\}_{\alpha,\beta\in I}\;,\;\{n_{\alpha\beta\gamma}\}_{\alpha,\beta,\gamma\in I}\bigr)\quad\text{by}\quad(A,\Lambda,n).
\end{equation*}

There are two types of gauge transformation for the element of Deligne--Beilinson 2-cocycle $(A,\Lambda,n)$. The ordinary gauge transformation with its parameter function
\begin{equation}\label{equ2050}
    \{\xi_\alpha\}_{\alpha\in I}\;,\;\xi_\alpha\in\Omega^0(U_\alpha)
\end{equation}
is described by
\begin{align}\label{equ2060}
    &A_{\alpha}\to A_{\alpha}+d\xi_{\alpha},&&\Lambda_{\alpha\beta}\to \Lambda_{\alpha\beta}+(\delta\xi)_{\alpha\beta}.
\end{align}
It is clear that this transformation sends a DB 2-cocycle to a DB 2-cocycle. On the other hand, a large gauge transformation with its parameter function
\begin{equation}\label{equ2070}
    \{m_{\alpha\beta}\}_{\alpha,\beta\in I}\;,\;\xi_\alpha\in\Omega^{-1}(U_\alpha)
\end{equation}
is defined by
\begin{align}\label{equ2080}
    &\Lambda\to \Lambda-m,&&n\to n+\delta m.
\end{align}
We can also see that this transformation sends a DB 2-cocycle to a DB 2-cocycle.

Let us define Deligne--Beilinson 2-cohomology $H^2_\mathrm{DB}(M)$ as a quotient of $Z^2_\mathrm{DB}(M)$ divided by the equivalence relation $\sim$ generated by these two gauge transformations. Written in mathematical form,
\begin{equation}\label{equ2090}
    H^2_\mathrm{DB}(M)=Z^2_\mathrm{DB}(M)/\sim.
\end{equation}
We adopt the notation $[(A,\Lambda,n)]$ to denote the element of $H^2_\mathrm{DB}(M)$ using the natural projection $[\bullet]$ of the quotient space, which is consistent with the addition operation on $Z^2_\mathrm{DB}(M)$.

$H^2_\mathrm{DB}(M)$ has a gauge-invariant line integral. For closed curve $\gamma:S^1\to M$, we can arbitrarily take a direct sum decomposition of $\gamma$
\begin{equation}\label{equ2100}
    \gamma=\gamma_{\alpha_1}+s_{\alpha_1\alpha_2}+\gamma_{\alpha_2}+s_{\alpha_2\alpha_3}+\dots+\gamma_{\alpha_N}+s_{\alpha_N\alpha_1}
\end{equation}
that satisfies
\begin{align}\label{equ2110}
    &\gamma_{\alpha_i}\subset U_{\alpha_i},&&\gamma_{\alpha_i}\cong \;]0,1[\,,&&s_{\alpha_j\alpha_k}\in U_{\alpha_j\alpha_k},
\end{align}
where the notation $]0,1[$ denotes the open interval. Then the line integral of $(A,\Lambda,n)$ along $\gamma$ can be well-defined by
\begin{equation}\label{equ2120}
    \int_\gamma (A,\Lambda,n)=\sum_{i=1}^N\int_{\gamma_i} A_{\alpha_i}+\sum_{i=1}^N\Lambda_{\alpha_i\alpha_{i+1}}(s_{\alpha_i\alpha_{i+1}}).
\end{equation}
The value of the line integration is modulo-$\mathbb{Z}$ invariant under the choice of the decomposition of $\gamma$ and the two gauge transformations of $(A,\Lambda,n)$. Therefore, the line integral can be said to be $\mathbb{R}/\mathbb{Z}$ valued gauge-invariant integral.

At the end of this section, let us mention the relation between $U(1)$ connection on classical field theory and DB $2$-cohomology. The patchwise $U(1)$ connection defined in common textbooks such as~\cite{Nakahara:2003nw} is exactly equivalent to the Deligne--Beilinson $2$-cocycle, and its gauge fixing is equivalent to the Deligne--Beilinson 2-cohomology. The only difference is that $\frac{1}{2\pi i}\log g_{\alpha\beta}=\Lambda_{\alpha\beta}$ is used instead of the usual $U(1)$ transition function $g_{\alpha\beta}$, and $n$ corresponds to the $2\pi i$ ambiguity in $\log$ that arises when defining $\Lambda$. The gauge-invariant line integral of the DB cocycle corresponds to the line integral of the $U(1)$ connection, i.e., the Wilson line. Similar to the definition of $F=dA$ in $U(1)$ gauge theory, the curvature $dA$ of the Deligne--Beilinson cocycle is defined as a global $2$-form by the gluing condition, and its period is always an integer.
\subsection{Star product in continuum DB cohomology}\label{sec220}
The star product is a bilinear mapping $\star:Z^2_\mathrm{DB}(M)\times Z^2_\mathrm{DB}(M)\to Z^4_\mathrm{DB}(M)$. Actually, $\star:H^2_\mathrm{DB}(M)\times H^2_\mathrm{DB}(M)\to H^4_\mathrm{DB}(M)$ also makes sense because the star product is consistent with gauge equivalence. $Z^4_\mathrm{DB}(M)$ is a certain set of $5$-tuples whose elements satisfy certain gluing conditions, and $H^4_\mathrm{DB}(M)$ can be defined by dividing $Z^4_\mathrm{DB}(M)$ by the equivalence relation given by gauge transformations and also has a gauge-invariant $\mathbb{R}/\mathbb{Z}$-valued integral along $M$. We omit further discussions about the definition of $H^4_\mathrm{DB}(M)$.

Let us consider two DB cocycles $\bm{\omega}=(\omega^{(1,0)},\omega^{(0,1)},\omega^{(-1,2)})$ and $\bm{\eta}=(\eta^{(1,0)},\eta^{(0,1)},\eta^{(-1,2)})$. The star product of these two DB cocycles $\bm{\omega}\star\bm{\eta}$ is defined by
\begin{equation}\label{equ2130}
    \begin{split}
        &\bm{\omega}\star\bm{\eta}\\
        =&(
        \omega^{(1,0)}_{\alpha}\wedge d\eta^{(1,0)}_{\alpha},
        \omega^{(0,1)}_{\alpha\beta}\wedge d\eta^{(1,0)}_{\beta},
        \omega^{(-1,2)}_{\alpha\beta\gamma}\wedge \eta^{(1,0)}_{\gamma},
        \omega^{(-1,2)}_{\alpha\beta\gamma}\wedge \eta^{(0,1)}_{\gamma\rho},
        \omega^{(-1,2)}_{\alpha\beta\gamma}\wedge \eta^{(-1,2)}_{\gamma\rho\sigma}).
    \end{split}
\end{equation}
Since the product is consistent with the projection map, 
\begin{equation}\label{equ2140}
    \begin{split}
        &[\bm{\omega}]\star[\bm{\eta}]=[\bm{\omega}\star\bm{\eta}]\\
        =&[(
        \omega^{(1,0)}_{\alpha}\wedge d\eta^{(1,0)}_{\alpha},
        \omega^{(0,1)}_{\alpha\beta}\wedge d\eta^{(1,0)}_{\beta},
        \omega^{(-1,2)}_{\alpha\beta\gamma}\wedge \eta^{(1,0)}_{\gamma},
        \omega^{(-1,2)}_{\alpha\beta\gamma}\wedge \eta^{(0,1)}_{\gamma\rho},
        \omega^{(-1,2)}_{\alpha\beta\gamma}\wedge \eta^{(-1,2)}_{\gamma\rho\sigma})]
    \end{split}
\end{equation}
also makes sense.

The star product on $H^2_\mathrm{DB}(M)$ has two important properties: (graded) commutativity\footnote{The product of $H^2_\mathrm{DB}(M)$ is simply commutative because $(-1)^{2\cdot 2}=1$.} and Pontrjagin duality. Pontrjagin duality is the property that there exists a Pontrjagin dual $\bm{\eta}_\gamma\in H^2_\mathrm{DB}(M)$ such that
\begin{equation}\label{equ2150}
    \int_\gamma \bm{\omega}\stackrel{\mathbb{Z}}{=}\int_M \bm{\omega}\star\bm{\eta}_\gamma\qquad(\text{for all $\bm{\omega}\in H^2_\mathrm{DB}(M)$})
\end{equation}
for any closed curve $\gamma$, where $\stackrel{\mathbb{Z}}{=}$ means modulo $\mathbb{Z}$ equality.

In the continuous DB formalism, it is said that the Pontrjagin dual $\bm{\eta}_\gamma$ ``exists'' for all closed curves $\gamma$; however, that is not actually an accurate claim. It shares a similar problem for the formulation of chainwise Poincar\'{e} duality on de Rham cohomology theory. The chainwise Poincar\'{e} duality is mathematically defined as a de Rham current, which is a counterpart of the Schwartz distribution in de Rham theory. For similar reasons, a distributional formulation should be applied to the definition of Pontrjagin duality, but we will skip a detailed discussion here.
\subsection{Continuum Chern--Simons theory via DB cohomology}\label{sec230}
We define level $2k$ Chern--Simons theory on $M$ by
\begin{align}\label{equ2160}
    &S_\mathrm{CS}[[\bm{a}]]=2\pi k\int_M [\bm{a}]\star[\bm{a}],&&\mathcal{Z}=\int_{[\bm{a}]\in H^2_\mathrm{DB}(M)}\mathcal{D}[\bm{a}]\exp\{iS_\mathrm{CS}[[\bm{a}]]\}.
\end{align}
The gauge invariance of the action and the $\mathbb{R}/\mathbb{Z}$-valuedness of the integral require $k\in\mathbb{Z}_{>0}$. There is no gauge redundancy in the path integral because we set the integral region as $[\bm{a}]\in H^2_\mathrm{DB}(M)$ and $H^2_\mathrm{DB}(M)$ is already divided by gauge equivalence. We assume path integral measure $\mathcal{D}[\bm{a}]$ is invariant under shift transformation $[\bm{a}]\to [\bm{a}+\bm{\omega}]$.

Next, let us discuss the properties of the Chern--Simons theory we defined. For simplicity, we only consider the case $M=\mathbb{T}^3$. We should formulate the $\mathbb{Z}_{2k}$ global $1$-form symmetry that Chern--Simons theory must have. Let each $\beta_{23},\beta_{31},\beta_{12}\in \Omega^1(\mathbb{T}^3)$ be a chainwise Poincar\'{e} dual of $e_{23},e_{31},e_{12},$ which is the basis of $H_2(\mathbb{T}^3;\mathbb{Z})$.\footnote{As mentioned in Section~\ref{sec220}, $\beta$ should be defined as a de Rham current, so it is actually inappropriate to write $\Omega^1(\mathbb{T}^3)$.} Defining $\delta_{-1}$ by
\begin{equation}\label{equ2170}
    \delta_{-1}:\Omega^1(M)\ni\omega\mapsto\{(\delta_{-1}\omega)_\alpha\}_{\alpha\in I}=\{\omega|_{U_\alpha}\}_{\alpha\in I},
\end{equation}
we find $(\delta_{-1}\omega,0,0)$ is a DB 2-cocycle on $M$ for all $\omega\in\Omega^1(M)$. We can define $\mathbb{Z}_{2k}$ global $1$-form symmetry by
\begin{equation}\label{equ2180}
    [\bm{a}]\to\left[\bm{a}+\frac{\bm{\beta}_{ij}}{2k}\right],
\end{equation}
where
\begin{equation}\label{equ2190}
    \frac{\bm{\beta}_{ij}}{2k}=\left(\frac{1}{2k}\delta_{-1}\beta_{ij},0,0\right)\in Z^2_\mathrm{DB}(M)
\end{equation}
and the level $2k$ Chern--Simons action is invariant under this transformation. In fact, from bilinearity of $\star$ and (graded) commutativity, the action changes as follows:
\begin{equation}\label{equ2200}
    k\int_{\mathbb{T}^3} \bm{a}\star\bm{a}\to k\int_{\mathbb{T}^3} \bm{a}\star\bm{a}+2k\int_{\mathbb{T}^3} \bm{a}\star\frac{\bm{\beta}_{ij}}{2k}+k\int_{\mathbb{T}^3} \frac{\bm{\beta}_{ij}}{2k}\star\frac{\bm{\beta}_{ij}}{2k}.
\end{equation}
Referring to $d\beta_{ij}=0$ and the definition of $\star$, we can see the second term is an integer, and the third term is $0$. Therefore, $S_\mathrm{CS}$ is certainly modulo-$2\pi\mathbb{Z}$ invariant under the shift $\bm{a}\to\bm{a}+\bm{\beta}/2k$.
\subsection{Expectation values of Wilson lines}\label{sec240}
The expectation value of the Wilson line $\gamma$, denoted $\langle W(\gamma)\rangle$, is defined by
\begin{equation}\label{equ2210}
    \begin{split}
        \langle W(\gamma)\rangle &=\mathcal{Z}^{-1}\int_{[\bm{a}]\in H^2_\mathrm{DB}(M)}\mathcal{D}[\bm{a}]\exp\left[iS_\mathrm{CS}[[\bm{a}]]+2\pi i\int_\gamma[\bm{a}]\right]\\
        &=\mathcal{Z}^{-1}\int_{[\bm{a}]\in H^2_\mathrm{DB}(M)}\mathcal{D}[\bm{a}]\exp\left[2\pi ik\int_{\mathbb{T}^3}[\bm{a}]\star[\bm{a}]+2\pi i\int_{\mathbb{T}^3}[\bm{a}]\star[\bm{\eta}_\gamma]\right].
    \end{split}
\end{equation}
The $\mathbb{Z}_{2k}$ $1$-form symmetry of the action leads the transformation of the path integral variable
\begin{equation}\label{equ2220}
    [\bm{a}]\to\left[\bm{a}+\frac{\bm{\beta}_{ij}}{2k}\right],
\end{equation}
and we obtain
\begin{equation}\label{equ2230}
    \langle W(\gamma)\rangle=\exp\left[2\pi i\int_{\mathbb{T}^3}\left[\frac{\bm{\beta}_{ij}}{2k}\right]\star[\bm{\eta}_\gamma]\right]\langle W(\gamma)\rangle.
\end{equation}
If there is $\bm{\beta}_{ij}$ that satisfies
\begin{equation}\label{equ2240}
    \exp\left[2\pi i\int_{\mathbb{T}^3}\left[\frac{\bm{\beta}_{ij}}{2k}\right]\star[\bm{\eta}_\gamma]\right]\neq 1,
\end{equation}
then, $\langle W(\gamma)\rangle=0$ is concluded. To require the Wilson line to have a nontrivial expectation value, we should impose a condition 
\begin{equation}\label{equ2250}
    \forall \bm{\beta}_{ij},\int_{\mathbb{T}^3}\left[\frac{\bm{\beta}_{ij}}{2k}\right]\star[\bm{\eta}_\gamma]\in \mathbb{Z}\;\biggl(\Longleftrightarrow\; [\gamma]\in H_1(\mathbb{T}^3;2k\mathbb{Z})\biggr)
\end{equation}
to the Wilson line $\gamma$. When the condition $[\gamma]\in H_1(\mathbb{T}^3;2k\mathbb{Z})$ is satisfied, a division $[\frac{\bm{\eta}_\gamma}{2k}]$ can be well-defined.\footnote{A more careful discussion is actually required for this argument; see also Section~\ref{sec960}.} Then we can complete the square as follows:
\begin{equation}\label{equ2260}
    \begin{split}
        &\langle W(\gamma)\rangle \\
        =&\mathcal{Z}^{-1}\int_{[\bm{a}]\in H^2_\mathrm{DB}(M)}\mathcal{D}[\bm{a}]\exp\Biggl[2\pi ik\int_{\mathbb{T}^3}\biggl\{\left([\bm{a}]+\left[\frac{\bm{\eta}_\gamma}{2k}\right]\right)\star\left([\bm{a}]+\left[\frac{\bm{\eta}_\gamma}{2k}\right]\right)
        -\left[\frac{\bm{\eta}_\gamma}{2k}\right]\star\left[\frac{\bm{\eta}_\gamma}{2k}\right]\biggr\}\Biggr]\\
        =&\exp\left[-2\pi ik\int_{\mathbb{T}^3}\left[\frac{\bm{\eta}_\gamma}{2k}\right]\star\left[\frac{\bm{\eta}_\gamma}{2k}\right]\right].
    \end{split}
\end{equation}
As a consequence, we find
\begin{equation}\label{equ2270}
    \langle W(\gamma)\rangle=\left\{\begin{array}{cl}
        \displaystyle\exp\left[-2\pi ik\int_{\mathbb{T}^3}\left[\frac{\bm{\eta}_\gamma}{2k}\right]\star\left[\frac{\bm{\eta}_\gamma}{2k}\right]\right]&(\gamma\in H_1(\mathbb{T}^3;2k\mathbb{Z}))\\
        &\\
        0&(\gamma\notin H_1(\mathbb{T}^3;2k\mathbb{Z}))
    \end{array}\right..
\end{equation}
The integral $k\int_{\mathbb{T}^3}\left[\frac{\bm{\eta}_\gamma}{2k}\right]\star\left[\frac{\bm{\eta}_\gamma}{2k}\right]$ is a $(\frac{1}{2k}\mathbb{Z})/\mathbb{Z}$-valued topological number, and is equal to the $\mathrm{mod}\,2k$ linking number\footnote{Actually, there is an ambiguity in the distributional definition of $\bm{\eta}_\gamma$, and all $\mathrm{mod}\,4k$ self-linking numbers that should appear are manually set to 0. See~\cite{Guadagnini:2008bh,Thuillier:2015vma} for details.} defined later in Section~\ref{sec960}.
\newpage
\section{Lattice Chern--Simons theory formulated using the modified Villain formalism}\label{sec300}
In this section, we will give a very brief review of the modified Villain formalism construction of lattice Chern--Simons theory based on the papers~\cite{Jacobson:2023cmr,Xu:2024hyo}.
\subsection{Modified Villain formalism}\label{sec310}
We consider a three-dimensional cubic lattice without boundary. In the Villain formalism, the $U(1)$ gauge field is constructed by $\mathbb{R}$-valued link variable $a$ and $\mathbb{Z}$-valued plaquette variable $n$. These variables transform under ordinary gauge transformations as
\begin{equation}\label{equ3010}
    \left\{\begin{array}{l}
        a\to a+d\lambda\\
        n\to n
    \end{array}\right.,
\end{equation}
and under large gauge transformations as
\begin{equation}\label{equ3020}
    \left\{\begin{array}{l}
        a\to a+2\pi m\\
        n\to n+dm
    \end{array}\right.,
\end{equation}
where $\mathbb{R}$-valued site variable $\lambda$ and $\mathbb{Z}$-valued link variable $m$ are the parameter functions of these gauge transformations.

The most na\"{i}ve lattice Chern--Simons action might be
\begin{equation}\label{equ3030}
    \sum_c \frac{ik}{4\pi}a\smallsmile da,
\end{equation}
where $\smallsmile$ is the cup product on cubic lattice, and $\sum_c$ means the sum over $3$d cubes on the lattice. This action changes by
\begin{equation}\label{equ3040}
    \sum_c\frac{ik}{2}[m\smallsmile da+a\smallsmile dm+2\pi m\smallsmile dm]
\end{equation}
under the large gauge transformation. To cancel the first and second term, let us try to add a term $-2\pi(a\smallsmile n+n\smallsmile a)$ to the Lagrangian. Then the improved action becomes
\begin{equation}\label{equ3050}
    \sum_c \frac{ik}{4\pi}[a\smallsmile da-2\pi(a\smallsmile n+n\smallsmile a)].
\end{equation}
While this action is not invariant under ordinary gauge transformation, in the case $dn=0$, this action is invariant under the ordinary gauge transformation. Thus, we keep the symmetry by imposing $dn = 0$. 
To enforce this condition, we introduce a Lagrange-multiplier site variable $\varphi$ and add the term
\begin{equation}\label{equ3060}
    i\sum_c \varphi \smallsmile dn
\end{equation}
to the action. The action then becomes
\begin{equation}\label{equ3070}
    \sum_c \left\{
        \frac{ik}{4\pi}\bigl[a\smallsmile da - 2\pi(a\smallsmile n + n\smallsmile a)\bigr]
        + i\,\varphi\smallsmile dn
    \right\}.
\end{equation}
Now that the action has the term $\varphi\smallsmile dn$, we can implement off-shell ordinary gauge invariance. The action changes by
\begin{equation}\label{equ3080}
    \sum_c\frac{ik}{4\pi}[d(\lambda\smallsmile da)-2\pi(d\lambda\smallsmile n+n\smallsmile d\lambda)]=\sum_c\frac{ik}{4\pi}[-2\pi(\lambda\smallsmile dn+dn\smallsmile \lambda)]
\end{equation}
under (ordinary) gauge transformation. We require that $\varphi$ changes under ordinary gauge transformation as follows:
\begin{equation}\label{equ3090}
    \varphi \;\to\; \varphi - k\lambda.
\end{equation}
Then the action changes by
\begin{equation}\label{equ3100}
    \sum_c \frac{ik}{4\pi}\bigl[-2\pi(-\lambda\smallsmile dn + dn\smallsmile \lambda)\bigr]
    \;=\;
    \sum_c -\frac{ik}{2}\, d\lambda \smallsmile_1 dn
\end{equation}
under ordinary gauge transformation. Adding $-\frac{ik}{2}a\smallsmile_1 dn$, the action becomes
\begin{equation}\label{equ3110}
    S[a,n,\varphi]=\sum_c \left\{\frac{ik}{4\pi}[a\smallsmile da-2\pi(a\smallsmile n+n\smallsmile a)]-\frac{ik}{2}a\smallsmile_1 dn+i\varphi\smallsmile dn\right\},
\end{equation}
and the (off-shell ordinary) gauge invariance is realized. Since this action changes by $i\pi k\mathbb{Z}$ under a large gauge transformation, $k\in2\mathbb{Z}$ guarantees the large gauge invariance of the action. This means that the Chern--Simons level $k$ is restricted to even integers in this formulation.\footnote{It is known that adding Majorana fermion action can make the odd level Chern--Simons action well-defined. See~\cite{Xu:2024hyo} for details.} We can thus define the lattice Chern--Simons theory
\begin{equation}\label{equ3120}
    \mathcal{Z}=(\mathrm{gauge}\;\mathrm{vol.})^{-1}\int\mathcal{D}a\mathcal{D}n\mathcal{D}\varphi \;e^{S[a,n,\varphi]}.
\end{equation}
We will not discuss the gauge fixing and the convergence of the path integral.\footnote{For the path integral with the introduction of Maxwell terms and appropriate gauge fixing, see Section~2.3 and Chapter~5 in~\cite{Xu:2024hyo}.}
\subsection{Staggered symmetry and Wilson line}\label{sec320}
We first consider a transformation $A\to A+\epsilon$ to explore $1$-form symmetries of the action. The change of the action \eqref{equ3110} equals to
\begin{equation}\label{equ3130}
    \sum_c\left[\epsilon\smallsmile\left(da-2\pi n+\frac{1}{2}d\epsilon\right)+\left(da-2\pi n+\frac{1}{2}d\epsilon\right)\smallsmile\epsilon-2\pi\epsilon\smallsmile_1dn\right].
\end{equation}
Integrating $\varphi$ yields $dn=0$ and the third term vanishes, and if there are several $1$-form $\epsilon$ which satisfy
\begin{equation}\label{equ3140}
    \sum_c\left(\epsilon\smallsmile X+X\smallsmile\epsilon\right)=0\qquad(\mathrm{for}\;\mathrm{all}\; X\in C^2(\mathbb{R})),
\end{equation}
the transformation $A\to A+\epsilon$ can be considered as a $1$-form symmetry of the action. Such an $\epsilon$ should satisfy
\begin{equation}\label{equ3150}
    \epsilon(x,\mu)=-\epsilon(x+(1,1,1),\mu)\qquad(\text{for all $x,\mu$}),
\end{equation}
and we can construct it. Let us call this type of symmetry staggered symmetry. 

The Wilson line has to be invariant under the staggered symmetry to have a nontrivial expectation value. The ordinary definition of the Wilson line
\begin{equation}\label{equ3160}
    \exp \left[i\int_\gamma A\right]
\end{equation}
is generally not invariant under staggered symmetries. Then we first prepare a closed line  $\tilde{\gamma}$ on the dual lattice, splitting one line into two half lines, and define the framed Wilson line operator 
\begin{equation}\label{equ3170}
    \exp \left[\frac{i}{2}\int_{\tilde{\gamma}_+} A+\frac{i}{2}\int_{\tilde{\gamma}_-} A\right]=\exp\left[\frac{i}{2}\sum_c(A\smallsmile\mathrm{PD}(\tilde{\gamma})+\mathrm{PD}(\tilde{\gamma})\smallsmile A)\right]
\end{equation}
instead of an ordinary one. Note that $\tilde{\gamma}_{+}$ (resp. $\tilde{\gamma}_-$) means $\frac{1}{2}\hat{1}+\frac{1}{2}\hat{2}+\frac{1}{2}\hat{3}$ (resp. $-\frac{1}{2}\hat{1}-\frac{1}{2}\hat{2}-\frac{1}{2}\hat{3}$) translation of $\tilde{\gamma}$, and $\mathrm{PD}(\tilde{\gamma})$ denotes a Poincar\'{e} dual of $\tilde{\gamma}$, defined by 
\begin{equation}\label{equ3171}
    \begin{split}
        &\mathrm{PD}(\tilde{\gamma})(x;\mu,\nu)\\
        =&(\text{the signed number of times that $\tilde{\gamma}$ passes through the plaquette $(x;\mu,\nu)$}).
    \end{split}
\end{equation}
However, \eqref{equ3170} is still imperfect because it lacks the large gauge invariance. Adding a higher cup term to the imperfect definition, we complete the definition
\begin{equation}\label{equ3180}
    W(\tilde{\gamma})=\exp\left[\frac{i}{2}\sum_c(A\smallsmile\mathrm{PD}(\tilde{\gamma})+\mathrm{PD}(\tilde{\gamma})\smallsmile A)-i\pi\sum_c\mathrm{PD}(\tilde{\gamma})\smallsmile_1n\right].
\end{equation}
This definition respects both staggered symmetry and gauge symmetry. As we have seen, staggered symmetry, which was thought to be removed, is reinterpreted by Jacobson and Sulejmanpasic as a framing of the Wilson line.

Let us briefly comment on the expectation value of the Wilson line. As seen in \eqref{equ3120}, the path integral of the partition function is difficult to calculate, and the convergence of the path integral seems ambiguous. Jacobson and Sulejmanpasic introduced a background field of $\mathbb{Z}_{k}$ symmetry to circumvent this problem, and they proved the two following properties of the Wilson line. First, consider a closed curve $\tilde{\gamma}$ on the dual lattice. Let $\tilde{\gamma}_\mathrm{twist}$ be a locally deformed curve whose self-linking number increases by $l$. The expectation values of these two lines have a relation
\begin{equation}\label{equ3190}
    \langle W(\tilde{\gamma}_\mathrm{twist})\rangle=\exp\left(\frac{2\pi il}{2k}\right)\langle W(\tilde{\gamma})\rangle.
\end{equation}
In addition, let $\tilde{\gamma},\tilde{\gamma}'$ be two closed curves which have linking number $l$. Then we have
\begin{equation}\label{equ3200}
    \langle W(\tilde{\gamma})W(\tilde{\gamma}')\rangle=\exp\left(\frac{2\pi il}{k}\right)\langle W(\tilde{\gamma})\rangle\langle W(\tilde{\gamma}')\rangle.
\end{equation}
These properties implemented on the lattice are the same as the corresponding properties of the Wilson line in continuum level $k$ Chern--Simons theory.
\newpage
\part{The lattice DB cohomology on $d$-dim cubic torus}\label{part2}
We move on to the main topic of this paper. In this part, we define the lattice Deligne--Beilinson cohomology on a $d$-dimensional cubic toroidal lattice. The method used to define the lattice DB cohomology in this part follows the approach outlined in~\cite{Norton:2021psm}, but various details of the construction are completed in the present work.
\section{Lattice DB cochain}\label{sec400}
Let $L$ be a positive integer. We consider the cubic toroidal lattice of volume $L^d$. We define $\mathbb{L}^d$ as a set of grid points
\begin{equation}\label{equ4010}
    \mathbb{L}^d=\{0,1,\ldots,L-1\}^d\cong (\mathbb{Z}/L\mathbb{Z})^d\cong\mathbb{Z}^d/L\mathbb{Z}^d,
\end{equation}
and $\tilde{\mathbb{L}}^d$ as a set of dual grid points
\begin{equation}\label{equ4020}
    \tilde{\mathbb{L}}^d=\left\{\frac{1}{2},\frac{3}{2},\ldots,\frac{2L-1}{2}\right\}^d\cong(\mathbb{Z}+1/2)^d/L\mathbb{Z}^d.
\end{equation}
\subsection{Lattice \v{C}ech--de Rham chain}\label{sec410}
Let $x_{+\dots +}$ be an abbreviation for $x+\frac{1}{2}\hat{1}+\dots +\frac{1}{2}\hat{d}$ for each point $x\in\mathbb{L}^d$, where $\hat{i}=(0,\ldots,1,\ldots,0)\;\;(\text{only the $i$-th element is $1$})$.

We define a $d$-dimensional cube with sides of length one and centered at point $x_{+\dots +}$, and denote it by $\mathrm{cube}(x_{+\dots +})$. This cube is regarded as a subset of the torus $\mathbb{T}^d=\mathbb{R}^d/L\mathbb{Z}^d$ and is given by
\begin{equation}\label{equ4030}
    \mathrm{cube}(x_{+\dots +}):=\mathrm{pr}\left\{y\in\mathbb{R}^d\middle|\|x_{+\dots +}-y\|_\mathrm{max}\leq \frac{1}{2}\right\},
\end{equation}
where $\mathrm{pr}:\mathbb{R}^d\to\mathbb{T}^d$ is natural projection on quotient space $\mathbb{R}^d/L\mathbb{Z}^d$.

\begin{figure}[H]\centering
\begin{minipage}[t]{0.45\textwidth}\centering
\includegraphics[scale=0.26]{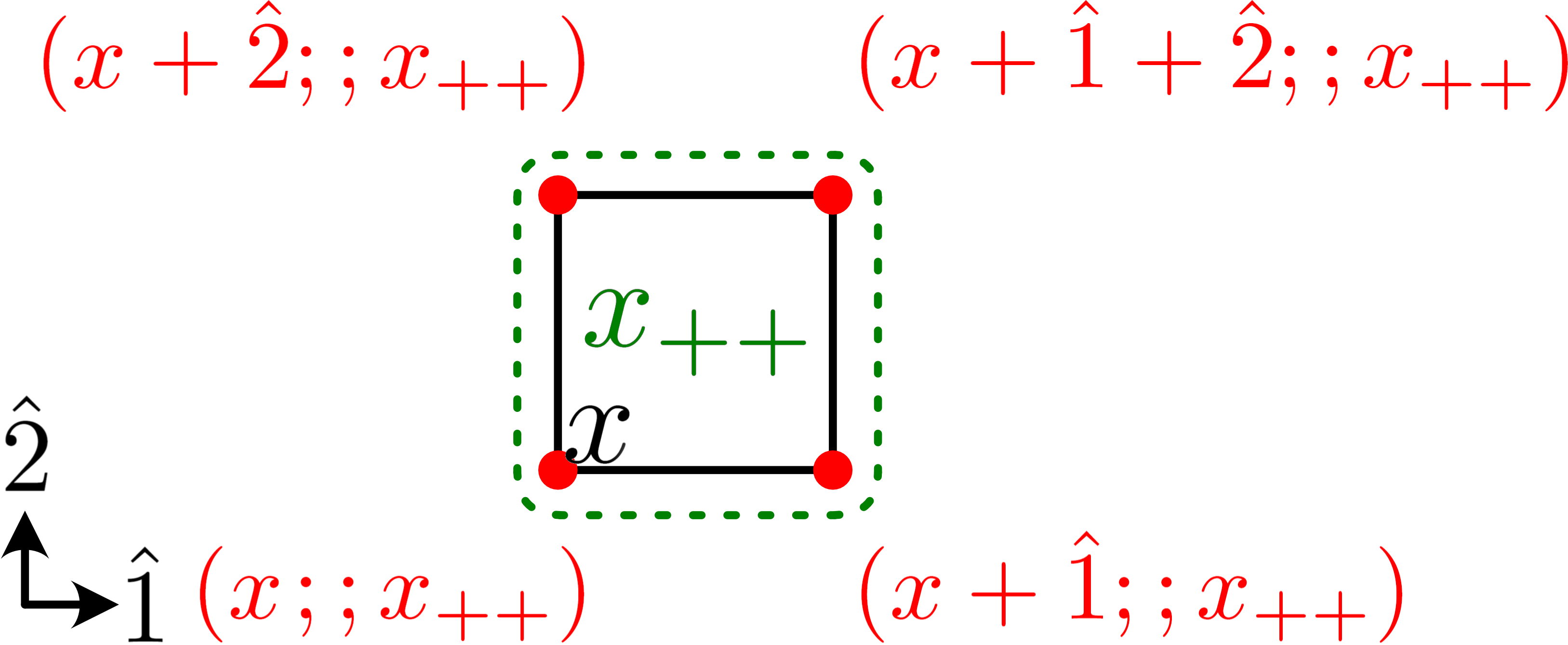}
\caption{Examples of elements of $\mathrm{cube}_{(0,0)}(\mathbb{L}^2)$}\label{fig:DB1}
\end{minipage}
\hspace{0.08\textwidth}
\begin{minipage}[t]{0.45\textwidth}\centering
\includegraphics[scale=0.26]{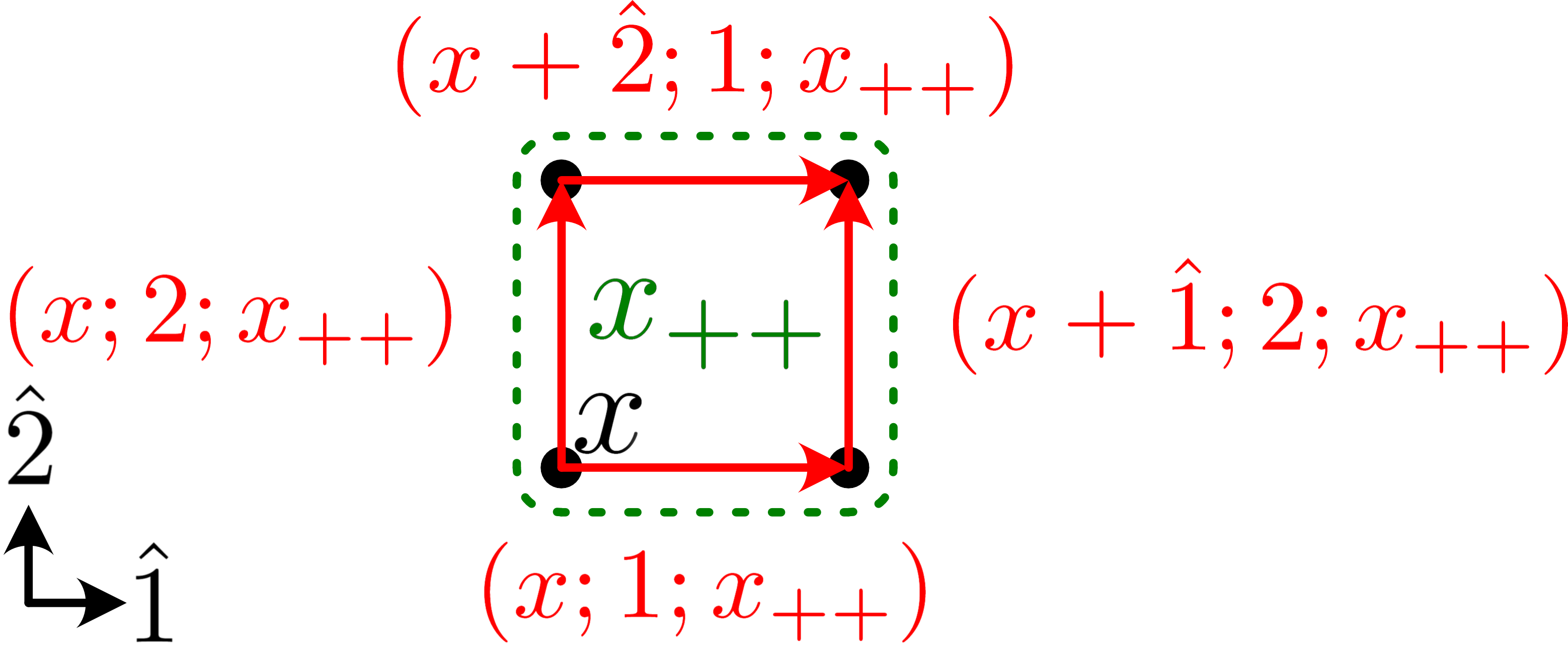}
\caption{Examples of elements of $\mathrm{cube}_{(1,0)}(\mathbb{L}^2)$}\label{fig:DB2}
\end{minipage}
\end{figure}

\begin{figure}[H]\centering
\begin{minipage}[t]{0.45\textwidth}\centering
\includegraphics[scale=0.26]{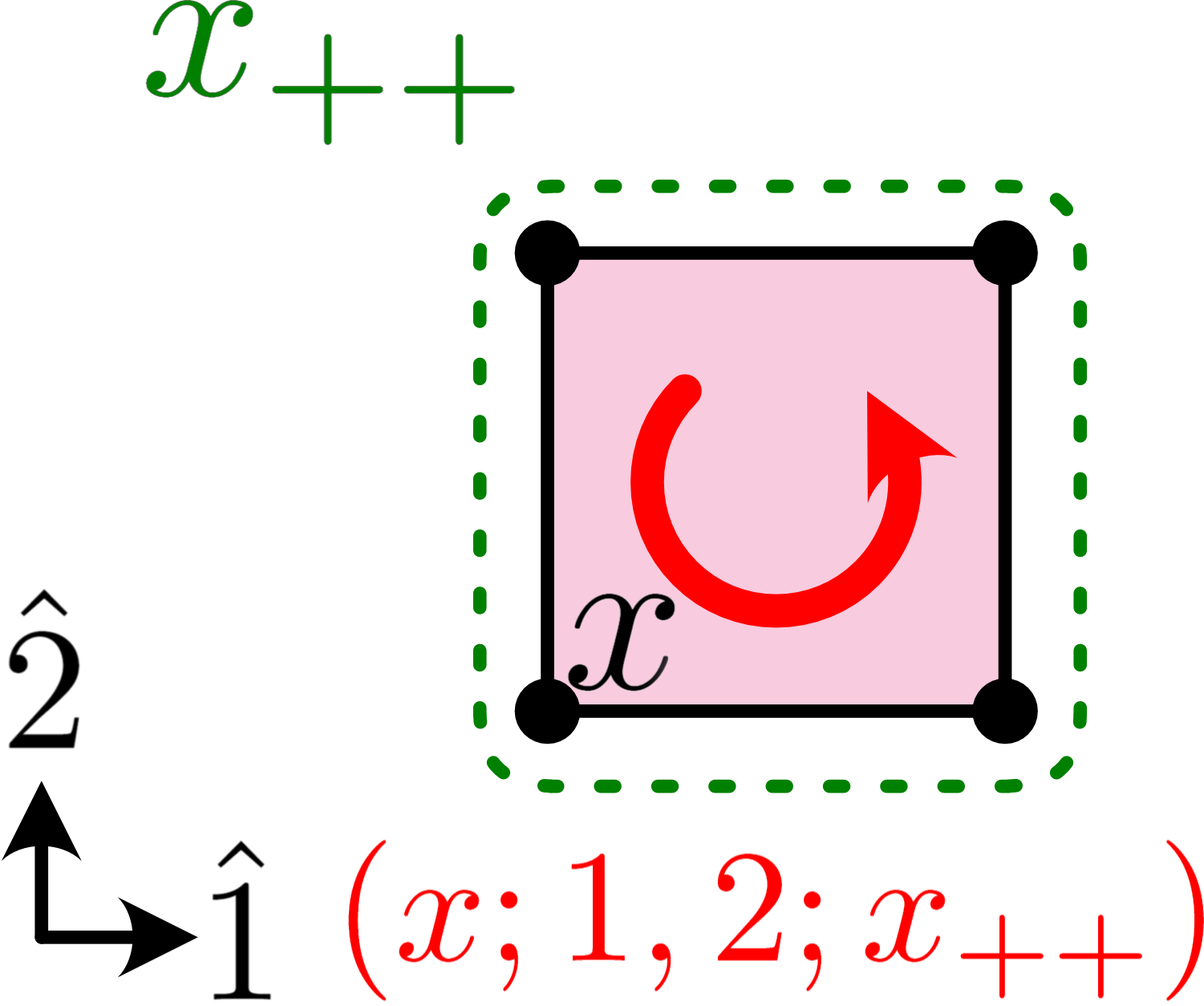}
\caption{An example of an element of $\mathrm{cube}_{(2,0)}(\mathbb{L}^2)$}\label{fig:DB3}
\end{minipage}
\hspace{0.08\textwidth}
\begin{minipage}[t]{0.45\textwidth}\centering
\includegraphics[scale=0.3]{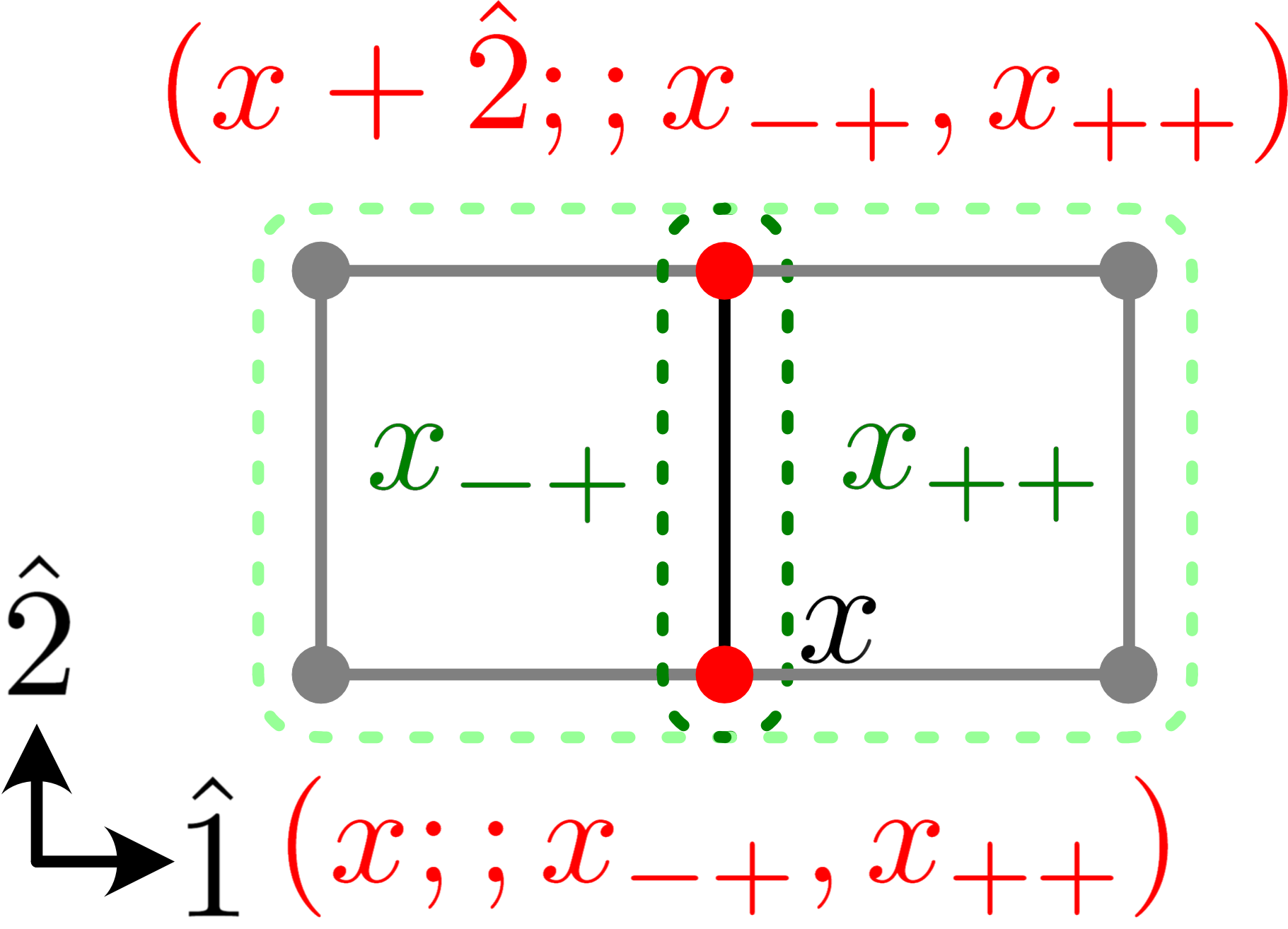}
\caption{Examples of elements of $\mathrm{cube}_{(0,1)}(\mathbb{L}^2)$}\label{fig:DB4}
\end{minipage}
\end{figure}

\begin{figure}[H]\centering
\begin{minipage}[t]{0.45\textwidth}\centering
\includegraphics[scale=0.3]{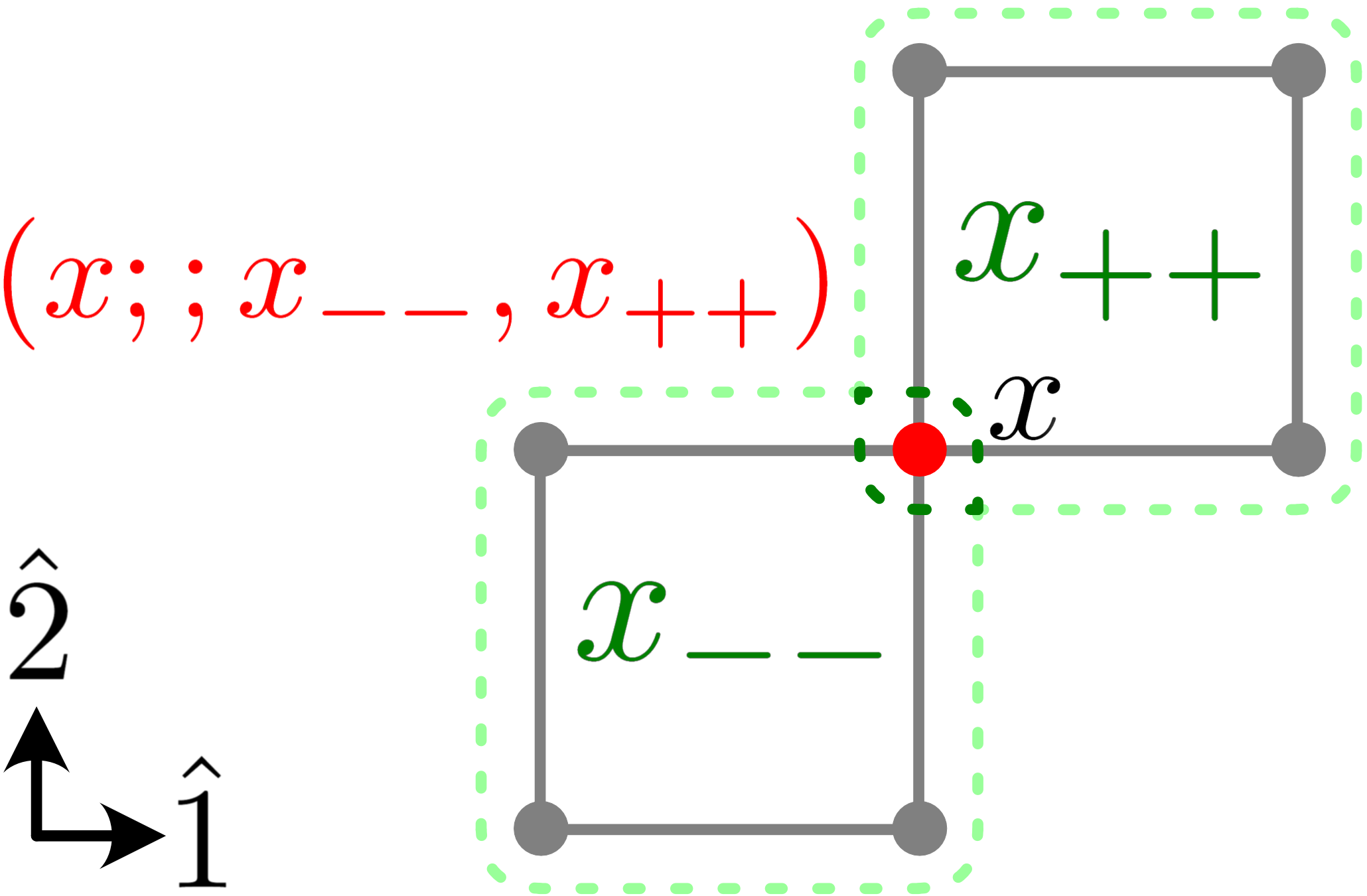}
\caption{Another example of an element of $\mathrm{cube}_{(0,1)}(\mathbb{L}^2)$}\label{fig:DB6}
\end{minipage}
\hspace{0.08\textwidth}
\begin{minipage}[t]{0.45\textwidth}\centering
\includegraphics[scale=0.3]{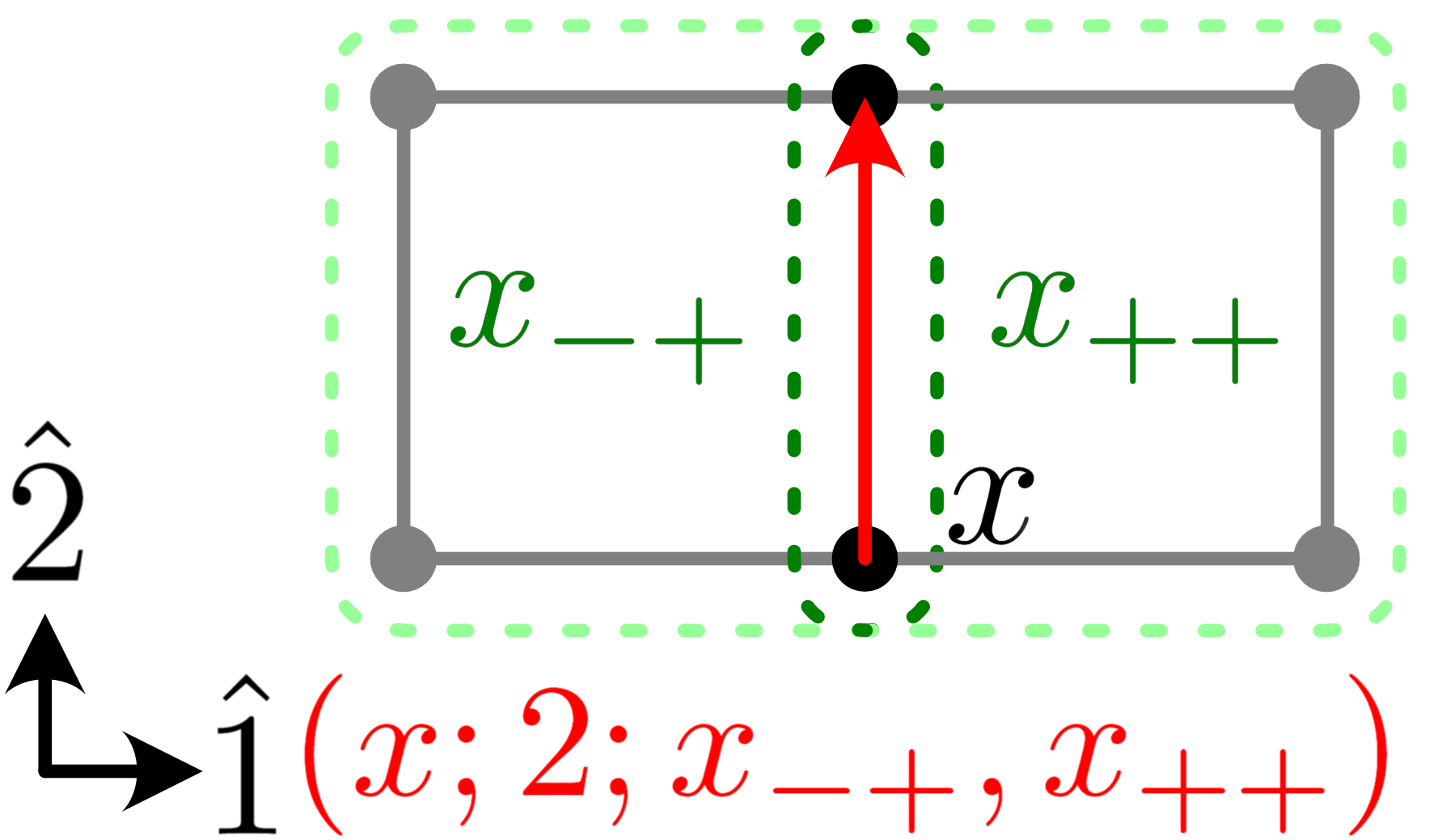}
\caption{An example of an element of $\mathrm{cube}_{(1,1)}(\mathbb{L}^2)$}\label{fig:DB5}
\end{minipage}
\end{figure}

\begin{figure}[H]\centering
\begin{minipage}[t]{0.45\textwidth}\centering
\includegraphics[scale=0.3]{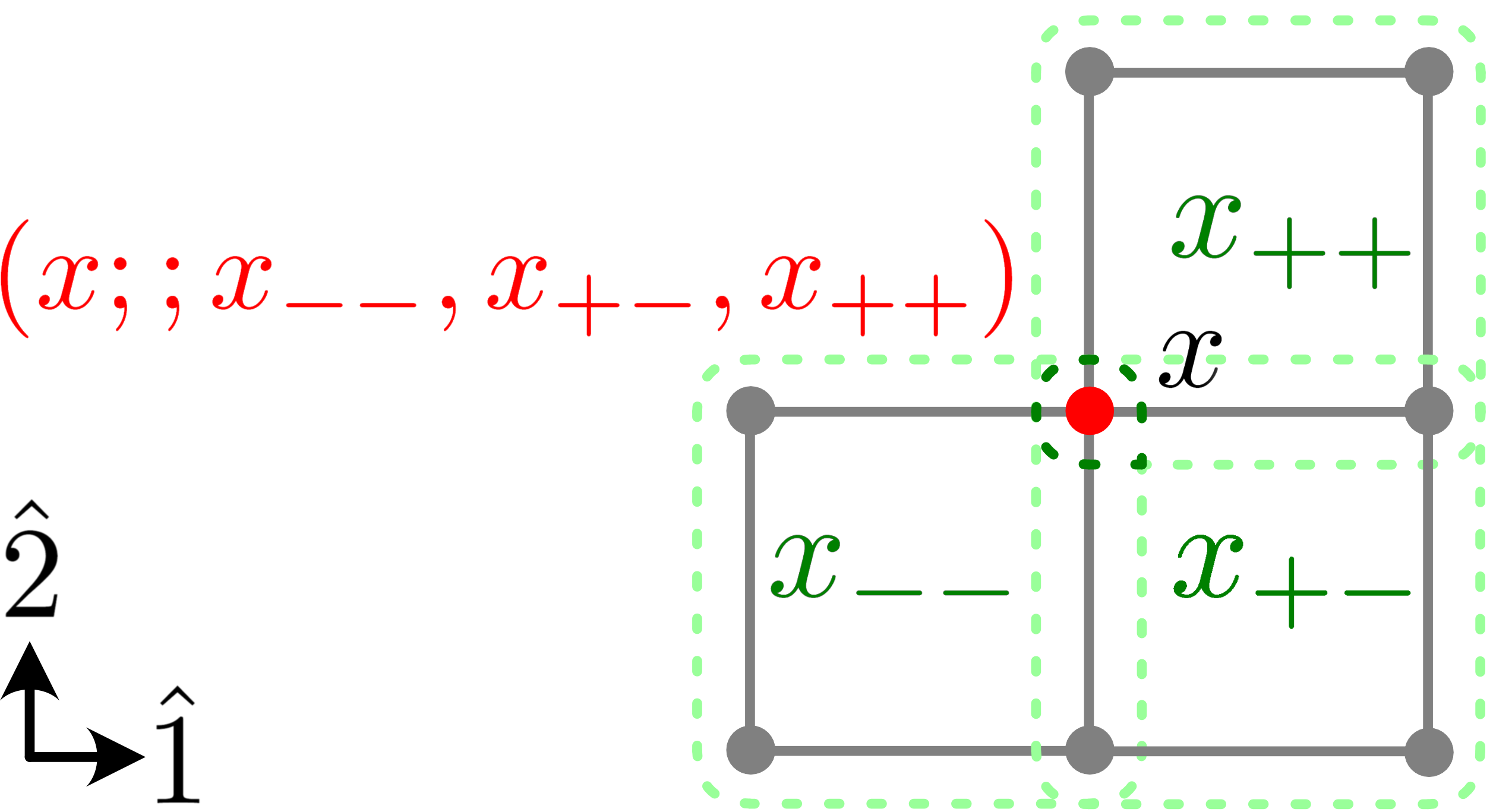}
\caption{An example of an element of $\mathrm{cube}_{(0,2)}(\mathbb{L}^2)$}\label{fig:DB7}
\end{minipage}
\hspace{0.08\textwidth}
\begin{minipage}[t]{0.45\textwidth}\centering
\includegraphics[scale=0.3]{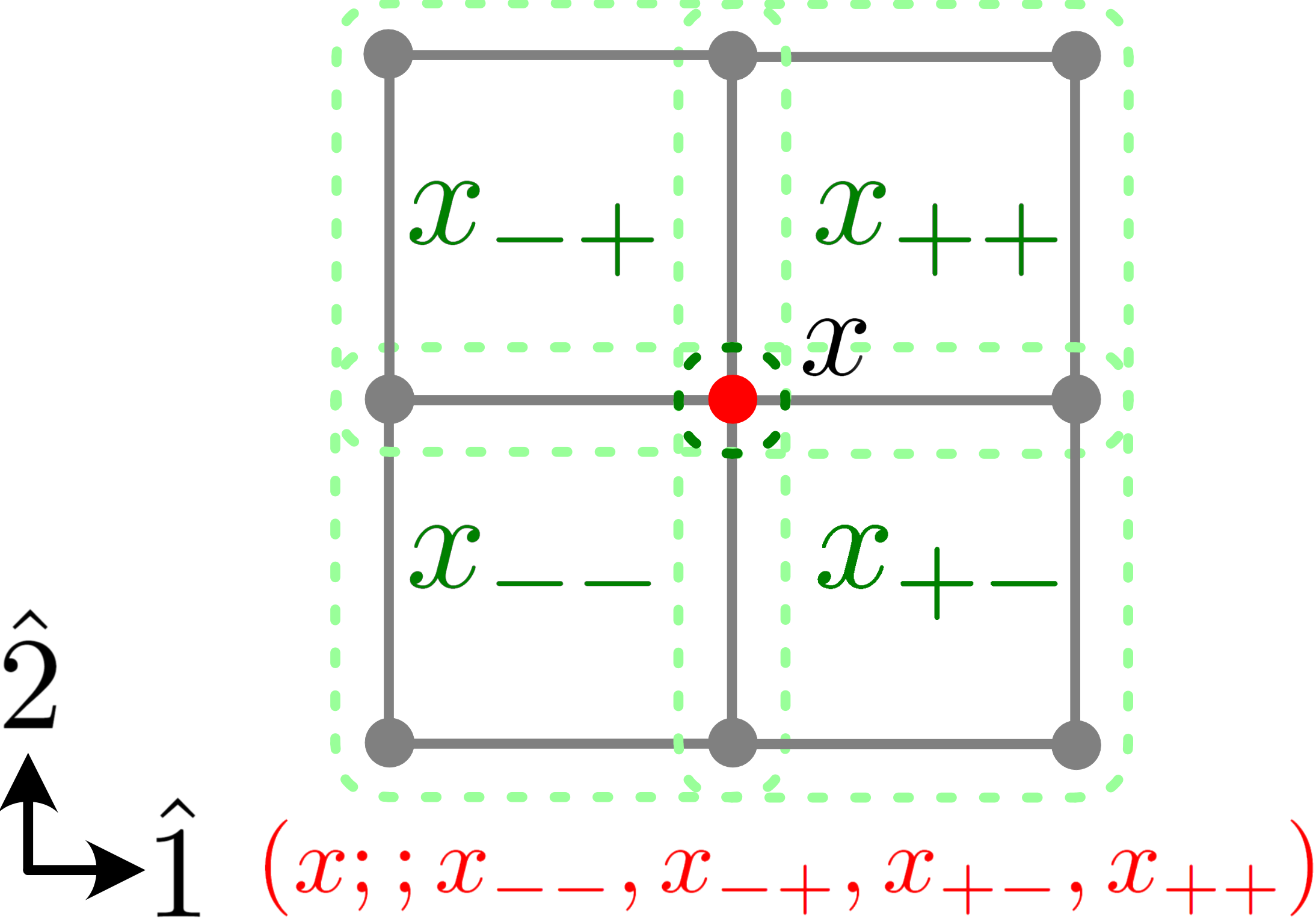}
\caption{An example of an element of $\mathrm{cube}_{(0,3)}(\mathbb{L}^2)$}\label{fig:DB8}
\end{minipage}
\end{figure}

For $0\leq i\leq d\;,\;-1\leq j\leq d$,\footnote{Note that when $j=-1$, $(\tilde{\mathbb{L}}^d)^{j+1}$ is the empty direct product and can be understood as a one point set $\{\ast\}$ and  $\mathrm{cube}(\tilde{x}^{(0)})\cap\cdots\cap\mathrm{cube}(\tilde{x}^{(j)})$ is identified with $\mathbb{T}^d$ itself. In any case, $\mathrm{cube}_{(i,j)}(\mathbb{L}^d)$ is well-defined even in $j=-1$.} we define a set of lattice \v{C}ech--de Rham cube of degree $(i,j)$, denoted $\mathrm{cube}_{(i,j)}(\mathbb{L}^d)$, by
\begin{equation}\label{equ4040}
    \begin{split}
        \mathrm{cube}_{(i,j)}(\mathbb{L}^d):=&\Bigl\{(x\;;\;\mu_1,\dots,\mu_i \;;\;\tilde{x}^{(0)},\dots,\tilde{x}^{(j)})\in\mathbb{L}^d\times\{1,\dots,d\}^i\times(\tilde{\mathbb{L}}^d)^{j+1}\\
        &\mid 0\leq\;\; \forall\lambda_1,\dots,\forall\lambda_i\;\;\leq 1,\\
        &(x+\lambda_1\hat{\mu}_1+\dots\lambda_i\hat{\mu}_i)/L\mathbb{Z}^d\in \mathrm{cube}(\tilde{x}^{(0)})\cap\cdots\cap\mathrm{cube}(\tilde{x}^{(j)})\Bigr\},
    \end{split}
\end{equation}
and we extend the definition for $i=-1\;,\;-1\leq j\leq d$ by
\begin{equation}\label{equ4050}
    \begin{split}
        \mathrm{cube}_{(-1,j)}(\mathbb{L}^d):=&\Bigl\{(\ast\;;\;\ast \;;\;\tilde{x}^{(0)},\dots,\tilde{x}^{(j)})\in\{\ast\}\times\{\ast\}\times(\tilde{\mathbb{L}}^d)^{j+1}\\
        &\mid\mathrm{cube}(\tilde{x}^{(0)})\cap\cdots\cap\mathrm{cube}(\tilde{x}^{(j)})\neq\emptyset\Bigr\},
    \end{split}
\end{equation}
where $\{\ast\}$ is the one point set. Figures~\ref{fig:DB1}--\ref{fig:DB8} provide concrete examples and illustrations of lattice \v{C}ech--de Rham cubes on a two-dimensional cubic lattice $\mathbb{L}^2$.

For $-1\leq i\leq d\;,\;-1\leq j\leq d$, we define lattice \v{C}ech--de Rham chain $C_{(i,j)}(\mathbb{L}^d)$ of degree $(i,j)$ by
\begin{equation}\label{equ4060}
    C_{(i,j)}(\mathbb{L}^d):=\left(\mathrm{Span}_{\mathbb{Z}}(\mathrm{cube}_{(i,j)}(\mathbb{L}^d))\right)/\sim,
\end{equation}
where $\sim$ is an equivalence relation of antisymmetry under permutations of the $\mu$'s
\begin{equation}\label{equ4070}
    \begin{array}{llllll}
        \sigma\in\mathfrak{S}_{i},&&&&&\\
        &(x\;;&\;\mu_1,\dots,\mu_i \;&;&\;\tilde{x}^{(0)},\dots,\tilde{x}^{(j)}&)\\
        \sim\mathrm{sgn}(\sigma)&(x\;;&\;\mu_{\sigma(1)},\dots,\mu_{\sigma(i)} \;&;&\;\tilde{x}^{(0)},\dots,\tilde{x}^{(j)}&),
    \end{array}
\end{equation}
and of antisymmetry  under permutations of the $\tilde{x}$'s
\begin{equation}\label{equ4080}
    \begin{array}{llllll}
        \sigma\in\mathfrak{S}_{j+1},&&&&&\\
        &(x\;;&\;\mu_1,\dots,\mu_i \;&;&\;\tilde{x}^{(0)},\dots,\tilde{x}^{(j)}&)\\
        \sim\mathrm{sgn}(\sigma)&(x\;;&\;\mu_1,\dots,\mu_i \;&;&\;\tilde{x}^{(\sigma(0))},\dots,\tilde{x}^{(\sigma(j))}&).
    \end{array}
\end{equation}
\newpage
\newpage
\subsection{Lattice \v{C}ech--de Rham cochain}\label{sec420}
For $0\leq i\leq d\;,\;-1\leq j\leq d$, lattice \v{C}ech--de Rham cochain of degree $(i,j)$, denoted $C^{(i,j)}(\mathbb{L}^d)$, is defined by
\begin{equation}\label{equ4090}
    C^{(i,j)}(\mathbb{L}^d):=\{\omega:C_{(i,j)}(\mathbb{L}^d)\to\mathbb{R}\mid \text{$\omega$ is $\mathbb{Z}$-linear}\},
\end{equation}
or equivalently,
\begin{equation}\label{equ4100}
    \begin{split}
    C^{(i,j)}(\mathbb{L}^d):=\{&\omega:\mathrm{cube}_{(i,j)}(\mathbb{L}^d)\to\mathbb{R}\mid\text{$\omega$ is antisymmetric under}\\
    &\text{ permutations of $\{1,\ldots,d\}^i$ and permutations of $(\tilde{\mathbb{L}}^3)^{j+1}$}
    \}.
    \end{split}
\end{equation}

We extend the definition of $C^{(i,j)}(\mathbb{L}^d)$ for $i=-1\;,\;-1\leq j\leq d$ by
\begin{equation}\label{equ4110}
    C^{(-1,j)}(\mathbb{L}^d):=\{\omega:C_{(-1,j)}(\mathbb{L}^d)\to\mathbb{Z}\mid \text{$\omega$ is $\mathbb{Z}$-linear}\},
\end{equation}
or equivalently,
\begin{equation}\label{equ4120}
    \begin{split}
    C^{(-1,j)}(\mathbb{L}^d):=\{&\omega:\mathrm{cube}_{(-1,j)}\to\mathbb{Z}
    \\ &\mid\text{$\omega$ is antisymmetric under permutations of $(\tilde{\mathbb{L}}^3)^{j+1}$}
    \}.
    \end{split}
\end{equation}
However, for $C^{(-1,j)}$, we take the range of $\omega$ to be $\mathbb{Z}$ for later convenience. It should be emphasized that this convention is crucial throughout the entire paper.

We will write the arguments of $\omega^{(i,j)}\in C^{(i,j)}(\mathbb{L}^d)$ with two semicolons separating the three sets of variables: for example,
\begin{align}\label{equ4130}
    &\omega^{(i,j)}(x\;;\;\mu_1,\ldots,\mu_i\;;\;\tilde{x}^{(0)},\ldots,\tilde{x}^{(j)}),&\omega^{(-1,j)}(\;;\;\;;\;\tilde{x}^{(0)},\ldots,\tilde{x}^{(j)}).
\end{align}

The cochain we have defined above has $\tilde{\mathbb{L}}^d$ as a variable, which corresponds to the indices of patches in the definition of a patchwise gauge field \eqref{equ2030}. That is, we consider each unit cube of the lattice as a patch, and such a method is already widely used~\cite{Luscher:1981zq,Abe:2023ncy}.
\subsection{Coboundary operations}\label{sec430}
\subsubsection{\v{C}ech coboundary operation}\label{sec431}
For $-1\leq j\leq d-1$, we introduce $j$-th \v{C}ech coboundary operation $\delta_j:C^{(i,j)}(\mathbb{L}^d)\to C^{(i,j+1)}(\mathbb{L}^d)$ by
\begin{equation}\label{equ4140}
    (\delta_j\omega)^{(i,j+1)}(\bullet\;;\;\bullet\;;\;\tilde{x}^{(0)},\ldots,\tilde{x}^{(j+1)})=\sum_{k=0}^{j+1}(-1)^k\omega^{(i,j)}(\bullet\;;\;\bullet\;;\;\tilde{x}^{(0)},\ldots,\check{\tilde{x}}^{(k)},\ldots,\tilde{x}^{(j+1)}),
\end{equation}
where $\check{\tilde{x}}^{(k)}$ means eliminating $\tilde{x}^{(k)}$.

Even for $j=-1$, $(-1)$-th \v{C}ech coboundary operation $\delta_{-1}:\Omega^{i}(\mathbb{L}^d)\to C^{(i,0)}(\mathbb{L}^d)$ can be defined by
\begin{equation}\label{equ4150}
    (\delta_{-1}\omega)(\bullet\;;\;\bullet\;;\;\tilde{x})=\omega^{(i,-1)}(\bullet\;;\;\bullet\;;\;),
\end{equation}
where $\Omega^i(\mathbb{L}^d)$ is the set of $i$-cochains on $\mathbb{L}^d$. This definition is the same as we defined in equation \eqref{equ2170}. We can see $\delta_{j+1}\circ\delta_{j}=0$ for $-1\leq j$.
\subsubsection{Lattice de Rham coboundary operation}\label{sec432}
For $0\leq i\leq d-1$, we define $i$-th de Rham coboundary operation $d_i:C^{(i,j)}(\mathbb{L}^d)\to C^{(i+1,j)}(\mathbb{L}^d)$ by
\begin{equation}\label{equ4160}
    \begin{split}
    (d_i\omega)^{(i+1,j)}(x\;;\; \mu_1,\ldots,\mu_{i+1}\;;\;\bullet)&\\
    =\frac{1}{(i+1)!}\sum_{\tau\in\mathfrak{S}_{i+1}}\mathrm{sgn}(\tau)\sum_{k=1}^{i+1}\Biggl[(-1)^{k-1}\Bigl(&\omega^{(i,j)}(x+\hat{\mu}_{\tau(k)}\;;\; \mu_{\tau(1)},\ldots,\check{\mu}_{\tau(k)},\ldots,\mu_{\tau(i+1)}\;;\;\bullet)\\
    -&\omega^{(i,j)}(x\;;\; \mu_{\tau(1)},\ldots,\check{\mu}_{\tau(k)},\ldots,\mu_{\tau(i+1)}\;;\;\bullet)\Bigr)\Biggr],
    \end{split}
\end{equation}
where $\check{\mu}_{k}$ means eliminating $\mu_k$.

Even for $i=-1$, $(-1)$-th de Rham coboundary operation $d_{-1}:C^{(-1,j)}(\mathbb{L}^d)\to C^{(0,j)}(\mathbb{L}^d)$ can be defined by
\begin{equation}\label{equ4170}
    (d_{-1}\omega)^{(0,j)}(x\;;\;\;;\;\bullet)=\omega^{(-1,j)}(\;;\;\;;\;\bullet).
\end{equation}

We can see $d_d=0$, and for $-1\leq i\leq d-1$, we find $d_{i+1}\circ d_{i}=0$.
In fact, in the case $i=d-1$, $d_d\circ d_{d-1}=0$ is obviously true because $d_d=0$, and in the case $i=-1$,
\begin{equation}\label{equ4180}
    \begin{split}
    (d_0d_{-1}\omega)^{(1,j)}(x\;;\;\mu\;;\;\bullet)&=(d_{-1}\omega)^{(0,j)}(x+\hat{\mu}\;;\;\;;\;\bullet)-(d_{-1}\omega)^{(0,j)}(x\;;\;\;;\;\bullet)\\
    &=\omega^{(-1,j)}(\;;\;\;;\;\bullet)-\omega^{(-1,j)}(\;;\;\;;\;\bullet)=0
    \end{split}
\end{equation}
holds, and in $0\leq i\leq d-2$, we can see $d_{i+1}\circ d_{i}=0$ from
\begin{equation}\label{equ4190}
    \begin{split}
    (d_{i+1}d_{i}\omega)^{(i+2,j)}(x\;;\;\mu_1,\ldots,&\mu_{i+2}\;;\;\bullet)\\
    =\text{antisym.}\sum_{k=1}^{i+2}\Biggl[
        (-1)^{k-1}\Bigl(&
            (d_{i}\omega)^{(i+1,j)}(x+\hat{\mu}_k\;;\;\mu_1,\ldots,\check{\mu}_k,\ldots,\mu_{i+2}\;;\;\bullet)\\-&(d_{i}\omega)^{(i+1,j)}(x\;;\;\mu_1,\ldots,\check{\mu}_k,\ldots,\mu_{i+2}\;;\;\bullet)
        \Bigr)
    \Biggr]\\
    =\text{antisym.}\sum_{k=1}^{i+2}\Biggl\{\sum_{1\leq l<k}\Biggl[
        (-1)^{l+k}\Bigl(&
            \omega^{(i,j)}(x+\hat{\mu}_l+\hat{\mu}_k\;;\;\mu_1,\ldots,\check{\mu}_l,\ldots,\check{\mu}_k,\ldots,\mu_{i+2}\;;\;\bullet)\\
            -&\omega^{(i,j)}(x+\hat{\mu}_k\;;\;\mu_1,\ldots,\check{\mu}_l,\ldots,\check{\mu}_k,\ldots,\mu_{i+2}\;;\;\bullet)\\
            -&\omega^{(i,j)}(x+\hat{\mu}_l\;;\;\mu_1,\ldots,\check{\mu}_l,\ldots,\check{\mu}_k,\ldots,\mu_{i+2}\;;\;\bullet)\\
            +&\omega^{(i,j)}(x\;;\;\mu_1,\ldots,\check{\mu}_l,\ldots,\check{\mu}_k,\ldots,\mu_{i+2}\;;\;\bullet)
        \Bigr)
    \Biggr]\\
    +\sum_{k<l\leq i+2}\Biggl[
        (-1)^{k+l-1}\Bigl(&
            \omega^{(i,j)}(x+\hat{\mu}_k+\hat{\mu}_l\;;\;\mu_1,\ldots,\check{\mu}_k,\ldots,\check{\mu}_l,\ldots,\mu_{i+2}\;;\;\bullet)\\
            -&\omega^{(i,j)}(x+\hat{\mu}_k\;;\;\mu_1,\ldots,\check{\mu}_k,\ldots,\check{\mu}_l,\ldots,\mu_{i+2}\;;\;\bullet)\\
            -&\omega^{(i,j)}(x+\hat{\mu}_l\;;\;\mu_1,\ldots,\check{\mu}_k,\ldots,\check{\mu}_l,\ldots,\mu_{i+2}\;;\;\bullet)\\
            +&\omega^{(i,j)}(x\;;\;\mu_1,\ldots,\check{\mu}_k,\ldots,\check{\mu}_l,\ldots,\mu_{i+2}\;;\;\bullet)
        \Bigr)
    \Biggr]\Biggr\}\\
    =\text{antisym.}\sum_{1\leq l<k\leq i+2}\Biggl[
        (-1)^{l+k}(1-1)\Bigl(&
            \omega^{(i,j)}(x+\hat{\mu}_l+\hat{\mu}_k\;;\;\mu_1,\ldots,\check{\mu}_l,\ldots,\check{\mu}_k,\ldots,\mu_{i+2}\;;\;\bullet)\\
            -&\omega^{(i,j)}(x+\hat{\mu}_k\;;\;\mu_1,\ldots,\check{\mu}_l,\ldots,\check{\mu}_k,\ldots,\mu_{i+2}\;;\;\bullet)\\
            -&\omega^{(i,j)}(x+\hat{\mu}_l\;;\;\mu_1,\ldots,\check{\mu}_l,\ldots,\check{\mu}_k,\ldots,\mu_{i+2}\;;\;\bullet)\\
            +&\omega^{(i,j)}(x\;;\;\mu_1,\ldots,\check{\mu}_l,\ldots,\check{\mu}_k,\ldots,\mu_{i+2}\;;\;\bullet)
        \Bigr)
    \Biggr]\\
    =0,&
    \end{split}
\end{equation}
where ``antisym.'' denotes antisymmetrization.
\subsubsection{Lattice DB coboundary operation}\label{sec433}
For integer $r$ satisfying $0\leq r\leq d+1$, lattice Deligne--Beilinson cochain $C_{lat.\mathrm{DB}}^r(\mathbb{L}^d)$ of degree $r$\footnote{It is by convention $i+j=r-1$ rather than $i+j=r$.} can be defined by
\begin{equation}\label{equ4200}
    C_{\mathrm{lat.DB}}^r(\mathbb{L}^d):=\bigoplus_{\substack{i+j=r-1\\0\leq j\leq r}}C^{(i,j)}(\mathbb{L}^d).
\end{equation}

For $0\leq r\leq d+1$, we define $C_{\mathrm{ev}}^r(\mathbb{L}^d)$ by
\begin{equation}\label{equ4210}
    C_{\mathrm{ev}}^r(\mathbb{L}^d):=\bigoplus_{j=1}^{r}C^{(r-j-1,j)}(\mathbb{L}^d).
\end{equation}

Finally, we introduce lattice DB coboundary operations $D_r^\mathrm{ev},D_r$, given by
\begin{align}
    D_r^\mathrm{ev}&=\delta_0+\left[\sum_{k=1}^{r}\bigl(\delta_k+(-1)^k d_{r-k-1}\bigr)\right]&&:C_\mathrm{lat.DB}^r(\mathbb{L}^d)\to C_{\mathrm{ev}}^{r+1}(\mathbb{L}^d),\label{equ4220}\\
    D_r&=\sum_{k=0}^{r-1}(\delta_k+(-1)^{k+1}d_{r-k-2})&&:C_\mathrm{lat.DB}^{r-1}(\mathbb{L}^d)\to C_{\mathrm{lat.DB}}^r(\mathbb{L}^d).\label{equ4230}
\end{align}
\newpage
\section{Lattice DB cohomology}\label{sec500}
\subsection{Lattice DB cycle/boundary}\label{sec510}
$D$ and $D^\mathrm{ev}$ have a property $D^\mathrm{ev}\circ D=0$. We can directly prove it using $\delta_{j+1}\circ \delta_{j}=0$, $d_{i+1}\circ d_{i}=0$, and commutativity between $\delta$ and $d$, as follows:
\begin{equation}\label{equ5010}
    \begin{split}
        &D_{r}^\mathrm{ev}\circ D_{r}=(D_{r}^\mathrm{ev}:C^{r}_\mathrm{lat.DB}\to C^{r}_\mathrm{ev})\circ (D_{r}:C^{r-1}_\mathrm{lat.DB}\to C^{r}_\mathrm{lat.DB})\\
        &=\left(\delta_0+\left[\sum_{k=1}^{r}\biggl(\delta_k+(-1)^k d_{r-k-1}\biggr)\right]\right)\left(\sum_{k=0}^{r-1}\biggl(\delta_k+(-1)^{k+1}d_{r-k-1}\biggr)\right)\\
        &=\sum_{k=0}^{r-1}\delta_{k+1}\delta_{k}+\sum_{k=-1}^{r-2}d_{k+1}d_{k}+\left[\sum_{k=0}^{r-1}\biggl((-1)^{k}d_{r-k-1}\delta_{k}+\delta_{k}(-1)^{k+1}d_{r-k-1}\biggr)\right]\\
        &=0.
    \end{split}
\end{equation}

As with the usual definition of cohomology theory, lattice Deligne--Beilinson cocycle/coboundary $Z_{\mathrm{lat.DB}}^r(\mathbb{L}^d),B_{\mathrm{lat.DB}}^r(\mathbb{L}^d)$ can be defined by
\begin{equation}\label{equ5020}
    \begin{split}
        Z_\mathrm{lat.DB}^{r}(\mathbb{L}^d):=&\{\bm{\omega}\in C_\mathrm{lat.DB}^{r}(\mathbb{L}^d)\mid D_{r}^\mathrm{ev}\bm{\omega}=0\},\\
        B_\mathrm{lat.DB}^{r}(\mathbb{L}^d):=&D_{r} C_\mathrm{lat.DB}^{r-1}(\mathbb{L}^d),
    \end{split}
\end{equation}
and we define lattice Deligne--Beilinson cohomology $H_{\mathrm{lat.DB}}^r(\mathbb{L}^d)$ by
\begin{equation}\label{equ5030}
    H_{\mathrm{lat.DB}}^r(\mathbb{L}^d):=\frac{Z_{\mathrm{lat.DB}}^r(\mathbb{L}^d)}{B_{\mathrm{lat.DB}}^r(\mathbb{L}^d)}.
\end{equation}

\subsection{Properties of lattice DB cocycle}\label{sec520}
In this section, we describe basic properties of the lattice DB cocycle. In the following, let $\bm{\omega},\bm{\omega}'$ be
\begin{equation*}
    \begin{split}
        \bm{\omega}&=(\omega^{(r-1,0)},\omega^{(r-2,1)},\dots,\omega^{(-1,r)})\in Z^r_\mathrm{lat.DB}(\mathbb{L}^d),\\
        \bm{\omega}'&=(\omega'^{(r-1,0)},\omega'^{(r-2,1)},\dots,\omega'^{(-1,r)})\in Z^r_\mathrm{lat.DB}(\mathbb{L}^d).
    \end{split}
\end{equation*}
Lattice DB cocycles have the following properties:
\begin{itemize}
    \item $(0,\dots,0)\in Z^r_\mathrm{lat.DB}(\mathbb{L}^d)$
    \item $\bm{\omega},\bm{\omega}'\in Z^r_\mathrm{lat.DB}(\mathbb{L}^d)$\\
    $\Longrightarrow \bm{\omega}+\bm{\omega}':=(\omega^{(r-1,0)}+\omega'^{(r-1,0)},\omega^{(r-2,1)}+\omega'^{(r-2,1)},\dots,\omega^{(-1,r)}+\omega'^{(-1,r)})\in Z^r_\mathrm{lat.DB}(\mathbb{L}^d)$
    \item $\omega^{(r-1,-1)}\in\Omega^{r-1}(\mathbb{L}^d)\Longrightarrow (\delta_{-1}\omega^{(r-1,-1)},0,0,\dots,0)\in Z^r_\mathrm{lat.DB}(\mathbb{L}^d)$
\end{itemize}
The proofs are simple and will be omitted.

An element of lattice DB cohomology can be multiplied by $\mathbb{Z}$, adding the same element an integer number of times, but in general it is not possible to multiply it by $\mathbb{R}$. However, as an exception, $\bm{\omega}$, realized as an image of $\Omega^{r-1}\hookrightarrow H_\mathrm{lat.DB}^{r}$, can be multiplied by $\mathbb{R}$ in $\Omega^{r-1}$.
\subsubsection{Global well-definedness of curvature}\label{sec521}
$d_{r-1}\omega^{(r-1,0)}$ can be considered as the curvature of $\bm{\omega}$. In the context of continuous Deligne--Beilinson $2$-cohomology, the curvature $d_{r-1}\omega^{(r-1,0)}$ corresponds to the curvature of the continuous principal $U(1)$-bundle theory. For the reason that connections on a principal $U(1)$-bundle do not necessarily have zero curvature, the condition $d_{r-1}\omega^{(r-1,0)}=0$ should not be imposed on lattice Deligne--Beilinson cochains. This is why we adopt $D^\mathrm{ev}_{r+1}\circ D_r=0$ instead of $D_{r+1}\circ D_r=0$ in the construction of the DB coboundary operation.

In the end, we describe that the curvature $d_{r-1}\omega^{(r-1,0)}$ can be interpreted as a global $r$-form on lattice. First, since $\bm{\omega}$ is a cocycle, it follows that
\begin{equation}\label{equ5040}
    \delta\omega^{(r-1,0)}-d\omega^{(r-2,1)}=0.
\end{equation}
Then we get
\begin{equation}\label{equ5045}
    \delta(d_{r-1}\omega^{(r-1,0)})=\delta d\omega^{(r-1,0)}=d\delta\omega^{(r-1,0)}=0.
\end{equation}
This means the values of $d_{r-1}\omega^{(r-1,0)}$ are independent of the choice of $\tilde{\mathbb{L}}^d$ indices. Then we conclude
\begin{equation}\label{equ5050}
    \exists\omega^{(r,-1)}\in \Omega^{r}(\mathbb{L}^d)\qquad\text{s.t.}d_{r-1}\omega^{(r-1,0)}=\delta_{-1}(\omega^{(r,-1)}).
\end{equation}
We sometimes simply write $d\bm{\omega}$ instead of $\omega^{(r,-1)}$.
\subsubsection{Properties of $\omega^{(-1,r)}$}\label{sec522}
Since $\delta\omega^{(0,r-1)}+(-1)^rd_{-1}\omega^{(-1,r)}=0$ holds for any DB $r$-cocycle $\bm{\omega}$, we find
\begin{equation}\label{equ5060}
    d_{-1}\delta_{r}\omega^{(-1,r)}=0,
\end{equation}
and we can see $\delta_r\omega^{(-1,r)}=0$ from the definition of $d_{-1}$.\footnote{For degrees other than $d_{-1}$, the condition $d\omega=0$ does not necessarily imply $\omega=0$, and thus such an argument cannot be applied.} Therefore, $[\omega^{(-1,r)}]$ can be considered to belong to \v{C}ech $r$-cohomology, and we can later observe that $[\omega^{(-1,r)}]$ and $[\omega^{(r,-1)}]$ are equivalent under the identification by isomorphism 
\begin{equation}\label{equ5070}
    (\text{\v{C}ech $r$-cohomology on $\mathbb{L}^3$})\cong H^r_\text{de Rham}(\mathbb{L}^3).
\end{equation}

The lattice DB coboundary operator $D_r^\mathrm{ev}:C^{r}_\mathrm{lat.DB}\to C^{r}_\mathrm{ev}$ defined in \eqref{equ4220}, has a $\delta_{r}$ term. Although the condition $D_{r}^\mathrm{ev}\bm{\omega}=0$ implies $\delta_{r}\omega^{(-1,r)}=0$ even when the $\delta_{r}$ term is omitted, we keep the $\delta_r$ term in $D^\mathrm{ev}_r$ because it leads to a more convenient definition of the star product introduced in Section~\ref{sec660}.
\newpage
\section{Products of cochain}\label{sec600}
\subsection{\v{C}ech cup product}\label{sec610}
\v{C}ech cup product of two \v{C}ech--de Rham cochains $\omega_1^{(-1,j_1)},\omega_2^{(-1,j_2)}$ is written by
\begin{equation}\label{equ6010}
    \omega_1^{(-1,j_1)}\smallsmile\omega_2^{(-1,j_2)}=(\omega_1\smallsmile\omega_2)^{(-1,j_1+j_2)}.
\end{equation}

In the following, we denote $\omega^{(-1,j)}(;;\tilde{x}^{(0)},\dotsc,\tilde{x}^{(j)})$ by $\omega(0,\dotsc,j)$ as abbreviation. Under this notation, let the product $\omega_1\smallsmile\omega_2$ be defined by
\begin{equation}\label{equ6020}
    (\omega_1\smallsmile\omega_2)(0,\dotsc,j_1+j_2)=\mathrm{sgn}(\sigma)\omega_1(\sigma(0),\dotsc,\sigma(j_1))\omega_2(\sigma(j_1),\dotsc,\sigma(j_1+j_2)).
\end{equation}
Note that $\sigma\in\mathfrak{S}_{j_1+j_2+1}$ is a unique permutation that satisfies $\tilde{x}^{(\sigma(0))}\prec \tilde{x}^{(\sigma(1))}\prec\cdots$, where $\prec$ denotes the total ordering\footnote{The total ordering of simplicial complex, sometimes called the branching structure, is defined by arbitrarily choosing and fixing the total order of $\tilde{\mathbb{L}}^d$. } of patch indices. It should be emphasized that $\omega_1\smallsmile\omega_2$ is defined antisymmetrically with respect to the patch indices.
\subsection{\v{C}ech cup-$1$ product}\label{sec620}
The \v{C}ech cup product defined in Section~\ref{sec610} is graded non-commutative at the cochain level. The non-commutativity is expressed by the \v{C}ech cup-$1$ product~\cite{Steenrod:1947}. In the following, we will use the same abbreviation as in Section~\ref{sec610}.

For $0\leq j_1\;,\;0\leq j_2\;,\;1\leq j_1+j_2$, \v{C}ech cup-$1$ product of $\omega_1^{(-1,j_1)}$ and $\omega_2^{(-1,j_2)}$ is defined by
\begin{equation}\label{equ6030}
    \begin{split}
        &(\omega_1\smallsmile_1\omega_2)(0,\dotsc,j_1+j_2-1)\\
        =&\mathrm{sgn}(\sigma)\sum_{0\leq k\leq j_1-1}\omega_1(\sigma(0),\dotsc,\sigma(k),\sigma(k+j_2),\dotsc,\sigma(j_1+j_2-1))\omega_2(\sigma(k),\dotsc,\sigma(k+j_2)),
    \end{split}
\end{equation}
where $\sigma\in\mathfrak{S}_{j_1+j_2}$ is the same as in Section~\ref{sec610}.

The coboundary formula, which is an important formula of the cup-$1$ product
\begin{equation}\label{equ6040}
    \begin{split}
        &\omega_1^{(-1,j_1)}\smallsmile\omega_2^{(-1,j_2)}-(-1)^{j_1j_2}\omega_2^{(-1,j_2)}\smallsmile\omega_1^{(-1,j_1)}\\
        =&\delta(\omega_1^{(-1,j_1)}\smallsmile_1\omega_2^{(-1,j_2)})+(\delta\omega_1^{(-1,j_1)})\smallsmile_1\omega_2^{(-1,j_2)}+(-1)^{j_1}\omega_1^{(-1,j_1)}\smallsmile_1(\delta\omega_2^{(-1,j_2)})
    \end{split}
\end{equation}
is known~\cite{Mosher:2008}.
\subsection{Lattice wedge product}\label{sec630}
Lattice wedge product of two \v{C}ech--de Rham cochains $\omega_1^{(i_1,-1)},\omega_2^{(i_2,-1)}$ is written by
\begin{equation}\label{equ6050}
    \omega_1^{(i_1,-1)}\wedge \omega_2^{(i_2,-1)}=(\omega_1\wedge\omega_2)^{(i_1+i_2,-1)}.
\end{equation}

In the following, we denote $\omega^{(i,-1)}(x;\mu_1,\dots,\mu_i;)$ by $\omega(x;1,\dots,i)$ as abbreviation. Under this notation, let the product $\omega_1\wedge\omega_2$ be defined by
\begin{equation}\label{equ6060}
    \begin{split}
        &(j_1!j_2!)(\omega_1\wedge\omega_2)(x;1,\dotsc,i_1+i_2)\\
        =&\sum_{\sigma\in\mathfrak{S}_{j_1+j_2}}\mathrm{sgn}(\sigma)\omega_1(x;\sigma(1),\dotsc,\sigma(j_1))\omega_2(x+\hat{\sigma}(1)+\cdots+\hat{\sigma}(j_1);\sigma(j_1+1),\dotsc,\sigma(j_1+j_2)).
    \end{split}
\end{equation}
Note that $\omega_1\wedge\omega_2$ is defined antisymmetrically with respect to the indices $1,\dotsc,i_1+i_2$.
\subsection{\v{C}ech--de Rham product}\label{sec650}
Using $\smallsmile$ and $\wedge$, we define products of \v{C}ech--de Rham chain $\stackrel{\smallsmile}{\wedge}$.
\subsubsection{Definition of $\stackrel{\smallsmile}{\wedge}$}\label{sec651}
For two \v{C}ech--de Rham cochains
\begin{align}\label{equ6080}
    &\omega^{(i,j)}(x\;;\;\mu_1,\dots,\mu_i\;;\;\tilde{x}^{(0)},\dots,\tilde{x}^{(j)})&
    &  \eta^{(k,l)}(x\;;\;\nu_1,\dots,\nu_k\;;\;\tilde{y}^{(0)},\dots,\tilde{y}^{(l)}),
\end{align}
let
\begin{equation}\label{equ6090}
    \omega\stackrel{\smallsmile}{\wedge}\;\eta\in C^{(i+k,j+l)}(\mathbb{L}^d)
\end{equation}
be
\begin{align}
    \begin{split}\label{equ6100}
        &(\omega\stackrel{\smallsmile}{\wedge}\,\eta)^{(i+k,j+l)}(x\;;\;\mu_1,\dots,\mu_{i+k}\;;\;\tilde{x}^{(0)},\dots,\tilde{x}^{(j+l)})\\
        :=&\frac{1}{i!k!}\sum_{\tau\in\mathfrak{S}_{i+k}}\mathrm{sgn}(\tau)\omega(x\;;\;\mu_{\tau(1)},\dots,\mu_{\tau(i)}\;;\;\tilde{x}^{\sigma(0)},\dots,\tilde{x}^{\sigma(j)})\\
        &\eta(x+\hat{\mu}_{\tau(1)}+\dots+\hat{\mu}_{\tau(i)}\;;\;\mu_{\tau(i+1)},\dots,\mu_{\tau(i+k)}\;;\;\tilde{x}^{\sigma(j)},\dots,\tilde{x}^{\sigma(j+l)})\\
        =&\frac{1}{i!k!}\sum_{\tau\in\mathfrak{S}_{i+k}}\mathrm{sgn}(\tau)\Biggl.\Bigl(\omega(x\;;\;\mu_{\tau(1)},\dots,\mu_{\tau(i)}\;;\;-)\\
        &\smallsmile\eta(x+\hat{\mu}_{\tau(1)}+\dots+\hat{\mu}_{\tau(i)}\;;\;\mu_{\tau(i+1)},\dots,\mu_{\tau(i+k)}\;;\;-)\Bigr)\Biggr|_{(\tilde{x}^{(0)},\dots,\tilde{x}^{(j+l-1)})}\\
        =&\mathrm{sgn}(\sigma)\Biggl.\Bigl(\omega(-\;;\;-\;;\;\tilde{x}^{\sigma(0)},\dots,\tilde{x}^{\sigma(j)})\wedge\eta(-\;;\;-\;;\;\tilde{x}^{\sigma(j)},\dots,\tilde{x}^{\sigma(j+l)})\Bigr)\Biggr|_{(x\;;\;\mu_1,\dots,\mu_{i+k-1})},
    \end{split}
\end{align}
where $\omega^{(i,j)}(-;-;\bullet)$ means to regard $\omega^{(i,j)}$ as $\omega^{(i,-1)}$ substituting only for the $\bullet$ part. Similarly, $\omega^{(i,j)}(\bullet;\bullet;-)$ means to regard $\omega^{(i,j)}$ as $\omega^{(-1,j)}$ substituting only for $\bullet$ part.
\subsubsection{Leibniz rule of $\stackrel{\smallsmile}{\wedge}$}\label{sec652}
$\stackrel{\smallsmile}{\wedge}$ is defined as a composite operation of $\smallsmile$, $\wedge$, and substituting. Therefore, Leibniz rule of $\delta,d$ can also be applied to $\stackrel{\smallsmile}{\wedge},$ then we obtain
\begin{align}
    &\delta(\omega^{(i,j)}\stackrel{\smallsmile}{\wedge}\eta^{(k,l)})=(\delta\omega^{(i,j)})\stackrel{\smallsmile}{\wedge}\eta^{(k,l)}+(-1)^j\omega^{(i,j)}\stackrel{\smallsmile}{\wedge}(\delta\eta^{(k,l)}),\label{equ6130}\\
    &d(\omega^{(i,j)}\stackrel{\smallsmile}{\wedge}\eta^{(k,l)})=(d\omega^{(i,j)})\stackrel{\smallsmile}{\wedge}\eta^{(k,l)}+(-1)^i\omega^{(i,j)}\stackrel{\smallsmile}{\wedge}(d\eta^{(k,l)}).\label{equ6140}
\end{align}
\subsection{Definition of the lattice DB star product}\label{sec660}
For integer $a,b$ satisfies $0\leq a\leq b\leq r$, we define the direct sum of \v{C}ech--de Rham cochain $C^{r}_{[a,b]}(\mathbb{L}^d)$ by
\begin{equation}\label{equ6180}
    C^{r}_{[a,b]}(\mathbb{L}^d):=\bigoplus_{0\leq a\leq k\leq b\leq r}C^{(r-k-1,k)}(\mathbb{L}^d).
\end{equation}
The relationship with the notation we have used is
\begin{align}\label{equ6190}
&C^r_\mathrm{lat.DB}(\mathbb{L}^d)=C^{r}_{[0,r]}(\mathbb{L}^d),&&C^r_\mathrm{ev}(\mathbb{L}^d)=C^{r}_{[1,r]}(\mathbb{L}^d).
\end{align}
Next, we will give the definition of $\bm{\omega}\star\bm{\eta}\in C_{[s+u,t+v]}^{p+q}(\mathbb{L}^d)$ for
\begin{equation}\label{equ6200}
    \begin{split}
        \bm{\omega}=(\omega^{(p-s-1,s)},\omega^{(p-s-2,s+1)},\dots,\omega^{(p-t,t-1)},\omega^{(p-t-1,t)})\in C^{p}_{[s,t]}(\mathbb{L}^d),\\
        \bm{\eta}=(\eta^{(q-u-1,u)},\eta^{(q-u-2,u+1)},\dots,\eta^{(q-v,v-1)},\eta^{(q-v-1,v)})\in C^{q}_{[u,v]}(\mathbb{L}^d)
    \end{split}
\end{equation}
by
\begin{equation}\label{equ6210}
    \begin{split}
        &\bm{\omega}\star\bm{\eta}=\\&
        \begin{array}{rrlcrll}
        (&&\omega^{(p-s-1,s)}&\stackrel{\smallsmile}{\wedge}&d_{q-u-1}&\eta^{(q-u-1,u)}&,\\
        &&\omega^{(p-s-2,s+1)}&\stackrel{\smallsmile}{\wedge}&d_{q-u-1}&\eta^{(q-u-1,u)}&,\\
        &&\vdots&\vdots&&\vdots&\\
        &&\omega^{(p-t,t-1)}&\stackrel{\smallsmile}{\wedge}&d_{q-u-1}&\eta^{(q-u-1,u)}&,\\
        &d_{p-t-1}&\omega^{(p-t-1,t)}&\stackrel{\smallsmile}{\wedge}&&\eta^{(q-u-1,u)}&,\\
        &d_{p-t-1}&\omega^{(p-t-1,t)}&\stackrel{\smallsmile}{\wedge}&&\eta^{(q-u-2,u+1)}&,\\
        &&\vdots&\vdots&&\vdots&\\
        &d_{p-t-1}&\omega^{(p-t-1,t)}&\stackrel{\smallsmile}{\wedge}&&\eta^{(q-v,v-1)}&,\\
        &d_{p-t-1}&\omega^{(p-t-1,t)}&\stackrel{\smallsmile}{\wedge}&&\eta^{(q-v-1,v)}&\;).
        \end{array}
    \end{split}
\end{equation}
From the definition, we can see
\begin{equation}\label{equ6220}
\begin{array}{l@{\hspace{2pt}}c@{\hspace{2pt}}lcccccl}
    C^a_\mathrm{lat.DB}(\mathbb{L}^d)&\star& C^b_\mathrm{lat.DB}(\mathbb{L}^d)&=&C^a_{[0,a]}(\mathbb{L}^d)\star C^b_{[0,b]}(\mathbb{L}^d)&=&C^{a+b}_{[0,a+b]}(\mathbb{L}^d)&=&C^{a+b}_\mathrm{lat.DB}(\mathbb{L}^d),\\
    C^a_\mathrm{ev}(\mathbb{L}^d)&\star& C^b_\mathrm{lat.DB}(\mathbb{L}^d)&=&C^a_{[1,a]}(\mathbb{L}^d)\star C^b_{[0,b]}(\mathbb{L}^d)&=&C^{a+b}_{[1,a+b]}(\mathbb{L}^d)&=&C^{a+b}_\mathrm{ev}(\mathbb{L}^d),\\
    C^a_\mathrm{lat.DB}(\mathbb{L}^d)&\star& C^b_\mathrm{ev}(\mathbb{L}^d)&=&C^a_{[0,a]}(\mathbb{L}^d)\star C^b_{[1,b]}(\mathbb{L}^d)&=&C^{a+b}_{[1,a+b]}(\mathbb{L}^d)&=&C^{a+b}_\mathrm{ev}(\mathbb{L}^d).
\end{array}
\end{equation}

Below, we will introduce the properties of $\star$ without proof, which will be provided in Section~\ref{sec670}. $\star$ satisfies
\begin{itemize}
    \item $\forall\bm{\omega},\forall\bm{\omega}'\in C^a_\mathrm{lat.DB}(\mathbb{L}^d),\forall\bm{\eta}\in C^b_\mathrm{lat.DB}(\mathbb{L}^d)\;\;,\;(\bm{\omega}+\bm{\omega}')\star\bm{\eta}=\bm{\omega}\star\bm{\eta}+\bm{\omega}'\star\bm{\eta}$
    \item $\forall\bm{\omega}\in C^a_\mathrm{lat.DB}(\mathbb{L}^d),\forall\bm{\eta}\in C^b_\mathrm{lat.DB}(\mathbb{L}^d)\;\;,\;D_{a+b}(\bm{\omega}\star\bm{\eta})=(D_a\bm{\omega})\star\bm{\eta}+(-1)^{a}\bm{\omega}\star(D_b\bm{\eta})$
\end{itemize}
on DB cochain, and satisfies
\begin{itemize}
    \item $\forall\bm{\omega}\in Z^a_\mathrm{lat.DB}(\mathbb{L}^d),\forall\bm{\eta}\in Z^b_\mathrm{lat.DB}(\mathbb{L}^d)\;\;,\;\bm{\omega}\star\bm{\eta}\in Z^{a+b}_\mathrm{lat.DB}(\mathbb{L}^d)$
    \item $\forall\bm{\omega},\forall\bm{\omega}'\in Z^a_\mathrm{lat.DB}(\mathbb{L}^d),\forall\bm{\eta}\in Z^b_\mathrm{lat.DB}(\mathbb{L}^d)\;\;,\;[\bm{\omega}]=[\bm{\omega}']\Longrightarrow [\bm{\omega}\star\bm{\eta}]=[\bm{\omega}'\star\bm{\eta}]$
\end{itemize}
on DB cocycle.

From the above properties, we can see that $\star$ is formulated as a product on the lattice DB cohomology; however, it is not graded commutative.
\subsection{Proofs of the properties of $\star$}\label{sec670}
Below, we consider the case
\begin{equation}\label{equ6230}
    \begin{split}
        \bm{\omega}=(\omega^{(a-1,0)},\omega^{(a-2,1)},\dots,\omega^{(0,a-1)},\omega^{(-1,a)},)\in C^a_\mathrm{lat.DB}(\mathbb{L}^d)\\
        \bm{\eta}=(\eta^{(b-1,0)},\eta^{(b-2,1)},\dots,\eta^{(0,b-1)},\eta^{(-1,b)})\in C^b_\mathrm{lat.DB}(\mathbb{L}^d).
    \end{split}
\end{equation}
\subsubsection{Proof of $(\boldsymbol{\omega}+\boldsymbol{\omega}')\star\boldsymbol{\eta}=\boldsymbol{\omega}\star\boldsymbol{\eta}+\boldsymbol{\omega}'\star\boldsymbol{\eta}$}\label{sec671}
It is obvious from the linearity of $d,\smallsmile,\wedge$.\newpage
\subsubsection{Proof of $D_{a+b}^\mathrm{ev}(\boldsymbol{\omega}\star\boldsymbol{\eta})=(D_a^\mathrm{ev}\boldsymbol{\omega})\star\boldsymbol{\eta}+(-1)^{a}\boldsymbol{\omega}\star(D_b^\mathrm{ev}\boldsymbol{\eta})$}\label{sec672}
This is proved by the following direct calculation:\footnote{The sizes of parentheses in DB cochain expressions are adjusted for typographical clarity and carry no additional meaning.}
\begin{equation}
    (D_a^\mathrm{ev}\bm{\omega})\star\bm{\eta}=\left(\begin{array}{rrl}
        (\delta_{0}\omega^{(a-1,0)}+(-1)^{1}d_{a-2}\omega^{(a-2,1)})&,\\
        \vdots&,\\
        (\delta_{a-2}\omega^{(1,a-2)}+(-1)^{a-1}d_{0}\omega^{(0,a-1)})&,\\
        (\delta_{a-1}\omega^{(0,a-1)}+(-1)^{a}d_{-1}\omega^{(-1,a)})&,\\
        d_{-1}\delta_{a}\omega^{(-1,a)}&
    \end{array}\right)\star\left(\begin{array}{rll}
        \eta^{(b-1,0)}&,\\
        \eta^{(b-2,1)}&,\\
        \vdots\;\;\;\;&,\\
        \eta^{(0,b-1)}&,\\
        \eta^{(-1,b)}&
    \end{array}\right)\notag
\end{equation}
\begin{equation}
    \begin{split}
    \begin{array}{rrcr}\label{equ6240}
        =\Bigl(&(\delta_{0}\omega^{(a-1,0)}+(-1)^{1}d_{a-2}\omega^{(a-2,1)})&\stackrel{\smallsmile}{\wedge}&d_{b-1}\eta^{(b-1,0)},\\
        &\vdots&&,\\
        &(\delta_{a-2}\omega^{(1,a-2)}+(-1)^{a-1}d_{0}\omega^{(0,a-1)})&\stackrel{\smallsmile}{\wedge}&d_{b-1}\eta^{(b-1,0)},\\
        &(\delta_{a-1}\omega^{(0,a-1)}+(-1)^{a}d_{-1}\omega^{(-1,a)})&\stackrel{\smallsmile}{\wedge}&d_{b-1}\eta^{(b-1,0)},\\
        &d_{-1}\delta_{a}\omega^{(-1,a)}&\stackrel{\smallsmile}{\wedge}&\eta^{(b-1,0)},\\
        &\vdots&&,\\
        &d_{-1}\delta_{a}\omega^{(-1,a)}&\stackrel{\smallsmile}{\wedge}&\eta^{(0,b-1)},\\
        &d_{-1}\delta_{a}\omega^{(-1,a)}&\stackrel{\smallsmile}{\wedge}&\eta^{(-1,b)}\Bigr),
    \end{array}
    \end{split}
\end{equation}
\begin{equation}
    \begin{split}
    \bm{\omega}\star(D_b^\mathrm{ev}\bm{\eta})=\left(\begin{array}{rll}
        \omega^{(a-1,0)}&,\\
        \omega^{(a-2,1)}&,\\
        \vdots&,\\
        \omega^{(0,a-1)}&,\\
        \omega^{(-1,a)}&
    \end{array}\right)\star\left(\begin{array}{rrl}
        (\delta_{0}\eta^{(b-1,0)}+(-1)^{1}d_{b-2}\eta^{(b-2,1)})&,\\
        \vdots&,\\
        (\delta_{b-2}\eta^{(1,b-2)}+(-1)^{a-1}d_{0}\eta^{(0,b-1)})&,\\
        (\delta_{b-1}\eta^{(0,b-1)}+(-1)^{a}d_{-1}\eta^{(-1,b)})&,\\
        d_{-1}\delta_{b}\eta^{(-1,b)}&
    \end{array}\right)\\
    \begin{array}{rrcr}
        =\Bigl(&\omega^{(a-1,0)}&\stackrel{\smallsmile}{\wedge}&d_{b-1}\delta_0\eta^{(b-1,0)},\\
        &\omega^{(a-2,1)}&\stackrel{\smallsmile}{\wedge}&d_{b-1}\delta_0\eta^{(b-1,0)},\\
        &\vdots&&,\\
        &\omega^{(0,a-1)}&\stackrel{\smallsmile}{\wedge}&d_{b-1}\delta_0\eta^{(b-1,0)},\\
        &d_{-1}\omega^{(-1,a)}&\stackrel{\smallsmile}{\wedge}&(\delta_{0}\eta^{(b-1,0)}+(-1)^{1}d_{b-2}\eta^{(b-2,1)}),\\
        &d_{-1}\omega^{(-1,a)}&\stackrel{\smallsmile}{\wedge}&(\delta_{0}\eta^{(b-2,1)}+(-1)^{2}d_{b-3}\eta^{(b-3,2)}),\\
        &\vdots&&,\\
        &d_{-1}\omega^{(-1,a)}&\stackrel{\smallsmile}{\wedge}&(\delta_{b-1}\eta^{(0,b-1)}+(-1)^{b}d_{-1}\eta^{(-1,b)}),\\
        &d_{-1}\omega^{(-1,a)}&\stackrel{\smallsmile}{\wedge}&\delta_{b}\eta^{(-1,b)}\Bigr),
    \end{array}
    \end{split}
\end{equation}
\begin{equation}
    \begin{array}{rrcr}
        D_{a+b}^\mathrm{ev}(\bm{\omega}\star\bm{\eta})=\Bigl(&\qquad\quad\delta_{0}(\omega^{(a-1,0)}&\stackrel{\smallsmile}{\wedge}&d_{b-1}\eta^{(b-1,0)})\\+&(-1)^{1}d_{a+b-2}(\omega^{(a-2,1)}&\stackrel{\smallsmile}{\wedge}&d_{b-1}\eta^{(b-1,0)}),\\
        &&\vdots&,\\
        &\delta_{a-2}(\omega^{(1,a-2)}&\stackrel{\smallsmile}{\wedge}&d_{b-1}\eta^{(b-1,0)})\\+&(-1)^{a-1}d_{b}(\omega^{(0,a-1)}&\stackrel{\smallsmile}{\wedge}&d_{b-1}\eta^{(b-1,0)}),\\
        &\delta_{a-1}(\omega^{(0,a-1)}&\stackrel{\smallsmile}{\wedge}&d_{b-1}\eta^{(b-1,0)})\\+&(-1)^{a}d_{b-1}(d_{-1}\omega^{(-1,a)}&\stackrel{\smallsmile}{\wedge}&\eta^{(b-1,0)}),\\
        &\delta_{a}(d_{-1}\omega^{(-1,a)}&\stackrel{\smallsmile}{\wedge}&\eta^{(b-1,0)})\\+&(-1)^{a+1}d_{b-2}(d_{-1}\omega^{(-1,a)}&\stackrel{\smallsmile}{\wedge}&\eta^{(b-2,1)}),\\
        &&\vdots&,\\
        &\delta_{a+b-1}(d_{-1}\omega^{(-1,a)}&\stackrel{\smallsmile}{\wedge}&\eta^{(0,b-1)})\\+&(-1)^{a+b}d_{-1}(d_{-1}\omega^{(-1,a)}&\stackrel{\smallsmile}{\wedge}&\eta^{(-1,b)}),\\
        &\delta_{a+b}(d_{-1}\omega^{(-1,a)}&\stackrel{\smallsmile}{\wedge}&\eta^{(-1,b)})\Bigr)
    \end{array}\notag
\end{equation}
\begin{equation}\label{equ6270}
    \begin{array}{rcr}
        =\Bigl((\delta_{0}\omega^{(a-1,0)}+(-1)^{1}d_{a-2}\omega^{(a-2,1)})&\stackrel{\smallsmile}{\wedge}&d_{b-1}\eta^{(b-1,0)}\\+(-1)^{0}\omega^{(a-2,1)}&\stackrel{\smallsmile}{\wedge}&\delta_{0}d_{b-1}\eta^{(b-1,0)},\\
        &\vdots&,\\
        (\delta_{a-2}\omega^{(1,a-2)}+(-1)^{a-1}d_{0}\omega^{(0,a-1)})&\stackrel{\smallsmile}{\wedge}&d_{b-1}\eta^{(b-1,0)}\\+(-1)^{a-2}\omega^{(1,a-2)}&\stackrel{\smallsmile}{\wedge}&\delta_{0}d_{b-1}\eta^{(b-1,0)},\\
        (\delta_{a-1}\omega^{(0,a-1)}+(-1)^ad_{-1}\omega^{(-1,a)})&\stackrel{\smallsmile}{\wedge}&d_{b-1}\eta^{(b-1,0)}\\+(-1)^{a-1}\omega^{(0,a-1)}&\stackrel{\smallsmile}{\wedge}&\delta_{0}d_{b-1}\eta^{(b-1,0)},\\
        \delta_{a}d_{-1}\omega^{(-1,a)}&\stackrel{\smallsmile}{\wedge}&\eta^{(b-1,0)}\\+d_{-1}\omega^{(-1,a)}&\stackrel{\smallsmile}{\wedge}&(-1)^{a}(\delta_{0}\eta^{(b-1,0)}+(-1)^{1}d_{b-2}\eta^{(b-2,1)}),\\
        &\vdots&,\\
        \delta_{a}d_{-1}\omega^{(-1,a)}&\stackrel{\smallsmile}{\wedge}&\eta^{(0,b-1)}\\+d_{-1}\omega^{(-1,a)}&\stackrel{\smallsmile}{\wedge}&(-1)^{a}(\delta_{b-1}\eta^{(0,b-1)}+(-1)^{b}d_{-1}\eta^{(-1,b)}),\\
        \delta_{a}d_{-1}\omega^{(-1,a)}&\stackrel{\smallsmile}{\wedge}&\eta^{(-1,b)}\\+d_{-1}\omega^{(-1,a)}&\stackrel{\smallsmile}{\wedge}&(-1)^a\delta_{b}\eta^{(-1,b)}\Bigr)\\
        =(D_a^\mathrm{ev}\bm{\omega})\star\bm{\eta}+(-1)^a\bm{\omega}\star(D_b^\mathrm{ev}\bm{\eta}).&&
    \end{array}
\end{equation}
\newpage
\subsubsection{Proof of $\forall\boldsymbol{\omega}\in Z_\mathrm{lat.DB}\;,\;\boldsymbol{\eta}\in Z_\mathrm{lat.DB}\Longrightarrow \boldsymbol{\omega}\star\boldsymbol{\eta}\in Z_\mathrm{lat.DB}$}\label{sec673}
Recall Section~\ref{sec672}. In the case $\bm{\omega}\in Z_\mathrm{lat.DB}\;,\;\bm{\eta}\in Z_\mathrm{lat.DB}$, the RHS of
\begin{equation}\label{equ6280}
    D^\mathrm{ev}(\bm{\omega}\star\bm{\eta})=(D^\mathrm{ev}\bm{\omega})\star\bm{\eta}\pm\bm{\omega}\star(D^\mathrm{ev}\bm{\eta})
\end{equation}
equals to $0$. Then $\bm{\omega}\star\bm{\eta}\in Z_\mathrm{lat.DB}$ is concluded.
\subsubsection{Proof of $\forall\boldsymbol{\omega}\in Z^a_\mathrm{lat.DB},\;\forall\boldsymbol{\omega}'\in Z^a_\mathrm{lat.DB},\;\forall\boldsymbol{\eta}\in Z^b_\mathrm{lat.DB}$,\;\;$[\boldsymbol{\omega}]=[\boldsymbol{\omega}']\Longrightarrow [\boldsymbol{\omega}\star\boldsymbol{\eta}]=[\boldsymbol{\omega}'\star\boldsymbol{\eta}]$}\label{sec674}
By definition,
\begin{equation}\label{equ6290}
    [\bm{\omega}]=[\bm{\omega}']\Longleftrightarrow \exists\bm{\xi}\in C^{a-1}_\mathrm{lat.DB}\;\mathrm{s.t.}\;D_{a}\bm{\xi}=\bm{\omega}-\bm{\omega}'
\end{equation}
holds. $\bm{\xi}$ in this formula holds
\begin{equation}\label{equ6300}
    \begin{split}
        &D_{a+b}(\bm{\xi}\star\bm{\eta})=(d_{a+b-1}+D^\mathrm{ev}_{a+b-1})(\bm{\xi}\star\bm{\eta})\\
        =&d_{a-2}\xi^{(a-2,0)}\stackrel{\smallsmile}{\wedge}d_{b-1}\eta^{(b-1,0)}+(D_{a-1}^\mathrm{ev}\bm{\xi})\star\bm{\eta}\pm\bm{\eta}\star(D_b^\mathrm{ev}\bm{\eta})\\
        =&(D_a\bm{\xi})\star\bm{\eta}\pm\bm{\eta}\star(D_b^\mathrm{ev}\bm{\eta})=\bm{\omega}\star\bm{\eta}-\bm{\omega}'\star\bm{\eta}.
    \end{split}
\end{equation}
Therefore, $[\bm{\omega}\star\bm{\eta}]=[\bm{\omega}'\star\bm{\eta}]$ holds.
\subsection{Two equivalent definitions of lattice DB star product}\label{sec680}
The first component of the definition of $\star$ discussed so far has the form $AdB$, but for future discussions, we would also like to define a product whose first component has the form $dAB$. We denote each different operations by $\bm{\omega}\overrightarrow{\underleftarrow{\star}}\bm{\eta},\bm{\omega}\overleftarrow{\underrightarrow{\star}}\bm{\eta}$. As we will see shortly, we can see $[\bm{\omega}\overrightarrow{\underleftarrow{\star}}\bm{\eta}]=[\bm{\omega}\overleftarrow{\underrightarrow{\star}}\bm{\eta}]$ at least the case of degree $2$, therefore these definitions are gauge equivalent within the framework of lattice DB $2$-cohomology.

We introduce notations $\overrightarrow{\underleftarrow{\star}}=\star,$ and $\overleftarrow{\underrightarrow{\star}}$ by
\begin{align}\label{equ6310}&
    \begin{split}
        &\bm{\omega}\overrightarrow{\underleftarrow{\star}}\bm{\eta}=\bm{\omega}\star\bm{\eta}=\\&
        \begin{array}{r@{\hspace{2pt}}r@{\hspace{2pt}}l@{\hspace{2pt}}c@{\hspace{2pt}}r@{\hspace{2pt}}l@{\hspace{2pt}}l}
        (&&\omega^{(a-1,0)}&\stackrel{\smallsmile}{\wedge}&d_{b-1}&\eta^{(a-1,0)}&,\\
        &&\omega^{(a-2,1)}&\stackrel{\smallsmile}{\wedge}&d_{b-1}&\eta^{(a-1,0)}&,\\
        &&\vdots&\vdots&&\vdots&\\
        &&\omega^{(0,a-1)}&\stackrel{\smallsmile}{\wedge}&d_{b-1}&\eta^{(a-1,0)}&,\\
        &d_{-1}&\omega^{(-1,a)}&\stackrel{\smallsmile}{\wedge}&&\eta^{(a-1,0)}&,\\
        &d_{-1}&\omega^{(-1,a)}&\stackrel{\smallsmile}{\wedge}&&\eta^{(a-2,1)}&,\\
        &&\vdots&\vdots&&\vdots&\\
        &d_{-1}&\omega^{(-1,a)}&\stackrel{\smallsmile}{\wedge}&&\eta^{(0,a-1)}&,\\
        &d_{-1}&\omega^{(-1,a)}&\stackrel{\smallsmile}{\wedge}&&\eta^{(-1,a)}&\;),
        \end{array}
    \end{split}&&
    \begin{split}
        &\bm{\omega}\overleftarrow{\underrightarrow{\star}}\bm{\eta}=\\&
        \begin{array}{r@{\hspace{2pt}}l@{\hspace{2pt}}l@{\hspace{2pt}}l@{\hspace{2pt}}c@{\hspace{2pt}}r@{\hspace{2pt}}l@{\hspace{2pt}}l}
        (&(-1)^{a}&d_{a-1}&\omega^{(a-1,0)}&\stackrel{\smallsmile}{\wedge}&&\eta^{(a-1,0)}&,\\
        &(-1)^{2a}&d_{a-1}&\omega^{(a-1,0)}&\stackrel{\smallsmile}{\wedge}&&\eta^{(a-2,1)}&,\\
        &\;\;\;\vdots&&\vdots&\vdots&&\vdots&\\
        &(-1)^{a^2}&d_{a-1}&\omega^{(a-1,0)}&\stackrel{\smallsmile}{\wedge}&&\eta^{(0,a-1)}&,\\
        &(-1)^{a^2}&&\omega^{(a-1,0)}&\stackrel{\smallsmile}{\wedge}&d_{-1}&\eta^{(-1,a)}&,\\
        &(-1)^{(a-1)a}&&\omega^{(a-2,1)}&\stackrel{\smallsmile}{\wedge}&d_{-1}&\eta^{(-1,a)}&,\\
        &&&\vdots&\vdots&&\vdots&\\
        &(-1)^{a}&&\omega^{(0,a-1)}&\stackrel{\smallsmile}{\wedge}&d_{-1}&\eta^{(-1,a)}&,\\
        &(-1)^{0}&&\omega^{(-1,a)}&\stackrel{\smallsmile}{\wedge}&d_{-1}&\eta^{(-1,a)}&\;).
        \end{array}
    \end{split}
\end{align}
In the case of $a=2$,
\begin{equation}\label{equ6320}
    \begin{split}
        &(A,\Lambda,n)\overrightarrow{\underleftarrow{\star}}(B,\Pi,m)-(A,\Lambda,n)\overleftarrow{\underrightarrow{\star}}(B,\Pi,m)\\
        =&\left(\begin{array}{c}
            A\stackrel{\smallsmile}{\wedge}dB-(-1)^2dA\stackrel{\smallsmile}{\wedge}B,\\
            \Lambda \stackrel{\smallsmile}{\wedge}dB-(-1)^4dA\stackrel{\smallsmile}{\wedge}\Pi,\\
            dn\stackrel{\smallsmile}{\wedge}B-(-1)^4A\stackrel{\smallsmile}{\wedge}dm,\\
            dn\stackrel{\smallsmile}{\wedge}\Pi-(-1)^2\Lambda\stackrel{\smallsmile}{\wedge} dm,\\
            dn\stackrel{\smallsmile}{\wedge}m-(-1)^0n\stackrel{\smallsmile}{\wedge}dm
        \end{array}\right)=
        \left(\begin{array}{c}
            A\stackrel{\smallsmile}{\wedge}dB-dA\stackrel{\smallsmile}{\wedge}B,\\
            \Lambda \stackrel{\smallsmile}{\wedge}dB-dA\stackrel{\smallsmile}{\wedge}\Pi,\\
            dn\stackrel{\smallsmile}{\wedge}B-A\stackrel{\smallsmile}{\wedge}dm,\\
            dn\stackrel{\smallsmile}{\wedge}\Pi-\Lambda\stackrel{\smallsmile}{\wedge} dm,\\
            dn\stackrel{\smallsmile}{\wedge}m-n\stackrel{\smallsmile}{\wedge}dm
        \end{array}\right)
    \end{split}
\end{equation}
\begin{equation}\label{equ6330}
    \begin{split}
        =&\left(\begin{array}{c}
            A\stackrel{\smallsmile}{\wedge}dB-dA\stackrel{\smallsmile}{\wedge}B,\\
            \Lambda \stackrel{\smallsmile}{\wedge}dB-dA\stackrel{\smallsmile}{\wedge}\Pi,\\
            dn\stackrel{\smallsmile}{\wedge}B-A\stackrel{\smallsmile}{\wedge}dm,\\
            dn\stackrel{\smallsmile}{\wedge}\Pi-\Lambda\stackrel{\smallsmile}{\wedge} dm,\\
            0
        \end{array}\right)
        =D_4\left(\begin{array}{c}
            -A\stackrel{\smallsmile}{\wedge}B,\\
            -\Lambda\stackrel{\smallsmile}{\wedge} B+A\stackrel{\smallsmile}{\wedge}\Pi,\\
            -\Lambda\stackrel{\smallsmile}{\wedge}\Pi,\\
            0
        \end{array}\right)
    \end{split}
\end{equation}
holds. The important thing here is that the fourth component of the RHS of the equation is $0$. As we will see later in \eqref{equ7260}, the integral of DB cohomology has a modulo $\mathbb{Z}$ ambiguity depending on the DB boundary. However, since the fourth component is $0$, formula
\begin{equation}\label{equ6340}
    \int_\mathrm{DB cycle}(A,\Lambda,n)\overrightarrow{\underleftarrow{\star}}(B,\Pi,m)=
    \int_\mathrm{DB cycle}(A,\Lambda,n)\overleftarrow{\underrightarrow{\star}}(B,\Pi,m)
\end{equation}
exactly hold. This is important in the discussion of linking numbers that appears in Section~\ref{sec900}.
\newpage
\section{Lattice DB cycle and integration}\label{sec700}
We move on to the integral of a lattice DB cocycle. The lattice DB cycle introduced in this section corresponds to the integral domain. As we already defined the lattice DB chain $C_{(i,j)}(\mathbb{L}^d)$ in \eqref{equ4060}, we are going to define the lattice DB cycle via a boundary operation on the DB chain, given as a dual of the coboundary operation. By defining the DB cycle, the gauge-invariant $\mathbb{R}/\mathbb{Z}$-valued integral is well-defined, and Stokes' theorem for lattice DB cohomology also holds. Finally, we will show concrete examples of the lattice DB cycle.
\subsection{Boundary operations}\label{sec710}
Let us review the coboundary operators $ d$ and $ \delta$ on lattice DB cochains:
\begin{equation}\label{equ7010}
    (\delta_j\omega)^{(i,j+1)}(\bullet\;;\;\bullet\;;\;\tilde{x}^{(0)},\ldots,\tilde{x}^{(j+1)})=\sum_{k=0}^{j+1}(-1)^k\omega^{(i,j)}(\bullet\;;\;\bullet\;;\;\tilde{x}^{(0)},\ldots,\check{\tilde{x}}^{(k)},\ldots,\tilde{x}^{(j+1)}),
\end{equation}
\begin{equation}\label{equ7020}
    \begin{split}
    (d_i\omega)^{(i+1,j)}(x\;;\; \mu_1,\ldots,\mu_{i+1}\;;\;\bullet)\qquad\qquad\qquad\qquad\qquad&\\
    =\text{antisym.}\sum_{k=1}^{i+1}\Biggl[(-1)^{k-1}\Bigl(\omega^{(i,j)}(x+\hat{\mu}_k\;;\;& \mu_1,\ldots,\check{\mu}_{k},\ldots,\mu_{i+1}\;;\;\bullet)\\
    -\omega^{(i,j)}(x\;;\;& \mu_1,\ldots,\check{\mu}_{k},\ldots,\mu_{i+1}\;;\;\bullet)\Bigr)\Biggr].
    \end{split}
\end{equation}
We define dual operations of $d,\delta$ by
\begin{equation}\label{equ7030}
    \Delta_{j+1}(\bullet\;;\;\bullet\;;\;\tilde{x}^{(0)},\ldots,\tilde{x}^{(j+1)})=\sum_{k=0}^{j+1}(-1)^k(\bullet\;;\;\bullet\;;\;\tilde{x}^{(0)},\ldots,\check{\tilde{x}}^{(k)},\ldots,\tilde{x}^{(j+1)}),
\end{equation}
\begin{equation}\label{equ7040}
    \begin{split}
    \partial_{i+1}(x\;;\; \mu_1,\ldots,\mu_{i+1}\;;\;\bullet)
    =\text{antisym.}\sum_{k=1}^{i+1}\Biggl[(-1)^{k-1}\Bigl((x+\hat{\mu}_k\;&;\; \mu_1,\ldots,\check{\mu}_{k},\ldots,\mu_{i+1}\;;\;\bullet)\\
    -(x\;&;\; \mu_1,\ldots,\check{\mu}_{k},\ldots,\mu_{i+1}\;;\;\bullet)\Bigr)\Biggr].
    \end{split}
\end{equation}
Expanding the operations $\mathbb{Z}$-linearly, we get
\begin{equation}\label{equ7050}
    \begin{split}
        \Delta_{j+1}:C_{(i,j+1)}(\mathbb{L}^d)\to C_{(i,j)}(\mathbb{L}^d),\\
        \partial_{i+1}:C_{(i+1,j)}(\mathbb{L}^d)\to C_{(i,j)}(\mathbb{L}^d).
    \end{split}
\end{equation}
\subsection{Natural pairing and duality formula}\label{sec720}
Let us define the integral as a natural pairing of a lattice DB cochain and a lattice DB chain.

Since we defined lattice \v{C}ech--de Rham cochain of degree $(i,j)$ by
\begin{equation}\label{equ7060}
    C^{(i,j)}(\mathbb{L}^d):=\{\omega:C_{(i,j)}(\mathbb{L}^d)\to\mathbb{R}\mid\text{$\omega$ is $\mathbb{Z}$-inear}\},
\end{equation}
\v{C}ech--de Rham cochain/chain has natural pairing
\begin{equation}\label{equ7070}
    \langle\bullet,\bullet\rangle :C^{(i,j)}(\mathbb{L}^d)\times C_{(i,j)}(\mathbb{L}^d)\to \mathbb{R}.
\end{equation}
The domain of the natural pairing can be extended $\mathbb{Z}$-linearly to lattice DB cochains 
$C^r_{\mathrm{lat.DB}}(\mathbb{L}^d)=\bigoplus_{j=0}^r C^{(r-j-1,j)}(\mathbb{L}^d)$ as
\begin{equation}\label{equ7080}
    \langle\bullet,\bullet\rangle :
    C^r_{\mathrm{lat.DB}}(\mathbb{L}^d)\times 
    \left(\bigoplus_{j=0}^r C_{(r-j-1,j)}(\mathbb{L}^d)\right)
    \to \mathbb{R}.
\end{equation}

By the way, since we are going to define $\mathbb{R}/\mathbb{Z}$-valued integral rather than $\mathbb{R}$-value, we can always omit the pairing between $C^{(-1,r)}(\mathbb{L}^d)$ and  $C_{(-1,r)}(\mathbb{L}^d)$ because it is always $\mathbb{Z}$-valued.\footnote{Recall we defined $C^{(-1,r)}(\mathbb{L}^d)$ as the set of $\mathbb{Z}$-valued linear functions. We cannot ignore degrees other than $(-1,r)$ because there may be $\mathbb{R}$-valued contributions from it.} So it is sufficient to define the integral by
\begin{equation}\label{equ7090}
    \begin{split}
        \langle\bullet,\bullet\rangle :&C^r_\mathrm{lat.DB}(\mathbb{L}^d)\times \left(\bigoplus_{j=0}^{r-1}C_{(r-j-1,j)}(\mathbb{L}^d)\right)\to \mathbb{R},\\
        &((\omega^{(r-1,0)},\omega^{(r-2,1)},\dots,\omega^{(0,r-1)},\omega^{(-1,r)}),(\sigma_{(r-1,0)},\sigma_{(r-2,1)},\dots,\sigma_{(0,r-1)}))\\
        \mapsto& \langle\omega^{(r-1,0)},\sigma_{(r-1,0)}\rangle+\langle\omega^{(r-2,1)},\sigma_{(r-2,1)}\rangle+\dots+\langle\omega^{(0,r-1)},\sigma_{(0,r-1)}\rangle.
    \end{split}
\end{equation}
In fact, in order to formulate holonomy and Stokes' theorem consistently in later discussions, it is convenient to define it as the $\mathbb{R}/\mathbb{Z}$-valued pairing $\langle\bullet,\bullet\rangle$ as above. We call the pairing $\langle\bullet,\bullet\rangle$ defined in this way the integral of the lattice DB cochain.

The duality of $\partial,\Delta$ described in \eqref{equ7030},\eqref{equ7040} can be written as
\begin{align}\label{equ7100}
    &\langle\delta\omega^{(i,j)},\sigma_{(i,j+1)}\rangle=\langle\omega^{(i,j)},\Delta\sigma_{(i,j+1)}\rangle,&&
    \langle d\omega^{(i,j)},\sigma_{(i+1,j)}\rangle=\langle\omega^{(i,j)},\partial\sigma_{(i+1,j)}\rangle,
\end{align}
where $\omega^{(i,j)}\in C^{(i,j)}(\mathbb{L}^d),\sigma_{(i,j)}\in C_{(i,j)}(\mathbb{L}^d)$.
\subsection{Lattice DB cycle}\label{sec730}
So far, we have na\"{i}vely defined the integral of lattice DB cochain as a natural pairing, but it is not well-defined as the integral of lattice DB cohomology. Since lattice DB cohomology $H^r_\mathrm{lat.DB}(\mathbb{L}^d)$ is defined modulo DB boundary $D_rC^{r-1}_\mathrm{lat.DB}(\mathbb{L}^d)$, it is necessary to find an integration domain $\bm{\sigma}$ such that the integral of $D_rC^{r-1}_\mathrm{lat.DB}(\mathbb{L}^d)$ over $\bm{\sigma}$ is always $0$ modulo $\mathbb{Z}$.

Let $\bm{\omega}$ and $\bm{\sigma}$ be
\begin{equation}\label{equ7110}
    \begin{split}
        \bm{\omega}&=(\omega^{(r-2,0)},\omega^{(r-3,1)},\dots,\omega^{(0,r-2)},\omega^{(-1,r-1)})\in C^{r-1}_\mathrm{lat.DB}(\mathbb{L}^d),\\
        \bm{\sigma}&=(\sigma_{(r-1,0)},\sigma_{(r-2,1)},\dots,\sigma_{(0,r-1)})\in\bigoplus_{j=0}^{r-1}C_{(r-j-1,j)}(\mathbb{L}^d).
    \end{split}
\end{equation}
Calculating $\langle D_r\bm{\omega},\bm{\sigma}\rangle$, we get 
\begin{equation}\label{equ7120}
    \begin{split}
        \langle D_r\bm{\omega},\bm{\sigma}\rangle&\\
        &\begin{array}{rr@{\hspace{2pt}}cc}
            =\langle&&(-1)^{0}d_{r-2}\omega^{(r-2,0)}&,\sigma_{(r-1,0)}\rangle\\
            +\langle&\delta_{0}\omega^{(r-2,0)}&+(-1)^{1}d_{r-3}\omega^{(r-3,1)}&,\sigma_{(r-2,1)}\rangle\\
            +\langle&\delta_{1}\omega^{(r-3,1)}&+(-1)^{2}d_{r-4}\omega^{(r-4,2)}&,\sigma_{(r-3,2)}\rangle\\
            &\vdots&\vdots&\vdots\\
            +\langle&\delta_{r-4}\omega^{(2,r-4)}&+(-1)^{r-3}d_{1}\omega^{(1,r-3)}&,\sigma_{(2,r-3)}\rangle\\
            +\langle&\delta_{r-3}\omega^{(1,r-3)}&+(-1)^{r-2}d_{0}\omega^{(0,r-2)}&,\sigma_{(1,r-2)}\rangle\\
            +\langle&\delta_{r-2}\omega^{(0,r-2)}&+(-1)^{r-1}d_{-1}\omega^{(-1,r-1)}&,\sigma_{(0,r-1)}\rangle
        \end{array}\\
        &\begin{array}{rr@{\hspace{2pt}}cc}
            =\langle&\omega^{(r-2,0)},&\Delta_{1}\sigma_{(r-2,1)}&+(-1)^{0}\partial_{r-1}\sigma_{(r-1,0)}\rangle\\
            +\langle&\omega^{(r-3,1)},&\Delta_{2}\sigma_{(r-3,2)}&+(-1)^{1}\partial_{r-2}\sigma_{(r-2,1)}\rangle\\
            +\langle&\omega^{(r-4,2)},&\Delta_{3}\sigma_{(r-4,3)}&+(-1)^{2}\partial_{r-3}\sigma_{(r-3,2)}\rangle\\
            &\vdots&\vdots&\vdots\\
            +\langle&\omega^{(2,r-4)},&\Delta_{r-3}\sigma_{(2,r-3)}&+(-1)^{r-4}\partial_{3}\sigma_{(3,r-4)}\rangle\\
            +\langle&\omega^{(1,r-3)},&\Delta_{r-2}\sigma_{(1,r-2)}&+(-1)^{r-3}\partial_{2}\sigma_{(2,r-3)}\rangle\\
            +\langle&\omega^{(0,r-2)},&\Delta_{r-1}\sigma_{(0,r-1)}&+(-1)^{r-2}\partial_{1}\sigma_{(1,r-2)}\rangle\\
            +\langle&d_{-1}\omega^{(-1,r-1)},&&(-1)^{r-1}\sigma_{(0,r-1)}\rangle.
        \end{array}
    \end{split}
\end{equation}
We used the duality formula \eqref{equ7100} at the second equal sign. While the last term is a pairing to which we cannot apply the duality formula, we ignore it because it takes values in $\mathbb{Z}$.

From the above discussion, we can see that the condition that integration domain $\bm{\sigma}=(\sigma_{(r-1,0)},\sigma_{(r-2,1)},\dots,\sigma_{(0,r-1)})\in\bigoplus_{j=0}^{r-1}C_{(r-j-1,j)}(\mathbb{L}^d)$ should satisfy is
\begin{equation}\label{equ7130}
    \begin{array}{cc}
        \Delta_{1}\sigma_{(r-2,1)}&+(-1)^{0}\partial_{r-1}\sigma_{(r-1,0)}=0\\
        \Delta_{2}\sigma_{(r-3,2)}&+(-1)^{1}\partial_{r-2}\sigma_{(r-2,1)}=0\\
        \Delta_{3}\sigma_{(r-4,3)}&+(-1)^{2}\partial_{r-3}\sigma_{(r-3,2)}=0\\
        \vdots&\vdots\\
        \Delta_{r-3}\sigma_{(2,r-3)}&+(-1)^{r-4}\partial_{3}\sigma_{(3,r-4)}=0\\
        \Delta_{r-2}\sigma_{(1,r-2)}&+(-1)^{r-3}\partial_{2}\sigma_{(2,r-3)}=0\\
        \Delta_{r-1}\sigma_{(0,r-1)}&+(-1)^{r-2}\partial_{1}\sigma_{(1,r-2)}=0
    \end{array}
\end{equation}
in order for the integral of $D_rC^{r-1}_\mathrm{lat.DB}(\mathbb{L}^d)$ to be $0\pmod{\mathbb{Z}}$. If we define $\mathscr{D}_r$ by
\begin{equation}\label{equ7140}
    \mathscr{D}_r(\sigma_{(r-1,0)},\sigma_{(r-2,1)},\dots,\sigma_{(0,r-1)})=\left(\begin{array}{cc}
        \Delta_{1}\sigma_{(r-2,1)}&+(-1)^{0}\partial_{r-1}\sigma_{(r-1,0)},\\
        \Delta_{2}\sigma_{(r-3,2)}&+(-1)^{1}\partial_{r-2}\sigma_{(r-2,1)},\\
        \Delta_{3}\sigma_{(r-4,3)}&+(-1)^{2}\partial_{r-3}\sigma_{(r-3,2)},\\
        \vdots&\vdots\\
        \Delta_{r-3}\sigma_{(2,r-3)}&+(-1)^{r-4}\partial_{3}\sigma_{(3,r-4)},\\
        \Delta_{r-2}\sigma_{(1,r-2)}&+(-1)^{r-3}\partial_{2}\sigma_{(2,r-3)},\\
        \Delta_{r-1}\sigma_{(0,r-1)}&+(-1)^{r-2}\partial_{1}\sigma_{(1,r-2)}
    \end{array}\right),
\end{equation}
the duality formula of DB cohomology
\begin{equation}\label{equ7150}
    \langle D_r\bm{\omega},\bm{\sigma}\rangle\stackrel{\mathbb{Z}}{=}\langle \bm{\omega},\mathscr{D}_r\bm{\sigma}\rangle
\end{equation}
holds. We call any $\bm{\sigma}$ satisfying $\mathscr{D}_r\bm{\sigma}=0$ a lattice DB cycle of degree $r$. Let us denote the set of lattice DB cycles of degree $r$ on $\mathbb{L}^d$ by $Z_r^\mathrm{lat.DB}(\mathbb{L}^d)$. As described above, we found that a lattice DB cohomology of degree $r$ can be integrated by natural pairing with a lattice DB cycle of degree $r$, and that the integral value is well-defined as an $\mathbb{R}/\mathbb{Z}$ value.
\subsection{Stokes' theorem on DB cohomology theory}\label{sec740}
Now that we have defined lattice DB cohomology, let us move on to Stokes' theorem. In this section, we consider the integrand, denoted
\begin{equation}\label{equ7160}
    \bm{\omega}=(\omega^{(r-1,0)},\omega^{(r-2,0)},\dots,\omega^{(0,r-2)},\omega^{(-1,r-1)})\in Z^{r}_\mathrm{lat.DB}(\mathbb{L}^d).
\end{equation}
If we perform the integral $\langle d\omega^{(r-1,0)},\Sigma_{(r,0)}\rangle$ using $\Sigma_{(r,0)}\in C_{(r,0)}(\mathbb{L}^d)$, we obtain a value corresponds to the integral of curvature. We will show that there exists $\bm{\sigma}\in Z^\mathrm{lat.DB}_{r}(\mathbb{L}^d)$ corresponding to the boundary of $\Sigma_{(r,0)}$, and that Stokes' theorem holds modulo $\mathbb{Z}$.

The DB cycle $\bm{\sigma}$ is constructed inductively on $\Sigma_{(r,0)}$ by finding $\Sigma_{(i,j)}$ and $\sigma_{(i,j)}\in C_{(i,j)}(\mathbb{L}^d)$ such that the equation 
\begin{equation}\label{equ7170}
    \partial\Sigma_{(i+1,r-i-1)}=\sigma_{(i,r-i-1)}+(-1)^{r-i-1}\Delta\Sigma_{(i,r-i)}
\end{equation}
holds for $r-1\geq i\geq 0$.
This construction is always possible because for any $\Sigma_{(r,0)}$, we can simply set $\partial_{r}\Sigma_{(r,0)}=\sigma_{(r-1,0)}$ and set all other $\Sigma,\sigma$ to 0. There is some arbitrariness in the configuration of $\Sigma,\sigma$, including this ``trivial configuration''.

Let us see why Stokes' theorem holds. First, we introduce
\begin{equation}\label{equ7180}
    \begin{split}
        \langle\omega^{(i,j)},(-1)^{j}\Delta\Sigma_{(i,j+1)}\rangle=&\langle(-1)^{j}\delta\omega^{(i,j)},\Sigma_{(i,j+1)}\rangle\\
        =&\langle d\omega^{(i-1,j+1)},\Sigma_{(i,j+1)}\rangle=\langle\omega^{(i-1,j+1)},\partial\Sigma_{(i,j+1)}\rangle
    \end{split}
\end{equation}
as a formula for $\bm{\omega}\in Z^{r}_\mathrm{lat.DB}(\mathbb{L}^d)$. From the usual duality formula (which is the usual lattice version of Stokes' theorem),
\begin{equation}\label{equ7190}
    \langle d\omega^{(r-1,0)},\Sigma_{(r,0)}\rangle=\langle \omega^{(r-1,0)},\partial\Sigma_{(r,0)}\rangle
\end{equation}
holds. Applying $\partial\Sigma_{(r,0)}=\sigma_{(r-1,0)}+(-1)^0\Delta\Sigma_{(r-1,1)}$ and the formula, we get
\begin{equation}\label{equ7200}
    \langle d\omega^{(r-1,0)},\Sigma_{(r,0)}\rangle=\langle \omega^{(r-1,0)},\sigma_{(r-1,0)}\rangle+\langle \omega^{(r-2,1)},\partial\Sigma_{(r-1,1)}\rangle.
\end{equation}
Focusing on $\langle \omega^{(r-2,1)},\partial\Sigma_{(r-1,1)}\rangle$ and using $\partial\Sigma_{(r-1,1)}=\sigma_{(r-2,1)}+(-1)^1\Delta\Sigma_{(r-2,2)}$ and the formula \eqref{equ7180},
\begin{equation}\label{equ7210}
    \langle d\omega^{(r-1,0)},\Sigma_{(r,0)}\rangle=\langle \omega^{(r-1,0)},\sigma_{(r-1,0)}\rangle+\langle \omega^{(r-2,1)},\sigma_{(r-2,1)}\rangle+\langle \omega^{(r-3,2)},\partial\Sigma_{(r-2,2)}\rangle
\end{equation}
holds. Repeat this procedure inductively and finally use the formula $\partial\Sigma_{(1,r-1)}=\sigma_{(0,r-1)}+(-1)^{r-1}\Delta\Sigma_{(0,r)}$, we get
\begin{equation}\label{equ7220}
    \begin{split}
        \langle d\omega^{(r-1,0)},\Sigma_{(r,0)}\rangle=&\langle \bm{\omega},\bm{\sigma}\rangle+\langle \omega^{(0,r-1)},(-1)^{r-1}\Delta\Sigma_{(0,r)}\rangle\\
        =&\langle \bm{\omega},\bm{\sigma}\rangle+\langle d_{-1}\omega^{(-1,r)},\Sigma_{(0,r)}\rangle\stackrel{\mathbb{Z}}{=}\langle \bm{\omega},\bm{\sigma}\rangle.
    \end{split}
\end{equation}
Then we complete the lattice DB version of Stokes' theorem
\begin{equation}\label{equ7230}
    \langle d\omega^{(r-1,0)},\Sigma_{(r,0)}\rangle\stackrel{\mathbb{Z}}{=}\langle \bm{\omega},\bm{\sigma}\rangle.
\end{equation}
We can also prove that $\bm{\sigma}$ is a lattice DB cycle. In fact, by using $\sigma_{(i,r-i-1)}=\partial\Sigma_{(i+1,r-i-1)}+(-1)^{r-i}\Delta\Sigma_{(i,r-i)}$, we obtain the equality
\begin{equation}\label{equ7240}
    \begin{split}
        &\Delta\sigma_{(r-i-2,i+1)}+(-1)^{i}\partial\sigma_{(r-i-1,i)}\\
        =&\Delta(\partial\Sigma_{(r-i-2,i+1)}+(-1)^{r-i}\Delta\Sigma_{(r-i-3,i+2)})\\
        +&(-1)^{i}\partial(\partial\Sigma_{(r-i-1,i)}+(-1)^{i+1}\Delta\Sigma_{(r-i-2,i+1)})\\
        =&\Delta\partial\Sigma_{(r-i-2,i+1)}-\partial\Delta\Sigma_{(r-i-2,i+1)}=0
    \end{split}
\end{equation}
for $0\leq i\leq r-2$. This equality guarantees that $\bm{\sigma}$ is a lattice DB cycle of degree $r$.
\subsection{Lattice DB boundary}\label{sec750}
The lattice DB cycle is constructed by requiring that the integral of $D_rC^{r-1}_\mathrm{lat.DB}(\mathbb{L}^d)$ is always $0$ modulo $\mathbb{Z}$, but there also exist integral domains such that the integral of $Z^{r}_\mathrm{lat.DB}(\mathbb{L}^d)$ is always $0$ modulo $\mathbb{Z}$. We call such an integral domain a lattice DB boundary.

Let us give a definition of lattice DB boundary. Consider $\Sigma_{(i,j)}\in C_{(i,j)}(\mathbb{L}^d)$. For $\bm{\Sigma}=(\Sigma_{(r-1,1)},\Sigma_{(r-2,2)},\dots,\Sigma_{(0,r)})$, we define $\partial^\mathrm{ev}(\Sigma_{(r-1,1)},\Sigma_{(r-2,2)},\dots,\Sigma_{(0,r)})$ by
\begin{equation}\label{equ7250}
    \partial^\mathrm{ev}\left(\begin{array}{c}
        \Sigma_{(r-1,1)},\\
        \Sigma_{(r-2,2)},\\
        \vdots\\
        \Sigma_{(0,r)}
    \end{array}\right)=
    \left(\begin{array}{lcr}
        \Delta_{1}\Sigma_{(r-1,1)}&&,\\
        \Delta_{2}\Sigma_{(r-2,2)}&+&(-1)^{1}\partial_{r-1}\Sigma_{(r-1,1)},\\
        \Delta_{3}\Sigma_{(r-3,3)}&+&(-1)^{2}\partial_{r-2}\Sigma_{(r-2,2)},\\
        &\vdots&\\
        \Delta_{r-1}\Sigma_{(1,r-1)}&+&(-1)^{r-2}\partial_{2}\Sigma_{(2,r-2)},\\
        \Delta_{r}\Sigma_{(0,r)}&+&(-1)^{r-1}\partial_{1}\Sigma_{(1,r-1)}
    \end{array}\right).
\end{equation}
We call $\partial^\mathrm{ev}$ the lattice DB boundary operation. We can see that the image of $\partial^\mathrm{ev}$ is a lattice DB cycle. If we calculate the pairing with $\bm{\omega}\in C^r_\mathrm{lat.DB}(\mathbb{L}^d)$, we obtain
\begin{equation}\label{equ7260}
    \begin{split}
        \langle\bm{\omega},\partial^\mathrm{ev}\bm{\Sigma}\rangle&\\
        &\begin{array}{lclcr}
            =\langle\omega^{(r-1,0)}&,&\Delta_{1}\Sigma_{(r-1,1)}&&\rangle\\
            +\langle\omega^{(r-2,1)}&,&\Delta_{2}\Sigma_{(r-2,2)}&+&(-1)^{1}\partial_{r-1}\Sigma_{(r-1,1)}\rangle\\
            +\langle\omega^{(r-3,2)}&,&\Delta_{3}\Sigma_{(r-3,3)}&+&(-1)^{2}\partial_{r-2}\Sigma_{(r-2,2)}\rangle\\
            &\vdots&&\vdots&\\
            +\langle\omega^{(1,r-2)}&,&\Delta_{r-1}\Sigma_{(1,r-1)}&+&(-1)^{r-2}\partial_{2}\Sigma_{(2,r-2)}\rangle\\
            +\langle\omega^{(0,r-1)}&,&\Delta_{r}\Sigma_{(0,r)}&+&(-1)^{r-1}\partial_{1}\Sigma_{(1,r-1)}\rangle
        \end{array}\\
        &\begin{array}{lclcl}
            =\langle\delta_{0}\omega^{(r-1,0)}&+&(-1)^{1}d_{1}\omega^{(r-2,1)}&,&\Sigma_{(r-1,1)}\rangle\\
            +\langle\delta_{1}\omega^{(r-2,1)}&+&(-1)^{2}d_{2}\omega^{(r-3,2)}&,&\Sigma_{(r-2,2)}\rangle\\
            +\langle\delta_{2}\omega^{(r-3,2)}&+&(-1)^{3}d_{3}\omega^{(r-4,3)}&,&\Sigma_{(r-3,3)}\rangle\\
            &\vdots&&\vdots&\\
            +\langle\delta_{r-3}\omega^{(2,r-3)}&+&(-1)^{r-2}d_{r-2}\omega^{(1,r-2)}&,&\Sigma_{(2,r-2)}\rangle\\
            +\langle\delta_{r-2}\omega^{(1,r-2)}&+&(-1)^{r-1}d_{r-1}\omega^{(0,r-1)}&,&\Sigma_{(1,r-1)}\rangle\\
            +\langle\delta_{r-1}\omega^{(0,r-1)}&&&,&\Sigma_{(0,r)}\rangle
        \end{array}\\
        &=\langle D_r^\mathrm{ev}\bm{\omega},\bm{\Sigma}\rangle-\langle(-1)^rd_{-1}\omega^{(-1,r)},\Sigma_{(0,r)}\rangle\stackrel{\mathbb{Z}}{=}\langle D_r^\mathrm{ev}\bm{\omega},\bm{\Sigma}\rangle
    \end{split}
\end{equation}
which shows lattice DB version of modulo $\mathbb{Z}$ duality formula between $D^\mathrm{ev}$ and $\partial^\mathrm{ev}$. If $\bm{\omega}\in Z^r_\mathrm{lat.DB}(\mathbb{L}^d)$, then this pairing is always $0$. Therefore, we find that the difference of the lattice DB boundary can be ignored when we integrate the lattice DB cocycle.
\subsection{Lattice DB cycle corresponding to closed curve $\gamma$ in $\mathbb{T}^3$}\label{sec760}
As we have defined $[(A,\Lambda,n)]\in H^2_\mathrm{lat.DB}(\mathbb{L}^d)$, let us consider a line integral of $(A,\Lambda,n)$, which is equivalent to the calculation of the Wilson line. When $\partial\Sigma = \gamma$ holds, then by using Stokes' theorem in the previous section to create $\bm{\sigma}$ from $\Sigma$, we can create the pairing $\langle\bullet,\sigma\rangle$ that corresponds to the line integral. On the other hand, if $[\gamma]\neq 0\in H^1(\mathbb{L}^d)$, then there is no $\Sigma$ such that $\partial\Sigma=\gamma$, and therefore a lattice DB cycle cannot be constructed using Stokes' theorem. Therefore, it is necessary to construct the lattice DB cycle $(\sigma^{(1,0)},\sigma^{(0,1)})$  directly from the closed curve $\gamma$ on $\mathbb{L}^d$.

The construction is almost the same as Section~\ref{sec210}. Since $\gamma$ is a closed curve, we can use the parameter $0\leq s\leq L-1$ to display $\gamma$ by
\begin{equation}\label{equ7270}
    \begin{split}
        \gamma:\{s\mid 0\leq s\leq L-1\}\to\mathbb{L}^d\times\{1,2,\dots,d\},s\mapsto \gamma(s)=(x(s),\mu(s)),\\
        x(s+1)=x(s)+\hat{\mu}(s)\,(\mathrm{for}\;0\leq s\leq L-2)\;,\;x(0)=x(L-1)+\hat{\mu}(L-1).
    \end{split}
\end{equation}
Let $\sigma_{(1,0)}$ be given by
\begin{equation}\label{equ7280}
    \sigma_{(1,0)}=\sum_{s=0}^{L-1}(x(s);\mu(s);x_{+\dots+}(s)),
\end{equation}
where
\begin{equation}\label{equ7290}
    x_{\pm \dots\pm }(s)=x(s)\pm \frac{1}{2}\hat{1}\pm \frac{1}{2}\hat{2}\pm \dots\pm \frac{1}{2}\hat{d}\in\tilde{\mathbb{L}}^d.
\end{equation}
In this case, we chose all patches to be $x_{+\dots+}$, but as we explain later, the difference in this selection only results in a difference in the lattice DB boundary, so as long as we select the cube to which the edge $(x(s);\mu(s))$ belongs, we can change it for each $s$. 

From the equality
\begin{equation}\label{equ7300}
    \begin{split}
        (-1)^0\partial\sigma_{(1,0)}=&\sum_{s=0}^{L-1}\left((x(s)+\hat{\mu}(s);;x_{+\dots+}(s))-(x(s);;x_{+\dots+}(s))\right)\\
        =&\sum_{s=0}^{L-1}\left((x(s+1);;x_{+\dots+}(s))-(x(s);;x_{+\dots+}(s))\right)\\
        =&\Delta\left(\sum_{s=0}^{L-1}(x(s);;x_{+\dots+}(s)x_{+\dots+}(s-1))\right),
    \end{split}
\end{equation}
$\sigma_{(0,1)}$ can be defined by
\begin{equation}\label{equ7310}
    \sigma_{(0,1)}=\sum_{s=0}^{L-1}(x(s);;x_{+\dots+}(s)x_{+\dots+}(s-1)).
\end{equation}

We give a definition of DB cycle $\gamma_\mathrm{DB.cyc}$ using $\sigma^{(1,0)},\sigma^{(0,1)}$ by
\begin{equation}\label{equ7320}
    \gamma_\mathrm{DB.cyc}=(\sigma^{(1,0)},\sigma^{(0,1)}),
\end{equation}
which is the DB cycle corresponding to $\gamma$.

Hereafter, we will use $\gamma_\mathrm{DB.cyc}$ in this sense. Also, we will write the pairing with $\gamma_\mathrm{DB.cyc}$ as
\begin{equation}\label{equ7330}
    \int_{\gamma}\bm{\omega}:=\langle\bm{\omega},\gamma_\mathrm{DB.cyc}\rangle\;\;(\mathrm{for}\;\bm{\omega}\in H^2_\mathrm{lat.DB}(\mathbb{L}^d)).
\end{equation}

Finally, let us consider changing the patch selection. For example, suppose we want to replace only the patch $(x(1);\mu(1);x_{+\dots+}(1))$ and change it to $(x(1),\mu(1),\tilde{x})$. $\tilde{x}\in\tilde{\mathbb{L}}^d$ can be chosen arbitrarily as long as it is the center point of a cube containing the edge $(x(1);\mu(1))$, but we will fix it from now on. In this situation, the parts of the lattice DB cycle that need to be changed are
\begin{equation}\label{equ7340}
    \begin{split}
        (x(1);\mu(1);x_{+\dots+}(1))&\to (x(1);\mu(1);\tilde{x}),\\
        (x(1);;x_{+\dots+}(0),x_{+\dots+}(1))&\to (x(1);;x_{+\dots+}(0),\tilde{x}),\\
        (x(2);;x_{+\dots+}(1),x_{+\dots+}(2))&\to (x(1);;\tilde{x},x_{+\dots+}(2)).
    \end{split}
\end{equation}
The difference of the lattice DB cycle due to the change of the patch is clarified by
\begin{equation}\label{equ7350}
    \begin{split}
        &-(x(1);\mu(1);x_{+\dots+}(1))+(x(1);\mu(1);\tilde{x})\\
        &-(x(1);;x_{+\dots+}(0),x_{+\dots+}(1))+(x(1);;x_{+\dots+}(0),\tilde{x})\\
        &-(x(2);;x_{+\dots+}(1),x_{+\dots+}(2))+(x(1);;\tilde{x},x_{+\dots+}(2))\\
        &=\Delta(x(1);\mu(1);x_{+\dots+}(1)\tilde{x})\\
        &+\Delta(x(1);;x_{+\dots+}(0),\tilde{x},x_{+\dots+}(1))+(x(1);;\tilde{x},x_{+\dots+}(1))\\
        &+\Delta(x(2);;x_{+\dots+}(1),\tilde{x},x_{+\dots+}(2))+(x(2);;x_{+\dots+}(1),\tilde{x})\\
        &=(\Delta+(-1)^1\partial)(x(1);\mu(1);x_{+\dots+}(1)\tilde{x})\\
        &+\Delta\Bigl((x(1);;x_{+\dots+}(0),\tilde{x},x_{+\dots+}(1))+(x(2);;x_{+\dots+}(1),\tilde{x},x_{+\dots+}(2))\Bigr).
    \end{split}
\end{equation}
We can see that it is a lattice DB boundary.
\subsection{Lattice DB cycle corresponding to $\mathbb{L}^3$}\label{sec770}
Let us consider the definition of a lattice DB cycle corresponding to $\mathbb{L}^3$ in a top-down manner, similar to the construction for $\gamma$.

We will set $\sigma_{(3,0)}$ corresponding to $\mathbb{L}^3$, and solve $\sigma_{(2,1)}, \sigma_{(1,2)}, \sigma_{(0,3)}$ satisfying the conditions of the lattice DB cycle. Clearly, $\sigma_{(3,0)}$ can only be defined as
\begin{equation}\label{equ7360}
    \sigma_{(3,0)}=\sum_{x\in\mathbb{L}^3}(x;1,2,3;x_{+++}).
\end{equation}
Then the equality
\begin{equation}\label{equ7370}
    \begin{split}
        (-1)^0\partial_3\sigma_{(3,0)}=\sum_{x\in\mathbb{L}^3}\biggl\{
        &(x+\hat{1};2,3;x_{+++})-(x;2,3;x_{+++})\\-
        &(x+\hat{2};1,3;x_{+++})+(x;1,3;x_{+++})\\+
        &(x+\hat{3};1,2;x_{+++})-(x;1,2;x_{+++})\biggr\}\\
        =\sum_{x\in\mathbb{L}^3}\biggl\{
        &(x;2,3;x_{-++})-(x;2,3;x_{+++})\\-
        &(x;1,3;x_{+-+})+(x;1,3;x_{+++})\\+
        &(x;1,2;x_{+-+})-(x;1,2;x_{+++})\biggr\}\\
        =\Delta_1\sum_{x\in\mathbb{L}^3}\biggl\{
        &(x;2,3;x_{+++}x_{-++})+
        (x;1,3;x_{+-+}x_{+++})+
        (x;1,2;x_{+++}x_{++-})\biggr\}
    \end{split}
\end{equation}
holds, and we get the definition
\begin{equation}\label{equ7380}
        \sigma_{(2,1)}=-\sum_{x\in\mathbb{L}^3}\biggl\{
        (x;2,3;x_{+++}x_{-++})+
        (x;1,3;x_{+-+}x_{+++})+
        (x;1,2;x_{+++}x_{++-})\biggr\}.
\end{equation}
Similarly,
\begin{equation}\label{equ7390}
    \begin{split}
        (-1)^1\partial_2\sigma_{(2,1)}=-(-1)^1\partial_2\sum_{x\in\mathbb{L}^3}\biggl\{
        &(x;2,3;x_{+++}x_{-++})\\+
        &(x;1,3;x_{+-+}x_{+++})\\+
        &(x;1,2;x_{+++}x_{++-})\biggr\}\\
        =\sum_{x\in\mathbb{L}^3}\biggl\{
        &(x+\hat{2};3;x_{+++}x_{-++})-(x;3;x_{+++}x_{-++})\\-
        &(x+\hat{3};2;x_{+++}x_{-++})+(x;2;x_{+++}x_{-++})\\+
        &(x+\hat{1};3;x_{+-+}x_{+++})-(x;3;x_{+-+}x_{+++})\\-
        &(x+\hat{3};1;x_{+-+}x_{+++})+(x;1;x_{+-+}x_{+++})\\+
        &(x+\hat{1};2;x_{+++}x_{++-})-(x;2;x_{+++}x_{++-})\\-
        &(x+\hat{2};1;x_{+++}x_{++-})+(x;1;x_{+++}x_{++-})\biggr\}\\
        =\sum_{x\in\mathbb{L}^3}\biggl\{
        &(x;3;x_{+-+}x_{--+})-(x;3;x_{+++}x_{-++})\\-
        &(x;2;x_{++-}x_{-+-})+(x;2;x_{+++}x_{-++})\\+
        &(x;3;x_{--+}x_{-++})-(x;3;x_{+-+}x_{+++})\\-
        &(x;1;x_{+--}x_{++-})+(x;1;x_{+-+}x_{+++})\\+
        &(x;2;x_{-++}x_{-+-})-(x;2;x_{+++}x_{++-})\\-
        &(x;1;x_{+-+}x_{+--})+(x;1;x_{+++}x_{++-})\biggr\}
    \end{split}
\end{equation}
\begin{equation}\label{equ7400}
    \begin{split}
        =\sum_{x\in\mathbb{L}^3}\biggl\{
        -(x;1;x_{+--}x_{++-})+(x;1;x_{+-+}x_{+++})-&(x;1;x_{+-+}x_{+--})+(x;1;x_{+++}x_{++-})\\
        -(x;2;x_{++-}x_{-+-})+(x;2;x_{+++}x_{-++})+&(x;2;x_{-++}x_{-+-})-(x;2;x_{+++}x_{++-})\\
        +(x;3;x_{+-+}x_{--+})-(x;3;x_{+++}x_{-++})+&(x;3;x_{--+}x_{-++})-(x;3;x_{+-+}x_{+++})\biggr\}\\
        =\Delta_2\sum_{x\in\mathbb{L}^3}\biggl\{
        (x;1;x_{+++}x_{++-}x_{+--})+&(x;1;x_{+++}x_{+--}x_{+-+})\\
        +(x;2;x_{+++}x_{-++}x_{++-})+&(x;2;x_{-+-}x_{++-}x_{-++})\\
        +(x;3;x_{+++}x_{+-+}x_{--+})+&(x;3;x_{+++}x_{--+}x_{-++})\biggr\}
    \end{split}
\end{equation}
holds, and we define $\sigma_{(1,2)}$ by
\begin{equation}\label{equ7410}
    \begin{split}
        \sigma_{(1,2)}=-\sum_{x\in\mathbb{L}^3}\biggl\{
        (x;1;x_{+++}x_{++-}x_{+--})+&(x;1;x_{+++}x_{+--}x_{+-+})\\
        +(x;2;x_{+++}x_{-++}x_{++-})+&(x;2;x_{-+-}x_{++-}x_{-++})\\
        +(x;3;x_{+++}x_{+-+}x_{--+})+&(x;3;x_{+++}x_{--+}x_{-++})\biggr\}.
    \end{split}
\end{equation}
By the same procedure, we can see
\begin{equation}\label{equ7420}
    \begin{split}
        (-1)^2\partial_1\sigma_{(1,2)}=-\sum_{x\in\mathbb{L}^3}\biggl\{
        &(x+\hat{1};;x_{+++}x_{++-}x_{+--})-(x;;x_{+++}x_{++-}x_{+--})\\
        +&(x+\hat{1};;x_{+++}x_{+--}x_{+-+})-(x;;x_{+++}x_{+--}x_{+-+})\\
        +&(x+\hat{2};;x_{+++}x_{-++}x_{++-})-(x;;x_{+++}x_{-++}x_{++-})\\
        +&(x+\hat{2};;x_{-+-}x_{++-}x_{-++})-(x;;x_{-+-}x_{++-}x_{-++})\\
        +&(x+\hat{3};;x_{+++}x_{+-+}x_{--+})-(x;;x_{+++}x_{+-+}x_{--+})\\
        +&(x+\hat{3};;x_{+++}x_{--+}x_{-++})-(x;;x_{+++}x_{--+}x_{-++})\biggr\}\\
        =-\sum_{x\in\mathbb{L}^3}\biggl\{
        &(x;;x_{-++}x_{-+-}x_{---})-(x;;x_{+++}x_{++-}x_{+--})\\
        +&(x;;x_{-++}x_{---}x_{--+})-(x;;x_{+++}x_{+--}x_{+-+})\\
        +&(x;;x_{+-+}x_{--+}x_{+--})-(x;;x_{+++}x_{-++}x_{++-})\\
        +&(x;;x_{---}x_{+--}x_{--+})-(x;;x_{-+-}x_{++-}x_{-++})\\
        +&(x;;x_{++-}x_{+--}x_{---})-(x;;x_{+++}x_{+-+}x_{--+})\\
        +&(x;;x_{++-}x_{---}x_{-+-})-(x;;x_{+++}x_{--+}x_{-++})\biggr\}\\
        =-\Delta_3\sum_{x\in\mathbb{L}^3}\biggl\{
        -&(x;;x_{---}x_{-++}x_{++-}x_{-+-})
        +(x;;x_{+++}x_{+--}x_{--+}x_{+-+})\\
        -&(x;;x_{---}x_{+--}x_{++-}x_{-++})
        +(x;;x_{+++}x_{-++}x_{--+}x_{+--})\\
        -&(x;;x_{---}x_{--+}x_{+--}x_{-++})
        +(x;;x_{+++}x_{++-}x_{-++}x_{+--})\biggr\},
    \end{split}
\end{equation}
and we define
\begin{equation}\label{equ7430}
    \begin{split}
        \sigma_{(0,3)}=\sum_{x\in\mathbb{L}^3}\biggl\{
        -&(x;;x_{---}x_{-++}x_{++-}x_{-+-})
        +(x;;x_{+++}x_{+--}x_{--+}x_{+-+})\\
        -&(x;;x_{---}x_{+--}x_{++-}x_{-++})
        +(x;;x_{+++}x_{-++}x_{--+}x_{+--})\\
        -&(x;;x_{---}x_{--+}x_{+--}x_{-++})
        +(x;;x_{+++}x_{++-}x_{-++}x_{+--})\biggr\}.
    \end{split}
\end{equation}
Finally, the lattice DB cycle corresponding to $\mathbb{L}^3$ can be constructed from $\sigma_{(3,0)},\sigma_{(2,1)},\sigma_{(1,2)},\sigma_{(0,3)}$ by
\begin{equation}\label{equ7440}
    \mathbb{L}^3_\mathrm{DB.cyc}=(\sigma_{(3,0)},\sigma_{(2,1)},\sigma_{(1,2)},\sigma_{(0,3)}).
\end{equation}

Hereafter, we will use $\mathbb{L}^3_\mathrm{DB.cyc}$ in this sense. Also, we will adopt the notation
\begin{equation}\label{equ7450}
    \int_{\mathbb{L}^3}\bm{\omega}:=\langle\bm{\omega},\mathbb{L}^3_\mathrm{DB.cyc}\rangle\;\;(\mathrm{for}\;\bm{\omega}\in H^4_\mathrm{lat.DB}(\mathbb{L}^3)).
\end{equation}
\subsection{Gauge-invariant half-integer integral}\label{sec780}
For $[\bm{\omega}],[\bm{\eta}]\in H^2_\mathrm{lat.DB}(\mathbb{L}^3)$, let us consider an integral
\begin{equation}\label{equ7460}
    \frac{1}{2}\int_{\mathbb{L}^3}(\bm{\omega}\star\bm{\eta}+\bm{\eta}\star\bm{\omega}).
\end{equation}
This integral is calculated without any problems, but the value of the integral can change by $\frac{1}{2}\mathbb{Z}$ under the large gauge transformations of $\bm{\omega}$ and $\bm{\eta}$. Therefore, this is not defined as a modulo-$\mathbb{Z}$ invariant integral. In this section, we confirm that this integral is not gauge-invariant, and then consider how to make the integral $\pmod{\mathbb{Z}}$ well-defined.

First, we calculate the change of this integral under the gauge transformation. Let
\begin{align}\label{equ7470}
    &\bm{\xi}=(\xi^{(0,0)},m^{(-1,1)})\in C^1_\mathrm{lat.DB}(\mathbb{L}^3),&&\bm{\Xi}=(\Xi^{(0,0)},M^{(-1,1)})\in C^1_\mathrm{lat.DB}(\mathbb{L}^3)
\end{align}
be gauge transformation parameters. Applying gauge transformation
\begin{align}
    &\bm{\omega}\to\bm{\omega}+D_2\bm{\xi}=\bm{\omega}+(d_0\xi,\delta_0\xi-d_{-1}m,\delta_1 m),\label{equ7480}\\
    &\bm{\eta}\to\bm{\eta}+D_2\bm{\Xi}=\bm{\eta}+(d_0\Xi,\delta_0\Xi-d_{-1}M,\delta_1 M),\label{equ7490}
\end{align}
$\bm{\omega}\star\bm{\eta}$ and $\bm{\eta}\star\bm{\omega}$ changes as follows:
\begin{align}
    &\bm{\omega}\star\bm{\eta}\to\bm{\omega}\star\bm{\eta}+D_4(\bm{\xi}\star\bm{\eta}+\bm{\omega}\star\bm{\Xi}+\bm{\xi}\star D_2\bm{\Xi}),\label{equ7500}\\
    &\bm{\eta}\star\bm{\omega}\to\bm{\eta}\star\bm{\omega}+D_4(\bm{\Xi}\star\bm{\omega}+\bm{\eta}\star\bm{\xi}+D_2\bm{\Xi}\star \bm{\xi}),\label{equ7510}
\end{align}
and the change of equation \eqref{equ7460} can be calculated by
\begin{equation}\label{equ7520}
    \begin{split}
        &\frac{1}{2}\langle d_{-1}(\bm{\xi}\star\bm{\eta}+\bm{\eta}\star\bm{\xi}+\bm{\omega}\star\bm{\Xi}+\bm{\Xi}\star\bm{\omega}+\bm{\xi}\star D_2\bm{\Xi}+D_2\bm{\Xi}\star \bm{\xi})^{(-1,3)}\;,\;\sigma_{(0,3)}\rangle\\
        =&\frac{1}{2}\langle d_{-1}( m\smallsmile \eta^{(-1,2)}+\eta^{(-1,2)}\smallsmile m+\omega^{(-1,2)}\smallsmile M+ M\smallsmile \omega^{(-1,2)}\\
        +& m\smallsmile \delta_1 M+\delta_1 M\smallsmile m)\;\;,\;\;\sigma_{(0,3)}\rangle\\
        \stackrel{\mathbb{Z}}{=}&\frac{1}{2}\langle d_{-1}\delta_2(m\smallsmile_1 \eta^{(-1,2)}+\omega^{(-1,2)}\smallsmile_1 M+ m\smallsmile_1\delta_1 M)\\
        +&d_{-1}(\delta_1 m\smallsmile_1 \eta^{(-1,2)}+\omega^{(-1,2)}\smallsmile_1\delta_1 M+\delta_1 m\smallsmile_1\delta_1 M)\;,\;\sigma_{(0,3)}\rangle\\
        =&\frac{1}{2}\langle d_0d_{-1}(m\smallsmile_1 \eta^{(-1,2)}+\omega^{(-1,2)}\smallsmile_1 M+ m\smallsmile_1\delta_1 M)\;,\;\sigma_{(1,2)}\rangle\\
        +&\frac{1}{2}\langle d_{-1}(\delta_1 m\smallsmile_1 \eta^{(-1,2)}+\omega^{(-1,2)}\smallsmile_1\delta_1 M+\delta_1 m\smallsmile_1\delta_1 M)\;,\;\sigma_{(0,3)}\rangle\\
        =&\frac{1}{2}\langle d_{-1}(\delta_1 m\smallsmile_1 \eta^{(-1,2)}+\omega^{(-1,2)}\smallsmile_1\delta_1 M+\delta_1 m\smallsmile_1\delta_1 M)\;,\;\sigma_{(0,3)}\rangle.
    \end{split}
\end{equation}
We used $d_{-1}(d_{-1}\alpha^{(-1,a)}\stackrel{\smallsmile}{\wedge}\beta^{(-1,b)})=d_{-1}(\alpha^{(-1,a)}\smallsmile\beta^{(-1,b)})$ in the first equal sign, and the cup-1 coboundary formula in $\stackrel{\mathbb{Z}}{=}$.

As we see above, equation \eqref{equ7460} can change by a half integer under large gauge transformations. To make the equation \eqref{equ7460} well-defined, we should add a term of the cup-1 product. Thus, we define $\frac{\overleftrightarrow{\star}}{2}$ by
\begin{equation}\label{equ7530}
    \int_{\mathbb{L}^3}[\bm{\omega}]\frac{\overleftrightarrow{\star}}{2}[\bm{\eta}]:=\frac{1}{2}\int_{\mathbb{L}^3}(\bm{\omega}\star\bm{\eta}+\bm{\eta}\star\bm{\omega}+(0,0,0,d_{-1}(\omega^{(-1,2)}\smallsmile_1\eta^{(-1,2)}),0)).
\end{equation}
Since the term $d_{-1}(\omega^{(-1,2)}\smallsmile_1\eta^{(-1,2)})$ changes the same as \eqref{equ7520}, this integral is defined as a gauge-invariant $\mathbb{R}/\mathbb{Z}$-valued integral.
\newpage
\section{Lattice DB cohomology as a differential cohomology}\label{sec800}
Deligne--Beilinson cohomology is one of the cohomology theories called differential cohomology. Other formulations of the differential cohomology
\begin{itemize}
\item Cheeger--Simons formulation~\cite{Cheeger:2006}
\item The formulation of spark, which is a special de Rham current~\cite{Harvey:2003}
\end{itemize}
are also known, and both formulations are said to be equivalent~\cite{Bar:2014}.

In this section, we explain how a short exact sequence that appears in the axiomatic definition of differential cohomology is realized in a lattice Deligne--Beilinson cohomology on $\mathbb{L}^d$. Although we will not discuss differential cohomology or its axiomatic aspects further, the short exact sequence leads to a decomposition of the lattice Deligne--Beilinson cohomology on $\mathbb{L}^d$, which plays an important role in defining the path integral over $H_\mathrm{lat.DB}^2(\mathbb{L}^3)$.
\subsection{Two short exact sequences}\label{sec810}
The axiomatic formulation of differential cohomology $H^\bullet_\mathrm{diff}(M)$ requires that a diagram called a hexagon diagram holds. The diagram contains two important exact sequences
\begin{align}
        0&\to&(\Omega^{n-1}(M)/\Omega^{n-1}_\mathrm{closed}(M)_\mathbb{Z})&\to&H^n_\mathrm{diff}(M)&\to&H^n(M;\mathbb{Z})&\to&0,\label{equ8010}\\
        0&\to&H^{n-1}(M;\mathbb{R}/\mathbb{Z})&\to&H^n_\mathrm{diff}(M)&\to&\Omega^{n}_\mathrm{closed}(M)_\mathbb{Z}&\to&0,\label{equ8020}
\end{align}
where $\Omega^{n}_\mathrm{closed}(M)_\mathbb{Z}$ denotes the set of closed differential $n$-forms on $M$ whose periods are all integers. In the DB cohomology on $\mathbb{L}^3$ that we consider, there is essentially no difference between the two short exact sequences. We therefore show below that (8.1) also holds as a short exact sequence of $H^\bullet_\mathrm{lat.DB}$. This result will be used in later chapters to define the gauge-fixed path integral.
\subsubsection{Definition of injective map $(\Omega^{n-1}(\mathbb{L}^d)/\Omega_\mathrm{closed}^{n-1}(\mathbb{L}^d)_\mathbb{Z})\hookrightarrow  H^{n}_\mathrm{lat.DB}(\mathbb{L}^d)$}\label{sec811}
$\Omega^{n-1}(\mathbb{L}^d)\to H^n_\mathrm{lat.DB}(\mathbb{L}^d)$ can be defined by $\omega\mapsto(\delta_{-1}\omega,0,0)$.

On the other hand, there exists $\bm{\xi}\in C^{n-1}_\mathrm{lat.DB}(\mathbb{L}^d)\,\mathrm{s.t.}\,D_1\bm{\xi}=(\delta_{-1}\beta,0,0)$ for all $[\beta]\in H^{n-1}(\mathbb{L}^d;\mathbb{Z})$. Similarly, $D_1(\delta_{-1}\alpha,0,0)=(\delta_{-1}d\alpha,0,0)$ for all $\alpha\in C^{n-2}(\mathbb{L}^d)$. Therefore, it is well-defined to take the domain of $\omega\mapsto(\delta_{-1}\omega,0,0)$ to be quotient space $\Omega^{n-1}(\mathbb{L}^d)/(H^{n-1}(\mathbb{L}^d;\mathbb{Z})\oplus B^{n-1}(\mathbb{L}^d))$ instead of $\Omega^{n-1}(\mathbb{L}^d)$. From the equality $H^{n-1}(\mathbb{L}^d;\mathbb{Z})\oplus B^{n-1}(\mathbb{L}^d)=\Omega_\mathrm{closed}^{n-1}(\mathbb{L}^d)_\mathbb{Z}$, if we show that this mapping is injective, we obtain the mapping $(\Omega^{n-1}(\mathbb{L}^d)/\Omega_\mathrm{closed}^{n-1}(\mathbb{L}^d)_\mathbb{Z})\hookrightarrow  H^{n}_\mathrm{lat.DB}(\mathbb{L}^d)$.

From the equality $\Omega^{n-1}(\mathbb{L}^d)/(H^{n-1}(\mathbb{L}^d;\mathbb{Z})\oplus B^{n-1}(\mathbb{L}^d))=\frac{H^{n-1}(\mathbb{L}^d;\mathbb{R})}{H^{n-1}(\mathbb{L}^d;\mathbb{Z})}\oplus\Omega^{n-1}_\mathrm{coexact}(\mathbb{L}^d)=H^{n-1}(\mathbb{L}^d;\mathbb{Z}/\mathbb{R})\oplus\Omega^{n-1}_\mathrm{coexact}(\mathbb{L}^d)$, we can see that we just need to show that the equation $[(\delta_{-1}\beta_1+\delta_{-1}\alpha_1,0,0)]\neq[(\delta_{-1}\beta_2+\delta_{-1}\alpha_2,0,0)]$ holds for all $[\beta_1]\oplus\alpha_1\neq[\beta_2]\oplus\alpha_2 \in H^{n-1}(\mathbb{L}^d;\mathbb{Z}/\mathbb{R})\oplus\Omega^{n-1}_\mathrm{coexact}(\mathbb{L}^d)$.

Let us think about it case by case. First, consider the case of $\alpha_1\neq\alpha_2$. From the equality $\Omega^{n-1}(\mathbb{L}^d)=Z^{n-1}(\mathbb{L}^d)\oplus\Omega^{n-1}_\mathrm{coexact}(\mathbb{L}^d)$, if $\alpha_1\neq\alpha_2$ holds, then $d\alpha_1\neq d\alpha_2$ holds. From $d\beta_1=d\beta_2=0$, we can see that $d[(\delta_{-1}\beta_1+\delta_{-1}\alpha_1,0,0)]\neq d[(\delta_{-1}\beta_2+\delta_{-1}\alpha_2,0,0)]$ holds, therefore $[(\delta_{-1}\beta_1+\delta_{-1}\alpha_1,0,0)]\neq [(\delta_{-1}\beta_2+\delta_{-1}\alpha_2,0,0)]$ holds.

Next, consider the case where $\alpha_1=\alpha_2$ and $\beta_1\neq\beta_2$. From $\beta_1\neq\beta_2$, we see there exists $i\in\{1,2,3,\dots ,d\}$ such that $\int_{\gamma_i}\beta_1\neq\int_{\gamma_i}\beta_2$, where $\gamma_i$ is a cycle of $\mathbb{L}^d$ which winds a torus $\mathbb{L}^3$ once in the $\hat{i}$ direction. Note that $\int_{\gamma_i}\alpha_1=\int_{\gamma_i}\alpha_2$, we can see $\int_{\gamma_i}[(\delta_{-1}\beta_1+\delta_{-1}\alpha_1,0,0)]\neq \int_{\gamma_i}[(\delta_{-1}\beta_2+\delta_{-1}\alpha_2,0,0)]$, therefore $[(\delta_{-1}\beta_1+\delta_{-1}\alpha_1,0,0)]\neq [(\delta_{-1}\beta_2+\delta_{-1}\alpha_2,0,0)]$ holds.

The above discussion shows that $(\Omega^{n-1}(\mathbb{L}^d)/\Omega_\mathrm{closed}^{n-1}(\mathbb{L}^d)_\mathbb{Z})\hookrightarrow  H^{n}_\mathrm{lat.DB}(\mathbb{L}^d)\;,\;\omega\mapsto(\delta_{-1}\omega,0,0)$ is a well-defined injective mapping.

\subsubsection{Definition of surjective map $H^{n}_\mathrm{lat.DB}(\mathbb{L}^d)\twoheadrightarrow H^{n}(\mathbb{L}^d;\mathbb{Z})$}\label{sec812}
The mapping can be defined by $H^{n}_\mathrm{lat.DB}(\mathbb{L}^d)\twoheadrightarrow H^{n}(\mathbb{L}^d;\mathbb{Z})\;,\;[(\omega^{(d-1,0)},\dots,\omega^{(-1,d)})]\mapsto[d\omega^{(d-1,0)}]$.\footnote{Usually, it is defined by $[(\omega^{(d-1,0)},\dots,\omega^{(-1,d)})]\mapsto \omega^{(-1,d)}\in \check{H}^n(\mathbb{L}^d;\mathbb{Z})$. Since $\mathbb{L}^d$ has a property $\check{H}^n(\mathbb{L}^d;\mathbb{Z})=H^n(\mathbb{L}^d;\mathbb{Z})\cong\Omega^{n}_\mathrm{closed}(\mathbb{L}^d)_\mathbb{Z}/B^{n-1}(\mathbb{L}^d)$, these two definitions are equivalent. In general, this argument does not hold if the cohomology of the base space has a torsion part.} This mapping is surjective because, for all $\tilde{\omega}\in \Omega^d_\mathrm{closed}(\mathbb{L}^d)_{\mathbb{Z}}$, there exists DB $d$-cocycle $(\omega^{(d-1,0)},\dots,\omega^{(-1,d)})$ that satisfies $\delta_{-1}\tilde{\omega}=d\omega^{(d-1,0)}$. This proposition can be shown as follows. First, let us decompose $\omega$ into the non-trivial $\mathbb{Z}$-cochain part and the trivial part $\tilde{\omega} = \omega' + \omega''$. Then, since $\omega''$ is trivial, there exists $(d-1)$-form $\eta''$ such that $d\eta'' = \omega''$. Furthermore, a DB $d$-cocycle $\bm{\eta}'$ such that $d\bm{\eta}' = \omega'$ can be constructed using the method of \eqref{equ9040}. Then we can define $(\omega^{(d-1,0)},\dots,\omega^{(-1,d)})=\bm{\eta}'+(\delta_{-1}\omega'',0,0\cdots).$

The completeness of $(\Omega^{n-1}(\mathbb{L}^d)/\Omega_\mathrm{closed}^{n-1}(\mathbb{L}^d)_\mathbb{Z})\hookrightarrow  H^{n}_\mathrm{lat.DB}(\mathbb{L}^d)\twoheadrightarrow H^{n}(\mathbb{L}^d;\mathbb{Z})$ follows from $[\beta]\oplus\alpha\hookrightarrow [(\delta_{-1}\beta+\delta_{-1}\alpha,0,\dots,0)]\twoheadrightarrow [d\alpha]=0$. 

\subsubsection{Decomposition of $H^2_\mathrm{lat.DB}(\mathbb{L}^3)$}\label{sec813}
From the above discussion of short exact sequences, we get
\begin{equation}\label{equ8030}
    H^2_\mathrm{lat.DB}(\mathbb{L}^3)=\frac{\Omega^1(\mathbb{L}^3)}{\Omega^1_\mathrm{closed}(\mathbb{L}^3)_\mathbb{Z}}\oplus H^2(\mathbb{L}^3;\mathbb{Z}),
\end{equation}
and considering the Hodge decomposition in the Appendix~\ref{secA10}, we find
\begin{equation}\label{equ8040}
    H^2_\mathrm{lat.DB}(\mathbb{L}^3)=\Omega_\mathrm{coexact}^1(\mathbb{L}^3)\oplus H^1(\mathbb{L}^3;\mathbb{R}/\mathbb{Z})\oplus H^2(\mathbb{L}^3;\mathbb{Z}).
\end{equation}
This expression can be interpreted as the space of gauge-fixed configurations of $U(1)$ connections on $\mathbb{L}^3$.
\newpage
\section{Pontrjagin duality and linking number on $\mathbb{L}^3$}\label{sec900}
In this section, we discuss the Pontrjagin duality on $H^2_\mathrm{lat.DB}(\mathbb{L}^3)$. In the continuum, defining the Pontrjagin dual in DB cohomology requires distributional formulations, but on the lattice, we can construct it without them. The wedge/cup product on a lattice regulates the canonical framing of the Wilson line, enabling a treatment of (self-) linking numbers on the lattice.
\subsection{DB Pontrjagin dual of $\tilde{\gamma}$ when $[\tilde{\gamma}]=0\in H_1(\tilde{\mathbb{L}}^3)$}\label{sec910}
\begin{figure}
\centering
\begin{minipage}[t]{0.45\textwidth}
\includegraphics[width=\linewidth]{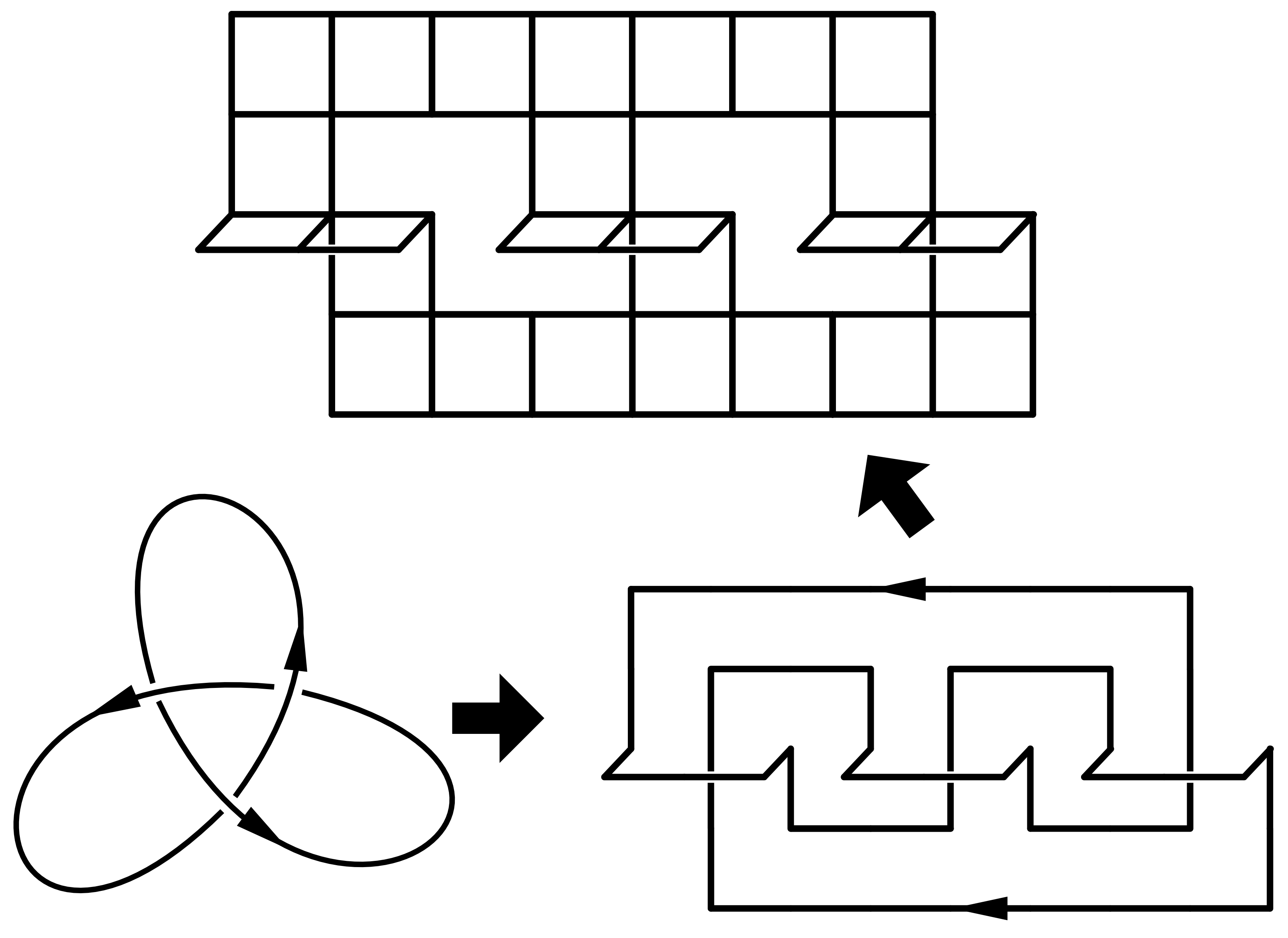}
\caption{Seifert surface of the trefoil knot on square lattice}\label{fig:Sei3}
\end{minipage}
\hspace{0.08\textwidth}
\begin{minipage}[t]{0.45\textwidth}
\includegraphics[width=\linewidth]{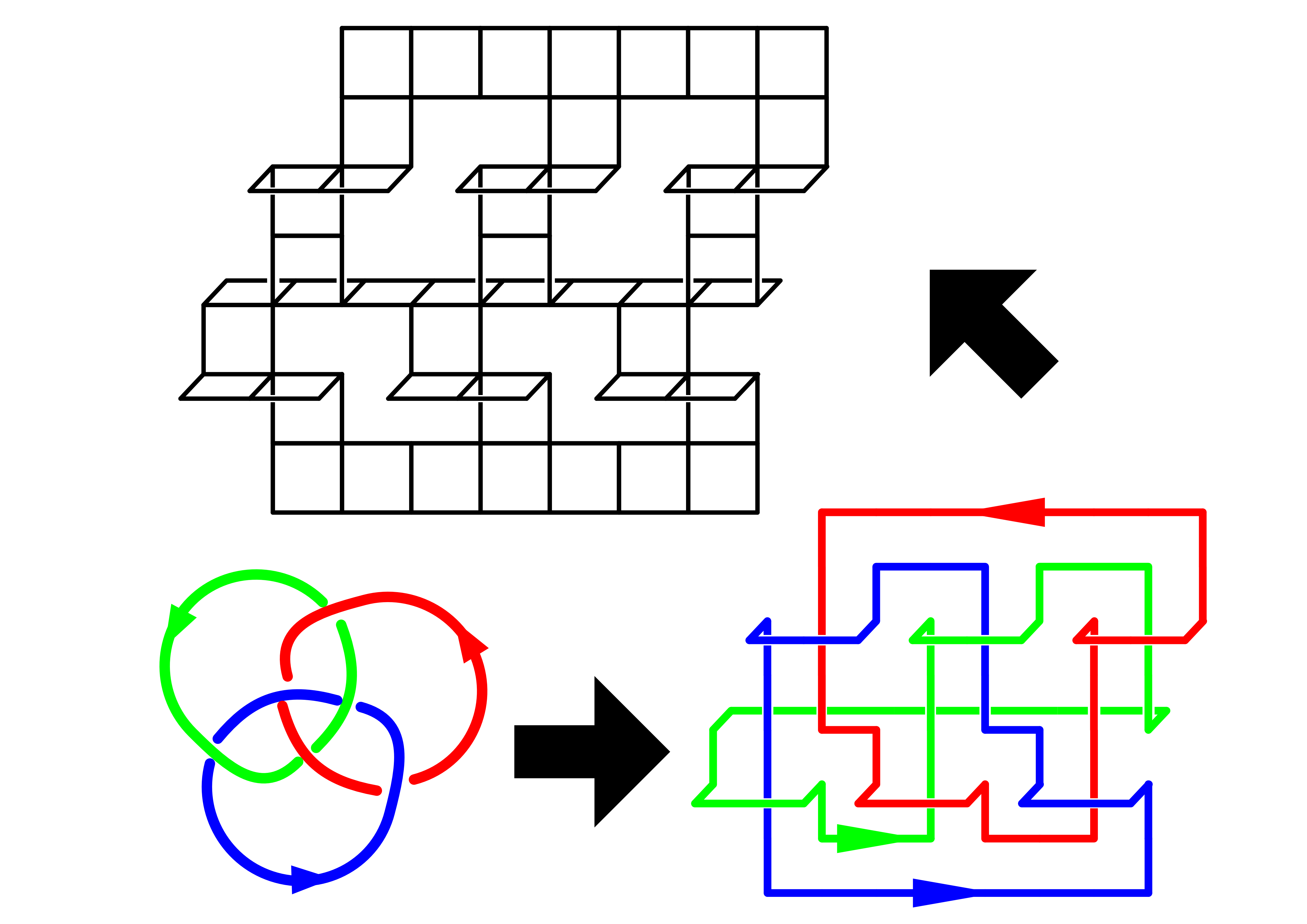}
\caption{Seifert surface of borromean ring on square lattice}\label{fig:Sei6}
\end{minipage}
\end{figure}

Let $\tilde{\gamma}$ be a 1-cycle of $\mathbb{Z}$-coefficient cohomology on $\tilde{\mathbb{L}}^3$. There exists $\tilde{\Sigma}\in C_2(\tilde{\mathbb{L}}^3;\mathbb{Z})$ such that $\tilde{\gamma}=\partial\tilde{\Sigma}$ holds. $\tilde{\Sigma}$ is called as a Seifert surface of $\gamma$. For explicit examples of Seifert surfaces on a square lattice and their illustrations, see Figures~\ref{fig:Sei3}~and~\ref{fig:Sei6}. There exists a Poincar\'{e} dual $\mathrm{PD}(\tilde{\Sigma})\in C^1(\mathbb{L}^3)$ of $\tilde{\Sigma}$, which is a $1$-cocycle on the lattice corresponding to the right-handed normal vector field on $\tilde{\Sigma}$. We can see that $d\mathrm{PD}(\tilde{\Sigma})=\mathrm{PD}(\tilde{\gamma})$ holds, where $\mathrm{PD}(\tilde{\gamma})$ is defined in \eqref{equ3171}. Let lattice DB cocycle $\bm{\eta}_{\tilde{\gamma}}\in H^2_\mathrm{lat.DB}(\mathbb{L}^3)$ be defined by
\begin{equation}\label{equ9010}
    \bm{\eta}_{\tilde{\gamma}}=(\delta_{-1}\mathrm{PD}(\tilde{\Sigma}),0,0).
\end{equation}
$\bm{\eta}_{\tilde{\gamma}}$ satisfies for all $\bm{A}=(A,\Lambda,n)\in H^2_\mathrm{lat.DB}(\mathbb{L}^3)$,
\begin{equation}\label{equ9020}
    \begin{split}
        &\int_{\mathbb{L}^3}\bm{\eta}_{\tilde{\gamma}}\star\bm{A}=\int_{\mathbb{L}^3}\bm{\eta}_{\tilde{\gamma}}\overleftarrow{\underrightarrow{\star}}\bm{A}\\
        =&\int_{\mathbb{L}^3}(d\delta_{-1}\mathrm{PD}(\tilde{\Sigma})\stackrel{\smallsmile}{\wedge}A,d\delta_{-1}\mathrm{PD}(\tilde{\Sigma})\stackrel{\smallsmile}{\wedge}\Lambda,\delta_{-1}\mathrm{PD}(\tilde{\Sigma})\stackrel{\smallsmile}{\wedge}dn,0,0)\\
        =&\int_{\mathbb{L}^3}(\delta_{-1}\mathrm{PD}(\tilde{\gamma})\stackrel{\smallsmile}{\wedge}A,\delta_{-1}\mathrm{PD}(\tilde{\gamma})\stackrel{\smallsmile}{\wedge}\Lambda,\delta_{-1}\mathrm{PD}(\tilde{\Sigma})\stackrel{\smallsmile}{\wedge}dn,0,0)\stackrel{\mathbb{Z}}{=}\int_{\tilde{\gamma}_+} \bm{A},
    \end{split}
\end{equation}
and
\begin{equation}\label{equ9030}
    \begin{split}
        &\int_{\mathbb{L}^3}\bm{A}\star\bm{\eta}_{\tilde{\gamma}}=\int_{\mathbb{L}^3}\bm{A}\overrightarrow{\underleftarrow{\star}}\bm{\eta}_{\tilde{\gamma}}\\
        =&\int_{\mathbb{L}^3}(A\stackrel{\smallsmile}{\wedge}d\delta_{-1}\mathrm{PD}(\tilde{\Sigma}),\Lambda\stackrel{\smallsmile}{\wedge}d\delta_{-1}\mathrm{PD}(\tilde{\Sigma}),dn\stackrel{\smallsmile}{\wedge}\delta_{-1}\mathrm{PD}(\tilde{\Sigma}),0,0)\\
        =&\int_{\mathbb{L}^3}(A\stackrel{\smallsmile}{\wedge}\delta_{-1}\mathrm{PD}(\tilde{\gamma}),\Lambda\stackrel{\smallsmile}{\wedge}\delta_{-1}\mathrm{PD}(\tilde{\gamma}),dn\stackrel{\smallsmile}{\wedge}\delta_{-1}\mathrm{PD}(\tilde{\Sigma}),0,0)\stackrel{\mathbb{Z}}{=}\int_{\tilde{\gamma}_-} \bm{A}.
    \end{split}
\end{equation}
Note that $\mathrm{PD}(\tilde{\Sigma})$ and $n$ is $\mathbb{Z}$-valued. $\tilde{\gamma}_\pm$ denotes the cycle of the lattice obtained by translating the cycle $\tilde{\gamma}$ by $\pm\frac{1}{2}\hat{1}\pm\frac{1}{2}\hat{2}\pm\frac{1}{2}\hat{3}$.

$\bm{\eta}_{\tilde{\gamma}}$ is a concept very similar to the Poincar\'{e} dual, but since the pairing is $\mathbb{R}/\mathbb{Z}$-valued, it is called the Pontrjagin dual of $\tilde{\gamma}$.

It should be noted that the above discussion does not essentially require the formulation of a lattice DB cocycle, but the essential argument lies in the construction of $\mathrm{PD}(\tilde{\Sigma})$ on the lattice cochain theory.
\subsection{DB Pontrjagin dual of $\tilde{\gamma}$ when $[\tilde{\gamma}]\neq 0\in H_1(\tilde{\mathbb{L}}^3)$}\label{sec920}
If $[\tilde{\gamma}]\neq 0\in H_1(\mathbb{L}^3)$, then there does not exist $\tilde{\Sigma}\in C^1(\mathbb{L}^3)$ such that $d\tilde{\Sigma}=\mathrm{PD}(\tilde{\gamma})$. The Seifert surface does not exist because it cannot cross a ``hole'' in the torus.

We consider constructing $\bm{\eta}_{\tilde{\gamma}}$ for any $\tilde{\gamma}$. The method employed here is called \v{C}ech--de Rham descent.

First, define the $\mathbb{Z}$-valued \v{C}ech--de Rham chain $\omega^{(1,0)}$ to be $d\omega^{(1,0)}=\delta_{-1}\mathrm{PD}(\tilde{\gamma})$. Such an $\omega^{(1,0)}$ always exists, because $d\mathrm{PD}(\tilde{\gamma})=0$ implies $d\delta_{-1}\mathrm{PD}(\tilde{\gamma})=0$, and $d\omega^{(1,0)}(\bullet;\bullet;\tilde{x})=\delta_{-1}\mathrm{PD}(\tilde{\gamma})(\bullet;\bullet;\tilde{x})$ has a solution whose existence is guaranteed by the triviality of $1$-cohomology of a $3$d cube.

Next, we define the $\mathbb{Z}$-valued \v{C}ech--de Rham chain $\omega^{(0,1)}$ to be $\delta_0\omega^{(1,0)}=d_{0}\omega^{(0,1)}$. Such an $\omega^{(0,1)}$ always exists because $d\left(\delta_0\omega^{(1,0)}\right)=\delta_0\delta_{-1}\mathrm{PD}(\tilde{\gamma})=0$, and $0$-cohomology of a $2$d cube is trivial.

Finally, we define the $\mathbb{Z}$-valued \v{C}ech--de Rham chain $\omega^{(-1,2)}$ to be $\delta_1\omega^{(0,1)}=-d_{-1}\omega^{(-1,2)}$. Such an $\omega^{(-1,2)}$ always exists because $d\left(\delta_1\omega^{(0,1)}\right)=\delta_1\delta_{0}\omega^{(1,0)}=0$, and the $0$-form whose exterior derivative is equal to $0$ is a constant function.

Using $\omega^{(1,0)},\omega^{(0,1)},\omega^{(-1,2)}$, we can give the definition of the lattice DB cocycle $\bm{\eta}_{\tilde{\gamma}}$ by
\begin{equation}\label{equ9040}
    \bm{\eta}_{\tilde{\gamma}}=(\omega^{(1,0)},\omega^{(0,1)},\omega^{(-1,2)}).
\end{equation}
$\bm{\eta}_{\tilde{\gamma}}$ satisfies an equality for all $\bm{A}=(A,\Lambda,n)\in H^2_\mathrm{lat.DB}(\mathbb{L}^3)$,
\begin{equation}\label{equ9050}
    \begin{split}
        &\int_{\mathbb{L}^3}\bm{\eta}_{\tilde{\gamma}}\star\bm{A}=\int_{\mathbb{L}^3}\bm{\eta}_{\tilde{\gamma}}\overleftarrow{\underrightarrow{\star}}\bm{A}\\
        =&\int_{\mathbb{L}^3}(d\omega^{(1,0)}\stackrel{\smallsmile}{\wedge}A,d\omega^{(1,0)}\stackrel{\smallsmile}{\wedge}\Lambda,\omega^{(1,0)}\stackrel{\smallsmile}{\wedge}dn,\omega^{(0,1)}\stackrel{\smallsmile}{\wedge}dn,\omega^{(-1,2)}\stackrel{\smallsmile}{\wedge}dn)\\
        =&\int_{\mathbb{L}^3}(\delta_{-1}\mathrm{PD}(\tilde{\gamma})\stackrel{\smallsmile}{\wedge}A,\delta_{-1}\mathrm{PD}(\tilde{\gamma})\stackrel{\smallsmile}{\wedge}\Lambda,\omega^{(1,0)}\stackrel{\smallsmile}{\wedge}dn,\omega^{(0,1)}\stackrel{\smallsmile}{\wedge}dn,\omega^{(-1,2)}\stackrel{\smallsmile}{\wedge}dn)\stackrel{\mathbb{Z}}{=}\int_{\tilde{\gamma}_+} \bm{A},
    \end{split}
\end{equation}
and
\begin{equation}\label{equ9060}
    \begin{split}
        &\int_{\mathbb{L}^3}\bm{A}\star\bm{\eta}_{\tilde{\gamma}}=\int_{\mathbb{L}^3}\bm{A}\overrightarrow{\underleftarrow{\star}}\bm{\eta}_{\tilde{\gamma}}\\
        =&\int_{\mathbb{L}^3}(A\stackrel{\smallsmile}{\wedge}d\omega^{(1,0)},\Lambda\stackrel{\smallsmile}{\wedge}d\omega^{(1,0)},dn\stackrel{\smallsmile}{\wedge}\omega^{(1,0)},dn\stackrel{\smallsmile}{\wedge}\omega^{(0,1)},dn\stackrel{\smallsmile}{\wedge}\omega^{(-1,2)})\\
        =&\int_{\mathbb{L}^3}(A\stackrel{\smallsmile}{\wedge}\delta_{-1}\mathrm{PD}(\tilde{\gamma}),\Lambda\stackrel{\smallsmile}{\wedge}\delta_{-1}\mathrm{PD}(\tilde{\gamma}),dn\stackrel{\smallsmile}{\wedge}\omega^{(1,0)},dn\stackrel{\smallsmile}{\wedge}\omega^{(0,1)},dn\stackrel{\smallsmile}{\wedge}\omega^{(-1,2)})\stackrel{\mathbb{Z}}{=}\int_{\tilde{\gamma}_-} \bm{A}.
    \end{split}
\end{equation}
Note again that $\mathrm{PD}(\tilde{\Sigma})$ and $n$ is $\mathbb{Z}$-valued, and $\tilde{\gamma}_\pm$ denotes the cycle obtained by translating the cycle $\tilde{\gamma}$ by $\pm\frac{1}{2}\hat{1}\pm\frac{1}{2}\hat{2}\pm\frac{1}{2}\hat{3}$.

The above discussion essentially requires the formulation of the lattice DB cocycle, and plays a crucial role in subsequent sections.
\subsection{A simple example of lattice DB Pontrjagin dual}\label{sec930}
For example, let us construct the Pontrjagin dual of a simple closed curve
\begin{equation}\label{equ9070}
    \tilde{\gamma}=\sum_{s=0}^{L-1}\left(\left(\frac{1}{2},\frac{1}{2},s+\frac{1}{2}\right);3\right)\in Z_1(\tilde{\mathbb{L}}^3).
\end{equation}
Since this cycle belongs to a nontrivial homology class on $\tilde{\mathbb{L}}^3$, we cannot apply the scheme of Seifert surface, so there is no solution $\omega\in\Omega^1(\mathbb{L}^3)$ for $d\omega=\mathrm{PD}(\tilde{\gamma})$, where
\begin{equation}\label{equ9080}
    (\mathrm{PD}(\tilde{\gamma}))(x;\mu,\nu)=\delta_{0,x}\epsilon_{\mu,\nu,3}.
\end{equation}

Let us first consider solving $d\eta^{(1,0)}=\delta_{-1}(\mathrm{PD}(\tilde{\gamma}))$ in each patch. We can find one of the solution
\begin{equation}\label{equ9090}
    \eta^{(1,0)}((x_1,x_2,x_3);\mu;(\tilde{x}_1,\tilde{x}_2,\tilde{x}_3))=\delta_{x_1,1}\delta_{x_2,0}\delta_{\mu,2}\delta_{\tilde{x}_1,\frac{1}{2}}\delta_{\tilde{x}_2,\frac{1}{2}}(\delta_{\tilde{x}_3,x_3-\frac{1}{2}}+\delta_{\tilde{x}_3,x_3+\frac{1}{2}}).
\end{equation}
If we compute $\delta_{0}\eta^{(1,0)}$ explicitly, we obtain
\begin{equation}\label{equ9100}
    \left\{\begin{array}{l}
        \delta_{0}\eta^{(1,0)}((1,0,s);2;(1/2,1/2,s+(1/2)),(3/2,1/2,s+(1/2)))=-1\;\;(0\leq s\leq L-1),\\
        \delta_{0}\eta^{(1,0)}((1,0,s+1);2;(1/2,1/2,s+(1/2)),(3/2,1/2,s+(1/2)))=-1\;\;(0\leq s\leq L-1),\\
        \delta_{0}\eta^{(1,0)}((1,0,s+1);2;(1/2,1/2,s+(1/2)),(3/2,1/2,s+(3/2)))=-1\;\;(0\leq s\leq L-1),\\
        \delta_{0}\eta^{(1,0)}((1,0,s);2;(1/2,1/2,s+(1/2)),(3/2,1/2,s-(1/2)))=-1\;\;(0\leq s\leq L-1),\\
        \delta_{0}\eta^{(1,0)}=0\;\;(\mathrm{otherwise}).
    \end{array}\right.
\end{equation}
We can find a solution of $d_0\eta^{(0,1)}=\delta_{0}\eta^{(1,0)}$ as follows:
\begin{equation}\label{equ9110}
    \left\{\begin{array}{l}
        \eta^{(0,1)}((1,0,s);;(1/2,1/2,s+(1/2)),(3/2,1/2,s+(1/2)))=1\;\;(0\leq s\leq L-1),\\
        \eta^{(0,1)}((1,0,s+1);;(1/2,1/2,s+(1/2)),(3/2,1/2,s+(1/2)))=1\;\;(0\leq s\leq L-1),\\
        \eta^{(0,1)}((1,0,s);;(1/2,1/2,s+(1/2)),(3/2,1/2,s+(3/2)))=1\;\;(0\leq s\leq L-1),\\
        \eta^{(0,1)}((1,0,s+1);;(1/2,1/2,s+(1/2)),(3/2,1/2,s-(1/2)))=1\;\;(0\leq s\leq L-1),\\
        \eta^{(0,1)}=0\;\;(\mathrm{otherwise}).
    \end{array}\right.
\end{equation}
Then we get 
\begin{equation}\label{equ9120}
    \left\{\begin{array}{l}
        \delta_1\eta^{(0,1)}((1,0,s);;(1/2,1/2,s\pm(1/2)),(3/2,1/2,s\pm(1/2)),((1/2)\pm 1,L-1/2,s\pm(1/2))),\\
        =1\;\;(0\leq s\leq L-1)\\
        \\
        \delta_1\eta^{(0,1)}=0\;\;(\mathrm{otherwise}),
    \end{array}\right.
\end{equation}
and can easily find a solution $\eta^{(-1,2)}$ of equation $d_{-1}\eta^{(-1,2)}=\delta_{1}\eta^{(0,1)}$. It is obvious that each $\eta^{(1,0)},\eta^{(0,1)},\eta^{(-1,2)}$ are $\mathbb{Z}$-valued. A direct calculation also shows that $\delta_2\eta^{(-1,2)}=0$. Then we complete the definition of $\bm{\eta}_{\tilde{\gamma}}$ by
\begin{equation}\label{equ9130}
    \bm{\eta}_{\tilde{\gamma}}=(\eta^{(1,0)},\eta^{(0,1)},\eta^{(-1,2)}).
\end{equation}
\subsection{Linking number of cycle}\label{sec940}
Next, let us discuss linking numbers. Consider $\gamma\in Z_1(\mathbb{L}^3;\mathbb{Z}),\tilde{\gamma}\in Z_1(\tilde{\mathbb{L}}^3;\mathbb{Z})$ which satisfy $[\gamma]=0\in H_1(\mathbb{L}^3;\mathbb{Z})$ and $[\tilde{\gamma}]=0\in H_1(\tilde{\mathbb{L}}^3;\mathbb{Z})$. A linking number of $\gamma$ and $\tilde{\gamma}$ should be well-defined because they never have common points.

According to knot theory, the linking number of $\gamma,\tilde{\gamma}$ is usually defined by the number of times $\gamma$ penetrates the Seifert surface $\tilde{\Sigma}$ of $\tilde{\gamma}$. Or equivalently, the linking number is defined as the number of times $\tilde{\gamma}$ penetrates the Seifert surface $\tilde{\Sigma}$ of $\gamma$. So, we should define linking number of $\gamma,\tilde{\gamma}$, denoted $\mathrm{link}(\gamma,\tilde{\gamma})$, by
\begin{equation}\label{equ9140}
    \mathrm{link}(\gamma,\tilde{\gamma})=\mathrm{link}(\tilde{\gamma},\gamma)=\int_\gamma \mathrm{PD}(\tilde{\Sigma})=\int_{\tilde{\gamma}} \mathrm{PD}(\Sigma).
\end{equation}
With this definition, as expected, a linking number is defined for any $[\gamma]=0\in H_1(\mathbb{L}^3;\mathbb{Z})$ and $[\tilde{\gamma}]=0\in H_1(\tilde{\mathbb{L}}^3;\mathbb{Z})$.

Next, let us consider the situation where both curves belong to the dual lattice which satisfies $[\tilde{\gamma}^1]=[\tilde{\gamma}^2]=0\in H_1(\tilde{\mathbb{L}}^3)$ and $(\mathrm{supp}\,\gamma^1)\cap(\mathrm{supp}\,\gamma^2)=\emptyset$. In this case, $\gamma^1$ and $\gamma^2$ are separated by an $L^\infty$-distance of at least $1$, and hence the linking number can be defined by
\begin{equation}\label{equ9150}
    \mathrm{link}(\tilde{\gamma}^1,\tilde{\gamma}^2):=\mathrm{link}(\tilde{\gamma}^1_\pm,\tilde{\gamma}^2)=\mathrm{link}(\tilde{\gamma}^1,\tilde{\gamma}^2_\pm).
\end{equation}
With this definition,
\begin{equation}\label{equ9160}
    \begin{split}
        &\mathrm{link}(\tilde{\gamma}^1,\tilde{\gamma}^2)\\
        =&\mathrm{link}(\tilde{\gamma}^1_+,\tilde{\gamma}^2)=\int_{\tilde{\gamma}^1_+}\mathrm{PD}(\tilde{\Sigma}^2)
        =\int_{\mathbb{L}^3}\bm{\eta}_{\tilde{\gamma}_1}\overleftarrow{\underrightarrow{\star}}\bm{\eta}_{\tilde{\gamma}_2}\\
        =&\mathrm{link}(\tilde{\gamma}^1_-,\tilde{\gamma}^2)=\int_{\tilde{\gamma}^1_-}\mathrm{PD}(\tilde{\Sigma}^2)
        =\int_{\mathbb{L}^3}\bm{\eta}_{\tilde{\gamma}_2}\overrightarrow{\underleftarrow{\star}}\bm{\eta}_{\tilde{\gamma}_1}\\
        =&\mathrm{link}(\tilde{\gamma}^1,\tilde{\gamma}^2_+)=\int_{\tilde{\gamma}^2_+}\mathrm{PD}(\tilde{\Sigma}^1)
        =\int_{\mathbb{L}^3}\bm{\eta}_{\tilde{\gamma}_2}\overleftarrow{\underrightarrow{\star}}\bm{\eta}_{\tilde{\gamma}_1}\\
        =&\mathrm{link}(\tilde{\gamma}^1,\tilde{\gamma}^2_-)=\int_{\tilde{\gamma}^2_-}\mathrm{PD}(\tilde{\Sigma}^1)
        =\int_{\mathbb{L}^3}\bm{\eta}_{\tilde{\gamma}_1}\overrightarrow{\underleftarrow{\star}}\bm{\eta}_{\tilde{\gamma}_2}
    \end{split}
\end{equation}
holds, then we obtain\footnote{Since $\star$ defined on the lattice is not graded commutative, this formula does not hold in general. In this case, it holds because $(\mathrm{supp}\,\tilde{\gamma}^1)\cap(\mathrm{supp}\,\tilde{\gamma}^2)=\emptyset$.}
\begin{equation}\label{equ9170}
        \mathrm{link}(\tilde{\gamma}^1,\tilde{\gamma}^2)
        =\int_{\mathbb{L}^3}\bm{\eta}_{\tilde{\gamma}_1}\star\bm{\eta}_{\tilde{\gamma}_2}
        =\int_{\mathbb{L}^3}\bm{\eta}_{\tilde{\gamma}_2}\star\bm{\eta}_{\tilde{\gamma}_1}.
\end{equation}

Finally, let us make some comments on the gauge transformation of the linking numbers we have defined so far. As mentioned above, $\star$ is modulo-$\mathbb{Z}$ invariant under gauge transformations, but the linking number defined here cannot allow the $\mathbb{Z}$-indeterminacy, and is therefore by no means invariant under gauge transformations. In fact, the linking number is invariant under $\mathbb{R}$-valued ordinary gauge transformations, but in general realizes all possible indeterminacy of integers under $\mathbb{Z}$-valued large gauge transformations. In other words, it is possible to change the linking number to any value by a large gauge transformation. Therefore, we should keep in mind that the above discussion is about a specific DB cocycle $\bm{\eta}_\gamma=(\mathrm{PD}(\Sigma),0,0)$. This problem is solved by introducing mod $2k$ linking numbers, which will be discussed in Section~\ref{sec960}.
\subsection{Change of Seifert surface and well-definedness of mod $2k$ linking number}\label{sec950}
First, we introduce notations
\begin{align}
    &H_1(\mathbb{L}^3;\mathbb{Z})=\mathrm{Span}_\mathbb{Z}\{[e_1],[e_2],[e_3]\},&&H_2(\mathbb{L}^3;\mathbb{Z})=\mathrm{Span}_\mathbb{Z}\{[e_{23}],[e_{31}],[e_{12}]\},\label{equ9180}\\
    &e_i=\sum_{s=0}^{L-1}(s\hat{i};i),&&e_{jk}=\sum_{s=0}^{L-1}\sum_{t=0}^{L-1}(s\hat{j}+t\hat{k};j,k),\label{equ9190}\\
    &H_1(\tilde{\mathbb{L}}^3;\mathbb{Z})=\mathrm{Span}_\mathbb{Z}\{[\tilde{e}_1],[\tilde{e}_2],[\tilde{e}_3]\},&&H_2(\tilde{\mathbb{L}}^3;\mathbb{Z})=\mathrm{Span}_\mathbb{Z}\{[\tilde{e}_{23}],[\tilde{e}_{31}],[\tilde{e}_{12}]\},\label{equ9200}\\
    &\tilde{e}_i=\sum_{s=0}^{L-1}((1/2,1/2,1/2)+s\hat{i};i),&&\tilde{e}_{jk}=\sum_{s=0}^{L-1}\sum_{t=0}^{L-1}((1/2,1/2,1/2)+s\hat{j}+t\hat{k};j,k).\label{equ9210}
\end{align}
Recall that the linking number we have discussed so far is restricted to the situation
\begin{align}\label{equ9220}
    &[\gamma]=0\in H_1(\mathbb{L}^3;\mathbb{Z}),&&[\tilde{\gamma}]=0\in H_1(\tilde{\mathbb{L}}^3;\mathbb{Z}).
\end{align}
This limitation is due to the ambiguity in choosing the Seifert surface.

For example, let us assume that $[\gamma]=0,[\tilde{\gamma}]=\tilde{e}_1$. In this case, a Seifert surface $\Sigma$ can be defined for $\gamma$, but if we define $\Sigma'=\Sigma+\partial V_{(3)}+e_{23}$, then $\partial\Sigma=\partial\Sigma'$ holds, and therefore $\Sigma'$ is also suitable as a Seifert surface of $\gamma$. The term $\partial V_{(3)}\in B_2(\mathbb{L}^3;\mathbb{Z})$ does not affect the calculation of the linking number, but $e_{23}\in H_2(\mathbb{L}^3;\mathbb{Z})$ causes changes in the linking number. In fact, we can see how the linking number changes from 
\begin{equation}\label{equ9230}
    \int_{\tilde{\gamma}}\mathrm{PD}(\Sigma')=\int_{\tilde{e}_1}(\mathrm{PD}(\Sigma)+\mathrm{PD}(e_{23}))=1+\int_{\tilde{\gamma}}\mathrm{PD}(\Sigma).
\end{equation}
In this case, replacing the Seifert surface changes the linking number by $1$. The reason why this happens is that $\tilde{\gamma}$ belongs to a nontrivial homology.

In order to give a well-defined linking number for closed curves belonging to non-trivial homology, we should consider providing a definition of linking number that is well-defined for any Seifert surface replacement, which leads to the idea of the mod $2k$ linking number.

From now on, we consider the situation
\begin{align}\label{equ9240}
    &[\gamma]=\sum_ic_i[e_i]\in H_1(\mathbb{L}^3;2k\mathbb{Z}),&&[\tilde{\gamma}]=\sum_j\tilde{c}_j[\tilde{e}_j]\in H_1(\tilde{\mathbb{L}}^3;2k\mathbb{Z}),
\end{align}
for $2k>0$.\footnote{It is also possible to consider mod $K$ linking numbers for general integers $K\in \mathbb{Z}_{>0}$ instead of $2k$. On the other hand, in the discussion of mod $4k$ self-linking numbers that appears later, it seems that mod $2K$ self-linking numbers cannot be defined in the same way. For this reason, in this paper, we will only consider mod $2k$ linking numbers.} By definition, it follows that $c_i,\tilde{c}_i\in 2k\mathbb{Z}$. There exists $\Sigma,\Sigma'$ such that
\begin{align}\label{equ9250}
    &\gamma=\partial \Sigma+\sum_ic_ie_i,&&\tilde{\gamma}=\partial\tilde{\Sigma}+\sum_j\tilde{c}_j\tilde{e}_j.
\end{align}
From $\mathbb{Z}$-linearlity of linking number and $c,\tilde{c}\in 2k\mathbb{Z}$, we would assume 
\begin{equation}\label{equ9260}
    \begin{split}
        \mathrm{link}(\gamma,\tilde{\gamma})
        &\;\;\;\text{``}=\text{''}\;\;\;
        \mathrm{link}\left(\partial \Sigma,\partial\tilde{\Sigma}\right)
        +\mathrm{link}\left(\partial \Sigma,\sum_j\tilde{c}_j\tilde{e}_j\right)\\
        &\qquad+\mathrm{link}\left(\sum_ic_ie_i,\partial\tilde{\Sigma}\right)
        +\mathrm{link}\left(\sum_ic_ie_i,\sum_j\tilde{c}_j\tilde{e}_j\right)\\
        &\;\;\;\text{``}\stackrel{2k\mathbb{Z}}{=}\text{''}\;\;\;
        \mathrm{link}\left(\partial \Sigma,\partial\tilde{\Sigma}\right)
    \end{split}
\end{equation}
as a formal discussion. Taking inspiration from here, we introduce the (geometric) definition of the mod $2k$ linking number of $\gamma,\tilde{\gamma}$ by
\begin{equation}\label{equ9270}
    \mathrm{link}^{\mathrm{mod}\,2k}(\gamma,\tilde{\gamma})\stackrel{2k\mathbb{Z}}{:=}\mathrm{link}\left(\partial \Sigma,\partial\tilde{\Sigma}\right)=\int_{\partial\Sigma}\mathrm{PD}(\tilde{\Sigma})=\int_{\partial\tilde{\Sigma}}\mathrm{PD}(\Sigma).
\end{equation}
This definition does not depend on the choice of the Seifert surfaces, since $\partial\Sigma,\partial\tilde{\Sigma}$ belong to trivial homology. Furthermore, $\mathrm{link}^{\mathrm{mod}\,2k}$ does not depend on the choice of the basis of $H_1$. In fact, if we change the choice of the basis of $H_1$ by
\begin{align}
    &H_1(\mathbb{L}^3;\mathbb{Z})=\mathrm{Span}([e'_1],[e'_2],[e'_3]),&&[e_1]=[e'_1],&&[e_2]=[e'_2],&&[e_3]=[e'_3],\label{equ9280}\\
    &H_1(\tilde{\mathbb{L}}^3;\mathbb{Z})=\mathrm{Span}([\tilde{e}'_1],[\tilde{e}'_2],[\tilde{e}'_3]),&&[\tilde{e}_1]=[\tilde{e}'_1],&&[\tilde{e}_2]=[\tilde{e}'_2],&&[\tilde{e}_3]=[\tilde{e}'_3],\label{equ9290}
\end{align}
there are $2$-cochains $\Sigma_i,\tilde{\Sigma}_i$ that satisfy
\begin{align}\label{equ9300}
    &e'_i=\partial\Sigma_i+e_i,&&\tilde{e}'_i=\partial\tilde{\Sigma}_i+\tilde{e}_i.
\end{align}
If we give decompositions of $\gamma,\tilde{\gamma}$ by
\begin{align}\label{equ9310}
    &\gamma=\partial \Sigma'+\sum_ic_ie'_i,&&\tilde{\gamma}=\partial\tilde{\Sigma}'+\sum_j\tilde{c}_j\tilde{e}_j',
\end{align}
the difference between $\partial\Sigma$ and $\partial\Sigma'$ is written as
\begin{align}\label{equ9320}
    &\partial\Sigma'-\partial\Sigma=-\sum_{i}c_i\partial\Sigma_i=:-2k\partial\sigma,&&\partial\tilde{\Sigma}'-\partial\tilde{\Sigma}=-\sum_{j}\tilde{c}_j\partial\tilde{\Sigma}_j=:-2k\partial\tilde{\sigma}.
\end{align}
Then we get
\begin{equation}\label{equ9330}
    \begin{split}
        \mathrm{link}(\partial\Sigma',\partial\tilde{\Sigma}')=&\mathrm{link}(\partial\Sigma,\partial\tilde{\Sigma})-\mathrm{link}(2k\partial\sigma,\partial\tilde{\Sigma})\\
        +&\mathrm{link}(\partial\Sigma,2k\partial\tilde{\sigma})+\mathrm{link}(2k\partial\sigma,2k\partial\tilde{\sigma})\\
        \stackrel{2k\mathbb{Z}}{=}&\;\mathrm{link}(\partial\Sigma,\partial\tilde{\Sigma}).
    \end{split}
\end{equation}

Another possible replacement is the replacement of the basis itself 
\begin{align}\label{equ9340}
    &e''_i=\sum_jM_{ij}\tilde{e}_j,&&\tilde{e}''_i=\sum_j\tilde{M}_{ij}\tilde{e}_j,
\end{align}
by $M,\tilde{M}\in GL(3,\mathbb{Z})$. $\partial\Sigma$ and $\partial\tilde{\Sigma}$ remain the same even after the replacement by $GL(3,\mathbb{Z})$. From
\begin{align}\label{equ9350}
    &c''_i=\sum_jM^{-1}_{ji}\tilde{c}_j,&&\tilde{c}''_i=\sum_j\tilde{M}^{-1}_{ji}\tilde{c}_j,
\end{align}
if $c_i,\tilde{c}_i$ are multiples of $2k$, $c''_i,\tilde{c}''_i$ are also multiples of $k$. Therefore, the value of $\mathrm{link}^{\mathrm{mod}\,2k}$ is invariant for all possible choices of $e_i,\tilde{e}_i$.
\subsection{Mod $2k$ linking number expressed as a star product}\label{sec960}
The mod $2k$ linking number can be calculated as the star product for a special gauge fixing. This is important for the discussion of Chern--Simons theory, which will be introduced later.

First, let us just give an overview, and we will later describe the detail. We introduce
\begin{align}\label{equ9360}
    &[\tilde{\gamma}_1],[\tilde{\gamma}_2]\in H_1(\tilde{\mathbb{L}}^3;2k\mathbb{Z}),&&(\mathrm{supp}\,\tilde{\gamma}_1)\cap(\mathrm{supp}\,\tilde{\gamma}_2)=\emptyset.
\end{align}
We consider DB cohomology class of DB Pontrjagin dual of $\tilde{\gamma}_1,\tilde{\gamma}_2$, denoted $[\bm{\eta}_{\tilde{\gamma}_1}],[\bm{\eta}_{\tilde{\gamma}_2}]$. The $\mathbb{Z}$-valued $1$-form gauge redundancy allows us to take $\bm{\eta}_{\tilde{\gamma}_1},\bm{\eta}_{\tilde{\gamma}_2}$ that satisfy
\begin{equation}\label{equ9370}
    \bm{\eta}_{\tilde{\gamma}_1},\bm{\eta}_{\tilde{\gamma}_2}\in Z_\mathrm{lat.DB}^1(\mathbb{L}^3;2k\mathbb{Z}),
\end{equation}
where $Z_\mathrm{lat.DB}^1(\mathbb{L}^3;2k\mathbb{Z})$ denotes the set of lattice DB cocycles $(\omega^{(1,0)},\omega^{(0,1)},\omega^{(-1,2)})$ with $2k\mathbb{Z}$-valued third element $\omega^{(-1,2)}$. There are several choices of $\bm{\eta}_{\tilde{\gamma}_1}, \bm{\eta}_{\tilde{\gamma}_2}$ that satisfy this condition, and the differences of these choices are described by $2k\mathbb{Z}$-valued $1$-form gauge transformations and $\mathbb{R}$-valued $0$-form gauge transformations.

Then we can divide each component of $\bm{\eta}$ by $2k$ and define $\frac{\bm{\eta}_{\tilde{\gamma}_1}}{2k}, \frac{\bm{\eta}_{\tilde{\gamma}_2}}{2k}$. An integral
\begin{equation}\label{equ9380}
    2k\int_{\mathbb{L}^3}\frac{\bm{\eta}_{\tilde{\gamma}_1}}{2k}\star\frac{\bm{\eta}_{\tilde{\gamma}_2}}{2k}
\end{equation}
is mod-$2k\mathbb{Z}$ invariant under $2k\mathbb{Z}$-valued $1$-form gauge transformations of $[\bm{\eta}]$ and $\mathbb{R}$-valued $0$-form gauge transformation of $\left[\frac{\bm{\eta}}{2k}\right]$. It can be shown that this integral corresponds to the $\mathrm{link}^{\mathrm{mod}\,2k}$ defined in \eqref{equ9270}.

Let us consider how we can choose $\bm{\eta}$ from $Z^2_\mathrm{lat.DB}(\mathbb{L}^3;2k\mathbb{Z})$ in more detail. First we can decompose $\tilde{\gamma}_1,\tilde{\gamma}_2\in Z_2(\tilde{\mathbb{L}}^3;2k\mathbb{Z})$ by
\begin{align}\label{equ9390}
    &\tilde{\gamma}_1=\partial\tilde{\Sigma}_1+\sum_j \tilde{b}_j\tilde{e}_j,&&\tilde{\gamma}_2=\partial\tilde{\Sigma}_2+\sum_j \tilde{c}_j\tilde{e}_j.
\end{align}
Next, we write $\bm{\eta}_{\tilde{\gamma}_1}$ and $\bm{\eta}_{\tilde{\gamma}_2}$ as
\begin{align}\label{equ9400}
    &\bm{\eta}_{\tilde{\gamma}_1}=\left(\delta_{-1}\mathrm{PD}(\tilde{\Sigma}_1),0,0\right)+\sum_i\tilde{b}_i\bm{\eta}_{\tilde{e}_i},&&\bm{\eta}_{\tilde{\gamma}_2}=\left(\delta_{-1}\mathrm{PD}(\tilde{\Sigma}_2),0,0\right)+\sum_j\tilde{c}_j\bm{\eta}_{\tilde{e}_j},
\end{align}
so that $\bm{\eta}_{\tilde{\gamma}_1},\bm{\eta}_{\tilde{\gamma}_2}\in Z_\mathrm{lat.DB}^1(\mathbb{L}^3;2k\mathbb{Z})$. Substituting these into the integral, we get
\begin{equation}\label{equ9410}
    \begin{split}
        &2k\int_{\mathbb{L}^3}\frac{\bm{\eta}_{\tilde{\gamma}_1}}{2k}\star\frac{\bm{\eta}_{\tilde{\gamma}_2}}{2k}\\
        =&2k\int_{\mathbb{L}^3}\left(\frac{1}{2k}\delta_{-1}\mathrm{PD}(\tilde{\Sigma}_1),0,0\right)\star\left(\frac{1}{2k}\delta_{-1}\mathrm{PD}(\tilde{\Sigma}_2),0,0\right)\\
        +&2k\int_{\mathbb{L}^3}\left(\sum_i\frac{\tilde{b}_i}{2k}\bm{\eta}_{\tilde{e}_i}\right)\star\left(\frac{1}{2k}\delta_{-1}\mathrm{PD}(\tilde{\Sigma}_2),0,0\right)\\
        +&2k\int_{\mathbb{L}^3}\left(\frac{1}{2k}\delta_{-1}\mathrm{PD}(\tilde{\Sigma}_1),0,0\right)\star\left(\sum_j\frac{\tilde{c}_j}{2k}\bm{\eta}_{\tilde{e}_j}\right)\\
        +&2k\int_{\mathbb{L}^3}\left(\sum_i\frac{\tilde{b}_i}{2k}\bm{\eta}_{\tilde{e}_i}\right)\star\left(\sum_j\frac{\tilde{c}_j}{2k}\bm{\eta}_{\tilde{e}_j}\right)\\
        \stackrel{2k\mathbb{Z}}{=}&2k\int_{\mathbb{L}^3}\left(\frac{1}{2k}\delta_{-1}\mathrm{PD}(\tilde{\Sigma}_1),0,0\right)\star\left(\frac{1}{2k}\delta_{-1}\mathrm{PD}(\tilde{\Sigma}_2),0,0\right)\\
        =&\frac{1}{2k}\sum_{\substack{\text{$3$d cubes}\\\text{in $\mathbb{L}^3$}}}\mathrm{PD}(\tilde{\Sigma}_1)\wedge d\mathrm{PD}(\tilde{\Sigma}_2)=\frac{1}{2k}\int_{(\partial\tilde{\Sigma}_2)_-}\mathrm{PD}(\tilde{\Sigma}_1)=\frac{1}{2k}\mathrm{link}^{\mathrm{mod}\,2k}(\tilde{\gamma}_1,\tilde{\gamma}_2),
    \end{split}
\end{equation}
where we used $\tilde{b}_i,\tilde{c}_j\in 2k\mathbb{Z}$ at the sign $\stackrel{2k\mathbb{Z}}{=}$. The gauge invariances allow us to write 
\begin{equation}\label{equ9420}
    2k\int_{\mathbb{L}^3}\left[\frac{\bm{\eta}_{\tilde{\gamma}_1}}{2k}\right]\star\left[\frac{\bm{\eta}_{\tilde{\gamma}_2}}{2k}\right]\stackrel{\mathbb{Z}}{=}\frac{1}{2k}\mathrm{link}^{\mathrm{mod}\,2k}(\tilde{\gamma}_1,\tilde{\gamma}_2).
\end{equation}
\subsection{Self-linking number}\label{sec970}
\subsubsection{Definition of self-linking number}\label{sec971}
In the case $[\tilde{\gamma}]=0\in H_1(\tilde{\mathbb{L}}^3;\mathbb{Z})$, we can give a simple definition of self-linking number of $\gamma$ by
\begin{equation}\label{equ9430}
    \mathrm{slk}(\tilde{\gamma}):=\int_{\mathbb{L}^3}\bm{\eta}_{\tilde{\gamma}}\star\bm{\eta}_{\tilde{\gamma}}=\mathrm{link}(\tilde{\gamma}_\pm,\tilde{\gamma})=\mathrm{link}(\tilde{\gamma},\tilde{\gamma}_\pm),
\end{equation}
where $\tilde{\gamma}_\pm$ are defined in \eqref{equ3170}. Normally, the self-linking number cannot be defined unless a framing is given to the closed curve, but in this case, the self-linking number is defined for $\gamma$ by giving the framing of $\gamma$ as $\gamma_+$. This argument is also pointed out in~\cite{Jacobson:2023cmr}, and our discussion proves that the same technique can be applied to lattice DB formulations.

Now, let us consider $\tilde{\gamma}$ that belongs to non-trivial homology. As we see in \eqref{equ9390}, $\frac{\bm{\eta}_{\tilde{\gamma}}}{2k}\in Z^2_\mathrm{lat.DB}(\mathbb{L}^3)$ can be chosen from arbitrary $\tilde{\gamma}\in Z_1(\tilde{\mathbb{L}}^3;2k\mathbb{Z})$. Then we introduce the definition of mod $4k$ self-linking number of $\gamma$ by
\begin{equation}\label{equ9440}
    \frac{1}{4k}\mathrm{slk}^{\mathrm{mod}\,4k}(\tilde{\gamma}):=k\int_{\mathbb{L}^3}\frac{\bm{\eta}_{\tilde{\gamma}}}{2k}\star\frac{\bm{\eta}_{\tilde{\gamma}}}{2k}.
\end{equation}

We discuss the geometric interpretation of the mod $4k$ self-linking number and why the definition of self-linking number is modulo $4k$. As in \eqref{equ9410}, we give the decomposition of $\tilde{\gamma}$ by
\begin{equation}\label{equ9450}
    \tilde{\gamma}=\partial\tilde{\Sigma}+\sum_i b_i\tilde{e}_i=\partial\tilde{\Sigma}+2k\tilde{\Gamma}\qquad(b_i\in 2k\mathbb{Z}),
\end{equation}
and we would assume
\begin{equation}\label{equ9460}
    \begin{split}
        \mathrm{slk}(\tilde{\gamma})\quad &\text{``}=\text{''}\quad
        \mathrm{link}\left(\partial\tilde{\Sigma},\partial\tilde{\Sigma}_+\right)
        +\mathrm{link}\left(2k\tilde{\Gamma}_-,\partial\tilde{\Sigma}\right)\\
        &\qquad+\mathrm{link}\left(\partial\tilde{\Sigma},2k\tilde{\Gamma}_+\right)
        +\mathrm{link}\left(2k\tilde{\Gamma}_-,2k\tilde{\Gamma}\right)\\
        \quad &\text{``}\stackrel{4k^2\mathbb{Z}}{=}\text{''}
        \quad\mathrm{link}\left(\partial\tilde{\Sigma},\partial\tilde{\Sigma}_+\right)
        +\mathrm{link}\left(2k\tilde{\Gamma}_-,\partial\tilde{\Sigma}\right)
        +\mathrm{link}\left(\partial\tilde{\Sigma},2k\tilde{\Gamma}_+\right)
    \end{split}
\end{equation}
as a formal discussion. Taking inspiration from here, we introduce the (geometric) definition of the mod $4k$ self-linking number by
\begin{equation}\label{equ9470}
    \begin{split}
        &\mathrm{slk}^{\mathrm{mod}\,4k}(\tilde{\gamma})\\
        \stackrel{4k\mathbb{Z}}{:=}\;&
        \mathrm{link}\left(\partial\tilde{\Sigma},\partial\tilde{\Sigma}_+\right)
        +\mathrm{link}\left(2k\tilde{\Gamma}_-,\partial\tilde{\Sigma}\right)
        +\mathrm{link}\left(\partial\tilde{\Sigma},2k\tilde{\Gamma}_+\right)\\
        =&\int_{\partial\tilde{\Sigma}_+}\mathrm{PD}(\tilde{\Sigma})
        +\int_{2k\tilde{\Gamma}_-}\mathrm{PD}(\tilde{\Sigma})
        +\int_{2k\tilde{\Gamma}_+}\mathrm{PD}(\tilde{\Sigma})
    \end{split}
\end{equation}
This definition has ambiguity due to the replacement of the Seifert surface of $\tilde{\Sigma}$
\begin{equation}\label{equ9480}
    \tilde{\Sigma}\to \tilde{\Sigma}+\tilde{e}_{ij},
\end{equation}
but since the change in slk when the replacement is made can be written as
\begin{equation}\label{equ9490}
    \int_{2k\tilde{\Gamma}_-}\mathrm{PD}(\tilde{e}_{ij})
    +\int_{2k\tilde{\Gamma}_+}\mathrm{PD}(\tilde{e}_{ij})\in 4k\mathbb{Z},
\end{equation}
we find that slk always changes by $4k\mathbb{Z}$. Let us also consider the replacement of representatives of the basis of $H_1(\tilde{\mathbb{L}}^3;\mathbb{Z})$, denoted
\begin{gather}
    H_1(\tilde{\mathbb{L}}^3;\mathbb{Z})=\mathrm{Span}([\tilde{e}'_1],[\tilde{e}'_2],[\tilde{e}'_3]),\qquad[\tilde{e}_1]=[\tilde{e}'_1],\quad[\tilde{e}_2]=[\tilde{e}'_2],\quad[\tilde{e}_3]=[\tilde{e}'_3],\label{equ9500}\\
    \tilde{e}'_1-\tilde{e}_1=\partial\sigma_1,\qquad\tilde{e}'_2-\tilde{e}_2=\partial\sigma_2,\qquad\tilde{e}'_3-\tilde{e}_3=\partial\sigma_3.\label{equ9510}
\end{gather}
The expression of the decomposition of $\tilde{\gamma}$ becomes
\begin{equation}\label{equ9520}
    \begin{split}
        \tilde{\gamma}=\partial\tilde{\Sigma}+2k\tilde{\Gamma}&=\partial\tilde{\Sigma}'+\sum_i \tilde{b}_i\tilde{e}'_i\\
        &=\partial\left(\tilde{\Sigma}-\sum_i \tilde{b}_i\sigma_i\right)+\sum_i \tilde{b}_i(\partial\sigma_i+\tilde{e}_i)=\partial\tilde{\Sigma}'+2k\tilde{\Gamma}',
    \end{split}
\end{equation}
therefore
\begin{equation}\label{equ9530}
    \quad\Gamma'-\Gamma=\sum_i \tilde{b}_i\sigma_i=:2k\sigma,\quad\Sigma'-\tilde{\Sigma}=-2k\sigma
\end{equation}
holds. The change of slk is given by
\begin{equation}\label{equ9540}
    \begin{split}
        &\mathrm{link}(\partial\tilde{\Sigma}',\partial\tilde{\Sigma}'_+)
        -\mathrm{link}(\partial\tilde{\Sigma},\partial\tilde{\Sigma}_+)\\
        +&\mathrm{link}(2k\Gamma_-,\partial\tilde{\Sigma}')
        -\mathrm{link}(2k\tilde{\Gamma}_-,\partial\tilde{\Sigma})\\
        +&\mathrm{link}(\partial\tilde{\Sigma}',2k\Gamma_+)
        -\mathrm{link}(\partial\tilde{\Sigma},2k\tilde{\Gamma}_+)\\
        =&-\mathrm{link}(2k\partial\sigma_-,\partial\tilde{\Sigma})
        -\mathrm{link}(\partial\tilde{\Sigma},2k\partial\sigma_+)
        +\mathrm{link}(2k\partial\sigma,2k\partial\sigma_+)\\
        +&\mathrm{link}(2k\partial\sigma_-,\partial\tilde{\Sigma})
        -\mathrm{link}(2k\tilde{\Gamma}_-,2k\partial\sigma)
        -\mathrm{link}(2k\partial\sigma_-,2k\partial\sigma)\\
        -&\mathrm{link}(2k\partial\sigma,2k\tilde{\Gamma}_+)
        +\mathrm{link}(\partial\tilde{\Sigma},2k\partial\sigma_+)
        -\mathrm{link}(2k\partial\sigma,2k\partial\sigma_+)\\
        =&\mathrm{link}(2k\partial\sigma,2k\partial\sigma_+)-\mathrm{link}(2k\tilde{\Gamma}_-,2k\partial\sigma)\\
        -&\mathrm{link}(2k\partial\sigma_-,2k\partial\sigma)-\mathrm{link}(2k\partial\sigma,2k\tilde{\Gamma}_+)
        -\mathrm{link}(2k\partial\sigma,2k\partial\sigma_+)\stackrel{4k^2\mathbb{Z}}{=}\;0,
    \end{split}
\end{equation}
then we see that the slk changes by $4k^2\mathbb{Z}$. From the above discussion, we conclude that the self-linking number is well-defined modulo $4k\mathbb{Z}$.

Finally, we show that the definition of slk using star products is equivalent to the geometric definition of slk. Using the notation in \eqref{equ9470}, we introduce the decomposition of the Pontrjagin dual by
\begin{equation}\label{equ9550}
    \bm{\eta}_{\tilde{\gamma}}:=\bm{\eta}_{\partial\tilde{\Sigma}}+2k\bm{\eta}_{\Gamma}\in Z^2_\mathrm{lat.DB}(\mathbb{L}^3;2k\mathbb{Z}).
\end{equation}
The equivalence of definitions of self-linking number can be proved by
\begin{equation}\label{equ9560}
    \begin{split}
        &k\int_{\mathbb{L}^3}\left(\frac{\bm{\eta}_{\partial\tilde{\Sigma}}}{2k}+\bm{\eta}_{\Gamma}\right)\star\left(\frac{\bm{\eta}_{\partial\tilde{\Sigma}}}{2k}+\bm{\eta}_{\Gamma}\right)\\
        =&k\int_{\mathbb{L}^3}\left(
            \frac{\bm{\eta}_{\partial\tilde{\Sigma}}}{2k}\star\frac{\bm{\eta}_{\partial\tilde{\Sigma}}}{2k}
            +\frac{\bm{\eta}_{\partial\tilde{\Sigma}}}{2k}\star\bm{\eta}_{\Gamma}
            +\bm{\eta}_{\Gamma}\star\frac{\bm{\eta}_{\partial\tilde{\Sigma}}}{2k}
            +\bm{\eta}_{\Gamma}\star\bm{\eta}_{\Gamma}
        \right)\\
        \stackrel{k\mathbb{Z}}{=}&k\int_{\mathbb{L}^3}\left(
            \frac{\bm{\eta}_{\partial\tilde{\Sigma}}}{2k}\star\frac{\bm{\eta}_{\partial\tilde{\Sigma}}}{2k}
            +\frac{\bm{\eta}_{\partial\tilde{\Sigma}}}{2k}\star\bm{\eta}_{\Gamma}
            +\bm{\eta}_{\Gamma}\star\frac{\bm{\eta}_{\partial\tilde{\Sigma}}}{2k}
        \right)\\
        =&\frac{1}{4k}\mathrm{link}\left(\partial\tilde{\Sigma},\partial\tilde{\Sigma}_+\right)
        +\frac{1}{2}\mathrm{link}\left(\tilde{\Gamma}_-,\partial\tilde{\Sigma}\right)
        +\frac{1}{2}\mathrm{link}\left(\partial\tilde{\Sigma},\tilde{\Gamma}_+\right)\\
        =&\frac{1}{4k}\left(
        \mathrm{link}\left(\partial\tilde{\Sigma},\partial\tilde{\Sigma}_+\right)
        +\mathrm{link}\left(2k\tilde{\Gamma}_-,\partial\tilde{\Sigma}\right)
        +\mathrm{link}\left(\partial\tilde{\Sigma},2k\tilde{\Gamma}_+\right)\right)=\frac{1}{4k}\mathrm{slk}^{\mathrm{mod}\,4k}(\tilde{\gamma}).
    \end{split}
\end{equation}
\subsubsection{Decomposition formula of self-linking number}\label{sec972}
As we see in \eqref{equ9150}-\eqref{equ9170}, or in later Section~\ref{sec973}, in the case of $(\mathrm{supp}\,\tilde{\gamma}^1)\cap(\mathrm{supp}\,\tilde{\gamma}^2)\neq\emptyset$, $\tilde{\gamma}^1,\tilde{\gamma}^2$ do not have well-defined linking number. This fact suggests that we should properly define the ``single'' Wilson line. In this section, we consider a suitable decomposition of general $\tilde{\gamma}\in H_1(\tilde{\mathbb{L}}^3;2k\mathbb{Z})$, and call the decomposed element of $\tilde{\gamma}$ a single knot. The aim of this section is to give an expression of $\mathrm{slk}(\tilde{\gamma})$ decomposed into the (self-) linking numbers of the single knots.

Decompose $\tilde{\gamma}$
\begin{equation}\label{equ9570}
    \tilde{\gamma}=\partial\tilde{\Sigma}+\sum_i b_i\tilde{e}_i
\end{equation}
as in \eqref{equ9410}. Next, decompose $2$-chain $\Sigma$ into connected components by
\begin{equation}\label{equ9580}
    \tilde{\Sigma}=\tilde{\Sigma}_1+\dots+\tilde{\Sigma}_{n-1}.
\end{equation}
Let $\gamma_i$'s be defined by
\begin{equation}\label{equ9590}
    \partial\tilde{\Sigma}=\partial\tilde{\Sigma}_1+\dots+\partial\tilde{\Sigma}_{n-1}=:\tilde{\gamma}_1+\dots+\tilde{\gamma}_{n-1},
\end{equation}
and then $\gamma_1,\dots,\gamma_{n-1}$ satisfy
\begin{equation}\label{equ9600}
    i\neq j\Longrightarrow (\mathrm{supp}\,\tilde{\gamma}_i)\cap(\mathrm{supp}\,\tilde{\gamma}_i)=\emptyset.
\end{equation}
$\tilde{\gamma}$ has the decomposition
\begin{equation}\label{equ9610}
    \tilde{\gamma}=\tilde{\gamma}^1+\dots+\tilde{\gamma}^{n-1}+\sum_i b_i\tilde{e}_i=\tilde{\gamma}^1+\dots+\tilde{\gamma}^{n-1}+2k\tilde{\gamma}^{n}.
\end{equation}
Note that $b_i\in 2k\mathbb{Z}$. Let us call this the decomposition of $\tilde{\gamma}$ into single knots.

Now, suppose that the decomposition of $\tilde{\gamma}\in Z_1(\tilde{\mathbb{L}}^3;2k\mathbb{Z})$ into single knots is given as $\tilde{\gamma}=\tilde{\gamma}^1+\dots+\tilde{\gamma}^{n-1}+2k\tilde{\gamma}^n$. Since $\bm{\eta}_{\tilde{\gamma}}$ has the decomposition
\begin{equation}\label{equ9620}
    \bm{\eta}_{\tilde{\gamma}}=\bm{\eta}_{\tilde{\gamma}^1}+\dots+\bm{\eta}_{\tilde{\gamma}^{n-1}}+\bm{\eta}_{2k\tilde{\gamma}^n},
\end{equation}
we get a formula
\begin{equation}\label{equ9630}
    \begin{split}
        &\frac{1}{4k}\mathrm{slk}^{\mathrm{mod}\,4k}(\tilde{\gamma})\\
        =&k\int_{\mathbb{L}^3}\left(
            \frac{\bm{\eta}_{\tilde{\gamma}^1}}{2k}+\cdots+\frac{\bm{\eta}_{\tilde{\gamma}^{n-1}}}{2k}+\frac{\bm{\eta}_{2k\tilde{\gamma}^{n}}}{2k}\right)
            \star
            \left(\frac{\bm{\eta}_{\tilde{\gamma}^1}}{2k}+\cdots+\frac{\bm{\eta}_{\tilde{\gamma}^{n-1}}}{2k}+\frac{\bm{\eta}_{2k\tilde{\gamma}^{n}}}{2k}\right)\\
        =&\sum_{i=1}^{n-1}k\int_{\mathbb{L}^3}\frac{\bm{\eta}_{\tilde{\gamma}^i}}{2k}\star\frac{\bm{\eta}_{\tilde{\gamma}^i}}{2k}+\sum_{1\leq i<j\leq n-1}2k\int_{\mathbb{L}^3}\frac{\bm{\eta}_{\tilde{\gamma}^i}}{2k}\star\frac{\bm{\eta}_{\tilde{\gamma}^j}}{2k}\\
        +&\sum_{i=1}^{n-1}\left[k\int_{\mathbb{L}^3}\left(\frac{\bm{\eta}_{\tilde{\gamma}^i}}{2k}\star\frac{\bm{\eta}_{2k\tilde{\gamma}^{n}}}{2k}+\frac{\bm{\eta}_{2k\tilde{\gamma}^{n}}}{2k}\star\frac{\bm{\eta}_{\tilde{\gamma}^i}}{2k}\right)\right]+k\int_{\mathbb{L}^3}\frac{\bm{\eta}_{2k\tilde{\gamma}^{n}}}{2k}\star\frac{\bm{\eta}_{2k\tilde{\gamma}^{n}}}{2k}\\
        \stackrel{k\mathbb{Z}}{=}&\sum_{i=1}^{n-1}\left[k\int_{\mathbb{L}^3}\left(\frac{\bm{\eta}_{\tilde{\gamma}^i}}{2k}\star\frac{\bm{\eta}_{\tilde{\gamma}^i}}{2k}+\frac{\bm{\eta}_{\tilde{\gamma}^i}}{2k}\star\frac{\bm{\eta}_{2k\tilde{\gamma}^{n}}}{2k}+\frac{\bm{\eta}_{2k\tilde{\gamma}^{n}}}{2k}\star\frac{\bm{\eta}_{\tilde{\gamma}^i}}{2k}\right)\right]+\sum_{1\leq i<j\leq n-1}2k\int_{\mathbb{L}^3}\frac{\bm{\eta}_{\tilde{\gamma}^i}}{2k}\star\frac{\bm{\eta}_{\tilde{\gamma}^j}}{2k}\\
        =&\sum_{i=1}^{n-1}\frac{1}{4k}\mathrm{slk}^{\mathrm{mod}\,4k}(\tilde{\gamma}^i+2k\tilde{\gamma}^n)+\sum_{1\leq i<j\leq n-1}\frac{1}{2k}\mathrm{link}(\tilde{\gamma}^i,\tilde{\gamma}^j)
    \end{split}
\end{equation}
from linearity of $\star$. Note that $i\neq j\Rightarrow \mathrm{supp}\,\tilde{\gamma}^i\cap\mathrm{supp}\,\tilde{\gamma}^j=\emptyset\Rightarrow \int\bm{\eta}_{\tilde{\gamma}^i}\star\bm{\eta}_{\tilde{\gamma}^j}=\int\bm{\eta}_{\tilde{\gamma}^j}\star\bm{\eta}_{\tilde{\gamma}^i}$ was used in the second equal sign. We finally conclude that if the decomposition of $\tilde{\gamma}$ into single knots is given, $\mathrm{slk}^{\mathrm{mod}\;4k}$ can be calculated from the linking number of each single knot and the mod $4k$ self-linking number of $\tilde{\gamma}^i+2k\tilde{\gamma}^n$.
\subsubsection{Change of self-linking number caused by local deformation}\label{sec973}
Consider again the closed curve belonging to trivial homology. From geometric discussions and the properties of $\star$, it is immediately clear that
\begin{align}
    &\int_{\mathbb{L}^3}\bm{\eta}_{\tilde{\gamma}^1}\overleftarrow{\underrightarrow{\star}}\bm{\eta}_{\tilde{\gamma}^2}=\mathrm{link}(\tilde{\gamma}^1_+,\tilde{\gamma}^2)=\mathrm{link}(\tilde{\gamma}^1,\tilde{\gamma}^2_-)=\int_{\mathbb{L}^3}\bm{\eta}_{\tilde{\gamma}^1}\overrightarrow{\underleftarrow{\star}}\bm{\eta}_{\tilde{\gamma}^2},\label{equ9640}\\
    &\int_{\mathbb{L}^3}\bm{\eta}_{\tilde{\gamma}^2}\overleftarrow{\underrightarrow{\star}}\bm{\eta}_{\tilde{\gamma}^1}=\mathrm{link}(\tilde{\gamma}^1,\tilde{\gamma}^2_+)=\mathrm{link}(\tilde{\gamma}^1_-,\tilde{\gamma}^2)=\int_{\mathbb{L}^3}\bm{\eta}_{\tilde{\gamma}^2}\overrightarrow{\underleftarrow{\star}}\bm{\eta}_{\tilde{\gamma}^1},\label{equ9650}
\end{align}
 but when $(\mathrm{supp}\,\tilde{\gamma})\cap(\mathrm{supp}\,\tilde{\gamma}')\neq\emptyset$ holds,
\begin{equation}\label{equ9660}
    \begin{split}
        \int_{\mathbb{L}^3}\bm{\eta}_{\tilde{\gamma}^1}\star\bm{\eta}_{\tilde{\gamma}^2}&=\mathrm{link}(\tilde{\gamma}^1_+,\tilde{\gamma}^2)=\mathrm{link}(\tilde{\gamma}^1,\tilde{\gamma}^2_-)\\
        &\neq\mathrm{link}(\tilde{\gamma}^1,\tilde{\gamma}^2_+)=\mathrm{link}(\tilde{\gamma}^1_-,\tilde{\gamma}^2)=\int_{\mathbb{L}^3}\bm{\eta}_{\tilde{\gamma}^2}\star\bm{\eta}_{\tilde{\gamma}^1}
    \end{split}
\end{equation}
holds in general.\footnote{Even if $\mathrm{supp}\,\tilde{\gamma}^1\cap\mathrm{supp}\,\tilde{\gamma}^2\neq\emptyset$, the inequality does not necessarily hold.}

Let us analyze the properties of slk based on this discussion. Consider the cycles $\tilde{\gamma}, \tilde{\gamma}'$ of the dual lattice, both of them belong to trivial cohomology. In this case, we can see that the self-linking number of $\tilde{\gamma}+\tilde{\gamma}'$ satisfies 
\begin{equation}\label{equ9670}
    \begin{split}
        \mathrm{slk}(\tilde{\gamma}+\tilde{\gamma}')-\mathrm{slk}(\tilde{\gamma})=\mathrm{slk}(\tilde{\gamma}')+\int_{\mathbb{L}^3}\left(\bm{\eta}_{\tilde{\gamma}}\star\bm{\eta}_{\tilde{\gamma}'}+\bm{\eta}_{\tilde{\gamma}'}\star\bm{\eta}_{\tilde{\gamma}}\right).
    \end{split}
\end{equation}
If $(\mathrm{supp}\,\tilde{\gamma})\cap(\mathrm{supp}\,\tilde{\gamma}')=\emptyset$ holds, then
\begin{equation}\label{equ9680}
    \int_{\mathbb{L}^3}\left(\bm{\eta}_{\tilde{\gamma}}\star\bm{\eta}_{\tilde{\gamma}'}+\bm{\eta}_{\tilde{\gamma}'}\star\bm{\eta}_{\tilde{\gamma}}\right)=2\mathrm{link}(\tilde{\gamma},\tilde{\gamma}')
\end{equation}
holds. In this case, $\tilde{\gamma}\to\tilde{\gamma}+\tilde{\gamma}'$ can be regarded as the addition of the closed curve $\gamma'$, and it can be seen that the self-linking number changes by twice their linking number.

On the other hand, if $(\mathrm{supp}\,\tilde{\gamma})\cap(\mathrm{supp}\,\tilde{\gamma}')\neq\emptyset$, then $\mathrm{link}(\tilde{\gamma},\tilde{\gamma}')$ is ill-defined, and we get
\begin{equation}\label{equ9690}
    \begin{split}
        \int_{\mathbb{L}^3}\left(\bm{\eta}_{\tilde{\gamma}}\star\bm{\eta}_{\tilde{\gamma}'}+\bm{\eta}_{\tilde{\gamma}'}\star\bm{\eta}_{\tilde{\gamma}}\right)=\mathrm{link}(\tilde{\gamma}_+,\tilde{\gamma}')+\mathrm{link}(\tilde{\gamma}_-,\tilde{\gamma}')
    \end{split}
\end{equation}
In this case, $\tilde{\gamma}\to\tilde{\gamma}+\tilde{\gamma}'$ can be regarded as a local deformation of the closed curve $\gamma$, and $\mathrm{link}(\tilde{\gamma}_+,\tilde{\gamma}')+\mathrm{link}(\tilde{\gamma}_-,\tilde{\gamma}')$ can be interpreted as the change in the self-linking number that occurs due to the deformation.
\newpage
\part{Lattice Chern--Simons theory on $3$D cubic torus formulated using lattice DB cohomology}\label{part3}
We move on to the second main topic of this paper. In this part, we define and analyze the lattice Chern--Simons theory formulated based on lattice Deligne--Beilinson cohomology on three-dimensional torus. Motivated by the approaches outlined in~\cite{Guadagnini:2008bh,Thuillier:2015vma} and~\cite[Section~6.4]{Norton:2021psm}, we present an analysis in the framework of lattice DB cohomology, which is one of the main results of this paper.

In the second half of this part, we discuss several topics, including applications to non-invertible defects of lattice massless QED. 
\section{Lattice Chern--Simons action on $\mathbb{L}^3$}\label{sec1000}
First we give the expression of level $2k$ lattice DB Chern--Simons action on $\mathbb{L}^3$ by
\begin{equation}\label{equ10010}
    \mathcal{CS}_\mathrm{lat.DB}^{2k}[\bm{a}]=\int_{\mathbb{L}^3}k\bm{a}\star\bm{a},
\end{equation}
where $k\in\mathbb{Z}_{\geq 1},[\bm{a}]\in H^2_\mathrm{lat.DB}(\mathbb{L}^3)$. The action coupling with external field $[\bm{A}]\in H^2_\mathrm{lat.DB}(\mathbb{L}^3)$ can be introduced by
\begin{equation}\label{equ10020}
    \mathcal{CS}_\mathrm{lat.DB}^{2k}[\bm{a},\bm{A}]=\int_{\mathbb{L}^3}\left(k\bm{a}\star\bm{a}+\bm{a}\frac{\overleftrightarrow{\star}}{2}\bm{A}\right).
\end{equation}
We define the Maxwell--Chern--Simons actions by
\begin{align}
    2\pi i\,\mathcal{MCS}_\mathrm{lat.DB}^{2k,\epsilon}[\bm{a}]&=2\pi i\,\mathcal{CS}_\mathrm{lat.DB}^{2k}[\bm{a}]-\sum_{\substack{\text{plaquettes}\\\text{in $\mathbb{L}^3$}}}\epsilon |d\bm{a}|^2,\label{equ10030}\\
    2\pi i\,\mathcal{MCS}_\mathrm{lat.DB}^{2k,\epsilon}[\bm{a},\bm{A}]&=2\pi i\,\mathcal{CS}_\mathrm{lat.DB}^{2k}[\bm{a},\bm{A}]-\sum_{\substack{\text{plaquettes}\\\text{in $\mathbb{L}^3$}}}\epsilon |d\bm{a}|^2\label{equ10040}
\end{align}
for coupling constant $0<\epsilon\ll 1$, where $d\bm{a}$ is defined in \eqref{equ5050}.

Since $\star$ product is modulo $\mathbb{Z}$ well-defined, and $d\bm{a}$ is well-defined as an operation on DB cohomology, i.e., independent of the choice of representative, the equalities
\begin{equation}\label{equ10050}
    \begin{split}
        [\bm{a}]=[\bm{a}']&\Longrightarrow 2\pi i\,\mathcal{CS}_\mathrm{lat.DB}[\bm{a}]\stackrel{2\pi i\mathbb{Z}}{=}2\pi i\,\mathcal{CS}_\mathrm{lat.DB}[\bm{a}'],\\ 
        [\bm{a}]=[\bm{a}']&\Longrightarrow 2\pi i\,\mathcal{MCS}_\mathrm{lat.DB}[\bm{a}]\stackrel{2\pi i\mathbb{Z}}{=}2\pi i\,\mathcal{MCS}_\mathrm{lat.DB}[\bm{a}']
    \end{split}    
\end{equation}
hold, and the same property for $\bm{A}$ holds on $\mathcal{CS}_\mathrm{lat.DB}^{2k}[\bm{a},\bm{A}]$ and $\mathcal{MCS}_\mathrm{lat.DB}^{2k,\epsilon}[\bm{a},\bm{A}]$.

Therefore, these four actions can be introduced as a functional on $H^2_\mathrm{lat.DB}(\mathbb{L}^3)$, denoted
\begin{align}\label{equ10060}
    &\mathcal{CS}_\mathrm{lat.DB}^{2k}[[\bm{a}]],&&\mathcal{CS}_\mathrm{lat.DB}^{2k}[[\bm{a}],[\bm{A}]],&&\mathcal{MCS}_\mathrm{lat.DB}^{2k,\epsilon}[[\bm{a}]],&&\mathcal{MCS}_\mathrm{lat.DB}^{2k,\epsilon}[[\bm{a}],[\bm{A}]].
\end{align}
However, to simplify notation, we will not use it in the future unless necessary.
\subsection{Global $\mathbb{Z}_{2k}$ $1$-form symmetry}\label{sec1010}
We introduce $1$-cocycle $\beta_{ij}\in Z^1(\mathbb{L}^3;\mathbb{Z})$ by
\begin{equation}\label{equ10070}
    \beta_{ij}(x,\mu)=\varepsilon_{ijk}\,\delta_{x_k,0}\,\delta_{\mu,k},
\end{equation}
where $i,j,k\in\{1,2,3\}$, and $\varepsilon_{ijk}$ is a Levi--Civita symbol, and $\delta$ is the Kronecker delta. We immediately find $d\beta=0$. By definition, 
\begin{equation}\label{equ10080}
    \begin{split}
        \sum_{\substack{\text{cubes}\\\text{in $\mathbb{L}^3$}}}\omega\wedge\beta_{ij}&=\left.\varepsilon_{ijk}\sum_{x_i=0}^{L-1}\sum_{x_j=0}^{L-1}\omega(x,k)\right|_{x_k=0},\\
        \sum_{\substack{\text{cubes}\\\text{in $\mathbb{L}^3$}}}\beta_{ij}\wedge\omega&=\left.\varepsilon_{ijk}\sum_{x_i=0}^{L-1}\sum_{x_j=0}^{L-1}\omega(x,k)\right|_{x_k=L-1}
    \end{split}
\end{equation}
hold for all $2$-cochain $\omega\in C^2(\mathbb{L}^3;\mathbb{R})$.

The global $\mathbb{Z}_{2k}$ $1$-form transformations are given by the shift transformation
\begin{equation}\label{equ10090}
    [\bm{a}]\to\left[\bm{a}+\left(\frac{1}{2k}\delta_{-1}\beta_{ij},0,0\right)\right]=:[\bm{a}+\bm{\beta}_{ij}/2k],
\end{equation}
for $i,j,k\in\{1,2,3\}$, and $\mathcal{CS}_\mathrm{lat.DB}^{2k}[\bm{a}]$ and $\mathcal{MCS}_\mathrm{lat.DB}^{2k,\epsilon}[\bm{a}]$ have symmetries of these transformations. In fact, we can confirm
\begin{equation}\label{equ10100}
    \begin{split}
        &\mathcal{CS}^{2k}_\mathrm{lat.DB}[\bm{a}+\bm{\beta}_{ij}/k]=\int_{\mathbb{L}^3}k\left(\bm{a}\star\bm{a}+\frac{\bm{\beta}_{ij}}{2k}\overrightarrow{\underleftarrow{\star}}\bm{a}+\bm{a}\overleftarrow{\underrightarrow{\star}}\frac{\bm{\beta}_{ij}}{2k}+\frac{\bm{\beta}_{ij}}{2k}\star\frac{\bm{\beta}_{ij}}{2k}\right)\\
        =&\int_{\mathbb{L}^3}k\bm{a}\star\bm{a}+\frac{1}{2}\sum_{\substack{\text{cubes}\\\text{in $\mathbb{L}^3$}}}(\beta_{ij}\wedge d\bm{a}+d\bm{a}\wedge\beta_{ij})\stackrel{\mathbb{Z}}{=}\mathcal{CS}^{2k}_\mathrm{lat.DB}[\bm{a}],
    \end{split}
\end{equation}
and
\begin{equation}\label{equ10110}
    \epsilon |d(\bm{a}+\bm{\beta}_{ij}/k)|^2=\epsilon |d\bm{a}|^2.
\end{equation}
Then we see that $\mathcal{CS}_\mathrm{lat.DB}^{2k}[\bm{a}]$ and $\mathcal{MCS}_\mathrm{lat.DB}^{2k,\epsilon}[\bm{a}]$ have global $\mathbb{Z}_{2k}$ $1$-form symmetry. 

On the other hand, since $\mathcal{CS}_\mathrm{lat.DB}^{2k}[\bm{a},\bm{A}]$ and $\mathcal{MCS}_\mathrm{lat.DB}^{2k,\epsilon}[\bm{a},\bm{A}]$ changes under $\bm{a}\to\bm{a}+\bm{\beta}/2k$ depending on the cohomology class of $d\bm{A}$, they do not have $\mathbb{Z}_{2k}$ $1$-form symmetry.
\subsection{Staggered symmetry}\label{sec1020}
According to~\cite{Jacobson:2023cmr}, or as we see in \eqref{equ3140}, there exist $1$-cochains $\chi$ that satisfy $\sum_{\substack{\text{cubes}\\\text{in $\mathbb{L}^3$}}}(X\wedge\chi+\chi\wedge X)=0$ for any $2$-cochain $X$.  Using $\chi$ to define $\bm{\chi}=(\delta_{-1}\chi,0,0)$, we find
\begin{equation}\label{equ10120}
    \begin{split}
        &\mathcal{CS}^{2k}_\mathrm{lat.DB}[\bm{a}+\bm{\chi}]=k\int_{\mathbb{L}^3}\left(\bm{a}\star\bm{a}+\bm{\chi}\overrightarrow{\underleftarrow{\star}}\bm{a}+\bm{a}\overleftarrow{\underrightarrow{\star}}\bm{\chi}+\bm{\chi}\star\bm{\chi}\right)\\
        =&\int_{\mathbb{L}^3}k\bm{a}\star\bm{a}+k\sum_{\substack{\text{cubes}\\\text{in $\mathbb{L}^3$}}}\left(\chi\wedge\left(d\bm{a}+\frac{1}{2}d\chi\right)+\left(d\bm{a}+\frac{1}{2}d\chi\right)\wedge\chi\right)=\mathcal{CS}^{2k}_\mathrm{lat.DB}[\bm{a}],
    \end{split}
\end{equation}
because
\begin{equation}\label{equ10130}
    \int_{\mathbb{L}^3}\left(\bm{a}\overleftarrow{\underrightarrow{\star}}\bm{\chi}+\bm{\chi}\overrightarrow{\underleftarrow{\star}}\bm{a}\right)=\sum_{\substack{\text{cubes}\\\text{in $\mathbb{L}^3$}}}(d\bm{a}\wedge\chi+\chi\wedge d\bm{a})=0.
\end{equation}
Thus, $\mathcal{CS}_\mathrm{lat.DB}^{2k}[\bm{a}]$ and $\mathcal{CS}_\mathrm{lat.DB}^{2k}[\bm{a},\bm{A}]$ have a symmetry of the shift $\bm{a}\to\bm{a}+\bm{\chi}/k$, which is called staggered symmetry. However, $\mathcal{MCS}$ does not have staggered symmetry because $|d\chi |^2\neq 0$.
\newpage
\section{Path integral and expectation value of Wilson line}\label{sec1100}
We define the path integral and Wilson line of the $\mathcal{CS},\mathcal{MCS}$ theory in this section. We formally calculate the expectation values of Wilson lines in the $\mathcal{CS}$ theory, assuming its path integral converges, and discuss how the self-linking numbers of Wilson lines arise. This formal argument is, of course, useful in analyzing $\mathcal{MCS}$ theory, since $\mathcal{MCS}$ has very similar dynamics for sufficiently small $\epsilon$.
\subsection{Path integral over $H^2_\mathrm{lat.DB}(\mathbb{L}^3)$}\label{sec1110}
We give definitions of the partition functions of lattice DB (Maxwell--) Chern--Simons theory by
\begin{equation}\label{equ11010}
    \begin{array}{ll}
    \mathcal{Z}_\mathrm{lat.\mathcal{CS}}^{2k}&\displaystyle=\int_{[\bm{a}]\in H^2_\mathrm{lat.DB}(\mathbb{L}^3)}\mathcal{D}[\bm{a}]\exp \left(2\pi i\,\mathcal{CS}^{2k}_\mathrm{lat.DB}[[\bm{a}]]\right),\\
    \mathcal{Z}_\mathrm{lat.\mathcal{CS}}^{2k}[[\bm{A}]]&\displaystyle=\int_{[\bm{a}]\in H^2_\mathrm{lat.DB}(\mathbb{L}^3)}\mathcal{D}[\bm{a}]\exp \left(2\pi i\,\mathcal{CS}^{2k}_\mathrm{lat.DB}[[\bm{a}],[\bm{A}]]\right),\\
    \mathcal{Z}_\mathrm{lat.\mathcal{MCS}}^{2k,\epsilon}&\displaystyle=\int_{[\bm{a}]\in H^2_\mathrm{lat.DB}(\mathbb{L}^3)}\mathcal{D}[\bm{a}]\exp \left(2\pi i\,\mathcal{MCS}^{2k,\epsilon}_\mathrm{lat.DB}[[\bm{a}]]\right),\\
    \mathcal{Z}_\mathrm{lat.\mathcal{MCS}}^{2k,\epsilon}[[\bm{A}]]&\displaystyle=\int_{[\bm{a}]\in H^2_\mathrm{lat.DB}(\mathbb{L}^3)}\mathcal{D}[\bm{a}]\exp \left(2\pi i\,\mathcal{MCS}^{2k,\epsilon}_\mathrm{lat.DB}[[\bm{a}],[\bm{A}]]\right).
    \end{array}
\end{equation}
Referring to Appendix~\ref{secA10} and \eqref{equ8040}, using isomorphisms
\begin{equation}\label{equ11020}
    \begin{split}
        H^2_\mathrm{lat.DB}(\mathbb{L}^3)&\cong\frac{\Omega^1(\mathbb{L}^3)}{\Omega_\mathrm{closed}^1(\mathbb{L}^3)_\mathbb{Z}}\oplus H^2(\mathbb{L}^3;\mathbb{Z})\cong\frac{\mathbb{R}^{3L^3}}{B^1(\mathbb{L}^3;\mathbb{R})\oplus H^1(\mathbb{L}^3;\mathbb{Z})}\oplus\mathbb{Z}^3\\
        &\cong\frac{\mathbb{R}^{3L^3}}{\mathbb{R}^{L^3-1}\oplus\mathbb{Z}^{3}}\oplus\mathbb{Z}^3\cong \mathbb{R}^{2L^3-2}\oplus(\mathbb{R}/\mathbb{Z})^{3}\oplus\mathbb{Z}^3,
    \end{split}
\end{equation}
we introduce path integral measure $\mathcal{D}[\bm{a}]$ by
\begin{equation}\label{equ11030}
    \begin{split}
        &\int_{[\bm{a}]\in H^2_\mathrm{lat.DB}(\mathbb{L}^3)}\mathcal{D}[\bm{a}]\exp(S[\bm{a}])\\
        =&\int_{\alpha\in\Omega^1_\mathrm{coexact}(\mathbb{L}^3)}d^{2L^3+1}\!\alpha\;\;\int_0^1dm_{12}\int_0^1dm_{23}\int_0^1dm_{31}\sum_{(n_1,n_2,n_3)\in\mathbb{Z}^3}\\
        &\exp\biggl\{S\Bigl[(\delta_{-1}\alpha,0,0)+m_{12}\bm{\beta}_{12}+m_{23}\bm{\beta}_{23}+m_{31}\bm{\beta}_{31}+n_1\bm{\eta}_{\tilde{e}_1}+n_2\bm{\eta}_{\tilde{e}_2}+n_3\bm{\eta}_{\tilde{e}_3}\Bigr]\biggr\},
    \end{split}
\end{equation}
where $\bm{\beta}_{ij}$ is defined in \eqref{equ10090}, and $\tilde{e}_i$, defined in \eqref{equ9210}, is a closed curve that winds the torus once in the $\hat{i}$ direction.

Notice the term $(\delta_{-1}\alpha,0,0)$ in the path integral defined here. We recall the $\chi$ discussed in Section~\ref{sec1020} and consider $\alpha=s\chi$, where $s$ is a real parameter. The successive integral 
\begin{equation}\label{equ11040}
    \int_{-\infty}^\infty ds \exp[2\pi i\mathcal{CS}[(\delta_{-1}(s\chi),0,0)+\cdots]]
\end{equation}
diverges because the action of $\mathcal{CS}$ degenerates that direction. In contrast, the same path integral of $\mathcal{MCS}$ converges because the action has the Maxwell term, which ensures that the action has a negative definite real part. 
\subsection{Definition of Wilson line and its expectation value}\label{sec1120}
We give the definition of Wilson line for $\gamma\in Z_1(\tilde{\mathbb{L}}^3;\mathbb{Z})$ by
\begin{equation}\label{equ11050}
    \begin{split}
        W(\gamma)=&\exp \left(2\pi i\int_{\mathbb{L}^3}\left([\bm{a}]\frac{\overleftrightarrow{\star}}{2}[\bm{\eta}_\gamma]\right)\right)\\
        =&\exp \left(\pi i\int_{\mathbb{L}^3}\left(\bm{a}\star\bm{\eta}_\gamma+\bm{\eta}_\gamma\star\bm{a}+(0,0,0,d_{-1}(a^{(-1,2)}\smallsmile_1\eta_\gamma^{(-1,2)}),0)\right)\right).
    \end{split}
\end{equation}
This is so similar to the method \eqref{equ3180}, including the property that framing of $\gamma$ is automatically given by the cup product on the lattice. From the definition of the partition function with an external field, we can introduce the expectation value of the Wilson line by
\begin{equation}\label{equ11060}
    \langle W(\gamma)\rangle_{\mathcal{CS}}^{2k}=\frac{\mathcal{Z}^{2k}_{\mathrm{lat.}\mathcal{CS}}[\bm{\eta}_\gamma]}{\mathcal{Z}^{2k}_{\mathrm{lat.}\mathcal{CS}}[0]}.
\end{equation}
The formal calculation of this expectation value, ignoring divergence, is almost the same as~\cite{Guadagnini:2008bh,Thuillier:2015vma}, and the expectation value is described by
\begin{equation}\label{equ11070}
    \langle W(\gamma)\rangle_\mathcal{CS}^{2k}=
    \left\{
    \begin{array}{ll}
    \displaystyle\exp\left[2\pi ik\int_{\mathbb{L}^3}\frac{\bm{\eta}_\gamma}{2k}\star\frac{\bm{\eta}_\gamma}{2k}\right]&(\gamma\in H_1(\tilde{\mathbb{L}}^3;2k\mathbb{Z}))\\
    &\\
    0&(\gamma\notin H_1(\tilde{\mathbb{L}}^3;2k\mathbb{Z}))
    \end{array}\right..
\end{equation}
It can be written using the mod $4k$ self-linking number by
\begin{equation}\label{equ11080}
    \langle W(\gamma)\rangle_\mathcal{CS}^{2k}=
    \left\{
    \begin{array}{ll}
    \displaystyle\exp\left[\frac{2\pi i}{4k}\mathrm{slk}^{\mathrm{mod}\,4k}(\gamma)\right]&(\gamma\in H_1(\tilde{\mathbb{L}}^3;2k\mathbb{Z}))\\
    &\\
    0&(\gamma\notin H_1(\tilde{\mathbb{L}}^3;2k\mathbb{Z}))
    \end{array}\right..
\end{equation}
However, as we have emphasized many times, this discussion is purely formal, since the path integral does not converge in $\mathcal{CS}$ theory.
\subsection{Wilson line as $\mathbb{Z}_{2k}$ $1$-form symmetry defect}\label{sec1130}
The global $\mathbb{Z}_{2k}$ $1$-form symmetry transformation $\bm{a}\to\bm{a}+\bm{\beta}/2k$ is a transformation that simultaneously applies a $1/2k$ shift to the normal link variables in the nontrivial 2-homology of the torus $\tilde{\mathbb{L}}^3$. If this shift transformation is performed locally, the action $\int k\bm{a}\star\bm{a}$ is not invariant, and Wilson lines appear at the boundaries of the support of the local shift transformation. This means that the Wilson line that belongs to a trivial homology class can be understood as a $\mathbb{Z}_{2k}$ $1$-form symmetry defect.

For example, let $\beta'\in\Omega^1(\mathbb{L}^3)$ be
\begin{equation}\label{equ11090}
    \beta'(x,\mu)=\delta_{x,(0,1,1)}\delta_{\mu,1}.
\end{equation}
Then $\mathrm{PD}(\beta')$ can be introduced as $2$-chain $((1/2,1/2,1/2);23)$, and its boundary $\partial\mathrm{PD}(\beta')$ is $1$-boundary $((1/2,1/2,1/2);2)+((1/2,3/2,1/2);3)-((1/2,1/2,3/2);2)-((1/2,1/2,1/2);3)$. From the equality $\mathrm{PD}(d\beta')=\partial\mathrm{PD}(\beta')$,
\begin{equation}\label{equ11100}
    (\delta_{-1}\beta',0,0)=\bm{\eta}_{\partial\mathrm{PD}(\beta')}
\end{equation}
holds. If we perform a transformation $\bm{a}\to\bm{a}+(1/2k)(\delta_{-1}\beta',0,0)$, we can find
\begin{equation}\label{equ11110}
    \mathcal{Z}^{2k}_{\mathrm{lat.}\mathcal{CS}}[\bm{0}]\to\mathcal{Z}^{2k}_{\mathrm{lat.}\mathcal{CS}}[\bm{\eta}_{\partial\mathrm{PD}(\beta')}],
\end{equation}
where we used $\beta'^{(-1,2)}=0$ and $\beta'd\beta'=0$.\footnote{If we consider $\beta'$ which does not satisfy the condition $\beta'd\beta'=0$, since $\partial\mathrm{PD}(\beta')$ itself has a non-vanishing self-linking number, the total self-linking number changes along the local deformation.} That is, we make the Wilson line appear along $\partial\mathrm{PD}(\beta')$ by performing a $\mathbb{Z}_{2k}$ transformation localized in the $\hat{1}$ direction at $(0,1,1)$.

Using this procedure in reverse, if we apply the path integral variable transformation $\bm{a}\to\bm{a}-(1/2k)(\delta_{-1}\beta',0,0)$ to $\mathcal{Z}^{2k}_{\mathrm{lat.}\mathcal{CS}}[\bm{\eta}_{\partial\mathrm{PD}(\beta')}]$, then we get
\begin{equation}\label{equ11120}
    \langle W(\partial\mathrm{PD}(\beta'))\rangle_\mathcal{CS}^{2k}=\mathcal{Z}^{2k}_{\mathrm{lat.}\mathcal{CS}}[\bm{\eta}_{\partial\mathrm{PD}(\beta')}]=\mathcal{Z}^{2k}_{\mathrm{lat.}\mathcal{CS}}[\bm{0}].
\end{equation}
This equation shows that a single Wilson line is a $\mathbb{Z}_{2k}$ symmetry defect. 

\begin{figure}
    \begin{center}
    \includegraphics[width=100mm]{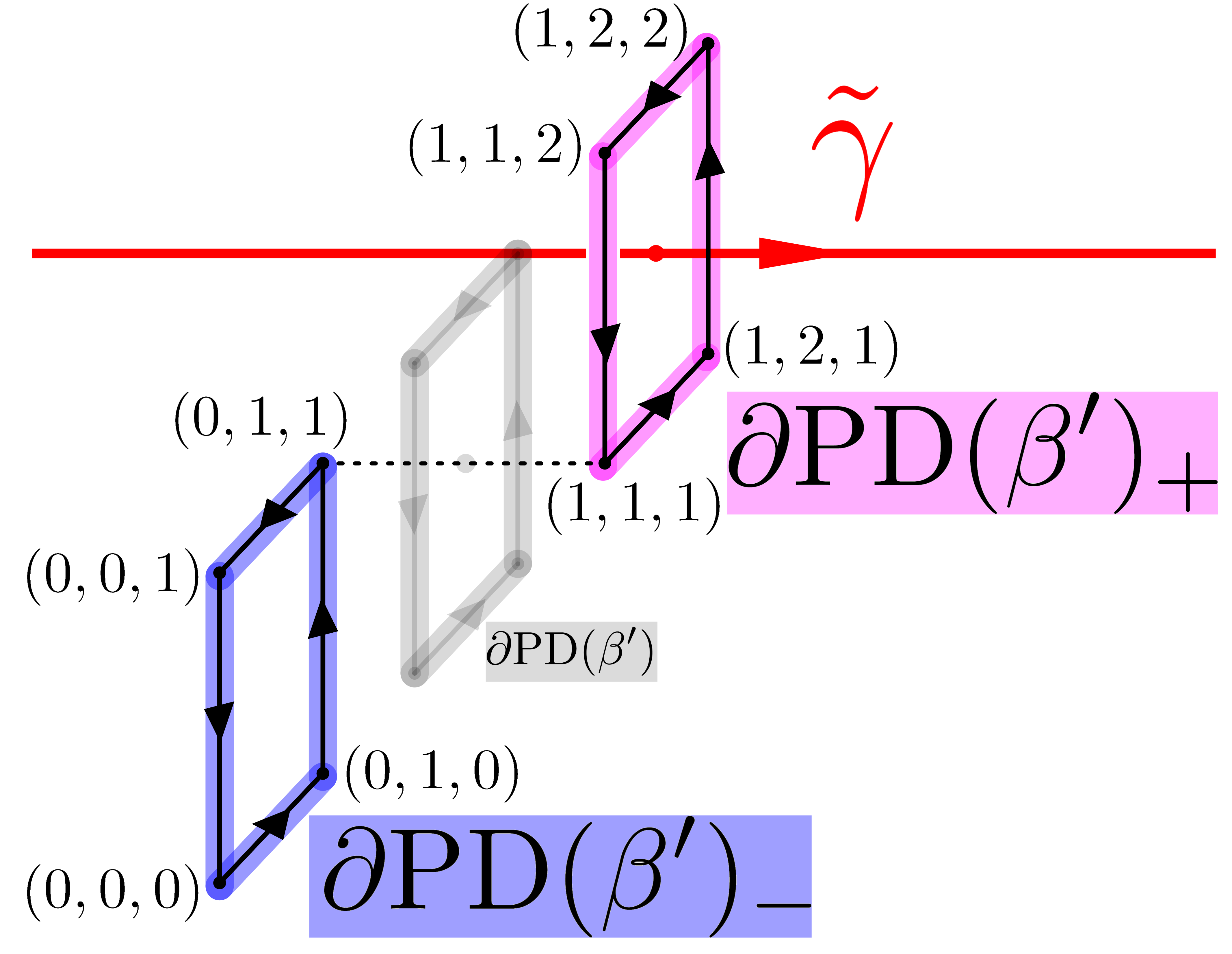}
    \caption{Illustration of $\beta'$, $\partial\mathrm{PD}(\beta')$, and the Wilson line $\tilde{\gamma}$. The dashed line denotes the link $((0,1,1);1)=\mathrm{supp}\,\beta'$}
    \label{fig:pPD}
    \end{center}
\end{figure}

\begin{figure}
    \begin{center}
    \includegraphics[width=100mm]{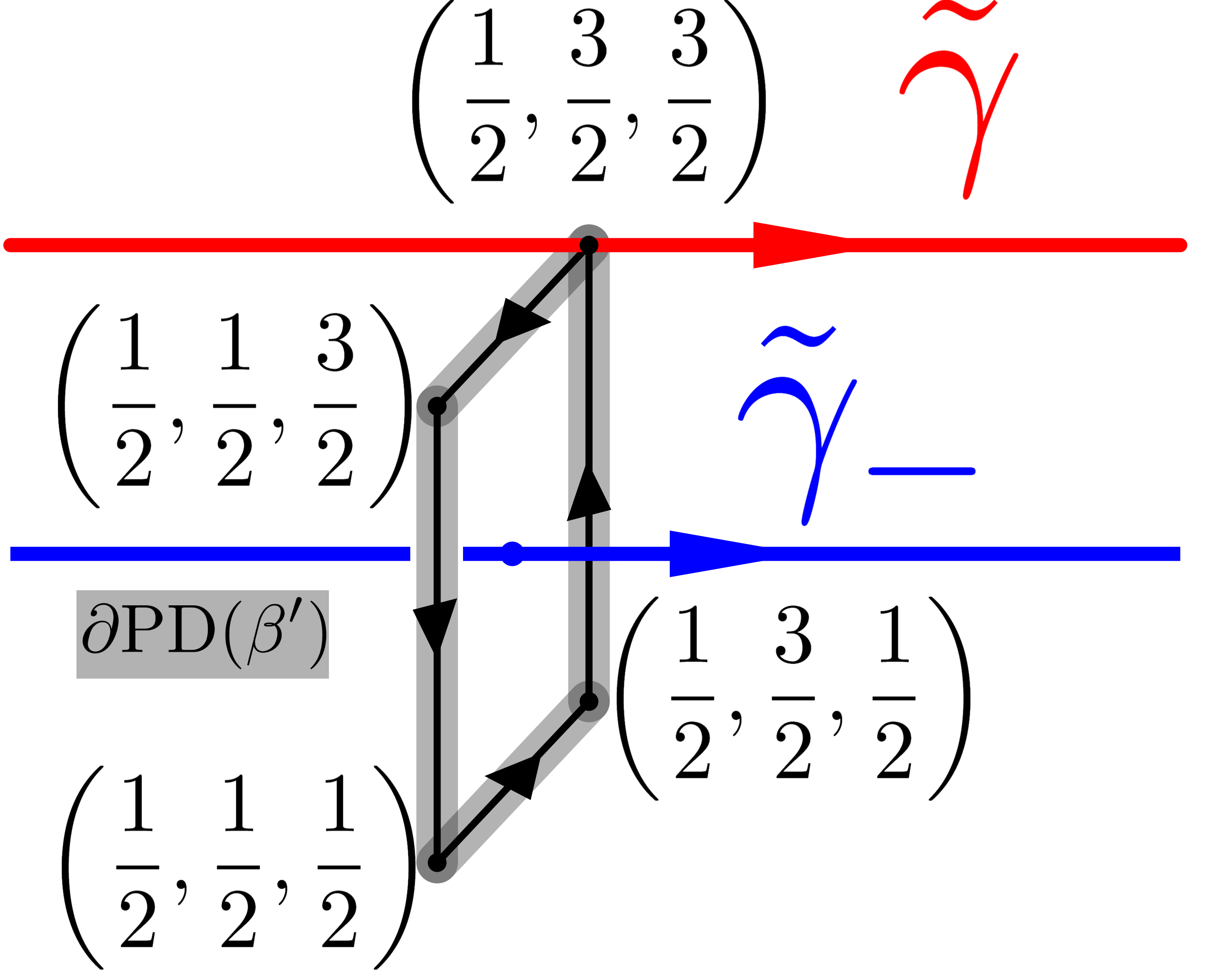}
    \caption{We can see that only $\partial\mathrm{PD}(\beta')$ and $\tilde{\gamma}_-$ are linked, but $\tilde{\gamma}_+$ is not. At the same time, $\tilde{\gamma}_-$ overlaps with the dashed line in Figure~\ref{fig:pPD}, and $\tilde{\gamma}_-$ is transformed by the path integral variable transformation $\bm{a}\to\bm{a}-(\beta'/2k,0,0)$.}
    \label{fig:beta}
    \end{center}
\end{figure}

\begin{figure}
    \begin{center}
    \includegraphics[width=110mm]{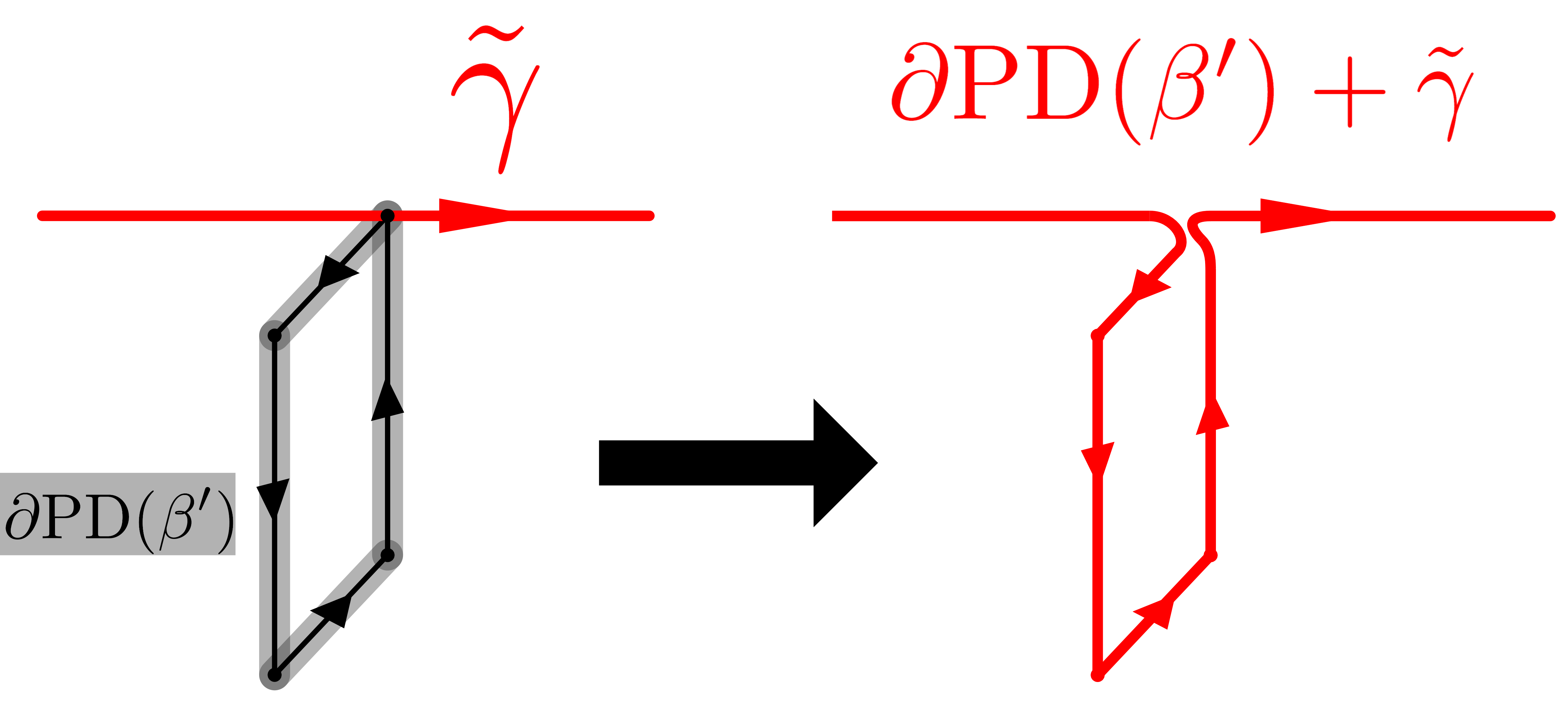}
    \caption{If $(\mathrm{supp}\,\partial\mathrm{PD}(\beta'))\cap(\mathrm{supp}\,\tilde{\gamma})\neq \emptyset$, additive transformation $\tilde{\gamma}\to\tilde{\gamma}+\partial\mathrm{PD}(\beta')$ introduces a local deformation that changes the self-linking number of $\tilde{\gamma}$.}
    \label{fig:slk}
    \end{center}
\end{figure}

What if, in addition to $\bm{\eta}_{\partial\mathrm{PD}(\beta')}$, there was also the Wilson line $\tilde{\gamma}$? If we assume that $\tilde{\gamma}_-$ contains one $((0,1,1),1)$ (see Figures \ref{fig:pPD} and \ref{fig:beta}), the variable transformation $\bm{a}\to\bm{a}-(1/k)(\delta_{-1}\beta',0,0)$ of the path integral changes the value of $W(\tilde{\gamma})$, and we see
\begin{equation}\label{equ11130}
    \begin{split}
        &\langle W(\partial\mathrm{PD}(\beta'))W(\tilde{\gamma})\rangle_\mathcal{CS}^{2k}=\frac{\mathcal{Z}^{2k}_{\mathrm{lat.}\mathcal{CS}}[\bm{\eta}_{\partial\mathrm{PD}(\beta')}+\bm{\eta}_{\tilde{\gamma}}]}{\mathcal{Z}^{2k}_{\mathrm{lat.}\mathcal{CS}}[\bm{0}]}\\
        =&e^{\frac{2\pi i}{4k}}\frac{\mathcal{Z}^{2k}_{\mathrm{lat.}\mathcal{CS}}[\bm{\eta}_{\tilde{\gamma}}]}{\mathcal{Z}^{2k}_{\mathrm{lat.}\mathcal{CS}}[\bm{0}]}=e^{\frac{2\pi i}{4k}}\langle W(\tilde{\gamma})\rangle_\mathcal{CS}^{2k}.
    \end{split}
\end{equation}
While paying attention to $(\mathrm{supp}\,\partial\mathrm{PD}(\beta'))\cap(\mathrm{supp}\,\tilde{\gamma})\neq \emptyset$, we can  find that this equality shows how the local $1$-form shift transformation changes the self-linking number of the whole Wilson line (see Figure \ref{fig:slk}). By a similar discussion, changes in the self-linking number can also be calculated from the shift transformation.

As discussed above, we can calculate the (self-) linking number of $\partial\mathrm{PD}(\beta')+\tilde{\gamma}$ from the shift transformation of the path integral. However, as pointed out so many times, this discussion is formal, since the path integral does not converge in $\mathcal{CS}$ theory. In the next subsection, we move away from formal discussion and move on to a discussion of convergent path integrals on $\mathcal{MCS}$.

\subsection{Expectation values in $\mathcal{MCS}^{2k,\epsilon}$ and error estimation}\label{sec1140}
The path integral does not converge in $\mathcal{CS}^{2k}$ theory, so the discussions of partition functions and expectation values so far have remained formal. Below we show that the path integral of $\mathcal{MCS}^{2k,\epsilon}$ is a finite complex Gaussian integral, and explicitly calculate the path integral of $\mathcal{Z}_{\mathrm{lat.}\mathcal{MCS}}^{2k,\epsilon}[\bm{A}]$. We also evaluate the difference, denoted $\mathrm{err}_{2k,\epsilon}(\bm{A})$, from the results ``derived'' from $\mathcal{CS}^{2k}$ theory.

\subsubsection{$\mathcal{Z}_{\mathrm{lat.}\mathcal{MCS}}[\boldsymbol{A}]$ and expectation value of Wilson line}\label{sec1141}
Let us start by calculating the partition function. The definition we introduced is
\begin{equation}\label{equ11140}
    \begin{split}
        &\mathcal{Z}_\mathrm{lat.\mathcal{MCS}}^{2k,\epsilon}[\bm{A}]\\
        &=\int_{[\bm{a}]\in H^2_\mathrm{lat.DB}(\mathbb{L}^3)}\mathcal{D}[\bm{a}]\exp \left(2\pi i\,\mathcal{MCS}^{2k,\epsilon}_\mathrm{lat.DB}[[\bm{a}],[\bm{A}]]\right)\\
        &=\int_{[\bm{a}]\in H^2_\mathrm{lat.DB}(\mathbb{L}^3)}\mathcal{D}[\bm{a}]\exp \biggl(\int_{\mathbb{L}^3}\pi i\Bigl(2k\bm{a}\star\bm{a}+\bm{a}\star\bm{A}+\bm{A}\star\bm{a}\\
        &\qquad\qquad\qquad\qquad\quad+(0,0,0,d_{-1}(a^{(-1,2)}\smallsmile_1A^{(-1,2)}),0)\Bigr)-\epsilon\sum_\text{plauettes}|d\bm{a}|^2\biggr).\\
    \end{split}
\end{equation}
Next we introduce the decompositions of $\bm{a}$ and $\bm{A}$
\begin{equation}\label{equ11150}
    \begin{split}
        \bm{a}=(\delta_{-1}\omega,0,0)+(\delta_{-1}\beta,0,0)+(\eta_\gamma^{(1,0)},\eta_\gamma^{(0,1)},\eta_\gamma^{(-1,2)})=:\bm{\omega}+\bm{\beta}+\bm{\eta}_\gamma,\\
        \bm{A}=(\delta_{-1}\Omega,0,0)+(\delta_{-1}B,0,0)+(\eta_\Gamma^{(1,0)},\eta_\Gamma^{(0,1)},\eta_\Gamma^{(-1,2)})=:\bm{\Omega}+\bm{B}+\bm{\eta}_\Gamma,\\
    \end{split}
\end{equation}
where
\begin{align}\label{equ11160}
    &\omega,\Omega\in\Omega_\mathrm{coexact}^1(\mathbb{L}^3),&&\beta,B\in H^1(\mathbb{L}^3;\mathbb{R}/\mathbb{Z}),&&[\gamma],[\Gamma]\in H_1(\mathbb{L}^3;\mathbb{Z}).
\end{align}
Then, let us assign the decomposition to the integral. From $\int\bm{\omega}\star\bm{\beta}=\int\bm{\beta}\star\bm{\omega}=\int\bm{\beta}\star\bm{\beta}=0$, $d\bm{\beta}=0$, and from $k\int \bm{\eta}_\gamma\star\bm{\eta}_\gamma\in\mathbb{Z}$, we get
\begin{equation}\label{equ11170}
    \begin{split}
        &\mathcal{Z}_\mathrm{lat.\mathcal{MCS}}^{2k,\epsilon}[\bm{A}]=\int_{\omega\in\Omega_\mathrm{coexact}^1(\mathbb{L}^3)} d\omega\sum_{\gamma\in H_1(\mathbb{L}^3;\mathbb{Z})}\\
        &\Biggl[\int_{\beta\in H^1(\mathbb{L}^3;\mathbb{R}/\mathbb{Z})} d\beta\exp \biggl(\pi i\int_{\mathbb{L}^3}(2k\bm{\eta}_\gamma+\bm{A})\star\bm{\beta}+\bm{\beta}\star(2k\bm{\eta}_\gamma+\bm{A})\biggr)\\
        &\exp\biggl(\int_{\mathbb{L}^3}\pi i\{2k(\bm{\omega}\star\bm{\omega}+\bm{\omega}\star\bm{\eta}_\gamma+\bm{\eta}_\gamma\star\bm{\omega})\\
        &+(\bm{\omega}+\bm{\eta}_\gamma)\star\bm{A}+\bm{A}\star(\bm{\omega}+\bm{\eta}_\gamma)+(0,0,0,d_{-1}(a^{(-1,2)}\smallsmile_1A^{(-1,2)}),0)\}\\
        &-\epsilon\sum_\text{plaquettes}|d\omega+d\bm{\eta}_\gamma|^2\biggr)\Biggr].
    \end{split}
\end{equation}
Using $\bm{A}\star\bm{\beta}=(\bm{\Omega}+\bm{B}+\bm{\eta}_\Gamma)\star\bm{\beta}=\bm{\eta}_\Gamma\star\bm{\beta}$ (resp. left and right reversed), and integrating $\bm{\beta}$, we find
\begin{equation}\label{equ11180}
    \begin{split}
        &=\int_{\omega\in\Omega_\mathrm{coexact}^1(\mathbb{L}^3)} d\omega\sum_{\gamma\in H_1(\mathbb{L}^3;\mathbb{Z})}\Biggl[\delta_{-2k\gamma,\Gamma}\\
        &\exp\biggl(\int_{\mathbb{L}^3}\pi i\{2k(\bm{\omega}\star\bm{\omega}+\bm{\omega}\star\bm{\eta}_\gamma+\bm{\eta}_\gamma\star\bm{\omega})\\
        &+(\bm{\omega}+\bm{\eta}_\gamma)\star\bm{A}+\bm{A}\star(\bm{\omega}+\bm{\eta}_\gamma)+(0,0,0,d_{-1}(a^{(-1,2)}\smallsmile_1A^{(-1,2)}),0)\}\\
        &-\epsilon\sum_\text{plaquettes}|d\omega+d\bm{\eta}_\gamma|^2\biggr)\Biggr]
    \end{split}
\end{equation}
\begin{equation}\label{equ11190}
        =\left\{\begin{array}{cl}
        &\\
        0&(\Gamma\notin H_1(\mathbb{L}^3;2k\mathbb{Z}))\\
        &\\
        &\\
        \displaystyle\int_{\omega\in\Omega_\mathrm{coexact}^1(\mathbb{L}^3)}d\omega\exp\biggl(\int_{\mathbb{L}^3}\pi i\{2k(\bm{\omega}\star\bm{\omega}-\bm{\omega}\star\bm{\eta}_{\Gamma/2k}-\bm{\eta}_{\Gamma/2k}\star\bm{\omega})&\\
        &\\
        +(\bm{\omega}-\bm{\eta}_{\Gamma/2k})\star\bm{A}+\bm{A}\star(\bm{\omega}-\bm{\eta}_{\Gamma/2k})&\\
        &(\Gamma\in H_1(\mathbb{L}^3;2k\mathbb{Z}))\\
        +(0,0,0,d_{-1}(a^{(-1,2)}\smallsmile_1A^{(-1,2)}),0)\}&\\
        &\\
        \displaystyle-\epsilon\sum_\text{plaquettes}|d\omega-d\bm{\eta}_{\Gamma/k}|^2\biggr)&\\
        &
        \end{array}\right..
\end{equation}
When $\Gamma\in H_1(\mathbb{L}^3;2k\mathbb{Z})$, it is possible to perform a gauge transformation such that $\bm{A}\in H_\mathrm{lat.DB}^2(\mathbb{L}^3;2k\mathbb{Z})$. In this case, we can express $\frac{1}{2k}\bm{A}$ as $\frac{1}{2k}(\bm{\Omega}+\bm{B})+\bm{\eta}_{\Gamma/2k}$. Furthermore, from the equality
\begin{equation}\label{equ11200}
    \int_{\mathbb{L}^3}(0,0,0,d_{-1}(a^{(-1,2)}\smallsmile_1A^{(-1,2)}),0)\in 2k\mathbb{Z},
\end{equation}
which derived from $\bm{A}\in H_\mathrm{lat.DB}^2(\mathbb{L}^3;2k\mathbb{Z})$, we complete the square using $\frac{1}{2k}\bm{A}$, and we get
\begin{equation}\label{equ11210}
        =\left\{\begin{array}{cl}
        &\\
        0&(\Gamma\notin H_1(\mathbb{L}^3;2k\mathbb{Z}))\\
        &\\
        &\\
        \displaystyle\int_{\omega\in\Omega_\mathrm{coexact}^1(\mathbb{L}^3)}d\omega\exp\Biggl[&\\
        &\\
        \displaystyle\int_{\mathbb{L}^3}2\pi ik\left(\bm{\omega}-\bm{\eta}_{\Gamma/2k}+\frac{1}{2k}\bm{A}\right)\star\left(\bm{\omega}-\bm{\eta}_{\Gamma/2k}+\frac{1}{2k}\bm{A}\right)&(\Gamma\in H_1(\mathbb{L}^3;2k\mathbb{Z}))\\
        &\\
        \displaystyle -2\pi ik\int_{\mathbb{L}^3}\frac{\bm{A}}{2k}\star\frac{\bm{A}}{2k}-\epsilon\sum_\text{plaquettes}|d\omega-d\bm{\eta}_{\Gamma/2k}|^2\Biggr]&\\
        &
        \end{array}\right..
\end{equation}
From $-\bm{\eta}_{\Gamma/k}+\frac{1}{k}\bm{A}=\frac{1}{k}(\bm{\Omega}+\bm{B})$, and $\int_{\mathbb{L}^3}(\bm{\omega}+\bm{\Omega})\star\bm{B}=\int_{\mathbb{L}^3}\bm{B}\star(\bm{\omega}+\bm{\Omega})=\int_{\mathbb{L}^3}\bm{B}\star\bm{B}=0$, we obtain
\begin{equation}\label{equ11220}
        =\left\{\begin{array}{cl}
        &\\
        0&(\Gamma\notin H_1(\mathbb{L}^3;2k\mathbb{Z}))\\
        &\\
        &\\
        \displaystyle\int_{\omega\in\Omega_\mathrm{coexact}^1(\mathbb{L}^3)}d\omega
        \exp\Biggl[\int_{\mathbb{L}^3}2\pi ik\left(\bm{\omega}+\frac{1}{2k}\bm{\Omega}\right)\star\left(\bm{\omega}+\frac{1}{2k}\bm{\Omega}\right)&\\
        &(\Gamma\in H_1(\mathbb{L}^3;2k\mathbb{Z}))\\
        \displaystyle -2\pi ik\int_{\mathbb{L}^3}\frac{\bm{A}}{2k}\star\frac{\bm{A}}{2k}-\epsilon\sum_\text{plaquettes}|d\omega-d\bm{\eta}_{\Gamma/2k}|^2\Biggr]&\\
        &
        \end{array}\right..
\end{equation}
Next, we change the path integral variable by $\omega'=\omega+\frac{1}{k}\Omega$. Since the measure $d\omega$ is invariant and $d\bm{A}=d\bm{\omega}+d\bm{\eta}_\Gamma$ holds, we conclude
\begin{equation}\label{equ11230}
        =\left\{\begin{array}{cl}
        &\\
        0&(\Gamma\notin H_1(\mathbb{L}^3;2k\mathbb{Z}))\\
        &\\
        &\\
        &\\
        \displaystyle\exp\left(-2\pi ik\int_{\mathbb{L}^3}\frac{\bm{A}}{2k}\star\frac{\bm{A}}{2k}\right)\int_{\omega\in\Omega_\mathrm{coexact}^1(\mathbb{L}^3)}d\omega\exp\Biggl[&\\
        &(\Gamma\in H_1(\mathbb{L}^3;2k\mathbb{Z}))\\
        \displaystyle\int_{\mathbb{L}^3}2\pi ik\bm{\omega}\star\bm{\omega}-\epsilon\sum_\text{plaquettes}\left|d\omega-\frac{1}{2k}d\bm{A}\right|^2\Biggr]&\\
        &
        \end{array}\right..
\end{equation}

Let us compare the consequence of the formal argument on $\mathcal{CS}_{\mathrm{lat.DB}}^{2k}$. By extracting the values that emerge from formal discussions on $\mathcal{CS}_{\mathrm{lat.DB}}^{2k}$ and summarizing the rest as errors, we write
\begin{equation}\label{equ11240}
    \mathcal{Z}_{\mathrm{lat.}\mathcal{MCS}}^{2k,\epsilon}[\bm{A}]=(1+\mathrm{err}_{2k,\epsilon}(\bm{A})) \exp\left[-2\pi ik\int_{\mathbb{L}^3}\frac{\bm{A}}{2k}\star\frac{\bm{A}}{2k}\right]\mathcal{Z}_{\mathrm{lat.}\mathcal{MCS}}^{2k,\epsilon}[0]\qquad(\mathrm{for}\;\Gamma\in H_1(\mathbb{L}^3;2k\mathbb{Z})),
\end{equation}
where $\mathrm{err}_{2k,\epsilon}$ is defined by
\begin{equation}\label{equ11250}
    1+\mathrm{err}_{2k,\epsilon}(\bm{A})=\frac{\displaystyle\int_{\omega\in \Omega^1_\mathrm{coexact}(\mathbb{L}^3)}\mathcal{D}\omega\exp\left[2\pi ik\int_{\mathbb{L}^3}\bm{\omega}\star\bm{\omega}+\sum_\mathrm{plaquettes}\left| d\omega-\frac{1}{2k}d\bm{A}\right|^2\right]}{\displaystyle\int_{\omega\in \Omega^1_\mathrm{coexact}(\mathbb{L}^3)}\mathcal{D}\omega\exp\left[2\pi ik\int_{\mathbb{L}^3}\bm{\omega}\star\bm{\omega}+\sum_\mathrm{plaquettes}\left| d\omega\right|^2\right]}.
\end{equation}

If we consider $\bm{\eta}_\gamma$ instead of $\bm{A}$, the expectation value becomes
\begin{equation}\label{equ11260}
    \begin{split}
        &\langle W(\tilde{\gamma})\rangle_{\mathrm{lat.}\mathcal{MCS}}^{2k,\epsilon}\\
        =&(1+\mathrm{err}_{2k,\epsilon}(\bm{\eta}_{\tilde{\gamma}})) \exp\left[-2\pi ik\int_{\mathbb{L}^3}\frac{\bm{\eta}_{\tilde{\gamma}}}{2k}\star\frac{\bm{\eta}_{\tilde{\gamma}}}{2k}\right]\\
        =&(1+\mathrm{err}_{2k,\epsilon}(\bm{\eta}_{\tilde{\gamma}}))\exp\left[-\frac{2\pi i}{4k}\mathrm{slk}^{\mathrm{mod}\,4k}(\tilde{\gamma})\right]\qquad(\mathrm{for}\;{\tilde{\gamma}}\in H_1(\mathbb{L}^3;2k\mathbb{Z})).
    \end{split}
\end{equation}
\subsubsection{Expression of coboundary operation and Chern--Simons action by basis}\label{sec1142}
To estimate the value of $1+\mathrm{err}_{2k,\epsilon}(\bm{\eta}_{\tilde{\gamma}})$, let us formulate the path integral strictly as a Gaussian integral and calculate it. $\Omega_\mathrm{coexact}^1(\mathbb{L}^3)$ is expressed using $2L^3-2$ bases, shown in \eqref{equA180}. $\{\bm{\eta}_\gamma\}_{[\gamma]\in H_1(\mathbb{L}^3;\mathbb{Z})}\cong H^2(\mathbb{L}^3;\mathbb{Z})\cong \mathbb{Z}^3$ can be represented by $3$ bases. Here, we give notations for these bases by
\begin{equation}\label{equ11270}
    \begin{split}
        \Omega_\mathrm{coexact}^1(\mathbb{L}^3)&=\mathrm{Span}\{\alpha_1,\alpha_2,\dots,\alpha_{2L^3-2}\},\\
        \{\bm{\eta}_\gamma\}_{\gamma\in H_1(\mathbb{L}^3;\mathbb{Z})}&=\mathrm{Span}_\mathbb{Z}\{\alpha_{2L^3-1},\alpha_{2L^3},\alpha_{2L^3+1}\}.
    \end{split}
\end{equation}

As we see in \eqref{equA080}, the kernel space of $d:\mathbb{R}^{3L^3}\cong\Omega^1(\mathbb{L}^3)\to\Omega^2(\mathbb{L}^3)\cong\mathbb{R}^{3L^3}$ is $(L^3-1)$-dimensional. Dividing the domain of $d\cong\mathbb{R}^{3L^3}$ by $\mathrm{Ker}\,d$ allows us to define an isomorphism $d_\mathrm{aut}$ by
\begin{equation}\label{equ11280}
    d_\mathrm{aut}:\Omega_\mathrm{coexact}^1(\mathbb{L}^3)\oplus\{\bm{\eta}_\gamma\}_{\gamma\in H_1(\mathbb{L}^3;\mathbb{Z})}\to\Omega^2_\mathrm{closed}(\mathbb{L}^3)_\mathbb{Z}.
\end{equation}
Compositing inclusion map
\begin{equation}\label{equ11290}
    \iota:\Omega^2_\mathrm{closed}(\mathbb{L}^3)_\mathbb{Z}\hookrightarrow\Omega^2(\mathbb{L}^3)\cong\mathbb{R}^{3L^3}
\end{equation}
to $d_\mathrm{Aut}$, we can construct
\begin{equation}\label{equ11300}
    \iota\circ d=d_\mathrm{mat}:\Omega_\mathrm{coexact}^1(\mathbb{L}^3)\oplus\{\bm{\eta}_\gamma\}_{\gamma\in H_1(\mathbb{L}^3;\mathbb{Z})}\to\mathbb{R}^{3L^3}.
\end{equation}
$d_\mathrm{mat}$ can be expressed as a matrix $d_\mathrm{mat}\in M_{(3L^3)\times (2L^3+1)}(\mathbb{R})$ by the basis $\{\alpha_i\}_{i=1}^{2L^3+1}$. From the regularity of $d_\mathrm{aut}$, $\mathrm{rank}\,d_\mathrm{mat}=2L^3+1$ holds.

Next we introduce a matrix $S\in M_{(2L^3-2)\times(2L^3-2)}(\mathbb{R})$ by
\begin{equation}\label{equ11310}
    S_{ij}=2\pi k\int_{\mathbb{L}^3} \alpha_i\star\alpha_j.
\end{equation}
For all $\alpha=\sum_j b_j\alpha_j\;,\;b=(b_1,\dots,b_{2L^2-2})\in \mathbb{R}^{2L^2-2}$, it follows that
\begin{equation}\label{equ11320}
    2\pi ik\int_{\mathbb{L}^3} \alpha\star\alpha=ib^\mathsf{T}Sb=ib^\mathsf{T}S_\mathrm{sym}b,
\end{equation}
where $S_\mathrm{sym}$ denotes $\frac{1}{2}(S+S^\mathsf{T})$.
\subsubsection{Error estimation}\label{sec1143}
Finally, we estimate the value of $1+\mathrm{err}_{2k,\epsilon}(\bm{A})$. Applying the basis expression to the definition of $1+\mathrm{err}_{2k,\epsilon}(\bm{A})$, we get the expression
\begin{equation}\label{equ11330}
    1+\mathrm{err}_{2k,\epsilon}(\bm{A})=\frac{\displaystyle\int_{\mathbb{R}^{2L^3-2}}\exp\left(ix^\mathsf{T}S_\mathrm{sym}x-\epsilon \left(x+\frac{v}{2k}\right)^\mathsf{T}d_\mathrm{mat}^\mathsf{T}d_\mathrm{mat}\left(x+\frac{v}{2k}\right)\right)dx}{\displaystyle\int_{\mathbb{R}^{2L^3-2}}\exp(ix^\mathsf{T}S_\mathrm{sym}x-\epsilon x^\mathsf{T}d_\mathrm{mat}^\mathsf{T}d_\mathrm{mat}x)dx},
\end{equation}
denoting the basis expression of $\frac{1}{2k}\bm{\Omega}+\frac{1}{2k}\bm{\eta}_\Gamma$ by $\frac{v}{2k}$, where $\frac{d\bm{A}}{2k}=\frac{1}{2k}\bm{\Omega}+\frac{1}{2k}\bm{B}+\frac{1}{2k}\bm{\eta}_\Gamma$ given in \eqref{equ11150}.

From the inequality \eqref{equB260}, $\mathrm{err}_{2k,\epsilon}(\bm{A})$ can be evaluated by
\begin{equation}\label{equ11340}
    |\mathrm{err}_{2k,\epsilon}(\bm{A})|\leq2\epsilon \|d_\mathrm{mat}\|_\infty\|d_\mathrm{mat}\|_1\left\|\frac{v}{2k}\right\|^2+O(\epsilon^2).
\end{equation}
In the case where $16<L$,
\begin{align}\label{equ11350}
    &\|d_\mathrm{mat}\|_\infty=\max_i\sum_j|(d_\mathrm{mat})_{ij}|\leq L,&&\|d_\mathrm{mat}\|_1=\max_j\sum_i|(d_\mathrm{mat})_{ij}|\leq17
\end{align}
hold, and we conclude
\begin{equation}\label{equ11360}
    |\mathrm{err}_{2k,\epsilon}(\bm{A})|\leq\frac{17\epsilon L}{2k^2}\left\|v\right\|^2+O(\epsilon^2).
\end{equation}
\subsection{Comparison between the modified Villain formalism and the lattice DB formalism}
In this subsection, we collect and compare the corresponding concepts of Chern--Simons theory in both the modified Villain formalism~\cite{Jacobson:2023cmr, Xu:2024hyo} and our lattice DB formalism. To match the conventions of~\cite{Jacobson:2023cmr} and make the comparison easier, we refrain from using the notation $2k$ in our lattice DB formalism and instead denote the level by $K$ used in Section~\ref{sec1310}. Let us begin by examining the following table.
\begin{longtable}{|c||c|c|}\caption{Comparison between the modified Villain formalism and the lattice DB formalism}\\\hline
    &modified Villain formalism&lattice DB formalism\\\endfirsthead\hline\hline
    gauge fields&\parbox[]{4cm}{\begin{align*}a\in C^1(\text{lattice};\mathbb{R}),\\n\in C^2(\text{lattice};\mathbb{Z})\end{align*}}&\parbox[]{4cm}{\begin{gather*}a\in C^{(1,0)}_\mathrm{lat.DB}(\mathbb{L}^3),\\\lambda\in C^{(0,1)}_\mathrm{lat.DB}(\mathbb{L}^3),\\n\in C^{(-1,2)}_\mathrm{lat.DB}(\mathbb{L}^3),\\\delta_0a=d_0\lambda,\;\delta_1\lambda=-d_{-1}n\end{gather*}}\\\hline
    line integral&$\displaystyle\sum_{s=0}^{l-1}a(x(s);\mu(s))$&\parbox[]{5.5cm}{\begin{gather*}\sum_{s=0}^{l-1}\Bigl(a(x(s);\mu(s);x_{+++}(s))\\+\lambda(x(s);;x_{+++}(s),x_{+++}(s+1))\Bigr)\end{gather*}}\\\hline
    curvature&$da-2\pi n$&$2\pi d\bm{a}$\\\hline
    Chern class&$[n]\in H^2(\text{lattice};\mathbb{Z})$&\parbox[]{5.5cm}{\begin{equation*}\begin{split}[d\bm{a}]&\in H^2(\mathbb{L}^3;\mathbb{Z})\\\cong [n]&\in H^2(\text{\v{C}ech nerve};\mathbb{Z})\end{split}\end{equation*}}\\\hline
    \begin{tabular}{c}ordinary $0$-form\\$\mathbb{R}$-gauge trans.\end{tabular}&\parbox[]{4cm}{\begin{equation*}\begin{split}a&\to a+d\xi,\\n&\to n,\\\xi&\in C^0(\text{lattice};\mathbb{R})\end{split}\end{equation*}}&\parbox[]{4cm}{\begin{equation*}\begin{split}a&\to a+d_0\xi^{(0,0)},\\\lambda&\to \lambda+\delta_0\xi^{(0,0)},\\\xi&\in C^{(0,0)}(\mathbb{L}^3)\end{split}\end{equation*}}\\\hline
    \begin{tabular}{c}large $1$-form\\$\mathbb{Z}$-gauge trans.\end{tabular}&\parbox[]{4cm}{\begin{equation*}\begin{split}a&\to a+2\pi m,\\n&\to n+dm,\\m&\in C^{1}(\text{lattice};\mathbb{Z})\end{split}\end{equation*}}&\parbox[]{4cm}{\begin{equation*}\begin{split}\lambda&\to \lambda-d_{-1}m^{(-1,1)},\\n&\to n+\delta_{1}m^{(-1,1)},\\m&\in C^{(-1,1)}(\mathbb{L}^3)\end{split}\end{equation*}}\\\hline
    CS action&\parbox[]{4cm}{\begin{equation*}\begin{split}&S[a,n,\phi]\\=&\sum_c \Bigl\{\frac{ik}{4\pi}[a\smallsmile da\\-&2\pi(a\smallsmile n+n\smallsmile a)]\\-&\frac{ik}{2}a\smallsmile_1 dn+i\varphi\smallsmile dn\Bigr\}\end{split}\end{equation*}}&$\displaystyle S[\bm{a}]=\pi i\int_{\mathbb{L}^3}K\bm{a}\star\bm{a}$\\\hline
    $\mathbb{Z}_k$ $1$-form trans.&\parbox[]{4cm}{\begin{align*}a\to&a+\frac{2\pi}{k}\omega,\\\omega\in&C^1(\text{lattice};\mathbb{Z}),\\d\omega=&0\pmod k\end{align*}}&\parbox[]{4cm}{\begin{align*}\bm{a}\to&\bm{a}+\left(\delta_{-1}\left(\frac{1}{K}\beta\right),0,0\right),\\[5pt]\beta\in&C^1(\mathbb{L}^3;\mathbb{Z}),\quad d\beta=0\end{align*}}\\\hline
    staggered sym. trans.&$a\to a+\chi$&$\bm{a}\to\bm{a}+\left(\delta_{-1}\chi,0,0\right)$\\\hline
    framed Wilson line&\parbox[]{4cm}{\begin{align*}\pi i&\sum_{c}\Bigl\{\mathrm{PD}(\tilde{\gamma})\cup a\\+&a\cup\mathrm{PD}(\tilde{\gamma})+n\cup_1\mathrm{PD}(\tilde{\gamma})\Bigr\}\end{align*}}&\parbox[]{4cm}{\begin{align*}\pi i&\int_{\mathbb{L}^3}\Bigl\{\bm{\eta}_{\tilde{\gamma}}\star\bm{a}+\bm{a}\star\bm{\eta}_{\tilde{\gamma}}\\&+(0,0,0,\delta_{-1}(n\smallsmile_1\eta^{(-1,2)}_{\tilde{\gamma}}),0)\Bigr\}\end{align*}}\\\hline
\end{longtable}
Based on the above comparison, let us give several comments on whether the modified Villain formalism and our lattice DB formalism are equivalent, and how they differ.

Let us point out that the gauge-fixed Degrees of Freedom (DoF) of the $U(1)$ gauge field in the modified Villain formalism seem to match the gauge-fixed DoF in the lattice DB formalism. In the modified Villain formalism, the field $a$ can be decomposed using the lattice Hodge decomposition discussed in Appendix~\ref{secA10}. This decomposition splits $a$ into the DoF of a coexact $1$-form and those of the global holonomies $H^1(\text{lattice};\mathbb{R}/\mathbb{Z})$. This decomposition is exactly the same as that of $\alpha$ and $\beta$ defined in Section~\ref{sec800}. The DoF of gauge fixed $n$ in the modified Villain formalism correspond the $2$-cohomology on the lattice, and this is isomorphic to the DoF carried by $\bm{\eta}_{\gamma}$ in the lattice DB formalism.

However, in the lattice DB formalism, $n$ is a $\mathbb{Z}$-valued 2-cohomology class on the \v{C}ech nerve, rather than on the square lattice. For this reason, proving equivalence at the level of cocycles seems difficult. Since the $n$ in the lattice DB formalism has a clear counterpart in the continuum theory, a more precise comparison should be possible if one constructs an interpolation to a continuum $U(1)$ gauge field starting from both formulations.

Thus, by looking at the table, we can see that the two theories are very similar, yet several subtle points remain where the correspondence is not immediately clear. For instance, it is not obvious why $\lambda$ can be omitted in the Villain formalism, or why the definitions of the curvature differ. Examining these issues in detail should deepen our understanding of $U(1)$ gauge theories on the lattice.
\newpage
\section{Non-invertible defect of lattice massless QED associated with rational angle chiral transformation}\label{sec1200}
As an application of lattice Chern--Simons theory, we present the construction of non-invertible defects in lattice massless QED. The definition of the chiral transformation defect of fractional rotation in continuous massless QED is discussed by~\cite{Choi:2022jqy,Cordova:2022ieu}. The chiral transformation defect discussed by~\cite{Choi:2022jqy,Cordova:2022ieu} is constructed by coupling the level $N$ Chern--Simons theory to the na\"{i}ve chiral transformation operator. On the other hand the construction of the non-invertible defect in massless QED on a lattice~\cite{Honda:2024yte}, is based on lattice BF theory, instead of Chern--Simons theory.

In this section, we follow the method of~\cite{Choi:2022jqy,Cordova:2022ieu} and show that the non-invertible chiral transformation defect of fractional rotation can be defined on a four-dimensional cubic torus $\mathbb{L}^4$ using the Chern--Simons theory defined so far.

\subsection{Construction of discrete non-invertible chiral transformation defect in continuum}\label{sec1210}

In this section, we provide a very brief review of the construction method for non-invertible defects proposed by~\cite{Choi:2022jqy,Cordova:2022ieu}.

Let us consider the Euclidean action of continuum massless QED:
\begin{equation}\label{equ12010}
    S_\mathrm{QED}[\Psi,\overline{\Psi},A]=\int d^4x \frac{1}{4e^2}F_{\mu\nu}F^{\mu\nu}+i\overline{\Psi}(\partial_\mu-iA_\mu)\gamma^\mu\Psi.
\end{equation}
$S_\mathrm{QED}[\Psi,\overline{\Psi},A]$ as a classical theory has the symmetry of chiral transformation $\Psi\to e^{i\alpha\gamma_5/2}$. Noether's theorem guarantees the conservation of axial current
\begin{equation}\label{equ12020}
    j_\mu^A=\overline{\Psi}\gamma_5\gamma_\mu\Psi,
\end{equation}
and the chiral transformation operator on $M$ can be defined by
\begin{equation}\label{equ12030}
    U_\alpha(M):=\exp\left(\frac{i\alpha}{2}\oint_M\ast j^A\right),
\end{equation}
where $\ast$ is the Hodge dual operator. On the other hand, massless QED as a QFT has the chiral anomaly
\begin{equation}\label{equ12040}
    d\ast j^A=\frac{1}{16\pi^2}\epsilon_{\mu\nu\rho\sigma}F_{\mu\nu}F_{\rho\sigma}=\frac{1}{4\pi^2}F\wedge F=d\left(\frac{1}{4\pi^2}A\wedge dA\right),
\end{equation}
and $U_\alpha(M)$ cannot be defined as a well-defined defect.

Then we na\"{i}vely introduce $\ast\hat{j}^A:=\ast j^A-\frac{1}{4\pi^2}A\wedge dA$ and
\begin{equation}\label{equ12050}
    \hat{U}_\alpha(M)=\exp\left[\frac{i\alpha}{2}\oint_M\left(\ast\hat{j}^A\right)\right]=\exp\left[\frac{i\alpha}{2}\oint_M\left(\ast j^A-\frac{1}{4\pi^2}A\wedge dA\right)\right]
\end{equation}
to cancel the anomaly and try to define $\hat{U}_\alpha(M)$ as a well-defined topological defect. But the Chern--Simons-like term in $\hat{U}_\alpha(M)$
\begin{equation}\label{equ12060}
    -\frac{i\alpha}{2}\frac{1}{4\pi^2}A\wedge dA
\end{equation}
does not satisfy the quantization condition of gauge transformation, and we find $\hat{U}_\alpha(M)$ is still ill-defined. Then we restrict the situation to the case where $\alpha=\frac{2\pi}{N}$, and the Chern--Simons-like term
\begin{equation}\label{equ12070}
    -\frac{2\pi i}{2k}\frac{1}{4\pi^2}A\wedge dA
\end{equation}
can have another expression using $U(1)$ gauge field $a$ on $M$ and level $N$ Chern--Simons theory
\begin{equation}\label{equ12080}
    \exp\left(-\oint_M\frac{2\pi i}{2N}\frac{1}{4\pi^2}A\wedge dA\right)=\int_{a\in\mathcal{A}(M)/\mathcal{G}} \mathcal{D}a\exp\left\{i\oint_M\left(\frac{N}{4\pi}a\wedge da+\frac{1}{2\pi}a\wedge dA\right)\right\}.
\end{equation}
This is known in the context of the $(2+1)$-dimensional fractional quantum Hall effect~\cite{Tong:2016kpv}. Finally, we complete the definition of chiral transformation defect of the $\frac{\pi}{N}$ rotation by
\begin{equation}\label{equ12090}
    \mathcal{D}_{\frac{1}{N}}(M)=\exp\left[i\oint_M\left(\frac{2\pi}{2N}\ast j^A+\frac{N}{4\pi}a\wedge da+\frac{1}{2\pi}a\wedge dA\right)\right].
\end{equation}
While we do not present a proof, it can be shown that this defect is non-invertible.
\subsection{Overlap fermion and admissible gauge field on $\mathbb{L}^4$}\label{sec1220}
To discuss chiral anomalies in the framework of lattice fermions, we use overlap fermions~\cite{Neuberger:1997fp,Neuberger:1998wv,Ginsparg:1981bj,Luscher:1998du} in this section. We briefly review the definition and properties of overlap fermions on $\mathbb{L}^4$, following~\cite{Aoki:2012}. We also provide a very brief review of the properties of the admissible $U(1)$ gauge fields on $\mathbb{L}^4$ based on~\cite{Luscher:1998du,Hernandez:1998et,Luscher:1998kn,Kadoh:2003ii,Ikeda:2022}.

\subsubsection{Overlap Dirac operator and admissible $U(1)$ link variable}\label{sec1221}
In this section, we consider the $U(1)$ link variable $U(x;\mu)$ on $\mathbb{L}^4$. Using covariant differential $\nabla_\mu$ on lattice, we introduce Wilson Dirac operator $D_W$ by
\begin{gather}
    D_W=\sum_\mu\left\{-\frac{1}{2}\gamma_\mu(\nabla_\mu-\nabla^\dagger_\mu)+\frac{1}{2}\nabla^\dagger_\mu\nabla_\mu\right\},\label{equ12100}\\
    \nabla_\mu\Psi(x)=U(x,\mu)\Psi(x+\hat{\mu})-\Psi(x),\quad\nabla^\dagger_\mu\Psi(x)=\Psi(x)-U^\dagger(x-\hat{\mu},\mu)\Psi(x-\hat{\mu}),\label{equ12110}
\end{gather}
and the overlap Dirac operator $D$ by
\begin{gather}
    D=\frac{1}{2}\left(1+X\frac{1}{\sqrt{X^\dagger X}}\right),\label{equ12120}\\
    X=D_W-M\;,\;(0<M<2).\label{equ12130}
\end{gather}
The overlap Dirac operator $D$ satisfies the Ginsparg-Wilson relation
\begin{equation}\label{equ12140}
    D\gamma_5+\gamma_5 D=D\gamma_5D\quad(\Longleftrightarrow D\gamma_5+(1-2D)\gamma_5 D=0)
\end{equation}
and as a result, $D$ falls outside the scope of the Nielsen--Ninomiya theorem and can provide a single Weyl fermion. From the Ginsparg-Wilson relation, the overlap fermion action $S_F[\Psi,\overline{\Psi},U]=\sum_{x\in\mathbb{L}^4}\overline{\Psi}D\Psi$ is invariant under modified infinitesimal chiral transformations:
\begin{align}\label{equ12150}
    &\delta_\alpha\Psi(x)=i\alpha\hat{\gamma}_5\Psi(x),&&\delta_\alpha\bar{\Psi}(x)=i\alpha\bar{\Psi}(x)\gamma_5,&&\hat{\gamma}_5=(1-2D)\gamma_5.
\end{align}

Since the expression of $D$ has the inverse matrix $(X^\dagger X)^{-1/2}$, the locality of the interaction is not necessarily guaranteed. What becomes important here is the admissibility condition for the gauge field $U$. Let the plaquette product of $U$ be
\begin{equation}\label{equ12160}
    P(x;\mu,\nu)=U(x;\mu)U(x+\hat{\mu};\nu)U^{-1}(x+\hat{\nu};\mu)U^{-1}(x;\nu).
\end{equation}
We introduce the admissibility condition of $U$ by
\begin{equation}\label{equ12170}
    |1-P(x;\mu,\nu)|<\varepsilon<\frac{1}{30}\qquad(\text{for all $x,\mu,\nu$}),
\end{equation}
and we call $U$ that satisfies this condition an admissible gauge field. It is known that the overlap Dirac operator constructed from an admissible gauge field is guaranteed to make the interactions local in certain sense~\cite{Hernandez:1998et}.

In addition, it is known that the admissible gauge field can be divided into topological sectors~\cite{Luscher:1981zq,Luscher:1998du,Kadoh:2003ii}. This corresponds to the classification of the $U(1)$ gauge field by the first Chern class $H^2(\mathbb{L}^4;\mathbb{Z})$. To explain this classification, we first introduce the curvature $F_{\mu\nu}(x)$ by
\begin{equation}\label{equ12180}
    F_{\mu\nu}(x)=\frac{1}{i}\ln P(x;\mu,\nu)\qquad\left(\text{$-\frac{1}{2}<F_{\mu\nu}(x)<\frac{1}{2}$ for all $x,\mu,\nu$}\right).
\end{equation}
$F$ can be used to impose the admissibility condition by
\begin{equation}\label{equ12190}
    |F_{\mu\nu}(x)|<\varepsilon<\frac{1}{30}\qquad(\text{for all $x,\mu,\nu$}),
\end{equation}
which is a stricter condition than \eqref{equ12170}. While $F$ can be defined even if $U$ is not admissible, if $U$ is an admissible gauge field, $F$ satisfies Bianchi identity
\begin{equation}\label{equ12200}
    \partial_{\mu}F_{\nu\rho}+\partial_{\rho}F_{\mu\nu}+\partial_{\nu}F_{\rho\mu}=0
\end{equation}
where $\partial$ is forward difference of $\mathbb{L}^4$. Then we can see that $F$ is a cocycle on $\mathbb{L}^4$ and can verify its cohomology class from the magnetic flux of $F$
\begin{equation}\label{equ12210}
    m_{\mu\nu}=\frac{1}{2\pi}\sum_{s,t=0}^{L-1}F_{\mu\nu}(x+s\hat{\mu}+t\hat{\nu}).
\end{equation}
\subsubsection{Definition of the lattice massless QED and anomalous Ward--Takahashi identity}\label{sec1222}
Let us move on to the construction of lattice massless QED. The fermion action can be described by $S_F[\Psi,\overline{\Psi},U]$, defined in Section~\ref{sec1221}. Giving a definition of the lattice (Euclidean) Maxwell action $S_G[U]$ by
\begin{gather}
    S_G[U]=\frac{1}{4e^2}\sum_{x,\mu,\nu}c_{\mu\nu}(x)(F_{\mu\nu}(x))^2,\label{equ12220}\\
    c_{\mu\nu}(x)=\left\{\begin{array}{ll}
        \displaystyle\left(1-\frac{(F_{\mu\nu}(x))^2}{\varepsilon^2}\right)^{-1}&(\mathrm{if}\;|F_{\mu\nu}(x)|<\varepsilon<1/30)\\[10pt]
        \infty&(\text{otherwise})
    \end{array}\right.,\label{equ12230}
\end{gather}
we can construct the action while imposing admissibility on $U$. Let us give the definition of lattice massless QED by
\begin{gather}
    S_{\mathrm{lat.QED}}[\Psi,\overline{\Psi},U]=S_F[\Psi,\overline{\Psi},U]+S_G[U],\label{equ12240}\\
    \mathcal{Z}_{\mathrm{lat.QED}}=\int\mathcal{D}\Psi\mathcal{D}\overline{\Psi}dU\exp(-S_{\mathrm{lat.QED}}[\Psi,\overline{\Psi},U]).\label{equ12250}
\end{gather}
According to~\cite{Kikukawa:1998py}, lattice massless QED has axial current $j^A$, and the anomalous Ward--Takahashi identity~\cite{Fujikawa:1983bg} 
\begin{equation}\label{equ12260}
    \begin{split}
        0=&\int\mathcal{D}\Psi\mathcal{D}\overline{\Psi}dU \Biggl[e^{-S_{\mathrm{lat.QED}}[\Psi,\overline{\Psi},U]}\\
        &\exp\left\{
     i\sum_{\substack{4\text{d cube}\\\text{in }\mathbb{L}^4}}\left(-(\partial_\mu\alpha(x))j^A_\mu(x)+\frac{\alpha(x)}{16\pi^2}\epsilon_{\mu\nu\rho\sigma}F_{\mu\nu}(x)F_{\rho\sigma}(x+\hat{\mu}+\hat{\nu})\right)\right\}\Biggr]
    \end{split}
\end{equation}
of finite chiral transformation
\begin{align}\label{equ12261}
    &\Psi(x)\to e^{i\hat{\gamma}_5\alpha(x)/2}\Psi(x),&&\overline{\Psi}(x)\to \overline{\Psi}(x)e^{i\hat{\gamma}_5\alpha(x)/2}
\end{align}
also holds. When we introduce $\alpha(x)$ using integers $a,b,N$ by
\begin{equation}\label{equ12270}
    \alpha(x)=\left\{\begin{array}{ll}
        \displaystyle\frac{2\pi}{N}&(0\leq x_0\leq a-1\text{ or }b\leq x_0\leq L-1)\\[10pt]
        0&(\text{otherwise})
    \end{array}\right.,
\end{equation}
the anomalous Ward--Takahashi identity is expressed by
\begin{equation}\label{equ12280}
    \begin{split}
        &\int\mathcal{D}\Psi\mathcal{D}\overline{\Psi}dU \left[e^{-S_{\mathrm{lat.QED}}}\exp\left\{\frac{2\pi i}{N}\sum_{\substack{x_0=a\\(x_1,x_2,x_3)\in\mathbb{L}^3}}j_0^A(x)\right\}\exp\left\{-\frac{2\pi i}{N}\sum_{\substack{x_0=b\\(x_1,x_2,x_3)\in\mathbb{L}^3}}j_0^A(x)\right\}\right]\\
        =&\int\mathcal{D}\Psi\mathcal{D}\overline{\Psi}dU \left[e^{-S_{\mathrm{lat.QED}}}\exp\left\{\frac{2\pi i}{N}\sum_{\substack{0\leq x_0\leq a-1\\\text{ or }b\leq x_0\leq L-1\\(x_1,x_2,x_3)\in\mathbb{L}^3}}\left(\frac{1}{16\pi^2}\epsilon_{\mu\nu\rho\sigma}F_{\mu\nu}(x)F_{\rho\sigma}(x+\hat{\mu}+\hat{\nu})\right)\right\}\right].
    \end{split}
\end{equation}
We write this equality as an operator equation by
\begin{equation}\label{equ12290}
    U_{\frac{2\pi}{N}}(\mathbb{L}^3|_{x_0=a})U^\dagger_{\frac{2\pi}{N}}(\mathbb{L}^3|_{x_0=b})=\exp\left\{\frac{2\pi i}{N}\sum_{\substack{0\leq x_0\leq a-1\\\text{ or }b\leq x_0\leq L-1\\(x_1,x_2,x_3)\in\mathbb{L}^3}}\frac{1}{4\pi^2}F\wedge F\right\}.
\end{equation}

Here, let us consider $\sum F\wedge F$. Recall $F$ is a curvature defined from $U(1)$ link variable $U$ on $\mathbb{L}^4$. First, we give a method to decompose the DoF of $U$ into $\alpha',\beta',m$ without lattice DB formalism. Let us focus on $\left[\frac{F}{2\pi}\right]\in H^2(\mathbb{L}^4;\mathbb{Z})$. The $1$-form obtained by subtracting the magnetic flux $m_{\mu\nu}$ from $\frac{F}{2\pi}$ is the exact form, so the equation
\begin{equation}\label{equ12300}
    d\alpha'(x;\mu,\nu)=\frac{1}{2\pi}F_{\mu\nu}(x)-m_{\mu\nu}\delta_{x_\mu,0}\delta_{x_\nu,0}
\end{equation}
has a unique solution $\alpha'\in \Omega^1_\mathrm{coexact}(\mathbb{L}^4)$. Next, we identify the global holonomy part of the DoF. We focus on 
\begin{equation}\label{equ12301}
    d(U\cdot\exp(-2\pi i\alpha'))(x;\mu,\nu)=\mathbbm{1},
\end{equation}
and find that the line integral of $U\cdot\exp(-2\pi i\alpha')$ over closed line only depends on the homology class of the integration contour. Then we identify the global holonomy part $\beta'$ by
\begin{equation}\label{equ12310}
    \beta'=\frac{1}{2}\epsilon_{\mu\nu\rho}\left(\beta_{\mu\nu}\,\frac{1}{2\pi i}\ln\int_{e_\rho}(U\cdot\exp(-2\pi i\alpha'))\right),
\end{equation}
where $\beta_{\mu\nu}$ is defined in \eqref{equ10070}. These $\alpha',\beta'$ satisfy
\begin{equation}\label{equ12320}
    \int_{\Gamma}U=\exp\left(2\pi i\int_{\Gamma}(\alpha'+\beta')\right),\qquad F_{\mu\nu}(x)=d\alpha'+m_{\mu\nu}\delta_{x_\mu,0}\delta_{x_\nu,0}.
\end{equation}

Finally, let us express $\sum \frac{1}{4\pi^2}F\wedge F$ using $\alpha',\beta',m$. Since we can express $\frac{1}{4\pi^2}F\wedge F$ by
\begin{equation}\label{equ12330}
    \begin{split}
        &\frac{1}{4\pi^2}(F\wedge F)_{\mu\nu\rho\sigma}(x)=\frac{1}{4\pi^2}F_{\mu\nu}(x)F_{\rho\sigma}(x+\hat{\mu}+\hat{\nu})\\
        =&((d\alpha')_{\mu\nu}(x)+m_{\mu\nu}\delta_{x_\mu,0}\delta_{x_\nu,0})
        ((d\alpha')_{\rho\sigma}(x+\hat{\mu}+\hat{\nu})+m_{\rho\sigma}\delta_{x_\rho,0}\delta_{x_\sigma,0}),
    \end{split}
\end{equation}
we get the equality
\begin{equation}\label{equ12340}
    \begin{split}
        &\sum_{\substack{0\leq x_0\leq a-1\\\text{ or }b\leq x_0\leq L-1\\(x_1,x_2,x_3)\in\mathbb{L}^3}}\frac{1}{4\pi^2}F\wedge F\\
        =&\left(\sum_{\substack{0\leq x_0\leq a-1\\\text{ or }b\leq x_0\leq L-1}}\delta_{x_0,0}\right)\frac{1}{8}\sum_{\mu,\nu,\rho,\sigma}m_{\mu\nu}m_{\rho\sigma}-\sum_{\substack{x_0=b\\(x_1,x_2,x_3)\in\mathbb{L}^3}}\alpha'\wedge d\alpha'
        +\sum_{\substack{x_0=a\\(x_1,x_2,x_3)\in\mathbb{L}^3}}\alpha'\wedge d\alpha'\\
        +&\frac{1}{2}\sum_{ijk}\epsilon_{ijk}\biggl(m_{ij}\sum_{\substack{0\leq s\leq a-1\\\text{ or }b\leq s\leq L-1}}\sum_{t=0}^{L-1}(d\alpha')_{0k}(s\hat{0}+t\hat{k})+m_{0i}\sum_{s=0}^{L-1}\sum_{t=0}^{L-1}(d\alpha')_{jk}(s\hat{j}+t\hat{k})\\
        +&m_{ij}\sum_{\substack{0\leq s\leq a-1\\\text{ or }b\leq s\leq L-1}}\sum_{t=0}^{L-1}(d\alpha')_{0k}(\hat{i}+\hat{j}+s\hat{0}+t\hat{k})+m_{0i}\sum_{s=0}^{L-1}\sum_{t=0}^{L-1}(d\alpha')_{jk}(\hat{0}+\hat{i}+s\hat{j}+t\hat{k})\biggr)\\
        =&\frac{1}{8}\sum_{\mu,\nu,\rho,\sigma}m_{\mu\nu}m_{\rho\sigma}
        -\sum_{\substack{x_0=b\\(x_1,x_2,x_3)\in\mathbb{L}^3}}\alpha'\wedge d\alpha'
        +\sum_{\substack{x_0=a\\(x_1,x_2,x_3)\in\mathbb{L}^3}}\alpha'\wedge d\alpha'\\
        +&\frac{1}{2}\sum_{ijk}\epsilon_{ijk}\sum_{s=0}^{L-1}\biggl(-\alpha'_k(b\hat{0}+s\hat{k})-\alpha'_k(\hat{i}+\hat{j}+b\hat{0}+s\hat{k})+\alpha'_k(a\hat{0}+s\hat{k})+\alpha'_k(\hat{i}+\hat{j}+a\hat{0}+s\hat{k})\biggr)\\
        =&\frac{1}{8}\sum_{\mu,\nu,\rho,\sigma}m_{\mu\nu}m_{\rho\sigma}
        -\sum_{\substack{x_0=b\\(x_1,x_2,x_3)\in\mathbb{L}^3}}\alpha'\wedge d\alpha'
        +\sum_{\substack{x_0=a\\(x_1,x_2,x_3)\in\mathbb{L}^3}}\alpha'\wedge d\alpha'\\
        -&\frac{1}{2}\sum_{ijk}\epsilon_{ijk}\sum_{s=0}^{L-1}\biggl(\alpha'_k(b\hat{0}+s\hat{k})+\beta'_k(b\hat{0}+s\hat{k})+\alpha'_k(\hat{i}+\hat{j}+b\hat{0}+s\hat{k})+\beta'_k(\hat{i}+\hat{j}+b\hat{0}+s\hat{k})\biggr)\\
        +&\frac{1}{2}\sum_{ijk}\epsilon_{ijk}\sum_{s=0}^{L-1}\biggl(\alpha'_k(a\hat{0}+s\hat{k})+\beta'_k(a\hat{0}+s\hat{k})+\alpha'_k(\hat{i}+\hat{j}+a\hat{0}+s\hat{k})+\beta'_k(\hat{i}+\hat{j}+a\hat{0}+s\hat{k})\biggr),
    \end{split}
\end{equation}
where we used \eqref{equ12301} and \eqref{equ12310} in the last equal sign. We can also find 
\begin{equation}\label{equ12341}
    \begin{split}
        &\sum_{\substack{0<a\leq x_0\leq b-1< L-1\\(x_1,x_2,x_3)\in\mathbb{L}^3}}\frac{1}{4\pi^2}F\wedge F\\
        =&
        +\sum_{\substack{x_0=b\\(x_1,x_2,x_3)\in\mathbb{L}^3}}\alpha'\wedge d\alpha'
        -\sum_{\substack{x_0=a\\(x_1,x_2,x_3)\in\mathbb{L}^3}}\alpha'\wedge d\alpha'\\
        +&\frac{1}{2}\sum_{ijk}\epsilon_{ijk}\sum_{s=0}^{L-1}\biggl(\alpha'_k(b\hat{0}+s\hat{k})+\beta'_k(b\hat{0}+s\hat{k})+\alpha'_k(\hat{i}+\hat{j}+b\hat{0}+s\hat{k})+\beta'_k(\hat{i}+\hat{j}+b\hat{0}+s\hat{k})\biggr)\\
        -&\frac{1}{2}\sum_{ijk}\epsilon_{ijk}\sum_{s=0}^{L-1}\biggl(\alpha'_k(a\hat{0}+s\hat{k})+\beta'_k(a\hat{0}+s\hat{k})+\alpha'_k(\hat{i}+\hat{j}+a\hat{0}+s\hat{k})+\beta'_k(\hat{i}+\hat{j}+a\hat{0}+s\hat{k})\biggr),
    \end{split}
\end{equation}
from the same discussion.
\subsection{Non-invertible discrete chiral transformation defects realized by lattice DB Chern--Simons theory}\label{sec1230}
For simplicity, we assume that $N$ is an even integer.\footnote{It is possible to set up a situation where $N$ is odd using the odd level DB Chern--Simons theory described below.} In this section, we first define the lattice DB cocycle $\bm{A}[U|_{x_0=a}]$ corresponding to $U|_{x_0=a}$, and observe that $2\pi i\frac{N}{2}\int\frac{\bm{A}[U|_{x_0=a}]}{N}\star\frac{\bm{A}[U|_{x_0=a}]}{N}$ is equal to the boundary term of $\sum F\wedge F$. Then, we show that a non-invertible chiral symmetry defect on a lattice can be constructed in the same way as in~\cite{Choi:2022jqy}.

We assume that $m_{ij}\in N\mathbb{Z}\;(\text{for all $1\leq i<j\leq 3$})$ holds. In the following, we will use the notation of equations \eqref{equ12300}, \eqref{equ12310}, and \eqref{equ12320} to denote $\alpha,\beta,\bm{\eta}$ as
\begin{align}
    &\alpha=\alpha'|_{(\mathbb{L}^3|x_0=a)},&&\beta=\beta'|_{(\mathbb{L}^3|x_0=a)},\notag\\
    &\bm{\eta}\in Z^2_{\mathrm{lat.DB}}(\mathbb{L}^3|_{x_0=a};\mathbb{Z}),&&d\bm{\eta}=\delta_{-1}\left((x;\mu,\nu)\mapsto\sum_{1\leq i<j\leq 3}m_{ij}\delta_{x_i,0}\delta_{x_j,0}\right).\label{equ12350}
\end{align}
The lattice DB cocycle $\bm{A}[U|_{x_0=a}]$ equivalent to $U|_{x_0=a}$ can be defined by
\begin{equation}\label{equ12360}
    \bm{A}[U|_{x_0=a}]=(\delta_{-1}(\alpha+\beta),0,0)+\bm{\eta},
\end{equation}
which satisfies
\begin{equation}\label{equ12370}
    \begin{split}
        &\frac{N}{2}\int\frac{\bm{A}[U|_{x_0=a}]}{N}\star\frac{\bm{A}[U|_{x_0=a}]}{N}\\
        \stackrel{\mathbb{Z}}{=}&\frac{1}{2N}\sum_{\substack{\text{3d cubes in}\\(\mathbb{L}^3|x_0=a)}}\alpha\wedge d\alpha\\
        +&\frac{N}{2}\int_{(\mathbb{L}^3|x_0=a)}(\delta_{-1}(\alpha+\beta),0,0)\overrightarrow{\underleftarrow{\star}}\bm{\eta}+\bm{\eta}\overleftarrow{\underrightarrow{\star}}(\delta_{-1}(\alpha+\beta),0,0)\\
        =&\frac{1}{2N}\sum_{\substack{x_0=a\\(x_1,x_2,x_3)\in\mathbb{L}^3}}\alpha\wedge d\alpha\\
        +&\frac{1}{2N}\sum_{ijk}\epsilon_{ijk}\sum_{s=0}^{L-1}\biggl(\alpha_k(b\hat{0}+s\hat{k})+\beta_k(b\hat{0}+s\hat{k})+\alpha_k(\hat{i}+\hat{j}+b\hat{0}+s\hat{k})+\beta_k(\hat{i}+\hat{j}+b\hat{0}+s\hat{k})\biggr).
    \end{split}
\end{equation}
Using \eqref{equ12360} and the formula
\begin{equation}\label{equ12380}
    m_{12},m_{23},m_{31}\in N\mathbb{Z}\Longrightarrow \frac{1}{8N}\sum_{\mu,\nu,\rho,\sigma}m_{\mu\nu}m_{\rho\sigma}\in\mathbb{Z},
\end{equation}
we can obtain the desired result
\begin{equation}\label{equ12390}
    \begin{split}
        &\frac{N}{2}\int_{(\mathbb{L}^3|x_0=a)}\frac{\bm{A}[U|_{x_0=a}]}{N}\star\frac{\bm{A}[U|_{x_0=a}]}{N}
        -\frac{N}{2}\int_{(\mathbb{L}^3|x_0=b)}\frac{\bm{A}[U|_{x_0=b}]}{N}\star\frac{\bm{A}[U|_{x_0=a}]}{N}\\
        \stackrel{\mathbb{Z}}{=}&\;\frac{1}{N}\sum_{\substack{0\leq x_0\leq a-1\\\text{ or }b\leq x_0\leq L-1\\(x_1,x_2,x_3)\in\mathbb{L}^3}}\frac{1}{4\pi^2}F\wedge F,
    \end{split}
\end{equation}
or from a similar discussion,
\begin{equation}\label{equ12391}
    \begin{split}
        -&\frac{N}{2}\int_{(\mathbb{L}^3|x_0=a)}\frac{\bm{A}[U|_{x_0=a}]}{N}\star\frac{\bm{A}[U|_{x_0=a}]}{N}
        +\frac{N}{2}\int_{(\mathbb{L}^3|x_0=b)}\frac{\bm{A}[U|_{x_0=b}]}{N}\star\frac{\bm{A}[U|_{x_0=a}]}{N}\\
        \stackrel{\mathbb{Z}}{=}&\;\frac{1}{N}\sum_{\substack{a+1\leq x_0\leq b-1\\(x_1,x_2,x_3)\in\mathbb{L}^3}}\frac{1}{4\pi^2}F\wedge F.
    \end{split}
\end{equation}
As we see, the condition $m_{ij}\in N\mathbb{Z}\Leftrightarrow\left[\frac{F}{2\pi}\right]\in H^2(\mathbb{L}^3;N\mathbb{Z})$ is essential to prove \eqref{equ12390}, which guarantees the sum of $\frac{1}{4\pi^2N}F\wedge F$ can be written as a Chern--Simons term on boundary. This is consistent with the result in level $2k$ lattice Chern--Simons theory we formulated, where $\mathcal{Z}[\bm{A}]\neq 0$ if and only if $[d\bm{A}]\in H^1(\mathbb{L}^3;2k\mathbb{Z})$.

Finally, let us construct a non-invertible defect for $U_{\frac{2\pi}{k}}(\mathbb{L}^3|_{x_0=a})$ using the method of~\cite{Choi:2022jqy,Cordova:2022ieu}. From the discussion so far, we can interpret the $U(1)$ gauge field $U$ on $\mathbb{L}^3|_{x_0=a}$ as a DB cohomology class $\bm{A}[U|_{x_0=a}]$ on $\mathbb{L}^3|_{x_0=a}$. Let us then define the lattice DB Chern--Simons theory on $\mathbb{L}^3|_{x_0=a}$ by
\begin{equation}\label{equ12400}
    \begin{split}
        &\mathcal{Z}_{\text{lat.$\mathcal{MCS}$ on $(\mathbb{L}^3|x_0=a)$}}^{N,\epsilon}[\bm{A}[U|_{x_0=a}]]\\
        =&\int_{[\bm{a}]\in H^2_{\text{lat.DB}}(\mathbb{L}^3|x_0=a)}\exp\left\{2i\pi\int_{(\mathbb{L}^3|x_0=a)}\left(\frac{N}{2}\bm{a}\star\bm{a}+\bm{a}\frac{\overleftrightarrow{\star}}{2}\bm{A}[U|_{x_0=a}]\right)-\epsilon\sum_{\substack{\text{plaquettes}\\\text{in $(\mathbb{L}^3|x_0=a)$}}}|d\bm{a}|^2\right\}.
    \end{split}
\end{equation}
Then the equation
\begin{equation}\label{equ12410}
    \begin{split}
        &\frac{\mathcal{Z}_{\text{lat.$\mathcal{MCS}$ on $(\mathbb{L}^3|x_0=a)$}}^{N,\epsilon}[\bm{A}[U|_{x_0=a}]]}{\mathcal{Z}_{\text{lat.$\mathcal{MCS}$ on $(\mathbb{L}^3|x_0=a)$}}^{N,\epsilon}[0]}\\
        =&\delta_{m_{ij},N\mathbb{Z}}\exp\left\{-2\pi i\frac{N}{2}\int_{(\mathbb{L}^3|x_0=a)}\frac{\bm{A}[U|_{x_0=a}]}{N}\star\frac{\bm{A}[U|_{x_0=a}]}{N}\right\}(1+\mathrm{err}_{N,\epsilon}(\bm{A}[U|_{x_0=a}]))
    \end{split}
\end{equation}
holds, and we can define the non-invertible chiral transformation defect $\mathcal{D}_{\frac{1}{N}}$ on $\mathbb{L}^4$ by
\begin{equation}\label{equ12420}
    \mathcal{D}_\frac{1}{k}(\mathbb{L}^3|_{x_0=a}):=U_{\frac{2\pi}{k}}(\mathbb{L}^3|_{x_0=a})\frac{\mathcal{Z}_{\text{lat.$\mathcal{MCS}$ on $(\mathbb{L}^3|x_0=a)$}}^{N,\epsilon}[\bm{A}[U|_{x_0=a}]]}{\mathcal{Z}_{\text{lat.$\mathcal{MCS}$ on $(\mathbb{L}^3|x_0=a)$}}^{N,\epsilon}[0]}.
\end{equation}
$\delta_{m_{ij},N\mathbb{Z}}$ in \eqref{equ12410} is a Kronecker delta that requires the curvature $F$ of lattice massless QED to belong to an $N\mathbb{Z}$-coefficient Chern class. Therefore, we find that $\mathcal{D}_\frac{1}{k}$ is non-invertible. The fusion of defects can be written as
\begin{equation}\label{equ12430}
    \mathcal{D}_\frac{1}{k}(\mathbb{L}^3|_{x_0=a})\mathcal{D}^\dagger_\frac{1}{k}(\mathbb{L}^3|_{x_0=a})=(1+\mathrm{err})\delta_{m_{ij},N\mathbb{Z}}\neq \mathbbm{1},\qquad \mathrm{err}\approx O(\epsilon).
\end{equation}
It should be emphasized that the error, which is bounded by $d\bm{A}[U|_{x_0=a}]=\frac{1}{2\pi i}\ln dU|_{x_0=a}$, is completely controlled by the admissibility condition on the $U(1)$ gauge field in lattice massless QED.
\newpage
\section{Other applications}\label{sec1300}
In this section, we will introduce some more applications.
\subsection{Odd level lattice Chern--Simons theory}\label{sec1310}
The $\mathcal{CS}_\mathrm{lat.DB}^{2k}[\bm{a}]=\int_{\mathbb{L}^3}\frac{2k}{2}\bm{a}\star\bm{a}$ discussed so far corresponds to the case where the Chern--Simons level is $2k$, and we have not discussed the case where the Chern--Simons level is odd.

For odd $K$, $\int_{\mathbb{L}^3}\frac{K}{2}\bm{a}\star\bm{a}$ is not invariant under large gauge transformations. J.~Y.~Chen and Z.~A.~Xu proposed a solution~\cite{Chen:2019mjw,Xu:2024hyo} to this problem by adding a path integral for Majorana fermions. In this chapter, we evaluate the changes that appear from $\int_{\mathbb{L}^3}\frac{K}{2}\bm{a}\star\bm{a}$ under large gauge transformations, and show that the argument can also be applied in the lattice DB formulation.

We can see that $\int_{\mathbb{L}^3}\frac{K}{2}\bm{a}\star\bm{a}$ is not invariant under large gauge transformations by performing $\bm{a}=(a,\lambda,n)\to (a,\lambda-dm,n+\delta m)$. From \eqref{equ7120}, we get
\begin{align}\label{equ13010}
    \bm{a}\star\bm{a}&\to\bm{a}\star\bm{a}+D_4\left(\bm{a}\star(0,m)+(0,m)\star\bm{a}+(0,m)\star(0,m)\right),\notag\\
    \frac{K}{2}\int_{\mathbb{L}^3}\bm{a}\star\bm{a}&\to\frac{K}{2}\int_{\mathbb{L}^3}\bm{a}\star\bm{a}+\frac{K}{2}\langle n\smallsmile m+m\smallsmile n+m\smallsmile \delta m,(-1)^3\sigma_{(0,3)}\rangle\\
    &\stackrel{\mathbb{Z}}{\neq}\frac{K}{2}\int_{\mathbb{L}^3}\bm{a}\star\bm{a},\notag
\end{align}
where $\sigma_{(0,3)}$ is an element of the DB cycle corresponding to $\mathbb{L}^3$ defined in \eqref{equ7430}. We can certainly confirm that the Chern--Simons action where $K\notin 2\mathbb{Z}_{>0}$ is not gauge-invariant.

The change in the Chern--Simons action obtained here has the same form as that appearing in ~\cite[eq.(2.9)]{Xu:2024hyo} and~\cite[eq.(27)]{Chen:2019mjw}. However, there is a difference: $n,m$ in the lattice DB cocycle are $\mathbb{Z}$-valued cocycles with patch subscripts, i.e., they correspond to simplicial complexes,\footnote{To be precise, we should call them cocycles on the \v{C}ech nerve, but it is an equivalent formulation when we use a good cover of $M$.} and are not 2-cocycles on a cubic lattice as in the modified Villain formalism.

As discussed in~\cite[Section~3]{Chen:2019mjw}, the following discussion can be applied to both simplicial complexes and cubic cell complexes without any problems, and it is possible to change $\int_{\mathbb{L}^3}\frac{K}{2}\bm{a}\star\bm{a}$ into a gauge-invariant action in exactly the same way.

The detailed method is given in~\cite{Chen:2019mjw}, but below we outline how it is possible to make $\int_{\mathbb{L}^3}\frac{K}{2}\bm{a}\star\bm{a}$ gauge-invariant. First, recall that for $\bm{a}=(a,\lambda,n)$, $n$ is a $\mathbb{Z}$-valued cocycle on a simplicial complex. $\mathbb{Z}\twoheadrightarrow \mathbb{Z}_2$ allows us to consider $n\twoheadrightarrow n'\in Z^2(\text{simplicial complex};\mathbb{Z}_2)$. On the other hand, we set a Majorana fermion operator in each $2$-simplex of the simplicial complex. We consider an operator obtained by combining the Majorana fermion operators at that location in a certain way according to the value $\{0,1\}$ in each 2-simplex of $n'$, and let $z[n]$ be the partition function obtained by performing a certain fermion path integral.

Using $z[n]$, we improve the action of odd level Chern--Simons theory by
\begin{equation}\label{equ13020}
    \mathcal{Z}_{\mathrm{lat.}\mathcal{CS}}^{K}=\int_{[\bm{a}]\in H^2_{\mathrm{lat.DB}}(\mathbb{L}^3)}\mathcal{D}[\bm{a}](z[n])^K\exp\left[2\pi i\frac{K}{2}\int_{\mathbb{L}^3}\bm{a}\star\bm{a}\right].
\end{equation}
It can be seen that the sign arising from \eqref{equ13010} cancels with the sign arising from $(z[n])^K$ and that $\mathcal{Z}_{\mathrm{lat.}\mathcal{CS}}^{K}$ is defined as a partition function of the gauge-invariant action.
\subsection{BF theory formulated using lattice DB cohomology}\label{sec1320}
In contrast to Chern--Simons theory, where the action is described as $AdA$, the theory where the action is described as $AdB$ is called BF theory. By using lattice DB formalism, it is possible to construct BF-theory in the same way as lattice Chern--Simons theory. We define the action $S^k_{\mathrm{BF}}$ and the Wilson lines $W_A$ and $W_B$ as follows:
\begin{align}
    2\pi i\;S_\mathrm{BF}^{k}[[\bm{A}],[\bm{B}]]=&2\pi ik\int_{\mathbb{L}^3}[\bm{A}]\star[\bm{B}],\label{equ13030}\\
    W_A(\gamma)=&\exp\left(2\pi i\int_\gamma\bm{A}\right)=2\pi i\int_{\mathbb{L}^3}\bm{A}\star\bm{\eta}_{\gamma_+},\label{equ13040}\\
    W_B(\gamma)=&\exp\left(2\pi i\int_\gamma\bm{B}\right)=2\pi i\int_{\mathbb{L}^3}\bm{\eta}_{\gamma_-}\star\bm{B}.\label{equ13050}
\end{align}
The path integral in BF theory without the Maxwell term does not seem to converge, as we see in the lattice DB Chern--Simons theory. However we formally assume that the expectation values would be
\begin{align}
    \langle W_A(\gamma)\rangle_\mathrm{BF}^{k}=&1,\label{equ13060}\\
    \langle W_B(\gamma)\rangle_\mathrm{BF}^{k}=&1,\label{equ13070}\\
    \langle W_A(\gamma^A)W_B(\gamma^B)\rangle_\mathrm{BF}^{k}=&\frac{1}{\mathcal{Z}}\int\mathcal{D}[\bm{A}]\mathcal{D}[\bm{B}]\exp\left[2\pi i\int_{\mathbb{L}^3}(k\bm{A}\star\bm{B}+\bm{A}\star\bm{\eta}_{\gamma_+^A}+\bm{\eta}_{\gamma_-^B}\star\bm{B})\right]\label{equ13080}\\
    =&\left\{\begin{array}{ll}
        \displaystyle\exp\left[2\pi ik\int_{\mathbb{L}^3}\frac{\bm{\eta}_{\gamma_-^B}}{k}\star\frac{\bm{\eta}_{\gamma_+^A}}{k}\right]&([\gamma^A],[\gamma^B]\in H_1(\mathbb{L}^3;k\mathbb{Z}))\\
        &\\
        0&(\mathrm{otherwise})\\
    \end{array}
    \right.\label{equ13090}\\
    =&\left\{
        \begin{array}{ll}
        \displaystyle\exp\left[\frac{2\pi i}{k}\mathrm{link}^{\mathrm{mod}\,k}(\gamma_+^A,\gamma^B)\right]&([\gamma^A],[\gamma^B]\in H_1(\mathbb{L}^3;k\mathbb{Z}))\\
        &\\
        0&(\mathrm{otherwise})\\
    \end{array}
    \right..\label{equ13100}
\end{align}
If we define Maxwell BF theory using the coupling constant $0<\epsilon\ll1$ by
\begin{equation}\label{equ13110}
    2\pi i\;S_\mathrm{BF}^{k,\epsilon}[[\bm{A}],[\bm{B}]]=2\pi ik\int_{\mathbb{L}^3}[\bm{A}]\star[\bm{B}]-\epsilon\sum_\mathrm{plaquettes}|d\bm{A}|^2-\epsilon\sum_\mathrm{plaquettes}|d\bm{B}|^2,
\end{equation}
$|\mathcal{Z}|<\infty$ holds, and we may obtain
\begin{equation}\label{equ13120}
    \langle \bullet\rangle_\mathrm{BF}^{k}\approx\langle \bullet\rangle_\mathrm{BF}^{k,\epsilon}.
\end{equation}
\newpage
\section{Summary and outlook}\label{sec1400}
\subsection{Summary}\label{sec1401}
We defined the Deligne--Beilinson (DB) cohomology on the $d$-dimensional cubic toroidal lattice, and we proved that lattice DB cohomology has gauge-invariant $\mathbb{R}/\mathbb{Z}$-valued integrals over lattice DB cycles. By defining the product $\star$ of DB cohomology, a gauge-invariant level $2k$ (Maxwell--) Chern--Simons action on the lattice is defined. Pontrjagin duality gives the torus a structure of mod $2k$ linking numbers and mod $4k$ self-linking numbers. The Wilson line is also defined by Pontrjagin duality, and the cup-$1$ product allows us to define a framing structure on the Wilson lines in the gauge-invariant way. The Hodge decomposition on the lattice gives a gauge-fixed path integral measure in lattice Chern--Simons theory, and the path integral in Maxwell--Chern--Simons theory is a strictly convergent complex Gaussian integral. The properties of the expectation value of the framed Wilson line are the same as those expected from continuum theory, but in Maxwell--Chern--Simons theory, it is accompanied by an error $\approx O(\epsilon)$ controlled by the coupling constant $\epsilon$ of the Maxwell term. By coupling the external field to the Maxwell--Chern--Simons theory, we can produce the boundary term of the chiral anomaly of the lattice massless QED, which can be applied to the construction of a chiral non-invertible defect on the lattice.
\subsection{Outlook}\label{sec1410}
It is important to compare this study with the modified Villain formalism by~\cite{Jacobson:2023cmr,Xu:2024hyo}. In the lattice DB formalism, $a$ and $n$ in $\bm{a} = (a, \lambda, n)$ play a similar role to $a$ and $n$ used in the modified Villain formalism. It is important to investigate whether the modified Villain formalism and the lattice DB formalism are equivalent theories, or, if not, how they differ. In~\cite{Chen:2019mjw,Xu:2024hyo}, the correspondence with continuum theory is discussed through the interpolation of lattice gauge fields, and it may be possible to consider interpolation in the lattice DB formalism as well. Furthermore, in the lattice DB formalism, we have shown that the expectation value of the Wilson line can be approximated and that the error is controlled. It would be interesting to investigate whether this can also be achieved within the modified Villain formalism.

It was pointed out in~\cite{Xu:2024hyo} that the framing of the Wilson line can be changed by replacing the definition of the cup product. A possible direction for future work is to examine whether the lattice DB cohomology theory can still be consistently defined when the definition of the cup product in $\mathbb{L}^3$ is modified. Moreover, since $\mathbb{R}P^3$ can be realized as a cubic lattice, it is interesting to examine whether the lattice DB cohomology theory remains well-defined in this setting. $\mathbb{R}P^3$ has a torsion part in cohomology, so it is not obvious whether $k\int\frac{\bm{A}}{2k}\star\frac{\bm{A}}{2k}$ still makes sense as a linking form or not. It remains to be examined whether a spin structure required for odd-level Chern--Simons theory can be consistently implemented on $\mathbb{R}P^3$ or on other lattices. It is also known that it is possible to discuss the generalization of Chern--Simons theory not only in three dimensions but also in general odd dimensions. Thus, we expect that our model will be generalized to various situations.

Our method should also be applicable to the K-matrix Chern--Simons theory. Because K-matrix Chern--Simons theory includes theories known as BF theory and minimal $\mathbb{Z}_N$ TQFT $\mathcal{A}^{N,p}$ and so on, formulating these theories on the lattice is important from both theoretical and practical perspectives.

Our lattice DB Chern--Simons theory is limited to the gauge group of $U(1)$ by definition, but Gomi mathematically discusses the construction of a Chern--Simons theory for general gauge groups in (continuous) DB cohomology theory~\cite{Gomi:2001ym}. It would be interesting to examine whether Chern--Simons theories for $SU(2)$ or other gauge groups can be constructed on lattices using Gomi's method. Developing the discussion of the Jones polynomial~\cite{Witten:1988hf} on the lattice is an interesting direction for future research, and it would be important to investigate whether the lattice DB formalism can be applied to this problem. In addition, if lattice DB cohomology can be constructed for general gauge groups, it will be possible to construct topologically nontrivial gauge configurations on the lattice for general gauge groups. The application of DB formalism is not necessarily limited to Chern--Simons theory. It is an important problem to research the effects of topological terms, such as the $\theta$-term, on the lattice for general gauge groups. In the DB cohomology formulation considered here, the matter field and gauge field are not directly coupled, but it is also important to investigate how covariant derivatives can be defined for matter fields on a patchwise lattice.
\section*{Acknowledgement}\label{secACK}
I am grateful to Masashi~Kawahira for introducing me to DB cohomology. I thank Yoshio~Kikukawa and Yuki~Furukawa for sincere discussions and advice. I also thank Shoto~Aoki and Toshinari~Takemoto for useful discussions. I am also grateful to Yuji~Okawa for helpful comments on the writing of this paper.

This work was supported by JST SPRING, Grant Number JPMJSP2108.
\newpage
\appendix
\section{Hodge decomposition of $\Omega^1(\mathbb{L}^3)$}\label{secA00}
In this section, we discuss the Hodge decomposition of $\Omega^1(\mathbb{L}^3)$.
\subsection{Hodge decomposition of lattice $1$-form}\label{secA10}
Let $M$ be a finite lattice, and let $\langle\bullet,\bullet\rangle$ be a positive definite inner product of $\Omega^1(M)$. In Section~\ref{secA00}, we will use the notation that $d$ is the exterior derivative of the lattice and $\delta$ is the coderivative of the lattice. $d$ and $\delta$ are in the relationship $\langle\delta\alpha,\beta\rangle=\langle\alpha,d\beta\rangle$. The following lemma can then be proved.
\begin{itemize}
    \item $(\mathrm{im}\,d_0)^\perp=\mathrm{ker}\,\delta_1$ and $(\mathrm{im}\,\delta_2)^\perp=\mathrm{ker}\,d_1$\\
    (proof) From $\langle d_0\alpha,\beta\rangle=\langle\alpha,\delta_1\beta\rangle$, we see that two conditions for $\beta$, $\forall\alpha\in\Omega^0(M),\langle d\alpha,\beta\rangle=0$ and $\forall\alpha\in\Omega^0(M),\langle\alpha,\delta\beta\rangle=0$ are equivalent. Then we see equivalence between conditions for $\beta$, $\beta\in(\mathrm{im}\,d_0)^\perp$ and $\beta\in\mathrm{ker}\,\delta_1$, so the equality $(\mathrm{im}\,d_0)^\perp=\mathrm{ker}\,\delta_1$ holds. By the same argument, we can also prove the equivalence between the two conditions for $\alpha$, $\forall\beta\in\Omega^1(M),\langle d\alpha,\beta\rangle=0$ and $\forall\beta\in\Omega^1(M),\langle\alpha,\delta\beta\rangle=0$. Thus, we conclude $(\mathrm{im}\,\delta_2)^\perp=\mathrm{ker}\,d_1$//
    \item $(\mathrm{im}\,d_0)\perp(\mathrm{im}\,\delta_2)$\\
    (proof) $\forall\alpha\in\Omega^0(M),\forall\beta\in\Omega^2(M),\langle d\alpha,\delta\beta\rangle=\langle \alpha,\delta\delta\beta\rangle=0$//
    \item For $\Delta:=d\delta+\delta d:\Omega^1\to\Omega^1$, $\mathrm{ker}\,\Delta=(\mathrm{ker}\,d_0)\cap(\mathrm{ker}\,\delta_2)$\\
    (proof) Let $\omega\in\mathrm{ker}\,\Delta$ be chosen arbitrarily and fixed. From $\langle\Delta\omega,\omega\rangle=\langle\delta\omega,\delta\omega\rangle+\langle d\omega,d\omega\rangle=0$, $\delta\omega=0,d\omega=0$, we find $\mathrm{ker}\,\Delta\subset(\mathrm{ker}\,d_0)\cap(\mathrm{ker}\,\delta_2)$.\\
    On the other hand, obviously $\mathrm{ker}\,\Delta\supset(\mathrm{ker}\,d_0)\cap(\mathrm{ker}\,\delta_2)$//
\end{itemize}

Based on the above lemma, let us discuss the Hodge decomposition. First, let us focus on $(\mathrm{im}\,d_0)\perp(\mathrm{im}\,\delta_2)$. Apply the Gram--Schmidt orthonormalization method to $\Omega^1(M)$ to perform the orthogonal decomposition $\Omega^1(M)=(\mathrm{im}\,d_0)\oplus\mathcal{H}\oplus(\mathrm{ker}\,\delta_2)$. On the other hand, there is also an orthogonal decomposition $\Omega^1=(\mathrm{im}\,d_0)\oplus(\mathrm{ker}\,\delta_1)=(\mathrm{ker}\,d_1)\oplus(\mathrm{im}\,\delta_2)$, so $\mathcal{H}=(\mathrm{ker}\,d_1)/(\mathrm{im}\,d_0)=(\mathrm{ker}\,\delta_1)/(\mathrm{im}\,\delta_2)=H^1(M;\mathbb{R})$. Furthermore, $\mathcal{H}=(\mathrm{ker}\,d_1)\cap(\mathrm{ker}\,\delta_1)=\mathrm{ker}\,\Delta$. Thus, we get an orthogonal decomposition
\begin{equation}\label{equA010}
    \Omega^1(M)=(\mathrm{im}\,d_0)\oplus H^1(M;\mathbb{R})\oplus(\mathrm{ker}\,\delta_2)=(\mathrm{im}\,d_0)\oplus(\mathrm{ker}\,\Delta)\oplus(\mathrm{ker}\,\delta_2).
\end{equation}

As an example, let us give an orthonormal Hodge decomposition of $\Omega^1(\mathbb{L}^3)$. $\mathrm{im}\,d_0$ has a representation in an overcomplete basis 
\begin{equation}\label{equA020}
    \mathrm{im}\,d_0=\mathrm{Span}\left\{(x,\mu)\mapsto \frac{1}{\sqrt{6}}\sum_{\nu=1}^3 \biggl((\delta_{x+\hat{\nu},y}-\delta_{x,y})\delta_{\mu,\nu}\biggr)\middle| y\in\mathbb{L}^3\right\}.
\end{equation}
The basis of $H^1(M;\mathbb{R})$ can be chosen as
\begin{equation}\label{equA030}
    H^1(M;\mathbb{R})=\mathrm{Span}\left\{(x,\mu)\mapsto \frac{1}{\sqrt{L^3}}\delta_{\mu,\nu}\middle| \nu\in\{1,2,3\}\right\},
\end{equation}
and $\mathrm{ker}\,\delta_2$ has an expression in an excess basis
\begin{equation}\label{equA040}
    \begin{split}
        \mathrm{ker}\,\delta_2=\mathrm{Span}\biggl\{&((x;\mu)\mapsto \frac{1}{2}(\delta_{y,x}\delta_{\nu,\mu}+\delta_{y+\hat{\nu},x}\delta_{\lambda,\mu}-\delta_{y+\hat{\lambda},x}\delta_{\nu,\mu}-\delta_{y,x}\delta_{\lambda,\mu}))\\&\bigg|y\in\mathbb{L}^3\;,\;(\nu,\lambda)\in\{(1,2),(2,3),(3,1)\}\biggr\}.
    \end{split}
\end{equation}
Looking at these expressions, it is clear that $(\mathrm{im}\,d_0)\oplus H^1(M;\mathbb{R})\oplus(\mathrm{ker}\,\delta_2)$ is an orthogonal decomposition of $\Omega^1(\mathbb{L}^3)$.
\subsubsection{CONS of $\Omega^1(\mathbb{L}^3)$}\label{secA11}
In the following, we discuss the complete orthonormal system (CONS) of $\Omega^1(\mathbb{L}^3)$. In this subsection, we will use the notation $\ast$ for the lattice Hodge dual:
\begin{equation}\label{equA041}
    \begin{split}
        \ast((x;\mu)\mapsto \delta_{x,x'}\delta_{\mu,\nu})=&\epsilon_{\nu\lambda\tau}((\tilde{x};\rho,\sigma)\mapsto \delta_{\tilde{x},x'+\frac{1}{2}\hat{\mu}-\frac{1}{2}\hat{\lambda}-\frac{1}{2}\hat{\tau}}\delta_{\rho,\lambda}\delta_{\sigma,\tau}),\\
        \ast((x;\mu\nu)\mapsto\delta_{x,x'}\delta_{\mu,\rho}\delta_{\nu,\sigma})=&\epsilon_{\rho\sigma\tau}((\tilde{x},\lambda)\mapsto\delta_{\tilde{x},x'+\frac{1}{2}\hat{\rho}+\frac{1}{2}\hat{\sigma}-\frac{1}{2}\hat{\tau}}\delta_{\lambda,\tau}),
    \end{split}
\end{equation}
and the coderivative operation can be expressed as $\delta=\ast d\ast$. Let the inner product of $\Omega^1(\mathbb{L}^3)$ be $\langle\alpha,\beta\rangle=\int\alpha\wedge\ast\beta$, i.e., the Euclidean inner product of the vectors enumerating the values for each link.

As mentioned above, $\Omega^1(\mathbb{L}^3)$ can be decomposed using Hodge theory as
\begin{equation}\label{equA050}
    \begin{split}
        \Omega^1(\mathbb{L}^3)&=d_0\Omega^0(\mathbb{L}^3)\oplus H^1(\mathbb{L}^3)\oplus (\ast d_1\ast \Omega^2(\mathbb{L}^3))\\
        &=\Omega^1_\mathrm{exact}(\mathbb{L}^3)\oplus H^1(\mathbb{L}^3)\oplus\Omega^1_\mathrm{coexact}(\mathbb{L}^3).
    \end{split}
\end{equation}
Since the dimensions are given by
\begin{align}\label{equA060}
        &\mathrm{dim}\,\Omega^0(\mathbb{L}^3)=L^3,&&\mathrm{dim}\,\Omega^1(\mathbb{L}^3)=3L^3,&&\mathrm{dim}\,\Omega^2(\mathbb{L}^3)=3L^3,&&\mathrm{dim}\,H^1(\mathbb{L}^3)=3,
\end{align}
we see
\begin{equation}\label{equA070}
    \mathrm{dim}\,\mathrm{ker}\,d_0+\mathrm{dim}\,\mathrm{ker}(\ast d_1\ast)=L^3+3.
\end{equation}
The goal below is to find bases for $\Omega^1_\mathrm{exact}(\mathbb{L}^3)$ and $\Omega^1_\mathrm{coexact}(\mathbb{L}^3)$ by giving bases for $\mathrm{ker}\,d_0$ and $\mathrm{ker}(\ast d_1\ast)$.
First, let us consider $\mathrm{ker}\,d_0$. Since $B^0(\mathbb{L}^3)=\{0\}$, $\mathrm{ker}\,d_0\cong Z^0(\mathbb{L}^3)\cong H^0(\mathbb{L}^3)\cong \{\mathrm{constant}\;1\mathrm{-form}\}\cong \mathbb{R}$. Then, we get $\mathrm{dim}\,\mathrm{ker}\,d_0=1$ and
\begin{equation}\label{equA080}
    \mathrm{dim}\,\Omega^1_\mathrm{exact}(\mathbb{L}^3)=\mathrm{dim}\,\Omega^0(\mathbb{L}^3)-\mathrm{dim}\,\mathrm{ker}\,d_0=L^3-1.
\end{equation}
Therefore, $\mathrm{dim}\,\Omega^1_\mathrm{coexact}(\mathbb{L}^3)=2L^3-2$, and $\mathrm{dim}\,\mathrm{ker}(\ast d_1\ast)=L^3+2$ holds.

Next, let us consider $(\ast d_1\ast):\Omega^2(\mathbb{L}^3)\to \Omega^1(\mathbb{L}^3)$. If we choose a basis for $\Omega^2(\mathbb{L}^3)$ as
\begin{equation}\label{equA090}
    \Omega^2(\mathbb{L}^3)=\mathrm{Span}_\mathbb{Z}\{(x;\mu\nu)\mapsto \delta_{x,y}\varepsilon_{\mu\nu\rho}\mid y\in\mathbb{L}^3,\rho\in\{1,2,3\}\},
\end{equation}
the operation of $\ast d_1\ast$ can be written as
\begin{equation}\label{equA100}
    (\ast d_1\ast)((x;\mu\nu)\mapsto \delta_{x,y}\varepsilon_{\mu\nu\rho})=((z;\lambda)\mapsto \delta_{y,z}\delta_{\mu,\lambda}+\delta_{y,z+\hat{\mu}}\delta_{\nu,\lambda}-\delta_{y,z+\hat{\nu}}\delta_{\mu,\lambda}-\delta_{y,z}\delta_{\nu,\lambda}).
\end{equation}
By considering this geometrically, we obtain $L^3$ equations
\begin{equation}\label{equA110}
    \begin{split}
        \ast d_1\ast\left(\sum_{(\mu,\nu)\in\{(1,2),(2,3),(3,1)\}}\biggl(((x;\mu\nu)\mapsto \delta_{x,y+\hat{\rho}}\varepsilon_{\mu\nu\rho})-((x;\mu\nu)\mapsto \delta_{x,y}\varepsilon_{\mu\nu\rho})\biggr)\right)=0\\
        (\mathrm{for}\;\mathrm{all}\;y\in\mathbb{L}^3)
    \end{split}
\end{equation}
for the kernel space of $\ast d_1\ast$, and three equations
\begin{equation}\label{equA120}
        \ast d_1\ast\left(\sum_{s=0}^{L-1}\sum_{t=0}^{L-1}((x;\mu\nu)\mapsto \delta_{x,s\hat{\mu}+t\hat{\nu}}\varepsilon_{\mu\nu\rho})\right)=0\;\;
        \mathrm{for}\;(\mu,\nu)\in\{(1,2),(2,3),(3,1)\}.
\end{equation}
However, the $L^3$ equations in the \eqref{equA110} are not independent, and adding any $L^3-1$ equations from \eqref{equA110}, we can derive the remaining $1$ equation. Therefore, we can see that the dimension of the image space of $(\ast d_1\ast)$ is $3L^3-(L^3-1)-3=2L^3-2$.

Next, we should choose $2L^3-2$ bases from $\mathrm{coim}(\ast d_1\ast)\subset \Omega^2(\mathbb{L}^3)$ and find $\Omega^1_\mathrm{coexact}(\mathbb{L}^3)$. Choosing a basis for $\mathrm{coim}(\ast d_1\ast)$ is not always easy, but below we will introduce a method that can be obtained by analogy with the discussion of the complete axial gauge.

First, using the $L-1$ equations in the \eqref{equA110} specified by $y$ such that $1\leq y_1 \leq L-1\;,y_2=y_3=0$, eliminate the $L-1$ excess elements of $\Omega^2(\mathbb{L}^3)$ expressed as $(x;\mu\nu)\mapsto\delta_{x,y}\varepsilon_{\mu\nu 1}$.

Next, using the $L(L-1)$ equations in the \eqref{equA110} specified by $y$ such that $0\leq y_1 \leq L-1\;,\;1\leq y_2 \leq L-1\;,\;y_3=0$, eliminate the $L(L-1)$ excess elements of $\Omega^2(\mathbb{L}^3)$ expressed as $(x;\mu\nu)\mapsto\delta_{x,y}\varepsilon_{\mu\nu 2}$.

Finally, using the $L^2(L-1)$ equations in the \eqref{equA110} specified by $y$ such that $0\leq y_1 \leq L-1\;,\;0\leq y_2 \leq L-1\;,\;1\leq y_3 \leq L-1$, eliminate the $L^2(L-1)$ excess elements of $\Omega^2(\mathbb{L}^3)$ expressed by $(x;\mu\nu)\mapsto\delta_{x,y}\varepsilon_{\mu\nu 3}$.

The number of elements eliminated so far is $L^2(l-1)+L(l-1)+(l-1)=L^3-1$, so we can see that we have used all the $L^3-1$ independent equations in \eqref{equA110}. However, $L^2$ of the elements of $\Omega^2(\mathbb{L}^3)$
\begin{align}
    &\{((x;\mu\nu)\mapsto\delta_{x,y}\varepsilon_{\mu\nu 1})\mid y\in\mathbb{L}^3,y_1=0\},\label{equA130}\\
    &\{((x;\mu\nu)\mapsto\delta_{x,y}\varepsilon_{\mu\nu 2})\mid y\in\mathbb{L}^3,y_2=0\},\label{equA140}\\
    &\{((x;\mu\nu)\mapsto\delta_{x,y}\varepsilon_{\mu\nu 3})\mid y\in\mathbb{L}^3,y_3=0\}\label{equA150}
\end{align}
still remain. Therefore, we can use the \eqref{equA120} to eliminate the three excess elements of $\Omega^2(\mathbb{L}^3)$
\begin{equation}\label{equA160}
    \begin{split}
        &(x;\mu\nu)\mapsto\delta_{x,0}\varepsilon_{\mu\nu 1},\\
        &(x;\mu\nu)\mapsto\delta_{x,0}\varepsilon_{\mu\nu 2},\\
        &(x;\mu\nu)\mapsto\delta_{x,0}\varepsilon_{\mu\nu 3}.
    \end{split}
\end{equation}
As a result of the discussion so far, we have been able to eliminate the $L^3+2$ bases of $\Omega^2(\mathbb{L}^3)$, and the remaining $2L^3-2$ elements 
\begin{equation}\label{equA170}
    \begin{split}
        &\{((x;\mu\nu)\mapsto\delta_{x,y}\varepsilon_{\mu\nu\rho})\mid\\
        &\begin{array}{llr}
            \rho=3,&0\leq y_1 \leq L-1\;,\;0\leq y_2 \leq L-1\;,\;y_3=0,(y_1,y_2)\neq 0&\mathrm{or}\\
            \rho=2,&0\leq y_1 \leq L-1\;,\;0\leq y_2 \leq L-1\;,\;1\leq y_3 \leq L-1&\mathrm{or}\\
            \rho=2,&1\leq y_1 \leq L-1\;,\;y_2=y_3=0&\mathrm{or}\\
            \rho=1,&0\leq y_1 \leq L-1\;,\;0\leq y_2 \leq L-1\;,\;1\leq y_3 \leq L-1&\mathrm{or}\\
            \rho=1,&0\leq y_1 \leq L-1\;,\;1\leq y_2 \leq L-1\;,\;y_3=0&\}
        \end{array}
    \end{split}
\end{equation}
are the bases of $\mathrm{coim}(\ast d_1\ast)$.\footnote{$(L^2-1)+L^2(L-1)+(L-1)+L^2(L-1)+L(L-1)=2L^3-2$, so there are indeed $2L^3-2$ bases.} Finally, we can choose a basis for $\Omega^1_\mathrm{coexact}(\mathbb{L}^3)$ as
\begin{equation}\label{equA180}
    \begin{split}
        &\{((z;\lambda)\mapsto \delta_{y,z}\delta_{\mu,\lambda}+\delta_{y,z+\hat{\mu}}\delta_{\nu,\lambda}-\delta_{y,z+\hat{\nu}}\delta_{\mu,\lambda}-\delta_{y,z}\delta_{\nu,\lambda})\mid\\
        &\begin{array}{llr}
            (\mu,\nu)=(1,2),&0\leq y_1 \leq L-1\;,\;0\leq y_2 \leq L-1\;,\;y_3=0,(y_1,y_2)\neq 0&\mathrm{or}\\
            (\mu,\nu)=(3,1),&0\leq y_1 \leq L-1\;,\;0\leq y_2 \leq L-1\;,\;1\leq y_3 \leq L-1&\mathrm{or}\\
            (\mu,\nu)=(3,1),&1\leq y_1 \leq L-1\;,\;y_2=y_3=0&\mathrm{or}\\
            (\mu,\nu)=(2,3),&0\leq y_1 \leq L-1\;,\;0\leq y_2 \leq L-1\;,\;1\leq y_3 \leq L-1&\mathrm{or}\\
            (\mu,\nu)=(2,3),&0\leq y_1 \leq L-1\;,\;1\leq y_2 \leq L-1\;,\;y_3=0&\}.
        \end{array}
    \end{split}
\end{equation}
\newpage
\section{Mathematical supplement}\label{secB00}
\subsection{Proof of $\|A\|_2^2\leq\|A\|_\infty\|A\|_1$}\label{secB10}
\subsubsection{Definition of matrix norm $\|\bullet\|_p$}\label{secB11}
Let $A$ be a general rectangular complex matrix. For $1\leq p\leq\infty$, the matrix $p$-norm $\|\bullet\|_p$ is defined by
\begin{equation}\label{equB010}
    \|A\|_p=\max_{\|x\|_p=1}\|Ax\|_p.
\end{equation}

In the case $p=1$,
\begin{equation}\label{equB020}
    \begin{split}
        \|A\|_1=\max_{\|x\|_1=1}\|Ax\|_1=\max_{\|x\|_1=1}\sum_{i,j}|A_{ij}x_j|\leq\max_{\|x\|_1=1}\sum_{j}\left(\left(\sum_{i}|A_{ij}|\right)|x_j|\right)\\
        \leq\max_{\|x\|_1=1}\sum_{j}\left(\left(\max_j\sum_{i}|A_{ij}|\right)|x_j|\right)=\max_j\sum_{i}|A_{ij}|
    \end{split}
\end{equation}
holds. Using $j$ maximizes $\sum_{i}|A_{ij}|$, we define $v=(v_1,v_2,\dots),v_i=\delta_{ij}$, which satisfies $\|v\|_1=1$ and $\|Av\|_1=\max_j\sum_{i}|A_{ij}|$, and we get
\begin{equation}\label{equB030}
    \|A\|_1=\max_j\sum_{i}|A_{ij}|.
\end{equation}

In the case $p=\infty$,
\begin{equation}\label{equB040}
    \begin{split}
        \|A\|_\infty=\max_{\|x\|_\infty=1}\|Ax\|_\infty=\max_{\|x\|_\infty=1}\max_{i}\left(\sum_j|A_{ij}x_j|\right)\leq\max_{\|x\|_\infty=1}\max_{i}\left(\sum_j|A_{ij}||x_j|\right)\\
        \leq\max_{\|x\|_\infty=1}\left(\max_{i}\sum_j|A_{ij}|\right)\left(\max_{j}|x_j|\right)=\max_{i}\sum_j|A_{ij}|
    \end{split}
\end{equation}
holds. We define $v=(1,1,\dots)$,$\|v\|_\infty=1$ and obtain $\|Av\|_\infty=\max_i\sum_{j}|A_{ij}|$, and
\begin{equation}\label{equB050}
    \|A\|_\infty=\max_i\sum_{j}|A_{ij}|
\end{equation}
holds.
\subsubsection{Proof of $\|A\|_2^2\leq\|A\|_\infty\|A\|_1$}\label{secB12}
Recall the definition of the matrix $2$-norm,
\begin{equation}\label{equB060}
    \begin{split}
        \|A\|_2^2=\left(\max_{\|x\|_2=1}\sqrt{\sum_i\left(\left|\sum_jA_{ij}x_j\right|^2\right)}\right)^2=\max_{\|x\|_2=1}\sum_i\left|\sum_jA_{ij}x_j\right|^2\\
        \leq\max_{\|x\|_2=1}\sum_i\left(\sum_j|A_{ij}x_j|\right)^2=\max_{\|x\|_2=1}\sum_i\left(\sum_j|A_{ij}|\cdot|x_j|\right)^2.
    \end{split}
\end{equation}
Using Cauchy--Schwartz inequality $(\sum_j y_jz_j)^2\leq(\sum_j y_j^2)(\sum_j z_j^2)$ and setting $y_j=\sqrt{|A_{ij}|},z_j=\sqrt{|A_{ij}|}|x_j|$, we get
\begin{equation}\label{equB070}
    \begin{split}
        \max_{\|x\|_2=1}\sum_i\sum_j\left(|A_{ij}|\cdot|x_j|\right)^2&\leq\max_{\|x\|_2=1}\sum_i\left(\left(\sum_j|A_{ij}|\right)\left(\sum_j|A_{ij}|\cdot|x_j|^2\right)\right)\\
        &\leq\max_{\|x\|_2=1}\sum_i\left(\left(\max_i\sum_j|A_{ij}|\right)\left(\sum_j|A_{ij}|\cdot|x_j|^2\right)\right)\\
        &=\max_{\|x\|_2=1}\left(\|A\|_\infty\left(\sum_i\sum_j|A_{ij}|\cdot|x_j|^2\right)\right)\\
        &=\max_{\|x\|_2=1}\left(\|A\|_\infty\sum_j\left(\left(\sum_i|A_{ij}|\right)\cdot|x_j|^2\right)\right)\\
        &\leq\max_{\|x\|_2=1}\left(\|A\|_\infty\sum_j\left(\left(\max_j\sum_i|A_{ij}|\right)\cdot|x_j|^2\right)\right)\\
        &=\max_{\|x\|_2=1}\left(\|A\|_\infty\left(\left(\max_j\sum_i|A_{ij}|\right)\cdot\sum_j|x_j|^2\right)\right)\\
        &=\|A\|_\infty\|A\|_1.
    \end{split}
\end{equation}
Then we find
\begin{equation}\label{equB080}
    \|A\|_2^2\leq\|A\|_\infty\|A\|_1.
\end{equation}
\subsection{Complex Gauss integral}\label{secB20}
In this section, we take $S\in M_{2L^3-2}(\mathbb{R})$ and $0<\epsilon\ll 1$. For
\begin{align}\label{equB090}
    &K\in \mathcal{L}(\mathbb{R}^{2L^3+1},\mathbb{R}^{3L^3})\cong M_{(3L^3)\times (2L^3+1)}(\mathbb{R}),&&\mathrm{rank}K=2L^3+1,
\end{align}
the Gram matrix $K^\mathsf{T}K$ is positive definite ($\because v^\mathsf{T}K^\mathsf{T}Kv=\|Kv\|^2\geq 0\;,\;Kv=0\Leftrightarrow v=0$), therefore there exists an upper triangular matrix $J\in GL(2L^3+1,\mathbb{R})$ satisfies $K^\mathsf{T}K=J^\mathsf{T}J$, which is known as Cholesky decomposition.

We decompose $J$ into
\begin{equation}\label{equB100}
    J=\left(\begin{array}{c|c}
        J_1&J_2\\\hline
        0&J_3
    \end{array}\right),
\end{equation}
where
\begin{align}\label{equB110}
    &J_1\in GL(2L^3-2;\mathbb{R}),&
    &J_2\in M_{(2L^3-2)\times3}(\mathbb{R}),&
    &J_3\in GL(3;\mathbb{R}).
\end{align}
In particular, for
\begin{align}
    &x=(x_1,\ldots,x_{2L^3-2},0,0,0),&&V_1=(v_1,\ldots,v_{2L^3-2}),\label{equB120}\\
    &V_2=(v_{2L^3-1},v_{2L^3},v_{2L^3+2}),&&v=(V_1,V_2),\label{equB130}
\end{align}
we obtain equations
\begin{align}\label{equB140}
    &x^\mathsf{T}J^\mathsf{T} Jx=x^\mathsf{T}J_1^\mathsf{T} J_1x,
    &v^\mathsf{T}J^\mathsf{T} Jx=(V_1^\mathsf{T}J_1^\mathsf{T}+V_2^\mathsf{T} J_2^\mathsf{T})J_1x.
\end{align}
Therefore,
\begin{equation}\label{equB150}
    (x+v)^\mathsf{T}J^\mathsf{T}J(x+v)=x^\mathsf{T} J_1^\mathsf{T}J_1x+2(V_1^\mathsf{T}J_1^\mathsf{T}+V_2^\mathsf{T} J_2^\mathsf{T})J_1x+v^\mathsf{T}J^\mathsf{T} Jv
\end{equation}
holds.

Here, consider the integral
\begin{align}\label{equB160}
    &I_0=\int_{\mathbb{R}^{2L^3-2}}\exp(ix^\mathsf{T} Sx-\epsilon x^\mathsf{T}J^\mathsf{T}Jx)dx_1\cdots dx_{2L^3-2},&&x=(x_1,\ldots,x_{2L^3-2},0,0,0).
\end{align}
From
\begin{equation}\label{equB170}
    x^\mathsf{T} Sx=x^\mathsf{T}\frac{S+S^\mathsf{T}}{2}x=x^\mathsf{T}S_\mathrm{sym}x,
\end{equation}
we are allowed to assume $S$ is a real symmetric matrix without loss of generality, and we obtain
\begin{equation}\label{equB180}
    \begin{split}
        &\int_{\mathbb{R}^{2L^3-2}}\exp(ix^\mathsf{T} Sx-\epsilon x^\mathsf{T}J_1^\mathsf{T}J_1x)d^{2L^3-2}x\\
        =&\int_{\mathbb{R}^{2L^3-2}}\exp(ix^\mathsf{T}J_1^\mathsf{T}(J_1^\mathsf{T})^{-1} SJ_1^{-1}J_1x-\epsilon (J_1x)^\mathsf{T}J_1x)\;d^{2L^3-2}(J_1x)\\
        =&(\mathrm{det}\,J_1)\int_{\mathbb{R}^{2L^3-2}}\exp(ix^\mathsf{T}(J_1^\mathsf{T})^{-1} SJ_1^{-1}x-\epsilon x^\mathsf{T}x)\;d^{2L^3-2}x.
    \end{split}
\end{equation}
Since $(J_1\mathsf{T})^{-1} SJ_1^{-1}=(J_1^{-1})^{\mathsf{T}} SJ_1^{-1}$ is a real symmetric matrix, it has only real eigenvalues and is diagonalizable by $U\in SO(2L^3-2)$. Denoting $(J_1^\mathsf{T})^{-1} SJ_1^{-1}=U^\mathsf{T}\Lambda U$ and $\Lambda=\mathrm{diag}(\lambda_1,\dots,\lambda_{2L^3-2})$, we find
\begin{equation}\label{equB190}
    \begin{split}
        &(\mathrm{det}\,J_1)\int_{\mathbb{R}^n}\exp(ix^\mathsf{T}(J_1^\mathsf{T})^{-1} SJ_1^{-1}x-\epsilon x^\mathsf{T}x)\;d^{2L^3-2}x\\
        =&(\mathrm{det}\,J_1)\int_{\mathbb{R}^n}\exp(ix^\mathsf{T}U^\mathsf{T}(i\Lambda-\epsilon I_{2L^3-2})Ux)\;d^{2L^3-2}x\\
        =&(\mathrm{det}\,J_1)\prod_{j}\left(\sqrt{\frac{2\pi}{-i\lambda_j+\epsilon}}\right),
    \end{split}
\end{equation}
where $\mathrm{Re}\sqrt{-i\lambda_j+\epsilon}>0$.

Next for $x=(x_1,\ldots,x_{2L^3-2},0,0,0)\;,\;v=(v_1,\ldots,v_{2L^3-1},v_{2L^3},v_{2L^3+2})=(V_1,V_2)$, we consider the integral
\begin{equation}\label{equB200}
    I_v=\int_{\mathbb{R}^n}\exp(ix^\mathsf{T} Sx-\epsilon (x+v)^\mathsf{T}J^\mathsf{T}J(x+v))d^{2L^3-2}x.
\end{equation}
Note that $v$ is a fixed arbitrary vector and not an integral variable. We obtain
\begin{equation}\label{equB210}
    \begin{split}
        &\int_{\mathbb{R}^n}\exp(ix^\mathsf{T} Sx-\epsilon (x+v)^\mathsf{T}J^\mathsf{T}J(x+v))d^{2L^3+1}x\\
        =&e^{-\epsilon\|Jv\|^2}\int_{\mathbb{R}^n}
        \exp((J_1x)^\mathsf{T}(i(J_1^\mathsf{T})^{-1}SJ_1^\mathsf{T}-\epsilon I_{2L^3+1})(J_1x)-2\epsilon ((J_1V_1+J_2V_2))^\mathsf{T}(J_1x))d^{2L^3-2}x\\
        =&e^{-\epsilon\|Jv\|^2}(\mathrm{det}\,J_1)\int_{\mathbb{R}^n}
        \exp(x^\mathsf{T}(i(J_1^\mathsf{T})^{-1}SJ_1^\mathsf{T}-\epsilon I_{2L^3+1})x-2\epsilon ((J_1V_1+J_2V_2))^\mathsf{T}x)d^{2L^3-2}x\\
        =&e^{-\epsilon\|Jv\|^2}(\mathrm{det}\,J_1)\left(\prod_j\sqrt{\frac{2\pi}{-i\lambda_j+\epsilon}}\right)e^{\epsilon^2(J_1V_1+J_2V_2)^\mathsf{T}(i(J_1^\mathsf{T})^{-1}SJ_1^\mathsf{T}-\epsilon I_{2L^3+1})^{-1}(J_1V_1+J_2V_2)}\\
        =&e^{-\epsilon\|Jv\|^2}(\mathrm{det}\,J_1)\left(\prod_j\sqrt{\frac{2\pi}{-i\lambda_j+\epsilon}}\right)e^{\epsilon^2(U(J_1V_1+J_2V_2))^\mathsf{T}(i\Lambda-\epsilon I_n)^{-1}(U(J_1V_1+J_2V_2))}\\
        =&e^{-\epsilon\|Jv\|^2}(\mathrm{det}\,J_1)\left(\prod_j\sqrt{\frac{2\pi}{-i\lambda_j+\epsilon}}\right)\exp\sum_j\left(\frac{\epsilon^2(U(J_1V_1+J_2V_2))_j^2}{i\lambda_j-\epsilon}\right),
    \end{split}
\end{equation}
where we used the formula $\int e^{-x^\mathsf{T}Ax+2b^\mathsf{T}x}dx=\sqrt{\frac{2\pi}{\mathrm{det}\,A}}e^{-b^\mathsf{T}A^{-1}b}$. Recall that $0<\epsilon\ll 1$, and we find
\begin{equation}\label{equB220}
    \begin{split}
        \frac{I_v}{I_0}&=\exp\left\{-\epsilon\|J_3v\|^2+\sum_j\left(\frac{\epsilon^2(U(J_1V_1+J_2V_2))_j^2}{i\lambda_j-\epsilon}\right)\right\}\\
        &= 1-\epsilon\|J_3v\|^2+\sum_j\left(\frac{\epsilon^2(U(J_1V_1+J_2V_2))_j^2}{i\lambda_j-\epsilon}\right)+O(\epsilon^2).
    \end{split}
\end{equation}
Therefore, we obtain
\begin{equation}\label{equB230}
    \begin{split}
        \left\|1-\frac{I_v}{I_0}\right\|&\leq \epsilon\|J_3v\|^2+\epsilon^2\|(J_1V_1+J_2V_2)^\mathsf{T}(i\Lambda-\epsilon I_n)^{-1}(J_1V_1+J_2V_2)\|+o(\epsilon^2)\\
        &\leq \epsilon\|J_3\|^2\|v\|^2+\left\|\frac{\epsilon^2}{i\lambda_\mathrm{min}-\epsilon}\right\|\|J\|^2\|v\|^2+o(\epsilon^2)\\
        &\leq 2\epsilon\|J\|^2\|v\|^2+O(\epsilon^2),
    \end{split}
\end{equation}
where $\lambda_\mathrm{min}=\min\{\lambda_j\}_{j=1}^{2L^3+1}$, and $\|J_j\|$ is the matrix $2$-norm $\displaystyle\max_{\|v\|=1}\|J_jv\|$.

Using
\begin{equation}\label{equB240}
    \|J\|^2=\max_{\|x\|=1}x^\mathsf{T}J^\mathsf{T}Jx=\max_{\|x\|=1}x^\mathsf{T}K^\mathsf{T}Kx=\|K\|^2
\end{equation}
and \eqref{equB080}, we can evaluate
\begin{equation}\label{equB250}
    2\|J\|^2=2\|K\|^2\leq2\|K\|_\infty\|K\|_1,
\end{equation}
then we conclude
\begin{equation}\label{equB260}
    \left\|1-\frac{I_v}{I_0}\right\|\leq2\epsilon\|K\|_\infty\|K\|_1\|v\|^2+O(\epsilon^2).
\end{equation}

\end{document}